\documentclass[12pt]{article}
\usepackage{cite,colordvi,amssymb,color,setspace,epsfig}
\usepackage{a4}
\ifx\epsfannounce\undefined \def\epsfannounce{\immediate\write16}\fi
 \epsfannounce{This is `epsf.tex' v2.7k <10 July 1997>}%
\newread\epsffilein    
\newif\ifepsfatend     
\newif\ifepsfbbfound   
\newif\ifepsfdraft     
\newif\ifepsffileok    
\newif\ifepsfframe     
\newif\ifepsfshow      
\epsfshowtrue          
\newif\ifepsfshowfilename 
\newif\ifepsfverbose   
\newdimen\epsfframemargin 
\newdimen\epsfframethickness 
\newdimen\epsfrsize    
\newdimen\epsftmp      
\newdimen\epsftsize    
\newdimen\epsfxsize    
\newdimen\epsfysize    
\newdimen\pspoints     
\def\epsfbox#1{\global\def\epsfllx{72}\global\def\epsflly{72}%
   \global\def\epsfurx{540}\global\def\epsfury{720}%
   \def\lbracket{[}\def\testit{#1}\ifx\testit\lbracket
   \let\next=\epsfgetlitbb\else\let\next=\epsfnormal\fi\next{#1}}%
\def\epsfgetlitbb#1#2 #3 #4 #5]#6{%
   \epsfgrab #2 #3 #4 #5 .\\%
   \epsfsetsize
   \epsfstatus{#6}%
   \epsfsetgraph{#6}%
}%
\def\epsfnormal#1{%
    \epsfgetbb{#1}%
    \epsfsetgraph{#1}%
}%
\newhelp\epsfnoopenhelp{The PostScript image file must be findable by
TeX, i.e., somewhere in the TEXINPUTS (or equivalent) path.}%
\def\epsfgetbb#1{%
    \openin\epsffilein=#1
    \ifeof\epsffilein
        \errhelp = \epsfnoopenhelp
        \errmessage{Could not open file #1, ignoring it}%
    \else                       
        {
            \chardef\other=12
            \def\do##1{\catcode`##1=\other}%
            \dospecials
            \catcode`\ =10
            \epsffileoktrue         
            \epsfatendfalse     
            \loop               
                \read\epsffilein to \epsffileline
                \ifeof\epsffilein 
                \epsffileokfalse 
            \else                
                \expandafter\epsfaux\epsffileline:. \\%
            \fi
            \ifepsffileok
            \repeat
            \ifepsfbbfound
            \else
                \ifepsfverbose
                    \immediate\write16{No BoundingBox comment found in %
                                    file #1; using defaults}%
                \fi
            \fi
        }
        \closein\epsffilein
    \fi                         
    \epsfsetsize                
    \epsfstatus{#1}%
}%
\def\epsfclipoff{\def\epsfclipstring{\ifepsfdraft\space clip\fi}}%
\epsfclipoff 
\def\epsfspecial#1{%
     \epsftmp=10\epsfxsize
     \divide\epsftmp\pspoints
     \ifnum\epsfrsize=0\relax
       \includegraphics{\ifepsfdraft}%
     \else
       \epsfrsize=10\epsfysize
       \divide\epsfrsize\pspoints
       \includegraphics{\ifepsfdraft}%
     \fi
}%
\def\epsfframe#1%
{%
  \leavevmode                   
  \setbox0 = \hbox{#1}%
  \dimen0 = \wd0                                
  \advance \dimen0 by 2\epsfframemargin         
  \advance \dimen0 by 2\epsfframethickness      
  \vbox
  {%
    \hrule height \epsfframethickness depth 0pt
    \hbox to \dimen0
    {%
      \hss
      \vrule width \epsfframethickness
      \kern \epsfframemargin
      \vbox {\kern \epsfframemargin \box0 \kern \epsfframemargin }%
      \kern \epsfframemargin
      \vrule width \epsfframethickness
      \hss
    }
    \hrule height 0pt depth \epsfframethickness
  }
}%
\def\epsfsetgraph#1%
{%
   \leavevmode
   \hbox{
     \ifepsfframe\expandafter\epsfframe\fi
     {\vbox to\epsfysize
     {%
        \ifepsfshow
            \vfil
            \hbox to \epsfxsize{\epsfspecial{#1}\hfil}%
        \else
            \vfil
            \hbox to\epsfxsize{%
               \hss
               \ifepsfshowfilename
               {%
                  \epsfframemargin=3pt 
                  \epsfframe{{\tt #1}}%
               }%
               \fi
               \hss
            }%
            \vfil
        \fi
     }%
   }}%
   \global\epsfxsize=0pt
   \global\epsfysize=0pt
}%
\def\epsfsetsize
{%
   \epsfrsize=\epsfury\pspoints
   \advance\epsfrsize by-\epsflly\pspoints
   \epsftsize=\epsfurx\pspoints
   \advance\epsftsize by-\epsfllx\pspoints
   \epsfxsize=\epsfsize{\epsftsize}{\epsfrsize}%
   \ifnum \epsfxsize=0
      \ifnum \epsfysize=0
        \epsfxsize=\epsftsize
        \epsfysize=\epsfrsize
        \epsfrsize=0pt
      \else
        \epsftmp=\epsftsize \divide\epsftmp\epsfrsize
        \epsfxsize=\epsfysize \multiply\epsfxsize\epsftmp
        \multiply\epsftmp\epsfrsize \advance\epsftsize-\epsftmp
        \epsftmp=\epsfysize
        \loop \advance\epsftsize\epsftsize \divide\epsftmp 2
        \ifnum \epsftmp>0
           \ifnum \epsftsize<\epsfrsize
           \else
              \advance\epsftsize-\epsfrsize \advance\epsfxsize\epsftmp
           \fi
        \repeat
        \epsfrsize=0pt
      \fi
   \else
     \ifnum \epsfysize=0
       \epsftmp=\epsfrsize \divide\epsftmp\epsftsize
       \epsfysize=\epsfxsize \multiply\epsfysize\epsftmp
       \multiply\epsftmp\epsftsize \advance\epsfrsize-\epsftmp
       \epsftmp=\epsfxsize
       \loop \advance\epsfrsize\epsfrsize \divide\epsftmp 2
       \ifnum \epsftmp>0
          \ifnum \epsfrsize<\epsftsize
          \else
             \advance\epsfrsize-\epsftsize \advance\epsfysize\epsftmp
          \fi
       \repeat
       \epsfrsize=0pt
     \else
       \epsfrsize=\epsfysize
     \fi
   \fi
}%
\def\epsfstatus#1{
   \ifepsfverbose
     \immediate\write16{#1: BoundingBox:
                  llx = \epsfllx\space lly = \epsflly\space
                  urx = \epsfurx\space ury = \epsfury\space}%
     \immediate\write16{#1: scaled width = \the\epsfxsize\space
                  scaled height = \the\epsfysize}%
   \fi
}%
{\catcode`\%=12 \global\let\epsfpercent=
\global\def\epsfatend{(atend)}%
\long\def\epsfaux#1#2:#3\\%
{%
   \def\testit{#2}
   \ifx#1\epsfpercent           
       \ifx\testit\epsfbblit    
            \epsfgrab #3 . . . \\%
            \ifx\epsfllx\epsfatend 
                \global\epsfatendtrue
            \else               
                \ifepsfatend    
                \else           
                    \epsffileokfalse
                \fi
                \global\epsfbbfoundtrue
            \fi
       \fi
   \fi
}%
\def\epsfempty{}%
\def\epsfgrab #1 #2 #3 #4 #5\\{%
   \global\def\epsfllx{#1}\ifx\epsfllx\epsfempty
      \epsfgrab #2 #3 #4 #5 .\\\else
   \global\def\epsflly{#2}%
   \global\def\epsfurx{#3}\global\def\epsfury{#4}\fi
}%
\def\epsfsize#1#2{\epsfxsize}%
\usepackage[a4paper]{geometry}
\sloppy
\oddsidemargin  -4mm              
\evensidemargin  4mm              

\headheight     13mm              
\headsep        21pt              
\footskip       30pt              

\textheight 212mm                 
\textwidth 160mm                  
\columnsep 10mm                   
\columnseprule 0pt                

\newsavebox{\fmbox}

\newcommand{\uxm}{U_{x,\mu}}
\newcommand{\be}{\begin{equation}}
\newcommand{\ee}{\end{equation}}
\newcommand{\ba}{\begin{eqnarray}}
\newcommand{\ea}{\end{eqnarray}}

\newcommand{\Tr}{\ensuremath{\mathop{\rm Tr}}}
\newcommand{\mps}{m_{PS}}
\newcommand{\ksea}{\kappa_{\mathrm{sea}}}
\newcommand{\kval}{\kappa_{\mathrm v}}

\newcommand{\Esim}{\ensuremath{E_{\mathrm sim}}}

\newcommand{\gsim}{\ensuremath{\mathrel
               {\raise2pt\hbox to 8pt{\raise -5pt\hbox{$\sim$}\hss{$>$}}}}}
\let\orighspace\hspace
\renewcommand{\hspace}{\vrule width0pt\relax\orighspace}

\newcommand{\delv}{\vec{\nabla}}
\newcommand{\Mbz}{M_0}
\newcommand{\Ev}{\vec{E}}
\newcommand{\Bv}{\vec{B}}
\newcommand{\nl}{\nonumber \\}
\newcommand{\delfour}{{\Delta^{(4)}}}
\newcommand{\delsq}{\Delta^{(2)}}
\newcommand{\sigmav}{\vec{\sigma}}

\newcommand{\lqcd}{\Lambda_{QCD}}

\thispagestyle{empty}

\begin{document}
\begin{center} 

\vspace{4.5cm}

{ \Huge{Effective Field Theories
   for Quantum Chromodynamics on the Lattice}}

\vspace{1cm}

{\Large{A.~Ali~Khan}}

\vspace{1cm}

 Institut f\"ur Theoretische Physik, Universit\"at Regensburg, \\
 93040 Regensburg,
 Germany
\footnote{previous address: Humboldt-Universit\"at zu Berlin, Newtonstr. 15,
 12489 Berlin, Germany.}
 
 \parbox{\hsize}{\vspace{-35em}
 \hbox to \hsize
 {\hss \normalsize HU-EP-05/34}}

 \abstract{\noindent Light and heavy-light ($b$) hadrons are among the most 
interesting and 
 among the most challenging quantities to calculate in lattice
 gauge theory. One would like to avoid discretization 
 effects from  very heavy quarks and to calculate chiral extrapolations and 
 calculate  finite volume effects from light quarks. For this one uses 
 effective theories: Chiral 
 Perturbation Theory and Nonrelativistic QCD or Heavy Quark Effective Theory. 
 Lattice results are reviewed on hadrons containing light quarks and light and
$b$ quarks, and discussed in the framework of effective theories.}
\end{center} 
 \newpage
\tableofcontents
\newpage
\parbox{153mm}{\vspace{-5em}
\hbox to \hsize
{\hss \normalsize TO MY PARENTS}}
\section{INTRODUCTION}
One of the most appealing ideas is that the fundamental forces of Nature
can be explained from mathematical symmetries. Indeed the known conservation laws result from
space-time and internal symmetries, and the hope is that it is possible 
to find a consistent theory describing the interactions between all elementary particles and
describing how the world is kept together. A large step forward on this way was the development of the 
Standard Model of elementary particle physics 
which includes Quantum Chromodynamics (QCD), the theory of strong interactions,
based on a local $SU(3)$ symmetry, and the Glashow-Salam-Weinberg model of the electromagnetic and
weak forces based on a local $U(1) \times SU(2)$ symmetry.
It gives a surprisingly accurate description of the presently known elementary particle interactions,
in experiments with particle collisions at energies up to several 100 GeV. 
One  now tries to understand as much of the implications of the Standard Model as possible, 
and to test the model by enhancing the precision of theoretical predictions 
and experimental measurements and by searching for contradictions, 
to be able to find indications for a broader and perhaps simpler and
more symmetric physical theory.

An obvious expectation from the Standard Model, and QCD in particular, is to be able to
explain the properties of mesons and baryons. 
For example, hadron masses and hadronic reactions should be predictable from QCD, and weak and
electromagnetic matrix elements of hadrons experience contributions from the
strong interactions whose magnitude has to be known to be able to interpret the matrix elements
theoretically.
Of great interest are weak decays of hadrons which depend on the Cabibbo-Kobayashi-Maskawa (CKM) 
matrix elements which are fundamental parameters of the Standard Model describing
the flavor changing weak currents of quarks and
play a special role by providing a mechanism for $CP$ violation within hadron physics.

Among the CKM matrix elements which are at present intensively studied in experiments and theory
and which are relevant to $CP$ violation are those related to heavy flavored quarks, in particular the 
$b$ quark.

At energies less than 1 GeV, the interaction strength increases  and non-perturbative are required.
 Placing space and time onto a four-dimensional lattice~\cite{wilson1974} it is possible to 
calculate  matrix elements numerically on a computer within a path integral formalism.
In the past few years the lattice has become a standard method in QCD calculations.

To avoid finite size effects, the lattice extent $L$ should be much larger than the wavefunction size or the  
Compton wavelength of the particles one would like to describe, and the inverse lattice spacing $a^{-1}$
should be much larger than the relevant masses and momenta  in order to avoid cutoff effects from 
the finite lattice spacing $a$. 
The lightest hadrons, the pions, have a mass of around 140 MeV, whereas the $B$ meson has a mass
of $5.28$ GeV and contains a heavy quark with a mass of 5 GeV. Lattice calculations of hadrons with 
heavy and light quarks have to be able to describe physics from this large range of scales.


An remarkable property of hadrons with heavy quarks is that the energy level splittings of $b$ hadrons are much 
smaller than their masses:
of the order of $\lqcd = 200-500$ MeV or smaller, where $\lqcd$ is the energy scale where QCD becomes 
non-perturbative. 
The dynamics of the heavy quarks can be accounted for in an expansion in powers of the
inverse heavy quark mass.
This is a basis for effective field theories such as Heavy Quark Effective Theory (HQET) (for reviews see
\cite{neubert1994})
and non-relativistic QCD (NRQCD)~\cite{thacker1991,lepage1992}, which can be
used to simulate heavy quarks directly on the lattice while avoiding large discretization errors 
proportional to powers of the large mass. 
Besides, heavy quark effective field theory predictions can be used to guide extrapolations and interpolations 
of lattice simulation data as a function of the heavy quark mass.

Simulations with light quarks are computationally expensive and sensitive to the finite lattice
volume. Therefore one often uses quark masses much heavier than $u$ and $d$ quark 
masses  and extrapolates the results to the physical values.
A formalism for this can be derived using chiral perturbation theory ($\chi PT$), an expansion around the chiral
(zero quark mass) limit describing the low-energy degrees of freedom of QCD, i.e.\ pions and other light mesons
and their interactions with other hadrons at low energy.

If one assumes  that chiral symmetry of QCD is  spontaneously broken, one expects massless Goldstone bosons in the
particle spectrum. The physical light mesons can be identified with the
corresponding Goldstone bosons if explicit breaking of chiral symmetry is added by including quark mass terms in the 
Lagrangian.
Hadron interactions give corrections to the relations between quark and hadron masses. At low momenta the interactions
e.g.\ in $\pi-\pi$ scattering and $\pi-N$ scattering are weak and can be treated using perturbative methods.

The aim of this article is to summarize lattice calculations of the light and heavy-light hadron spectrum and decay 
constants the author performed within various collaborations, to discuss the
systematics of these calculations in some detail, and to compare the results to recent other work. 
To make the article more self-contained, a short introduction to some formalisms relevant to the calculations is also given.

In Section~\ref{sec:QCD} we give a review of the  basic theory and phenomena of QCD. 
 A brief introduction to the lattice formalism is given in 
Section~\ref{sec:lattice}. The concepts of effective theories which are relevant in the context of QCD 
calculations, namely NRQCD, HQET and Chiral Perturbation Theory, are discussed 
in Section~\ref{sec:eff}. 
Having made acquaintance with the lattice and the chiral effective field theory for the analysis of the light 
quark mass dependence, we discuss results on the light hadron spectrum and decay constants in 
Section~\ref{sec:lresults}. In Section~\ref{sec:heavy} we discuss lattice formalisms  for
heavy quarks using non-relativistic effective theories. In 
Sections~\ref{sec:hresults} and~\ref{sec:fB} lattice results on the $b$-light spectrum  
and decay constants are presented. Section~\ref{sec:concl} contains some conclusions.

\section{QCD AND THE STANDARD MODEL\label{sec:QCD}}
In this section  we give a brief overview of QCD emerging from nuclear physics and the
Standard Model of Elementary Particle Physics. For a recent review about hadron physics and methods
in QCD see e.g.~\cite{khodja2004}.
\subsection{Hadrons and the Quark Model}
In the early 20th century it was discovered that atoms possess nuclei which are much
smaller than the atom and which consist of smaller particles, protons 
and neutrons, which in turn are tied together by the strong nuclear force. 
We are interested in studying the strong force to understand what keeps all matter together.

Yukawa postulated in 1935 that the interactions between nucleons are due to
exchange of mesons~\cite{yukawa1935}. From the short range of the nuclear force (interaction
radius $r \sim 1$ fm), he concluded that the mesons should have a mass of $\sim
1/r \simeq 100$ MeV. 
In 1947, the $\pi$ meson was discovered experimentally with a mass of $\sim 140$ MeV. $K$ mesons
were discovered soon afterwards.

Since the 1970's indications began to accumulate  that hadrons, i.e.\
nuclear particles and other particles interacting with the strong 
force consist of even smaller particles: baryons such as protons ($p$), neutrons ($n$), hyperons and their like 
consist of three quarks and mesons such as pions, kaons, the $J/\psi$ and the
$\Upsilon$ of a quark and an antiquark.

The masses of proton and neutron are very closely degenerate, and they can be regarded as 
transforming as an approximately degenerate doublet of `nucleons' in the fundamental
representation of a global $SU(2)$ symmetry group. 
The corresponding quantum number is isospin ($I$).
Within the quark model, the nucleons and pions are composed of 
$u$ and $d$ quarks, both almost massless, which sit in the fundamental representation of $SU(2)$.

The discovery of hadrons with a new quantum number, the strangeness $S$, led to the introduction 
of the strange ($s$) quark. The spectra of strange and lighter hadrons can be 
described using an approximate symmetry between hadrons with strangeness and those without.
In  the 'Eightfold Way' model of Gell-Mann and Ne'eman~\cite{gellm1964}, the three quarks can be 
identified with the generators of an approximate $SU(3)$ flavor symmetry in the 
fundamental representation. The symmetry group contains isospin as well as the quantum number
 hypercharge $Y = B + S$, where $B$ is the baryon number.
The low-lying mesons composed of a quark and an antiquark can be grouped
into irreducible representations contained in the $3 × 3^\ast$ product representation
 of $SU(3)$, the singlet and the octet, and baryons consisting of three quarks can be grouped into the
octet or decuplet representations contained in the product $3 × 3 × 3$. 
The light baryon states, for which lattice results will be discussed later, and their quark content are 
shown in Table~\ref{tab:bary_qm}.
\begin{table}[thb]
\begin{center}
\begin{tabular}{|ccl|}
\hline
Quark content & Octet & Decuplet \\
$uuu$         &       & $\Delta^{ ++}$ \\
$uud$         &  $p$  & $\Delta^{ +}$ \\
$udd$         &  $n$  & $\Delta^{ 0}$ \\
$ddd$         &       & $\Delta^{ -}$ \\
\hline
$uus$         & $\Sigma^+$ & $\Sigma^{\ast +}$            \\
$uds$         & $\Sigma^0,\Lambda^0$ & $\Sigma^{\ast +}$  \\
$dds$         & $\Sigma^+$ & $\Sigma^{\ast +}$           \\
$uss$         & $\Xi^0$    & $\Sigma^{\ast 0}$            \\
$dss$         & $\Xi^-$    & $\Sigma^{\ast -}$            \\
$sss$         &            & $\Omega^-$                   \\
\hline
\end{tabular}
\end{center}
\caption{Baryons and their quark content within the quark model. Electric 
charges are indicated by superscripts. The names of strange 
baryons can be extended to those containing heavy quarks as discussed in sections
\protect\ref{sec:bbary} and \protect\ref{sec:bbbary}. }
\label{tab:bary_qm}
\end{table}

The discovery of the $\Delta^{++}$ gave an indication for the existence of an additional
degree of freedom, the color. According to the quark model
it consists of three $u$ quarks with a totally symmetric wave function.
If there was only one species of $u$ quark, this would violate the Pauli principle.

High energy experiments also showed that nucleons have constituents.
The existence of three colors of quarks showed up in
measurements of the cross section for $e^+ e^-$ decaying into hadrons.

In studies of the structure of the nucleon in high-energy scattering experiments, 
constituents (`partons') were found which behaved as almost free
particles in the high energy limit. Those can be identified with the quarks previously
suggested by the quark model, and the gluons.

Important for the understanding of QCD as the theory of the strong interactions was the
discovery of mesons consisting of heavier quarks. 
In 1974, a bound state of a heavier quark flavor (charm) and its antiquark, the $J/\psi$ meson, 
was discovered. A fifth species of quark ($b$) was found in 1977 with the discovery of the 
$\Upsilon$, a $\overline{b}b$ bound state. 
It had been postulated earlier by Kobayashi and Maskawa to explain $CP$ violation 
in the weak interactions, which had been discovered in 1964 by 
Christensen {\it et al} in decays of neutral $K$ mesons into pions.
Heavy-light mesons with $b$ quarks are $B^+ (\overline{b}u)$, $B^0 (\overline{b}d)$,
 $\overline{B}^0 (\overline{d}b)$, and $B^- (\overline{u}b)$.
The sixth quark postulated by the Standard Model, the top ($t$) quark, was
discovered in 1995.
\subsection{Gauge Theories: QED and QCD}
Interactions between elementary particles can be explained as resulting from local, i.e.\
space-time dependent, mathematical symmetries.

For example, the Lagrangian of free Dirac particles is invariant under global unitary transformations
\ba
q & \to & V q, \nonumber \\
\overline{q} & \to & \overline{q} V^\dagger,
\ea
where $q$ are Dirac spinors and $V \in U(1)$ or  $V \in SU(N)$. 
The symmetry group of the strong interactions is $SU(3)$ corresponding to three colors of
quarks, which sit in the fundamental representation of the gauge group.
Postulating invariance of the Lagrangian under local $U(1)$ or $SU(3)$
transformations requires  the introduction of gauge fields coupling to the quarks with the covariant
derivative $D_\mu$, and the Lagrangian takes the following form:
\be
{\cal L} = \overline{q}\left(i\gamma_\mu D^\mu -m\right) q + {\cal L}_g, \label{eq:qcd}
\ee
where ${\cal L}_g$ is the gauge field Lagrangian. The different features of QED and QCD at tree level
are compared in Table~\ref{tab:gauge}.
\begin{table}[thb]
\begin{center}
\begin{tabular}{ll}
\hline
  Quantum Electrodynamics & Quantum Chromodynamics \\
\hline
  $U(1)$: $V = \exp[i\alpha(x)]$  & $SU(3)$: $V = \exp[i\alpha^c(x)t^c]$ \\
      & $t^c$ $SU(3)$ generators. $[t^a,t^b] = if^{abc} t^c$ \\
  $D_\mu = \partial_\mu +ie \Black{A_\mu}$ & 
$D_\mu \equiv \partial_\mu - ig \Black{A_\mu^c} t^c$ \\
 electric charge $e$ & $g$ strong coupling $g$ \\
\Black{$A_\mu$ photon,  neutral} & \Black{$A_\mu^c$ gluon, color charged} 
\\
${\cal L}_g=-\frac{1}{4}  F_{\mu\nu}F^{\mu\nu}$ & 
${\cal L}_g=-\frac{1}{4}  G^c_{\mu\nu}G^{c\mu\nu}$  \\
$F_{\mu\nu} = \partial_\mu {A_\nu}-\partial_\nu \Black{A_\mu}$ & 
$G^a_{\mu\nu} = \partial_\mu {A_\nu^a}-\partial_\nu 
{A_\mu^a} +gf^{abc}\Black{A_\mu^b A_\nu^c}$  \\
\hline
\end{tabular}
\end{center}
\caption{Comparison of QED and QCD tree level Lagrangians.}
\label{tab:gauge}
\end{table}

Proof of the mathematical consistency of quantum theories with non-abelian
gauge symmetries~\cite{thooft1971} in the early 1970's prepared the stage for a description of 
the strong interaction as a gauge theory with a local $SU(3)$ color symmetry, namely QCD.
\subsection{QCD coupling }
If a field theory of point particles is quantized divergent mathematical expressions can occur.
If the divergences can be absorbed by a redefinition of the parameters of the Lagrangian the 
theory is (perturbatively) renormalizable. For non-abelian gauge theories renormalizability has been 
proven  at any order of perturbation theory~\cite{thooft1971}.
In general, the quantum corrections to a given physical process depend on the momenta of the
external particles. A dependence of physical couplings is introduced on
the momentum scales used to probe, i.e.\ look at, the physical system. 
For dimensional reasons the momentum dependence introduces an additional parameter, namely
a renormalization scale. Then one can derive
renormalization group equations,  which determine how couplings, masses
and Green functions evolve as functions of the  renormalization scale parameters or, alternatively, on the 
momentum scales. 

Renormalization can actually be defined in more general terms than in perturbation theory only.
The most interesting case occurs when limits exist where physical quantities such as dimensionless
couplings and mass ratios become scale invariant.
Renormalizability can then be viewed as the possibility to vary the bare parameters such that physics can be
kept fixed while the distance scales at which the system is probed are changed. 
In perturbative QED, an electric charge appears weaker if it is probed at a larger distance because of 
screening of the charge due to generation of virtual electron-positron pairs. 
Since the QCD gauge group is non-abelian, the gauge fields themselves carry charges of the 
strong interactions and interact not only with the color-charged matter fields, but also 
among themselves. Perturbation theory predicts that the QCD coupling becomes weaker due to the 
self-interactions if it is probed at a smaller distance, i.e.\ at higher momentum. 
At infinite momentum the coupling vanishes. In 1973 it was shown that QCD has a free-field theory
asymptotic behavior up to logarithmic corrections~\cite{asymp1973}.

The scale dependence of the coupling $\alpha_s = g^2/(4\pi)$ is described by the $\beta$ function
\be
\mu \frac{d\alpha_s}{d\mu} = 2\beta(\alpha_s), \label{eq:beta}
\ee
which  in perturbation theory can be expanded in powers of $\alpha_s$:
\be
2\beta(\alpha_s) = -b_0 \frac{\alpha_s^2}{2\pi}  - b_1 \frac{\alpha_s^3}{4\pi^2} + ...
\ee
At one loop, the evolution  of $\alpha_s$ from a scale $\mu_0$ to a scale $\mu$ is given by
\be
\alpha_s (\mu) = \frac{\alpha_s(\mu_0)}{1+\frac{\alpha_s(\mu_0)}{2\pi}
b_0 \ln\frac{\mu}{\mu_0}},
\ee
with $b_0 = 11-\frac{2}{3}N_f$, where $N_f$ is the number of flavors.  
If Eq.~(\ref{eq:beta}) is integrated, one can re-express the scale dependence
(or momentum dependence) in terms of the integration constant $\lqcd$:
\be
\alpha_s (\mu^2) = \frac{2\pi}{b_0 \ln\frac{\mu}{\lqcd}}.  \label{eq:alph}
\ee
At tree-level QCD without quark masses has  no scale parameter. The renormalized 
dimensionless gauge coupling can be re-expressed using the renormalization scale or the 
momentum and  the dimensionful parameter $\lqcd$.
This is called dimensional transmutation~\cite{coleman1973}.

Eq.~(\ref{eq:alph}) indicates that at larger distance, the coupling becomes strong and 
diverges at $\mu = \lqcd$. Of course, at strong couplings perturbation theory is not valid.
For example, the physics giving rise to masses of light hadrons is related to energy scales
$O(\lqcd)$ and has to be described using non-perturbative methods. 

\subsection{Chiral symmetry}
If the quark mass vanishes, i.e.\ $m = 0$, the Lagrangian~(\ref{eq:qcd}) is invariant under 
chiral transformations
 \ba 
q & \to & \exp i \gamma_5 \alpha^a t^a q ,                      \nonumber \\
\overline{q} & \to & \overline{q} \exp i \gamma_5 \alpha^a t^a ,
\ea
where $t^a$ are the generators of the symmetry group. The effect of chiral symmetry on the particle 
spectrum would be degeneracy of positive and negative parity states.  This is however not found in
nature. Instead, the spectrum contains pseudoscalar mesons which are extraordinarily light
compared to the other hadrons.
The lightest mesons are the pions $\pi^+, \pi^0, \pi^-$ which correspond in this picture to the 
Goldstone bosons of an $SU(2)$ chiral symmetry between the $u$ and $d$ quarks broken spontaneously 
and explicitly due to small mass terms.
If the strange quark is also considered as almost massless, the chiral symmetry is $SU(3)$ and the 
would-be Goldstone bosons are an octet of $\pi, K$ and $\eta$.

The approximate conservation of the axial vector current $A_\mu$, the Noether current of the chiral
symmetry, can be expressed using the PCAC relation. For $SU(2)$, this is
\be
\partial_\mu A_\mu^a = 2m_q \delta^{ab}\pi^b, \label{eq:pcac}
\ee
where $\pi^a$ is an interpolating field for the pion, $A_\mu^a(x) = \overline{q}(x)\gamma_\mu\gamma_5 
t^aq(x)$, $t^a$ being a generator of $SU(2)$, and $m_q$ is the quark mass. In this section, Euclidean formulae
are used for the discussion of the topology. 
PCAC denotes the partially conserved axial vector current. Eq.~(\ref{eq:pcac}) can be considered as an 
operator identity in the sense that it can be extended to a relation for matrix elements of $A_\mu^a$.

The spontaneous symmetry breaking leads to a vacuum expectation value of $\overline{q}q$, the chiral
condensate. This can be calculated e.g.\ in lattice simulations. 
In the sum rule method of Shifman, Vainshtein and Zakharov~\cite{svz1979} the chiral 
condensate, the gluon condensate $\langle \scriptsize{\frac{\alpha_s}{\pi}}G^a_{\mu\nu}
G^{a,\mu\nu}\rangle $, and condensates of higher dimensional operators are used to
parameterize nonperturbative 
contributions to the hadron spectrum and matrix elements and to make further predictions using
these parameters. For a recent review see \cite{ioffe2005}.

The tree-level QCD Lagrangian also has a global chiral $U(1)$ symmetry which has to be
broken since there are no degenerate states of opposite parity in the hadron spectrum.
The $\eta'$ meson has the correct quantum numbers for being the Goldstone boson
of spontaneous chiral $U(1)$ symmetry breaking, however its mass is too large. 
Experimentally this additional breaking of chiral symmetry is 
visible e.g. in the decay rate of $\pi^0 \to 2\gamma$.
The symmetry is broken due to  an anomaly which can be
explained from the noninvariance of the measure of the path integral under
chiral transformations~\cite{fujikawa1980}. The anomaly implies that the divergence 
of the flavor singlet axial vector current $A_\mu = \overline{q}\gamma_\mu\gamma_5 q$
is non-vanishing:
\be
\partial_\mu A_\mu(x) = N_f q(x), \label{eq:ano}
\ee
where $q$, the topological charge density, is related to the color-electromagnetic field
strengths by 
\be
q(x) = -\frac{1}{16\pi^2} G^a_{\mu\nu}\tilde{G}^a_{\mu\nu},
\ee
where  $\tilde{G}^{a}_{\mu\nu} = \scriptsize{\frac{1}{2}}\epsilon_{\mu\nu\alpha\beta}
G^{a}_{\alpha\beta}$ is the dual field strength. With the anomaly, a close relation between the topology
of the gauge field and the chiral symmetry of the lattice fermions is established.
An index theorem relates the topological charge to the difference between the number of 
right and left handed zero modes of the Dirac operator $D$:
\be
n_- - n_+ = N_f Q.
\ee
If the axial current vanishes at spatial infinity, one finds after integration of Eq.~(\ref{eq:ano}) also
that the topological charge is related to the difference of the axial charges at positive and negative 
temporal infinity by:
\be
\int d^3x A_0(t \to \infty) - \int d^3x A_0(t \to - \infty) = N_f Q.
\ee
The topological charge density can also be expressed as the divergence of a current which 
depends entirely on the gauge fields:
\ba
q(x) &=& \partial_\mu K_\mu(x) \nonumber \\
K_\mu &=& -\frac{1}{16\pi^2}\epsilon_{\mu\alpha\beta\gamma}(A^a_\alpha\partial_\beta A^a_\gamma
+\frac{1}{3}f^{abc} A_\alpha^a A_\beta^b A_\gamma^c ),
\ea
where $f^{abc}$ are the structure constants of $SU(3)$.
If this again decays sufficiently fast at spatial infinity, the topological charge can also be expressed
as the difference of the Chern-Simons numbers $N_{CS}$ of the gauge field at $t \to \infty$ and 
$t \to -\infty$, where
\be
N_{CS}(t) = \int d^3x K_0(t).
\ee 
The classical action of the Yang-Mills field is minimized if the gauge fields are
self-dual. The lower limit to the action of the classical field is proportional to the difference of
the Chern-Simons numbers. Instantons which are self-dual field configurations  with 
finite action corresponding to a change of  Chern-Simons numbers thus provide a possible classical or
semiclassical scenario for the occurrence of chiral symmetry breaking in QCD.
For reviews about the physics of instantons in QCD see~\cite{tschaefer1998,diakonov2003}.
 
There are indications that chiral symmetry in QCD is  restored for a larger number of quark flavors $N_f$.
The restoration is predicted to occur  for $N_f \geq 7$ by an
instanton model calculation~\cite{shuryak1994} and by lattice results~\cite{iwasaki2004}.
\subsection{Confinement}
In nature there are no free quarks and gluons. Color charges seem to be confined.
Already perturbation theory predicts a growing of the coupling constant at decreasing 
momentum scale. The one-loop relation is 
Eq.~(\ref{eq:alph}), which has a pole at the  scale  $\lqcd$. Perturbation theory is of
course not valid at very low scales, but it already indicates that there is an energy
scale where particles are very strongly bound and confinement occurs. Determining $\alpha_s$
at higher energy scales it is however possible to determine $\lqcd$ using 
perturbation theory, using a higher-loop equivalent to Eq.(\ref{eq:alph}).

The spectrum of hadrons  actually fits quite well into the picture of an inter-quark potential 
which rises linearly at large distances, corresponding to a flux string with a constant string 
tension. Using a space-time lattice it could be shown numerically that
the potential between static color sources, whose world lines are just parallel to 
the time axis, indeed increases at larger separation of the color sources, 
which implies that QCD without dynamical quarks is confining at low 
momentum scales or large distances.  
The theory with fermions remains confining, although there is a probability for
the string to break due to the generation of quark-antiquark pairs. 
The authors of Ref.~\cite{iwasaki2004} find with lattice calculations that for $N_f \leq 6$ QCD 
remains confining at low energies, while for a larger number of flavors confinement is lost 
and there are indications for a complex phase structure of QCD at large $N_f$.

It seems that confinement can be related to a color-electric flux
tube between color charges. A physical picture which could explain 
this on a classical level is the dual superconductor model~\cite{nambu1974}
 where the 
flux tube enclosed in the vacuum is characterized by the condensation  of color-magnetic 
abelian monopoles.  The condensation of monopoles is indeed found
in lattice simulations. One model which could potentially describe the physics of the  monopoles
is the Ginzburg-Landau model. 
Lattice results indicate that the monopoles in lattice QCD may rather be arising from center vortex 
structures which imply a different monopole dynamics from the picture realized in the  dual
Ginzburg-Landau superconductor. A dependence 
of the abelian monopoles on the
method used for the abelian projection is investigated in \cite{belavin1,belavin2,bornyakov05}.
A recent review on lattice results on the physics of confinement is given in~\cite{engel2004}.

The previous discussion is related to zero temperature. 
If the temperature is increased, QCD becomes non-confining. In 
QCD without dynamical fermions the corresponding transition is a first order
phase transition, while with more than two species of light fermions the
transition is found to be a crossover. Recent results for the  transition temperatures from
lattice simulations are~\cite{petreczky2004}
\ba
265(1)-296(2) \,\mbox{MeV},&\;& N_f = 0  \label{eq:q} \\
171(4)-173(7) \,\mbox{MeV},&\;& N_f = 2,  \\
164(2)-169(12)(4)\,\mbox{MeV},&\;& N_f = 2_{light} + 1_{strange},
\ea
where $N_f$ is the number of quarks flavors present in the lattice simulation.
The range of values quoted corresponds to different simulations. The difference of 
the $N_f = 0$ results is due to different ways of determining the lattice spacing.
For quenched gauge fields the transition temperature is thus approximately of the  order of magnitude of 
the scale $\lqcd$.
\subsection{The Standard Model}
The Standard Model of the strong and electroweak interactions 
is based on an 
$SU(3)\times SU(2)\times U(1)$ gauge symmetry with three generations of quarks and leptons as fermionic
matter fields and a scalar field, the Higgs, which is responsible for the masses of the weak $SU(2)$ gauge
bosons and the fermions, assuming neutrinos to be massless.
The $SU(3)$ gauge group gives rise to the QCD interactions, with gluons as gauge bosons. 
$SU(2) \times U(1)$ is the gauge group of the Glashow-Salam-Weinberg model of the electroweak interactions.
The gauge bosons are photons related to QED and the massive $W^±$ and $Z$ bosons related to the weak
interactions. 
The fermionic matter fields are three generations of quarks $(u,d,c,s,t,b)$, and leptons
$(e,\nu_e, \mu,\nu_\mu,\tau, \nu_\tau)$.

For a recent review about the status of the Standard Model and new physics
see e.g.~\cite{wilczek2003,pokorski2005}.
Although providing a remarkbly good description of elementary particle physics, 
the Standard Model has some unsatisfactory features. On the experimental side, recent discoveries 
such as neutrino mixings, the possibility of the existence of undiscovered matter in the cosmos and perhaps 
new results from accelerator experiments~\cite{Bhadr2004,g-22004} indicate possibilities where
physics beyond the Standard Model may appear.
The Higgs particle has not yet been found. Further, there are theoretical motivations to search for 
physics beyond the Standard Model which contains a relatively 
large set of coupling constants and masses as input parameters.
The values of typical energy scales such as the masses of the weak gauge bosons are input
parameters of the Standard Model and require new physics for explanations.

A strategy in the search for new physics is to simultaneously measure as many physical quantities as 
possible, test
the results for self-consistency using Standard Model formulae and search for indications for new theories 
which can imply the presently known phenomena.

The Standard Model parameters relevant to the interactions of quarks are
\begin{itemize}
\item $\alpha_s$ 
\item Quark masses 
\item Parameters of the weak  currents (CKM matrix elements)
\end{itemize}
For a review about recent results on quark masses from the lattice and sum rules see 
Ref.~\cite{gupta2003}.
A recent comprehensive review on the Standard Model is given by Ref.~\cite{altarelli2005}.
\subsection{CKM Matrix and Heavy Quark Decays\label{sec:CKM}}
\begin{figure}[thb]
\begin{center}
\vspace{-0.3cm}
\epsfysize=2.8cm \epsfbox[100 20 500 320]{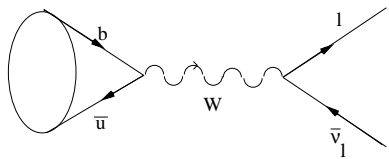}
\end{center}
\caption{Leptonic $B^+$ decay in the Standard Model.}
\label{fig:Bdecayplot}
\end{figure}

\begin{figure}[bht]
\begin{center}
\vspace{-1cm}
\centerline{
\hspace{2.8cm}
\epsfysize=2.8cm \epsfbox[300 50 500 400]{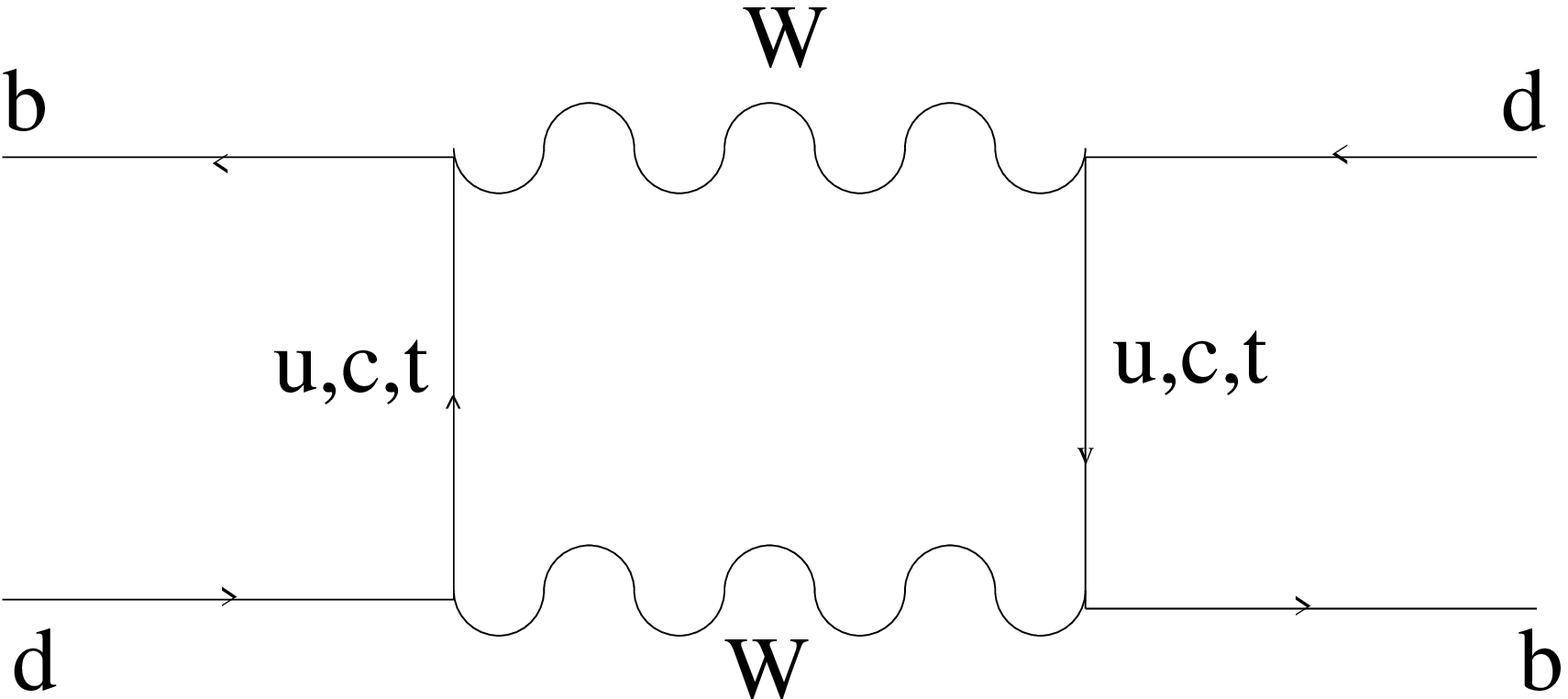}
\hspace{1cm}
\epsfysize=2.8cm \epsfbox[50 50 220 400]{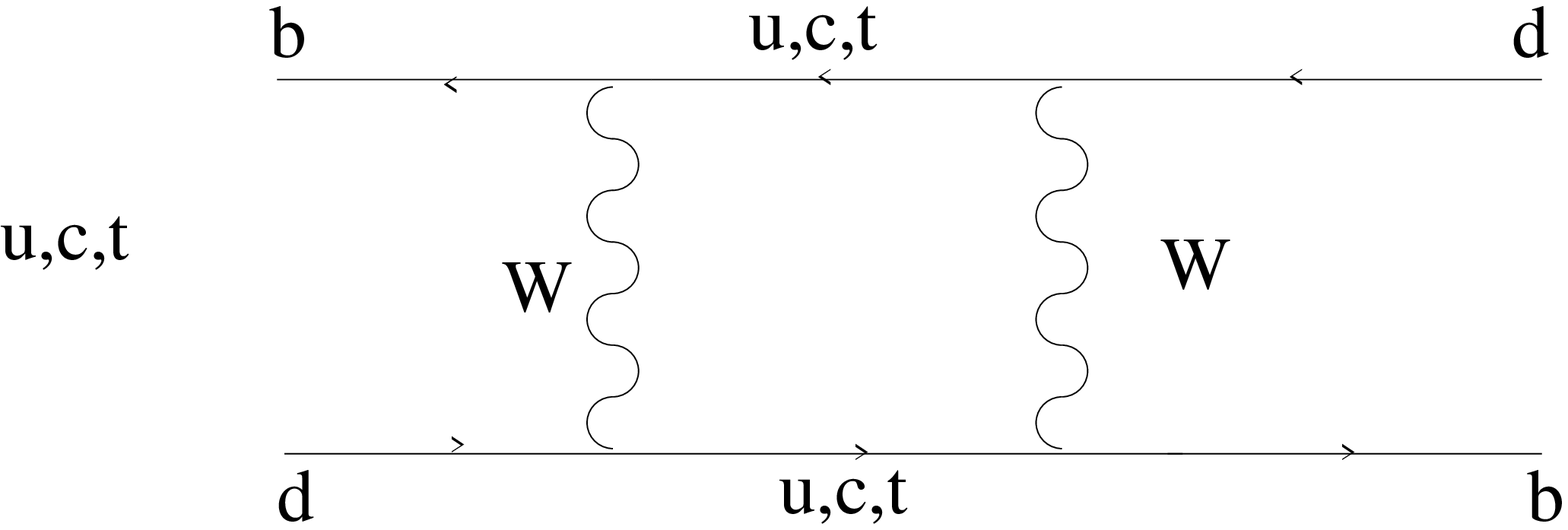}
\hspace{4.5cm}
\epsfysize=2.8cm \epsfbox[0 20 400 270]{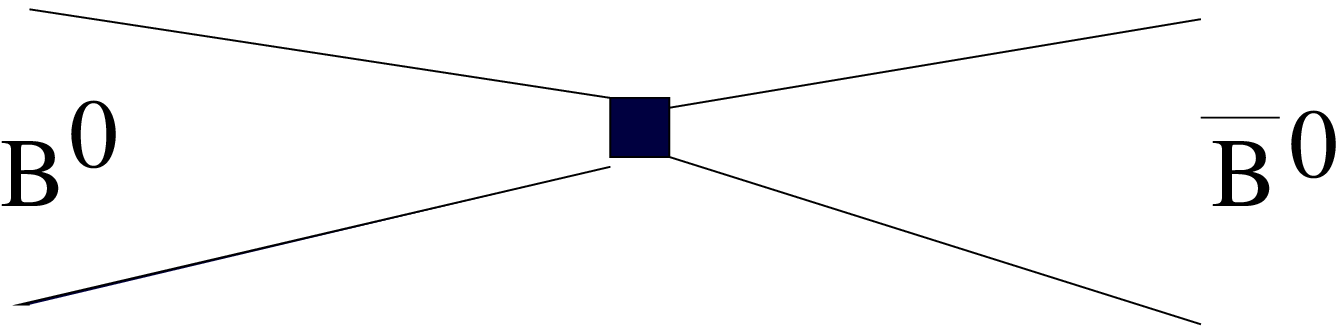}
}
\end{center}
\caption{Box diagrams describing $B^0-\overline{B^0}$ mixing in the electroweak theory (left and middle) and the 
weak effective theory (right).}
\label{fig:box}
\end{figure}

Weak flavor changing decays of quarks are investigated to determine elements of the CKM 
matrix~\cite{cabibbo63,KM73} which parameterizes 
the mixings of quark generations in the Standard Model. The corresponding Lagrangian density is 
given by:
\be
{\cal L} = -\frac{g_2}{\sqrt{2}} \left(\overline{u}_L\overline{c}_L\overline{t}_L\right)
\gamma^\mu (V_{q_1q_2}) \left(
\begin{array}{c}
d_L \\
s_L \\
b_L \\
\end{array}
\right) W^\dagger_\mu + \mathrm{h.c.}\;,
\ee
where $u_L,d_L,c_L,s_L,t_L,b_L$ are left-handed quark spinors, $W_\mu$ a charged weak gauge boson 
and $g_2$ the weak gauge coupling. If there are three generations of quarks, the CKM matrix $V$ should
be unitary. Its entries are usually parameterized with:
\be
V = \left(
\begin{array}{ccc}
V_{ud} & V_{us} & V_{ub} \\ 
V_{cd} & V_{cs} & V_{cb} \\ 
V_{td} & V_{ts} & V_{tb} \\ 
\end{array}
\right).
\ee
The CKM matrix elements most relevant to decays or mixings 
of the $b$ quark are $V_{cb}$, $V_{ub}$ and $V_{td}$. Since quarks are bound
their decays take place on the level of hadron decays. 
The decays of interest for lattice calculations are those to a definite hadron state,
the exclusive decays. In parallel one also uses inclusive decays, where one sums over 
all hadrons in the final state which can be the end products in a certain quark decay.
The high precision of measurements of inclusive decay rates is in contrast to some 
model dependence in the theoretical interpretation of the results.
$|V_{ub}|^2$ is e.g.\ proportional to the rate for the leptonic meson decay $B^- 
\to l^- \nu_l$ (see Fig.~\ref{fig:Bdecayplot},
where $l$ is a lepton ($e$, $\mu$ or $\tau$) and $\nu_l$ the corresponding neutrino.
Since $V_{ub}$ parameterizes the quark decay $b \to u$, its modulus can also measured by
the semileptonic decays of the $B$ meson $B \to (\pi, \rho, \omega)  l\overline{\nu}_l$.
$|V_{cb}|^2$ can be determined e.g.\ from the semileptonic decay $B \to D$. 
$|V_{td}|^2$ is proportional to the oscillation frequency between
the mass eigenstates of the $B^0-\overline{B^0}$ mixing, which is described by the 
left and middle diagrams in Figure~\ref{fig:box} in the electroweak theory.

\begin{figure}[thb]
\begin{center}
\epsfysize=4cm \epsfbox{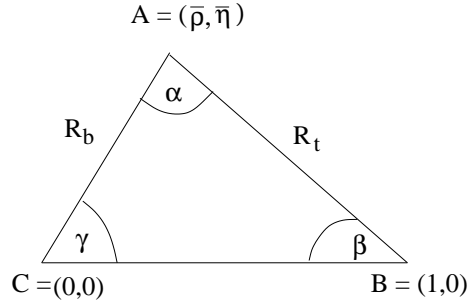}
\end{center}
\caption{Unitarity triangle.}
\label{fig:ut}
\end{figure}

If there are only three flavors of quarks and no new physics
the CKM matrix has to be unitary which leads to 9 relations of the matrix elements.
Two of the relations relevant 
to $b$ decays are given by:
\ba
V_{ud}V_{ub}^\ast+V_{cd}V_{cb}^\ast+V_{td}V_{tb}^\ast &=& 0 \label{eq:tri1} \\
V_{ub}V_{tb}^\ast+V_{us}V_{ts}^\ast+V_{ud}V_{td}^\ast &=& 0.\label{eq:tri2}
\ea
To describe $CP$ violation which has been discovered in $K^0-\overline{K}^0$ oscillations
in 1964 and has been established in $B^0$ meson decays in 2001~\cite{babarbelle2001}, 
within the Standard Model,
it is necessary that some CKM matrix elements have an imaginary part. 
Eqs.~(\ref{eq:tri1}) and (\ref{eq:tri2}) describe triangles in the
complex plane. Eq.~(\ref{eq:tri1}) is in particular used in the so-called `standard 
analysis' of the unitarity triangle shown in Fig.~\ref{fig:ut}
which focuses on the determination of the complex phase of
$V_{td}$, related to $CP$ violation within $B^0-\overline{B}^0$ oscillations via $t$ 
quark production. The procedure is e.g.\ reviewed in~\cite{buras2003}.

The CKM matrix parameterizes mixing between quark families. 
The relative smallness of the off-diagonal elements can be taken into
account with the  Wolfenstein parametrisation:
\be
V = \left(
\begin{array}{ccc}
1-\frac{\lambda^2}{2} &\lambda & A\lambda^3(\rho - i \eta) \\
  -\lambda    & 1-\frac{\lambda^2}{2} & A\lambda^2         \\
A\lambda^3(1-\rho-i\eta)& -A\lambda^2 & 1 \\
\end{array}
\right) + O(\lambda^4)
\ee
with $\lambda = |V_{us}|$.

With more exact measurements of the matrix elements, a generalized Wolfenstein 
parametrization~\cite{buras1994} has become useful:
\be
V = \left(
\begin{array}{ccc}
1-\frac{\lambda^2}{2} -\frac{\lambda^4}{8} &\lambda + O(\lambda^7) & 
A\lambda^3(\rho - i \eta) \\
  -\lambda + A^2\frac{\lambda^5}{2} [1-2(\rho + i\eta)]  
& 1-\frac{\lambda^2}{2} -\frac{\lambda^4}{8}(1+4A^2) & A\lambda^2 + O(\lambda^8) \\ 
A\lambda^3(1-\overline{\rho}-i\overline{\eta})& -A\lambda^2+A\frac{\lambda^4}{2}
[1-2(\rho+i\eta)] & 1-A^2\frac{\lambda^4}{2} \\
\end{array}
\right) + O(\lambda^6).
\ee
The parameters $\overline{\rho}$ and $\overline{\eta}$ differ from the 
not barred parameters by $O(\lambda^2)$:
\be
\overline{\rho} = \rho\left(1 -\frac{\lambda^2}{2}\right), 
\overline{\eta} = \eta\left(1-\frac{\lambda^2}{2}\right).
\ee
$|V_{ub}|, |V_{cb}$ and $|V_{td}|$ parameterize the sides of the triangle in the
complex plane shown in Fig.~\ref{fig:ut}:
\ba
R_b &=& \sqrt{\overline{\rho}^2 + \overline{\eta}^2} =
\left(1-\frac{\lambda^2}{2}\right) \frac{1}{\lambda}|V_{ub}/V_{cb}| \nonumber \\
R_t &=& \sqrt{(1-\overline{\rho})^2 + \overline{\eta}^2} 
\frac{1}{\lambda}|V_{td}/V_{cb}|.
\ea
One of its angles directly measures the phase of $V_{td}$:
\be
V_{td} = |V_{td}|e^{-i\beta}.
\ee
Lectures reviewing the CKM matrix and $CP$ violation in particular in the context of $B$ physics and
discussing future directions of research on $CP$ violation in $B$ decays
can be found in Ref.\cite{fleischer2004}.
A comprehensive review on the status of the elements of the CKM matrix is given in~\cite{ali2003}.

Processes at energy scales much less than the $W$ boson mass can be calculated within
the weak effective theory where interactions mediated by the $W$ or $Z$ particles can be
described  by point interactions. In other words, if the momentum of an exchanged
$W,Z$ boson is small with respect to its mass $M$, the propagator can be approximated by
\be
\frac{-i g_{\mu\nu}}{q^2-M^2} \simeq \frac{-ig_{\mu\nu}}{M^2} 
\ee

The leptonic decay of $B$ mesons (e.g. Fig.~\ref{fig:Bdecayplot}) can be defined from
matrix elements of the heavy-light axial vector current going into the vacuum:
\be
\langle 0|A_\mu(x)|B(p) \rangle = -if_B p_\mu e^{-ipx}, \label{eq:axial}
\ee
where $f_B$ is the $B$ decay constant.
The branching ratio for the decay $B^+ \to l^+\nu_l$  is 
\ba
BR(B^+ \to l^+ \nu_l) &=& \frac{G_F^2 m_B m_l^2}{8\pi}\left(1-
\frac{m_l^2}{m_B^2} \right)^2  f_B^2 |V_{ub}|^2 \tau_B, 
\ea
where $G_F= g_2^2/(8M_W^2)$ is the Fermi constant and $\tau_B$ the $B$ lifetime. If $f_B$ is known, 
$|V_{ub}|$ can in principle be determined experimentally from this decay. 
There exist only experimental upper bounds on $f_B$ and $f_{B_s}$. 
Recent experimental measurements of $f_D$ from  $D^+ \to \mu^+ \nu_\mu$ exist from CLEO-c~\cite{CLEO2005}
with the result
\ba
f_D & = & 223(17)(3) \mbox{ MeV},
\ea 
and from BES~\cite{BES2004}.
Experimental results on  $f_{D_s}$ are:
\ba
f_{D_s} &=& 280(17)(25)(35) \mbox{ MeV~\protect\cite{cleo_fDS}, and} \nonumber \\
f_{D_s}  &=& 285(19)(40) \mbox{ MeV~\protect~\cite{aleph_fDS}}. \nonumber \label{eq:fDs}
\ea 

In the weak effective theory exclusive semileptonic decays are described by matrix elements  of currents
between hadronic states. They can be parameterized in terms of form factors which
depend on the momenta of the hadron states. 
The form factors receive  contributions from long-distance QCD interactions,
and therefore have to be evaluated nonperturbatively. This can be done from first principles
using the lattice.
$B^0-\overline{B^0}$ mixing is described by the matrix element of
a current-current interaction, see the third diagram in Fig.~\ref{fig:box}. It is common to
parameterize it as $f_B^2B_B$, where the 'bag parameter' $B_B$ quantifies to what extent the matrix 
element is described by $B$-to-vacuum currents (see e.g.~\cite{hiorth2003}):
\be
B_B = \frac{3}{2}\frac{\langle \overline{B^0} | (\overline{b}_L\gamma_\mu d_L)(\overline{b}_L\gamma_\mu d_L)|B^0
\rangle}{\langle \overline{B^0} |\overline{b}\gamma_\mu\gamma_5 d|0\rangle
\langle 0|\overline{b}\gamma_\mu\gamma_5 d|B^0\rangle }.
\ee
In the vacuum saturation approximation, $B_B = 1$.
The oscillation frequency of the mass eigenstates is proportional to the mass difference and
related to the current matrix element by
\be
\Delta M_d \propto  |V_{tb}^\ast V_{td}|^2 f_B^2B_B.
\ee

\begin{figure}[thb]
\begin{center}
\vspace{-0.3cm}
\epsfysize=5cm \epsfbox{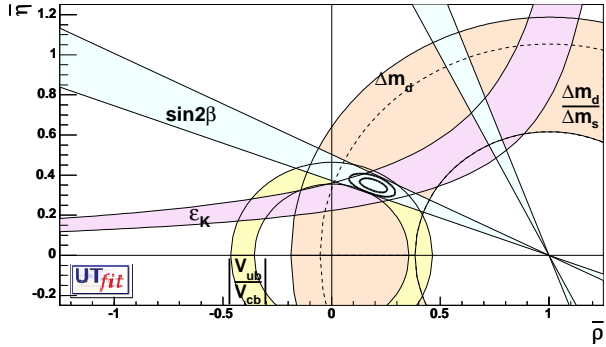}
\end{center}
\caption{Status of the CKM matrix after the CKM05 Workshop~\protect\cite{utfit2004}. The shaded regions
denote the values which are allowed at 95\% probability. The dotted line denotes the bound on the mass 
difference of the $B_s^0$ $CP$ eigenstates. For recent updates see 
Refs.~\protect\cite{utfit2004,ckmfitter}}
\label{fig:CKMfitter}
\end{figure}

The status of the constraints on the unitarity triangle in the $\overline{\rho}-
\overline{\eta}$ plane including results on $|V_{ub}/V_{cb}|$, $\epsilon_K$ and $B^0-
\overline{B}^0$ and $B_s^0-\overline{B}^0_s$ mixings from 
lattice~\cite{utfit2004} at the CKM05 Workshop 
is shown in Fig.~\ref{fig:CKMfitter}. The parameter $\epsilon_K$ measures $CP$ violation in the
neutral $K$ meson system. Again, the flavor eigenstates are not eigenstates to $CP$, 
rather they are superpositions of the $CP$ eigenstates and of the mass eigenstates. 
Denoting the heavier (or longer-lived) one as $K_L$ and the 
lighter (or shorter-lived) one as $K_S$, and normalizing $\sqrt{|p|^2+|q|^2}=1$, they can be parameterized as:
\ba
|K_S\rangle & = & e^{-\gamma_St}(p |K^0\rangle + q |\overline{K}^0\rangle) \nonumber \\
|K_L\rangle & = & e^{-\gamma_Lt}(p |K^0\rangle - q |\overline{K}^0\rangle),
\ea
where $\gamma_{L,S}$ are complex.
$\epsilon_K$ is the ratio of amplitudes of decays of $K_L$ and $K_S$ respectively into two
pions with isospin 0, and is related to the mixing parameters by
\be
\frac{|p|^2-|q|^2}{|p|^2+|q|^2} = \frac{2\mathrm{Re} \epsilon_K}{1+|\epsilon_K|^2}.
\ee
The existence of an imaginary part in CKM matrix elements is now well 
established, and apparently the Standard Model can explain the presently known
$CP$ violation in hadron physics. 
For recent reviews of the impact of lattice results on the determination of 
parameters of the CKM matrix see~\cite{lubicz2004}.
\section{QCD ON THE LATTICE\label{sec:lattice}}
The 4-dimensional Minkowski space-time can be substituted with a 4-dimensional 
Euclidean lattice by Wick-rotating real time into imaginary time. The Minkowski vacuum 
expectation values of products of fields can be reconstructed from the Euclidean Green
functions~\cite{osterwalder1973}.
Matter fields, e.g. quarks, sit on the lattice sites  separated by 
a distance $a$, the lattice spacing. Gauge fields sit on links connecting the lattice sites.
Here, a brief introduction to the features of the lattice formalism which appear relevant to
present calculations is given. For detailed introductions see e.g.\ the textbooks~\cite{latreviews} 
and the review articles~\cite{sharpe1994,davies2001,kronfeld2002,mcneile2003,degrand2004}.

A lattice can be considered as a gauge-invariant regulator. 
QCD has been shown to be renormalizable to all orders
in lattice perturbation theory~\cite{reisz1989}. It is believed that continuum QCD can 
be constructed as a continuum limit of lattice QCD also beyond perturbation theory.
\subsection{Gauge Fields}
We discuss an introduction into the discretization of a gauge theory, following 
the discussion of e.g.~\cite{sharpe1994}.
To arrive at gauge invariant products of fields at different space-time points as they
occur in actions containing derivatives,  
one connects them with path-ordered integrals of gauge fields
\be
L(y,x) = P\exp\left[ig\int_x^y dz_\mu A^\mu(z) \right].
\ee
The $L(x,y)$ are parallel transporters which transform under $V(x) \in SU(N_c)$
(where $N_c$ is the number of colors) as
\be
L(y,x) \to V(y) L(y,x) V^{-1}(x).
\ee
Then  $\bar{q}(y)L(y,x)q(x)$ is gauge 
invariant. Line integrals over closed loops, $L(x,x)$, are called  Wilson loops. 
In particular, $\Tr L(x,x)$ is gauge invariant.

The gauge fields on the lattice are represented by `link' variables
\ba
U_{x,\mu} &=& P \exp[ig a \int_{x+a\mu}^xdz_\mu A^c_\mu(z)t^c],
\ea
where the line integration starts at the lattice site which is shifted by one lattice
spacing in the $\mu$ direction from the site $x$.
In the continuum limit this is approximately equal to
\ba
 &=& \exp[-ig a A^c_\mu(x+an_\mu/2)t^c],
\ea
where $A^c_\mu(x)$ are the continuum gauge fields. $n_\mu$ is a unit vector in the
$\mu$ direction.
The $U$ fields  are parallel transporters between neighboring lattice sites transforming as
\be
U_{x,\mu} \to V_x \uxm V^\dagger_{x+\mu}, V(x) \in SU(N_c).
\ee
We use the short form $x+\mu$ for $x+an_\mu$.

\begin{figure}[thb]
\begin{center}
\epsfysize=2.5cm \epsfbox{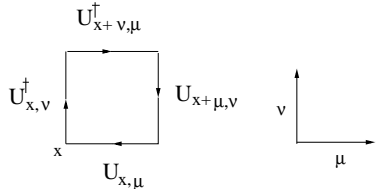}
\caption{Plaquette $W_{\mu\nu}^{1\times 1}$ in the $(\mu,\nu)$ plane. The corners are separated by the
lattice spacing $a$. }
\label{fig:plaq}
\end{center}
\end{figure}

The most elementary product of links along a closed path is the  ($1\times1$) Wilson loop,
or  the plaquette,
\be
W_{\mu\nu}^{1\times 1} = \Tr\uxm U_{x+\mu,\nu}U^\dagger_{x+\nu,\mu}U^\dagger_{x,\nu}, 
\ee
sketched in Fig.~\ref{fig:plaq}. It is used to define lattice actions for the gauge fields.

By expanding the gauge links around $a \simeq 0$, 
\be
U_{x,\mu} = 1 -  igaA_\mu(x+{\scriptsize{\frac{1}{2}}}an_\mu) + O(a^2),
\ee
with $A_\mu(x) = t^c A_\mu^c(x)$, 
one finds for the tree-level relation between plaquettes and the continuum field 
strengths:
\be
Re W_{\mu\nu}^{1\times 1} = N_c -\frac{g^2}{2}a^4\Tr G_{\mu\nu}G^{\mu\nu}
+O(a^6).
\ee 
Thus one obtains the Wilson (or `plaquette') lattice action as a discretization of the continuum gauge
field action of Eq.~(\ref{eq:qcd}),
\be
S_g = -\beta\sum_{x,\mu<\nu}\left(1 - \frac{1}{N_c}Re W_{\mu\nu}^{1\times 1}\right), \,\, 
\beta = \frac{2N_c}{g^2}. \label{eq:lataction}
\ee
The constant term is sometimes dropped for simplicity. The action~(\ref{eq:lataction}) has 
lattice spacing errors at $O(a^2)$. To reduce discretization effects further, actions
can be improved. For gauge field actions, this consists in adding larger Wilson 
loops. Improvement can be done  by removing discretization effects 
order by order in $a$ (originally suggested by Symanzik for scalar field theory~\cite{symanzik1980}, and
developed into an improvement program for on-shell quantities by Ref.~\cite{luscher1985}), or with a
renormalization group method. The perfect lattice gauge action~\cite{hasenfratz1994} is given as the
fixed point of a renormalization group trajectory. 
For the renormalization group improved (RG) action~\cite{iwasaki1985}, the 
coefficients are
determined using block spin transformations along a renormalized trajectory. 

Some of the lattice calculations whose results are discussed in the following use improved actions with a
$1\times 2$ Wilson loop ($W_{\mu\nu}^{1\times 2}$) added to the plaquette term:
\be
S_{imp} = \frac{\beta}{N_c}\sum_{x}\left( c_0
\sum_{\mu < \nu}Re W_{\mu\nu}^{1\times 1}
+c_1 \sum_{\mu \neq \nu}Re W_{\mu\nu}^{1\times 2}\right).
\ee
The coefficients are given by $c_1 = -0.331$  for the renormalization group improved
action, by  $-1.4088$ for the DBW2 action~\cite{dbw2}, and 
by $-1/12$ for tree-level Symanzik improvement~\cite{weisz83}.
The other coefficient is fixed by the condition $c_0 = 1 - 8c_1$.

At typical values of $\beta$ in lattice simulations,
there are large  coefficients in perturbation theory due to tadpole diagrams on the lattice.
The perturbative corrections can be reduced with `mean-field' (or `tadpole') improvement~\cite{lepage1992}: 
the gauge links $U$ are divided by their expectation value 
\be
U_\mu(x) \to U_\mu(x)/u_0,
\ee
where $u_0$ can 
be calculated in perturbation theory and determined in simulations. 
A gauge invariant definition is the fourth root of the plaquette 
\be
u_0 = \langle W_{\mu\nu}^{1\times 1} \rangle^{1/4}.
\ee
The expectation value of the gauge link $U_\mu$ in e.g.\ in the Landau gauge has been used in many 
lattice calculations of NRQCD for the tadpole improvement of the heavy quark. 
NRQCD quarkonium data were found to have a better scaling behavior than using the 
plaquette~\cite{shake1998}.
\subsection{Lattice Fermions\label{sec:ferm}}
One can discretize the Euclidean continuum Dirac action by substituting the covariant
derivatives by covariant lattice differences 
\be
\begin{array}{lccl}
\mbox{forward} &  \nabla_\mu q_x &  = & \scriptsize{\frac{1}{a}}(U_{x,\mu} 
q_{x+\mu} - q_x), \nonumber \\
\mbox{backward} &  \nabla^\ast_\mu q_x & = & \scriptsize{\frac{1}{a}}
(q_x - U^\dagger_{x-\mu,\mu} q_{x-\mu}),
\nonumber \\                           
\mbox{symmetric} &  \scriptsize{\frac{1}{2}}(\nabla_\mu q_x + \nabla^\ast_\mu q_x )
& = &  \scriptsize{\frac{1}{2a}}(U_{x,\mu} q_{x+\mu} - U^\dagger_{x-\mu,\mu} q_{x-\mu}).
\end{array}
\ee
Using the symmetric derivative, one can define the 'naive' lattice fermion action by
\ba
S_f &= &a^4\sum_{x,y}\overline{q}_x D_{xy} q_y, \nonumber \\
D_{xy} &=& \sum_\mu\frac{1}{2a}\gamma_\mu( U_{x,\mu}\delta_{x+\mu,y}
- U^\dagger_{x-\mu,\mu}\delta_{x-\mu,y})+ m \delta_{xy},
\ea
which is chirally symmetric if $m\to 0$ and has $O(a^2)$ errors.
However, the naive discretization leads to a flavor multiplication, the so-called 
'doublers'. 
The fermion propagator is a periodic function of the momentum $k_\mu$ 
Considering for example the case $m=0$, one finds that the free propagator has
next to the pole at $k_\mu = 0$ which also exists in the continuum, poles at $k_\mu = \pi/a$.
The states corresponding to the poles are degenerate and have finite energies. So one has   
16 species of fermions, which occur in pairs of opposite chirality. The occurrence of doublers 
is a problem with a quite general class of fermion actions. According to the Nielsen-Ninomiya 
theorem~\cite{nielsen1985}, for any lattice fermion action which is  translation invariant, 
local and has a hermitian Hamilton operator, there is an equal number of right- and lefthanded 
fermions corresponding to each quantum number. 

Wilson's solution to the doubling problem is to add a term of the form 
\ba
S_W & = & a^4 \sum_{x} \overline{q}_x D_{W,xy} q_y, \nonumber \\
D_{W,xy} & = & -\sum_\mu\frac{r}{2a}( U_{x,\mu}\delta_{x+\mu,y}
 + U^\dagger_{x-\mu,\mu}\delta_{x-\mu,y}- 2 \delta_{xy}), \label{eq:wilson}
\ea
to the action. 
With the Wilson term the doublers obtain masses $m_d$ which remain finite in lattice units: 
$m_d a \neq 0$, i.e.\ $m_d \sim 1/a$ if $a\to 0$.
The exact behavior of the doubler masses depends on the choice of $r$.
The most common value in lattice simulations is $r = 1$.

To normalize the term proportional to $\delta_{x,y}$ in the action to one, it is 
common to parameterize the mass $m$ in terms of the hopping parameter $\kappa$. Setting
$r = 1$, the action is:
\ba
S_f &=& \sum_x\overline{\psi}_x D^{hopp}_{xy} \psi_y, \nonumber \\
D^{hopp}_{xy} &=&  \delta_{xy} - \kappa \sum_\mu \left(
[1-\gamma_\mu]U_{x,\mu}\delta_{x+\mu,y} + [1+\gamma_\mu]
U^\dagger_{x-\mu,\mu}\delta_{x-\mu,y}\right)
\ea
with
\ba
am & = & \frac{1}{2\kappa}- 4
\ea
and 
\be
 q_x =  a^{-3/2} \sqrt{2\kappa}\psi_x ; \; \overline{q}_x  = 
 \overline{\psi}_x a^{-3/2} \sqrt{2\kappa}. \label{eq:contnorm}
\ee
The Wilson term introduces a shift in the quark mass parameter such that the massless free 
Lagrangian corresponds to $\kappa = \kappa_c = 0.125$.

In the interacting case, the critical hopping parameter is renormalized and the bare
quark mass is given by
\be
am = \frac{1}{2}\left(\frac{1}{\kappa}-\frac{1}{\kappa_c}\right),\label{eq:vwi}
\ee
also called the vector ward identity (VWI) quark mass. Now chiral symmetry is broken at 
$a \neq 0$. Discretization errors of $O(a)$ occur and manifest themselves on
dimensional grounds as $O(am), O(ap)$  effects, where $m$ is the quark mass and $p$ the
quark momentum which is in a hadron typically of $O(\lqcd)$.
The $\kappa$ value corresponding to zero quark mass, $\kappa_c$, can be determined using
the hadron spectrum. The pion mass is usually associated with explicit chiral symmetry breaking, 
therefore one defines $\kappa_c$ as the hopping parameter where the pseudoscalar meson mass vanishes. 
With a tadpole-improved Wilson action, $U_\mu \to U_\mu/u_0$ the hopping parameter is
redefined to  $\tilde{\kappa} = \kappa u_0$.
In the tree-level mean field improved theory the critical hopping parameter is 
\be
\kappa_c = 1/(8u_0),
\ee
such that $\tilde{\kappa} = 0.125$.

Another definition of quark mass used in lattice calculations~\cite{bochicchio1985,itoh1986} 
is to use the axial vector ward identity (AWI)~Eq.(\ref{eq:pcac}):
\be
2 m_q^{AWI} \langle 0|P|\pi(\vec{p}=0) \rangle = 
 \langle 0|\nabla_4 A_4(x)|\pi(\vec{p}=0) \rangle,
\ee
where $A_\mu(x)$ is a lattice version of the axial vector current, which  includes 
improvement terms where necessary.
$O(a)$ errors can be removed from the quark action using the clover term
\ba
S_c & = & - a^4 \frac{\kappa c_{SW}}{2} \sum_x \sum_{\mu < \nu} \overline{\psi}_x
\sigma_{\mu\nu}  G_{\mu\nu} \psi_x,
\ea
where  $\sigma_{\mu\nu} = \frac{i}{2}[\gamma_\mu,\gamma_\nu]$, and $G_{\mu\nu}$ denotes now
a discretized version of the field strength tensor using four neighboring plaquettes~\cite{SW1985}:
\be
G_{\mu\nu}^{ab}(x) = \frac{1}{8i}\sum \left((\tilde{W}^{1\times 1})^{ab}_{\mu\nu}-
(\tilde{W}^{1\times 1 \dagger})^{ab}_{\mu\nu}\right),
\ee
where the sum goes over all four plaquettes in the $(\mu,\nu)$ directions which are connected to the
lattice site $x$. The $\tilde{W}$ denote the untraced plaquettes, and $a,b$ are color indices.
The coefficient of the clover term can  be calculated in
perturbation theory. In many of the calculations whose results are discussed later,
$c_{SW}$ is at tree-level with tadpole-improvement. 
Most recent calculations use a non-perturbative determination of the clover coefficient~\cite{cSW},
i.e.\ the action is $O(a)$ improved to all orders in perturbation theory.

For the calculation of matrix elements of currents, improvement of the field is also
necessary. For $O(a)$ improvement this can be done by changing the quark fields by a term of
the form
\ba
q(x) & \to & \sqrt{1+b am} q(x), \label{eq:impr}
\ea
where $b$ can be determined nonperturbatively. The Fermilab normalization for massive quarks,
Eq.~(\ref{eq:z2}) of Section \ref{sec:fnal},  corresponds to $b = 1$ at tree-level. In our lattice
calculations discussed in Section~\ref{sec:fB} we used tadpole improvement, i.e.\ $b = u_0$.
The current operator itself has to be improved as well by adding a higher dimensional operator 
having the same symmetries with the appropriate coefficient. The improvement of currents
due to higher dimensional operators in effective theories is very similar and discussed in
Section~\ref{sec:cNRQCD}.

A spontaneous breaking of flavor and parity symmetries in lattice calculations with dynamical 
Wilson fermions has been conjectured by Aoki~\cite{s.aoki1984}. With two flavors, the $SU(2)$ 
symmetry breaks down to a residual $U(1)$, and there are two massless Goldstone bosons and
one Goldstone boson with mass proportional to the lattice spacing. 
Ref.~\cite{sharpe1998} finds, using chiral perturbation theory, that for certain 
parameter values of the chiral Lagrangian an Aoki phase might exist  close to the
continuum limit, while 
the authors of Ref.~\cite{ilgenfritz04} find indications with a dynamical lattice calculation that 
the Aoki phase is contained only within a region of relatively strong coupling.

Another standard technique to reduce the number of doublers is to use staggered fermions,
which is the method chosen in some of the lattice calculations discussed in the
sections~\ref{sec:lresults},~\ref{sec:hresults} and~\ref{sec:fB}.
Staggered fermions are obtained from a spin-diagonalization of naive four-component fermions:
\be
q_x = \gamma_x \chi_x \, , \;\; \overline{q}_x = \overline{\chi}_x \gamma_x^\dagger\;,
\ee
with $\gamma_x = \gamma_1^{x_1}\gamma_2^{x_2}\gamma_3^{x_3}\gamma_4^{x_4}$, where $x = 
(x_1,x_2,x_3,x_4)$. In the tree-level Lagrangian the spinor components of the $\chi$ fields are 
disentangled, and chosing one component one obtains the staggered fermions. The staggered
action is  at finite lattice spacing invariant under a vector $U(1)$ symmetry, 
and for zero mass there is an axial lattice $U(1)$ symmetry which is in the continuum limit flavor 
non-singlet. For a discussion see e.g.\ the review  article~\cite{sharpe1994}, and for a detailed 
explanation of staggered fermions and their symmetries~\cite{golterman1985}.
In the continuum limit there is a $SU(4)_L\times SU(4)_R\times U(1)_L\times U(1)_R$ flavor 
symmetry, where the axial part of the chiral $U(1)$ has the usual anomaly.
Discretization errors of staggered fermions are $O(a^2)$. Recent calculations use staggered quarks 
which are improved by adding higher dimensional operators, see e.g.~\cite{asqtad1999}.

\begin{table}[thb]
\begin{center}
\begin{tabular*}{\textwidth}{ll}
\hline
\multicolumn{1}{l}{Continuum symmetry:} &
\multicolumn{1}{l}{Lattice realization} \\
\hline
\Black{Gauge symmetry  & exact} \\
Euclidean  invariance:     & 
recovered for $a \to 0, V \to \infty$ \\
\begin{minipage}[t]{6.5cm}Fermion chiral symmetry: \end{minipage}& {\small \begin{tabular}{ll}
                                 naive fermions:           & \begin{minipage}[t]{5cm}
remaining, $O(a^2)$ errors, 15 doublers  \end{minipage} \vspace{0.2cm}  \\
                                 Wilson:          & \begin{minipage}[t]{5cm}
broken at $O(a)$, doubler masses lifted  \end{minipage} \vspace{0.2cm} \\
                                 staggered:          & \begin{minipage}[t]{5cm}
                                remaining, $O(a^2)$ errors, 3 doublers 
                                  \end{minipage} \vspace{0.2cm} \\
                                 clover:          & \begin{minipage}[t]{5cm}
                               $O(a)$ breaking terms removed, doubler masses lifted  
\end{minipage}\vspace{0.2cm} \\
                                 Ginsparg-Wilson:  & \begin{minipage}[t]{5cm}
                                $O(a^2)$ errors, single chirality possible, 
                               invariant under lattice chiral transformation at $a\neq 0$ 
                                    \end{minipage} \vspace{0.2cm} \\
                                 \end{tabular} }\\
\begin{minipage}[t]{6cm} $U(1)_L\times U(1)_R$ anomaly: \end{minipage}   & \begin{minipage}[t]{9cm}
{\small exists at finite lattice spacing for Ginsparg-Wilson }  \end{minipage} \\
       & \begin{minipage}[t]{8.5cm}
{\small continuum anomaly reproduced for $a\to 0$ for all five formulations}  \end{minipage} 
\vspace{0.2cm}\\
\hline
\end{tabular*}
\caption{Symmetries of lattice fermions.}
\label{tab:chiferm}
\end{center}
\end{table}

The Ginsparg-Wilson relation~\cite{gw1982} expresses a modified chiral symmetry.
If a lattice fermion Dirac operator $D$  obeys the equation
\be
\gamma_5 D + D \gamma_5 = aD\gamma_5 D,
\ee
the corresponding action is invariant under a modified chiral transformation~\cite{luscher1998}
\ba
\psi & \to & \exp\left[i\epsilon^a T^a\gamma_5 (1-\frac a 2 D )\right]\psi,
\nonumber \\
\overline{\psi} & \to & \overline{\psi}
\exp\left[i\epsilon^a T^a(1-\frac a 2 D ) \gamma_5 \right],
\ea
where $D$ is the massless gauge covariant lattice Dirac operator, and 
one  speaks of lattice chiral fermions or Ginsparg-Wilson fermions.
For  $a \to 0$, it corresponds to the usual chiral symmetry.

Under quantization, the measure in the path integral 
is not invariant under this transformation, and the lattice axial $U(1)$ symmetry 
has an anomaly with an index theorem similar to the chiral $U(1)$ anomaly known from 
continuum QCD. 

Chiral lattice fermions are e.g. domain wall fermions in the limit of infinite 5th
dimension \cite{domainwall}, overlap \cite{overlap} and 
fixed point fermions \cite{hasenfratz1998}. Their popularity in QCD lattice simulations
is increasing. Renormalization effects are simplified due to the chiral symmetry. 
Ginsparg-Wilson fermions are on-shell $O(a)$ improved, and due to the difference in the 
eigenvalue distribution of the Ginsparg-Wilson Dirac matrix compared to the clover Dirac
matrix it is possible to simulate at small quark masses.  

A brief summary of lattice fermions and their symmetries is given in Table~\ref{tab:chiferm}.
An important remark is that chiral fermions are automatically $O(a)$ 
improved, see the discussion in~\cite{niedermayer1999}. 
Chiral symmetry and chiral fermions on the lattice are reviewed in~\cite{luscher2001,chandra2004}. 

The Schr\"odinger functional~\cite{SF} is the partition function of a Euclidean 
system with certain boundary conditions at the initial and final time. It provides a scheme to calculate 
renormalized quantities as a function of the lattice extent, which is a natural length
scale in lattice calculations.
In practical calculations one often uses periodic boundary conditions  in space and Dirichlet boundary 
conditions in time~\cite{SF}.

\subsection{Hadron Correlation Functions \label{sec:corr}}
Green functions can be calculated by  evaluating the path integral over the lattice degrees of
freedom numerically. For example, a
two-point function of a field $O(x)$, which can for example be composed of gauge and fermion fields, 
is given by 
\be
\langle 0| O^\dagger(x) O(0)|0 \rangle = \frac{1}{Z}\int {\cal D}U{\cal D}q
{\cal D}\overline{q} O^\dagger(x)
O(0) e^{-S_g[U]-S_f[q,\overline{q}]}
\ee
Ideally, these calculations are done at various values of the lattice spacing, and the
continuum estimate is obtained by extrapolating as a function of $a$ to
$a \to$ 0. In practice, some lattice calculations are performed only at one or two
values of $a$, in which case a continuum limit cannot be taken, and the discretization effects
have to be included into the estimate of systematic errors. With NRQCD calculations, higher dimensional 
operators are included as discussed in Section~\ref{sec:NRQCD}. Therefore an $a \to 0$ 
extrapolation cannot be done.
Calculations at several values of $a$ then serve to determine the discretization uncertainty.

Correlation functions are expectation values of time ordered products of hadron 
operators:
\ba
 C(t,\vec{x}) 
   &=&\langle 0|T[O^\dagger(t,\vec{x}) O(0)]|0 \rangle.
\ea
One can now derive how masses can be extracted from lattice correlation functions
(see e.g.~\cite{sharpe1994}). 
We are interested in correlation functions of hadron operators for
states with definite spatial momentum $\vec{p}$, 
\ba
 C(t,\vec{p}) 
   &=&\langle 0|T[O^\dagger(t,\vec{p}) O(0)]|0 \rangle.
\ea
The momentum projection is obtained by Fourier transforming, i.e.
summing over all points of the spatial volume at the 'sink':
\ba
C(t,\vec{p}) &=& \sum_{\vec{x}}\langle 0|T[ O^\dagger(t,\vec{x}) \exp(-i\vec{p}\cdot
\vec{x}) O(0)| 0 \rangle \nonumber \\
& \equiv & \langle 0|T[ O^\dagger(t,\vec{p})  O(0)| 0 \rangle
\ea
For the moment we consider a Euclidean 4-dimensional lattice with spatial volume $V$ and 
infinite extent in time.
The time evolution of the operator is in the Heisenberg picture
described by exponentials of the Hamiltonian $H$, and the correlation function
can be re-written as
\ba
C(t,\vec{p})    &=& \langle  0 |e^{Ht}O^\dagger(0,\vec{p}) e^{-Ht}O(0)|0\rangle ,  
\, t>0 \nonumber \\
    & = & \langle 0|O^\dagger e^{-Ht}O|0\rangle.
\ea
Inserting a complete set of energy eigenstates $|n\rangle$ with energy $E_n$ and using relativistic
normalization, this gives
\be
C(t) = \sum_n|\langle 0|O|n\rangle|^2\frac{e^{-E_nt}}{2E_nV},
\ee
i.e.\ for sufficiently large times the correlation function decays exponentially with
the ground state energy. For $\vec{p} = 0$ this is given by the mass.
If the lattice is finite in the time direction, and depending on the fermion action,
there may be a modification to the exponential behavior.

To reduce discretization errors and to improve the resolution in the time direction,
some calculations use anisotropic lattices with a finer temporal than spatial lattice spacing.

Local interpolating operators for relativistic quarks to project onto pseudoscalar, scalar, vector and axial vector
 mesons are
\be
O^{ab}(x) = \overline{q}^a(x)\Gamma q^b(x) \label{eq:op}
\ee
where $\Gamma = 1, \gamma_5, \gamma_\mu, \gamma_5\gamma_\mu$, and $\alpha$ and $\beta$
are flavor indices. A correspondence of lattice operators to
continuum quantum numbers is given in Table~\ref{tab:qnum}.

\begin{table}[thb]
\begin{center}
\begin{tabular}{|ccccc|}
\hline
$J^P$    & $0^-$ & $0^+$ & $1^-$ & $1^+$ \\
$\Gamma$ & $\gamma_5,\gamma_0\gamma_5$ & $1,\gamma_0$ & $\gamma_k $ & $\gamma_k\gamma_5$ \\
\hline
\end{tabular}
\end{center}
\caption{Meson operators corresponding to a set of ground state quantum numbers.}
\label{tab:qnum}
\end{table}

Orbitally excited mesons (e.g. $J^P = 0^+, 1^+$ and $2^+$), can also be constructed using a 
combination of spin and derivative operators, see e.g.\ Table~\ref{tab:ops}.
In Table~\ref{tab:ops} also examples for baryon operators are given. Further 
examples are in Sec.~\ref{sec:lresults}.

The correlators can contain contributions from all states with the same quantum 
numbers.
To improve overlap with the ground state, one uses smearing, i.e.\  spatially
extended meson operators. 
For example, working in a fixed gauge, one can use wave functions motivated from  quark model 
considerations. For \cite{sgo1997,alikhan2000}, hydrogen-like smearing functions were used.
\ba
\phi(r) & = & \exp \left(-r/r_0\right) \\
\phi(r) & = & \frac{1}{2^{3/2}}\left(1-\frac{r}{2r_0}\right)\exp \left(-r/(2r_0)\right) 
\ea
to optimize overlap onto the ground and excited state $S$ wave state respectively. A gauge invariant 
technique is e.g. Jacobi smearing where the meson shape is approximated with the Green
function of a Klein-Gordon equation for a scalar field coupled to gluons in a hopping parameter
expansion. For an explanation of the method see e.g.~\cite{best1997}.

A measure for the quality of the projection to a single state is the `effective mass'. If the correlation 
function in Euclidean time is $C(t)$, then the effective mass is given by
\be
-\ln\left[\frac{C(at+an_4)}{C(at)}\right].
\ee
Examples of effective masses of $B$ meson correlators are given in  Fig.~\ref{fig:effmass}.
%
%
\subsection{Dynamical Fermions and Quenching}
In full QCD, the  partition function is an integral over 
gauge and fermionic fields $ \int {\cal D}U{\cal D}\overline{q}{\cal D}q$.
To decrease computational expenses, lattice calculations are often done in the 
quenched approximation, i.e.\ the path integral is calculated setting
the determinant of the quark matrix $D\!\!\!\slash +m$ to one.

Simulations with two flavors of light fermions have been performed for a range of lattice
spacings and various gluon and quark actions. Simulations in particular at small 
quark masses can be computationally very expensive.
An estimate of how  the computer time $T$ in
simulations including two flavors of clover fermions scales with quark mass, lattice
size $L$ and lattice spacing $a$ based on CP-PACS calculations is given by~\cite{ukawa2001}:
\ba
T \propto \left(\frac{L}{3\mbox{ fm}}\right)^5\left(
\frac{a^{-1}}{2\mbox{ GeV}}
\right)^{7}\left(\frac{m_\rho}{\mps}\right)^{6},
\ea
where $(\mps/m_\rho)^2$ is a rough measure for the quark mass (see the section on chiral
perturbation theory).

It is computationally significantly less prohibitive to simulate dynamical staggered instead of Wilson fermions
at relatively light quark masses. 
Because of the flavor doubling, dynamical simulations with staggered fermions use the square root
(for $N_f = 2$) or the fourth root (for $N_f = 1$) of the fermion determinant. 
Although Ref.~\cite{bunk2004} finds that the square root of the free staggered fermion operator is
non-local in the continuum limit, Ref.~\cite{shamir2005} finds a representation of the fourth root of the 
free staggered fermion matrix in terms of a local operator and discusses generalization to interacting fermions.
\subsection{Systematic Errors}
To use lattice results for phenomenology, it is necessary to estimate systematical errors 
as accurately as possible. The most important sources are:
\begin{itemize}
\item Finite lattice spacing 
\item Finiteness of lattice volume
\item Quenching (unphysical number of dynamical quarks)
\item Extrapolation to physical quark masses
\end{itemize}
As mentioned before, if the momentum scales are sufficiently smaller than the inverse lattice
spacing, in QCD with relativistic actions one can apply improvement in an expansion in powers of $a$. 
The Schr\"odinger  functional~\cite{SF} method  utilizes the finite lattice extent as a
length scale of the theory, and has been proven to have a well-defined continuum limit.
The results in the infinite volume are obtained using a step-scaling technique, i.e.\ rescaling the
lattice size at fixed $a$ in a first step and rescaling $L/a$, making the lattice larger and finer
in such a way that the renormalized coupling is kept fixed. For an explanation see 
e.g.~\cite{sommer2002}.
Quarks whose  masses are larger than one in lattice units, and where discretization errors proportional 
to powers of the mass are expected to be large with conventional formalisms, can be simulated using lattice 
formulations of
effective theories. If these are non-renormalizable, a continuum limit is not possible but instead one has
to find a lattice spacing with minimal systematic error. The reliability is greater if there is an
interval of lattice spacings where the results scale as a function of the lattice spacing. 
A theoretical handle to calculate extrapolations from the quark masses used in the lattice simulations to the  
$u,d$ quark masses is provided by chiral perturbation theory, discussed in the next section.
Chiral perturbation theory also provides a theoretical handle to calculate  finite volume effects.
Extrapolations of heavy masses are usually supported using predictions from HQET. Since the formulae
partly depend on unknown parameters, the most reliable way to reduce the extrapolation uncertainty 
is to simulate at quark masses which are as close as possible to the physical quark masses, which requires,
as mentioned above, effective theories if the quark mass is very large. This will be discussed in 
Section~\ref{sec:eff}. Before this, we look at various methods to set the scale, which is in 
particular closely associated with the discussion of quenching effects.
\subsection{The Static Quark Potential \label{sec:pot}}
The potential between a static quark-antiquark pair 
(see also Section~\ref{sec:lhqet}) parameterizes confinement at large 
distances as well as the strong
coupling $\alpha_s$ at short distances. Much interesting phyiscs is associated with it. For a review
see e.g.~\cite{bali2001}. Here we are just interested in  the practical aspects.
The potential contains length scales, and since it has small 
discretization errors ($O(a^2)$ or higher for quenched gauge fields), it is important for determinations of
the lattice scale. Further it is related to the strong coupling $\alpha_s$ which is a fundamental 
constant of nature and of interest here in particular for use in the perturbative matching calculations of
matrix elements. Some general features and results are described in the following.

Flux tube like field configurations in QCD give rise to a linearly growing part of the potential.
The zero point energy of the string vibrations gives rise to a $1/r$
contribution (see Refs.~[14,15] in~\cite{luscher2002}). Using the string picture, the QCD potential
is  predicted  to be of the form
\be
V(r) = \sigma r + \mu - \frac{\pi}{12r} + O\left(\frac{1}{r^2}\right),
\ee
where $r$ is the distance between the quarks, $\sigma$ the string tension and $\mu$ a mass parameter.
A usual parameterization of the potential is 
\be
V(r) = V_0 - \frac{\alpha}{r} + \sigma r. \label{eq:pot}
\ee
If the quarks mass is heavy but finite, the force between them also depends on the spins of the quarks 
and their  angular momentum. Derivations of the spin-dependent correction terms to the static potential are 
given in Refs.~\cite{eichten1981,chen1995}.

\begin{figure}[htb]
\vspace{0.1cm}
\begin{center}
\centerline{
\epsfysize=4.9cm \epsfbox{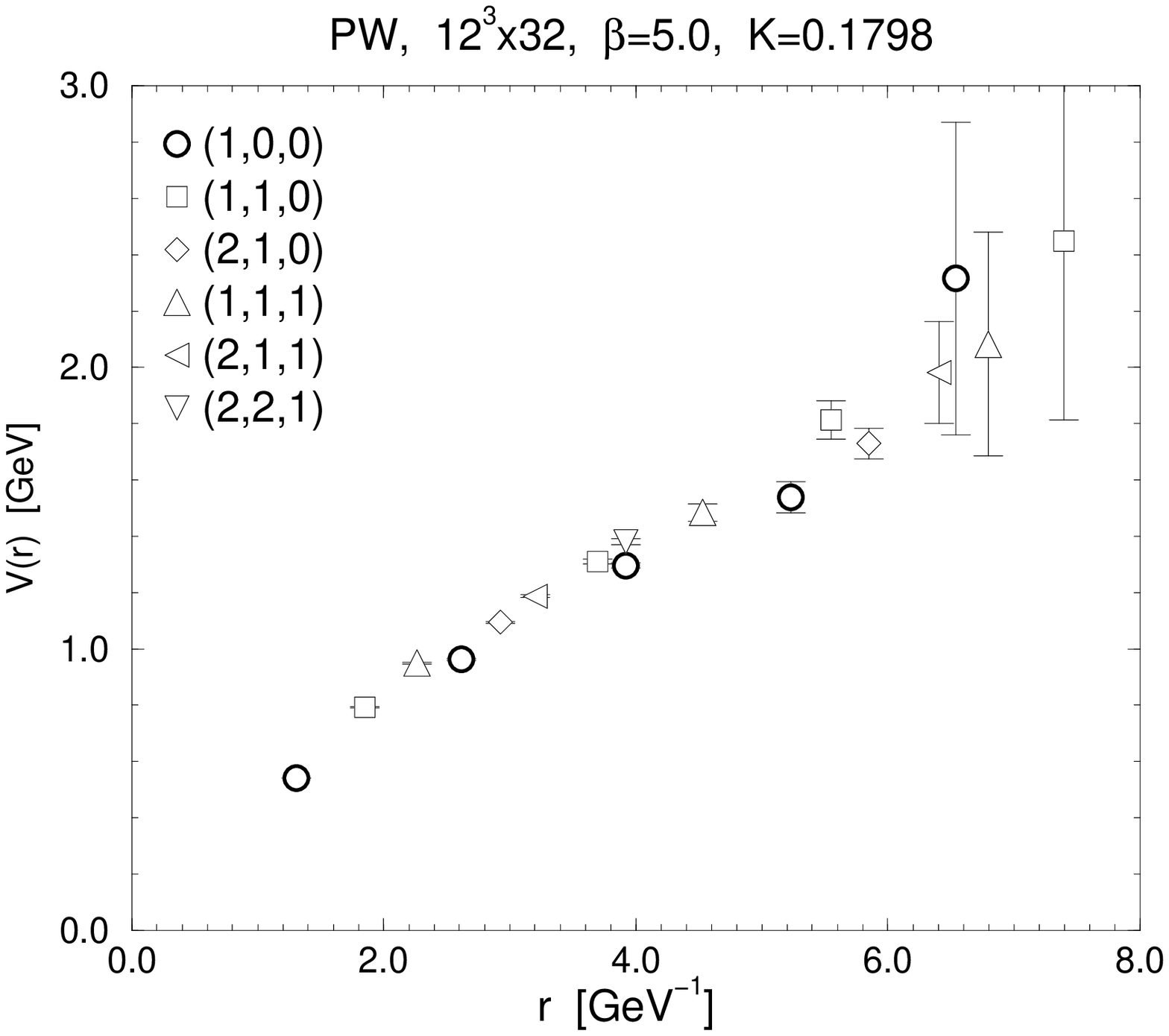}
\epsfysize=4.9cm \epsfbox{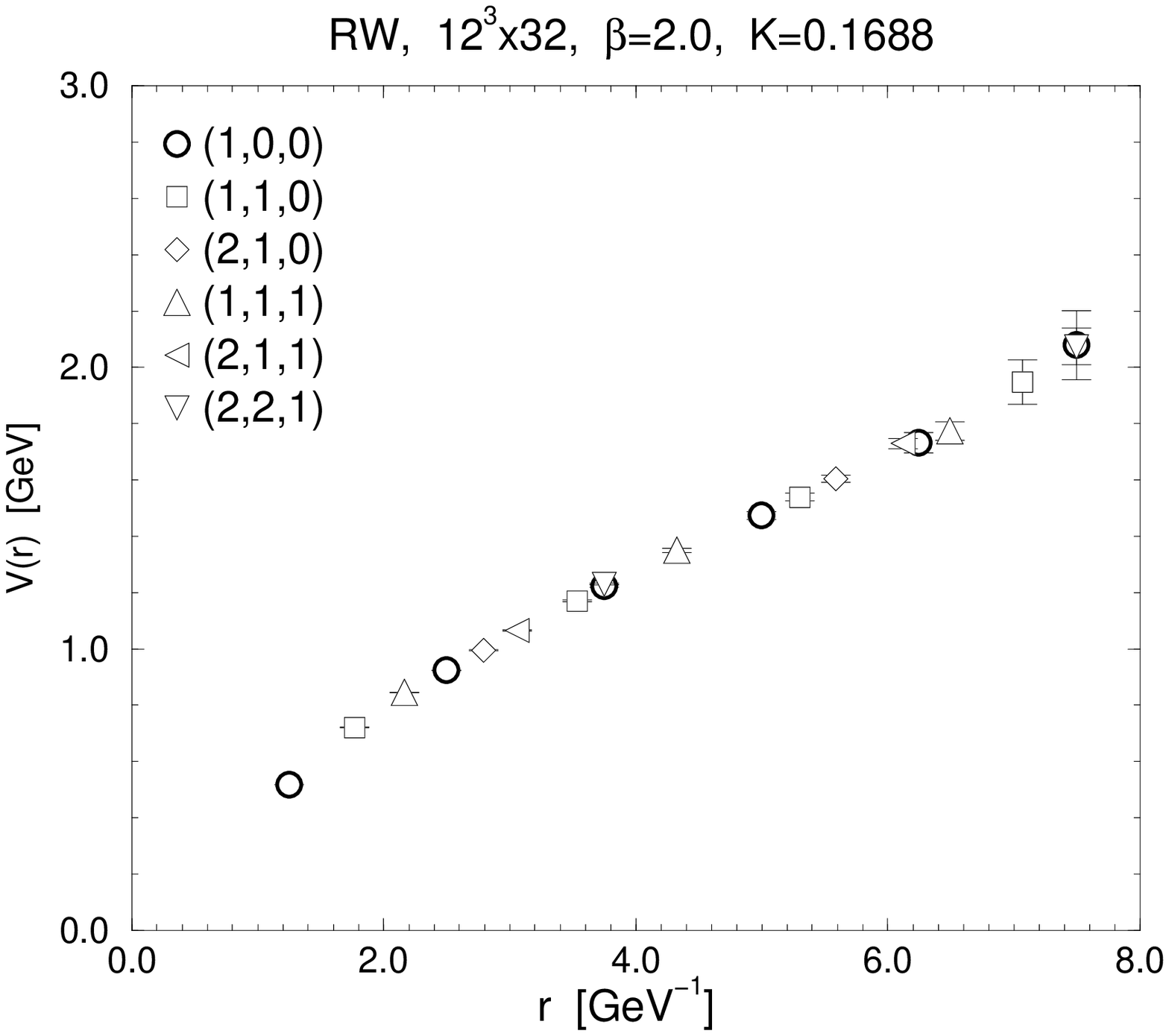}
}
\centerline{
\epsfysize=4.9cm \epsfbox{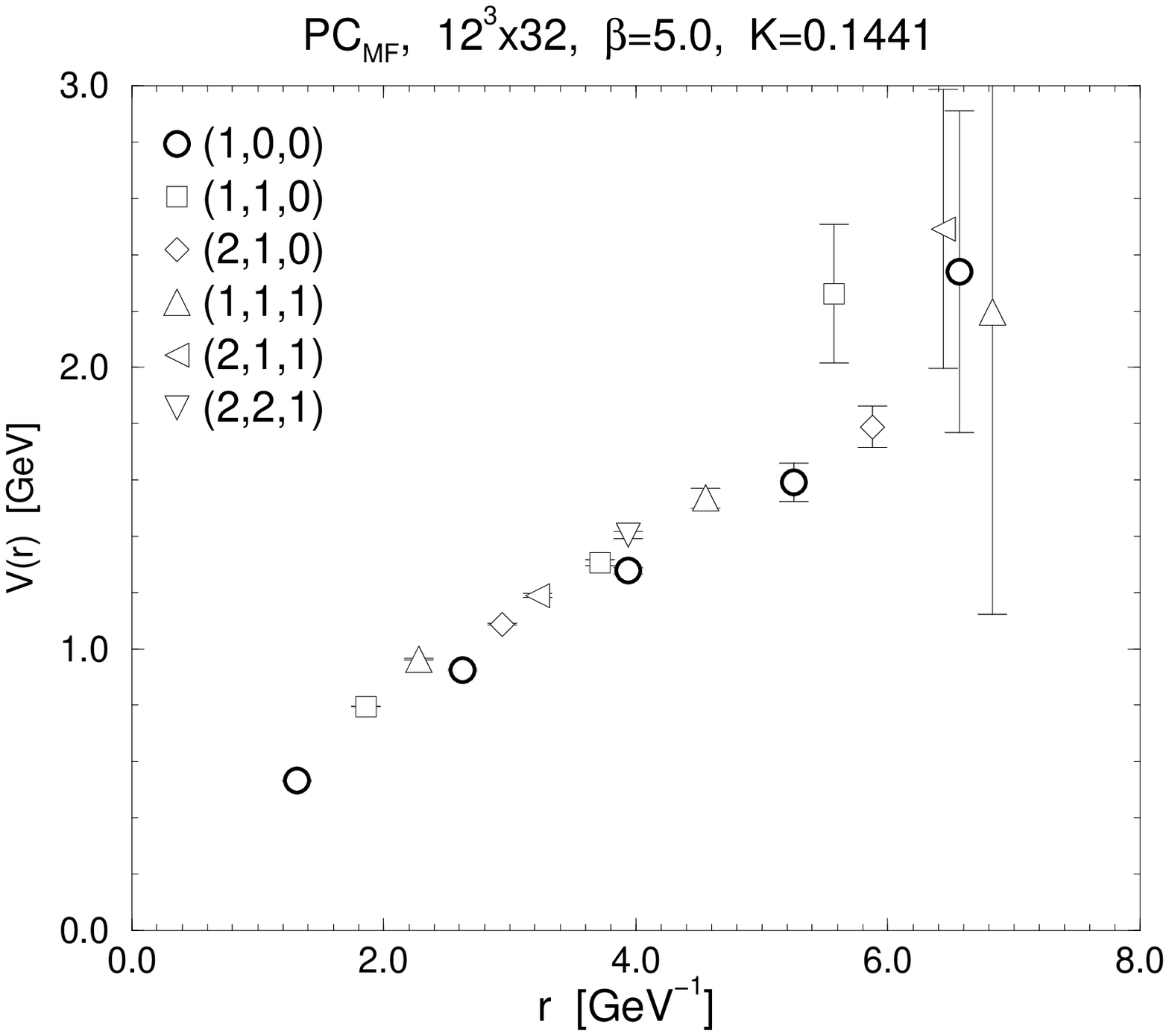}
\epsfysize=4.9cm \epsfbox{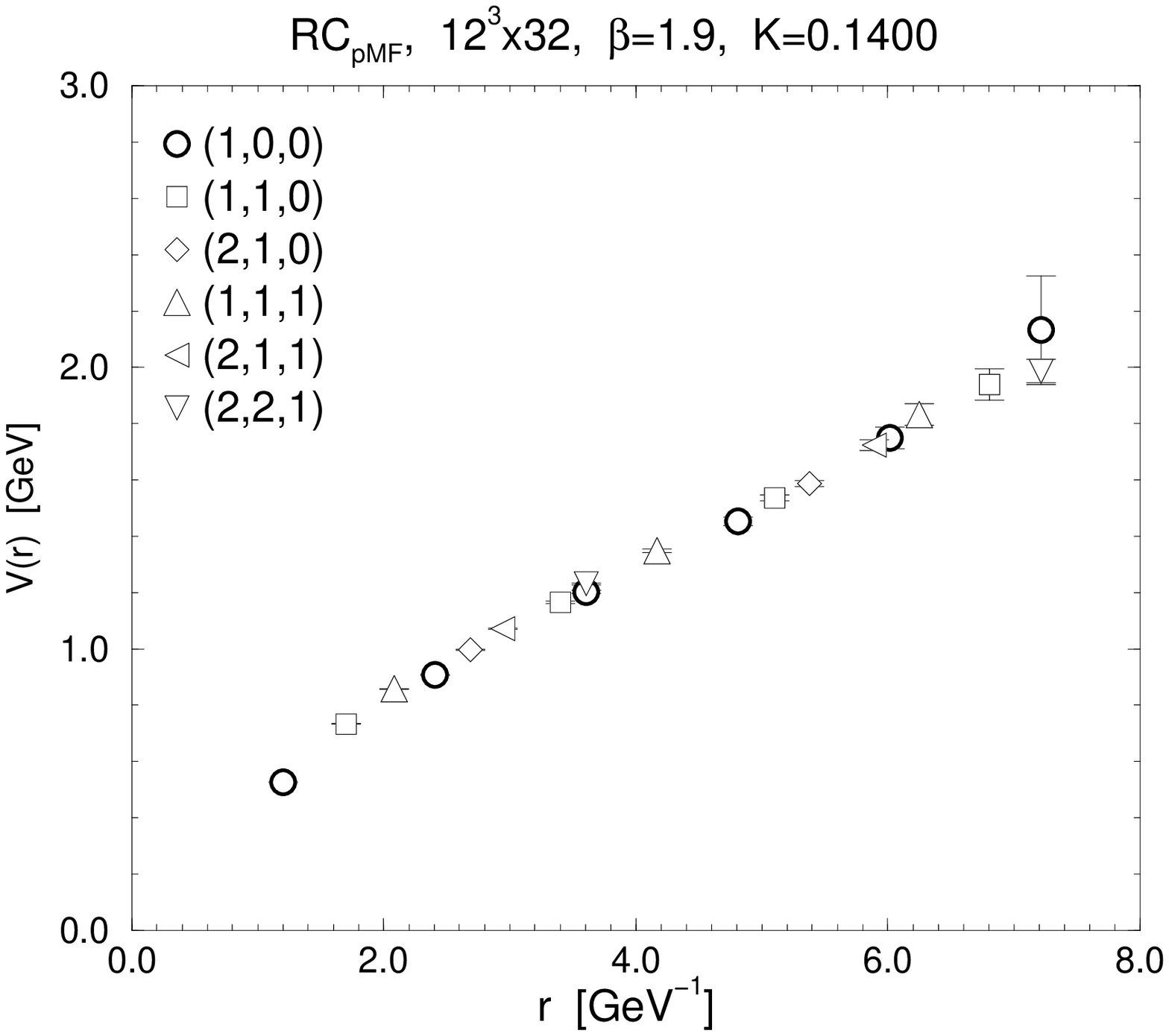}
}
\end{center}
\vspace{-0.5cm}
\caption{Heavy quark potential for $N_f = 2$ with 
Wilson gauge and quark actions (upper left), 
RG gauge and Wilson quark actions (upper right), Wilson gauge and clover quark
actions (lower left) and RG gauge and clover quark actions from~\protect\cite{aoki99}.
The clover quarks are mean-field improved. Different symbols denote the axes in units
of the lattice spacing along which the potential has been measured.}
\label{fig:rot_inv} 
\end{figure}

On the lattice, discretization errors of the static potential  can be reduced using 
improved actions. Breaking of the rotational invariance of the 
potential is a signal of discretization errors. Comparison of the rotational invariance with
 a renormalization group improved gauge action with respect to the Wilson gauge action is
demonstrated on dynamical lattices at an inverse lattice spacing of $a^{-1} \sim 1$ GeV  
in~\cite{aoki99}. The result is also given in Fig.~\ref{fig:rot_inv}. With dynamical gauge fields 
one expects a flattening of
the potential at larger distances since the string can break due to quark-antiquark pair creation. 
An explanation why flattening is not visible in the CP-PACS result of Fig.~\ref{fig:rot_inv} could be 
insufficient overlap of the operator projecting onto the static quark-antiquark pair to calculate the potential 
with the static-light meson-antimeson pair which is created as an end product of string breaking.

The Coulomb part can be used to define the coupling in the $V$ scheme~\cite{blm1983,lepage1993}
via the relation
\be
V(q^2) = -4\pi\frac{ C_F \alpha_V(q)}{q^2},
\ee
at large $q^2$.  $C_F = 4/3$ for $SU(3)$. Their definition relates the strong coupling directly 
to the potential, while higher order corrections do not appear. The procedure is expected to 
minimize higher order corrections to other relevant physical quantities as well~\cite{morningstar1996}. 
The scale of the running coupling is then also directly related to the momentum transfer between the
static quarks.  

It can be argued that the plaquette is dominated by short distance effects and
non-perturbative contributions are small. Therefore it can be expanded in a power series in
the coupling. E.g.\ using the $V$ scheme one has:
\be
-\ln W^{1\times 1}(N_f) = \sum_{i=1}^2 c_i^{(N_f)}[\alpha_V^{(N_f)}(q_{1\times 1}^\ast)]^i,
\ee
where $q^\ast_{1\times 1}$ is the average gluon momentum in the perturbative calculation of the
plaquette, explained below.
Truncating after the second term one obtains the definition for the coupling in the
plaquette or $P$ scheme~\cite{davies1995},
such that the plaquette has contributions only up to second order in perturbation 
theory. The $P$ scheme has been used e.g.\ in the perturbative calculations for the decay
constant calculation presented in~\cite{alikhan1998}.

To calculate perturbative corrections to a given physical process one can use the averaged exchanged 
gluon momentum, $q^\ast$, as a scale for $\alpha_s$~\cite{blm1983,hornbostel2003} 
to minimize higher order perturbative corrections. $q^\ast$ depends upon the
physical quantity $I$ which is supposed to be calculated.  
$\delta I$, the one-loop correction to $I$, is an integral over a function $\xi$ of the momentum $q$.
One can choose the scale for the running coupling such that the one-loop correction is equal to the value 
with $\alpha_V$ for the dressed gluon integrated over the momenta~\cite{hornbostel2003}:
\be
\delta I =  \alpha_V(q^\ast) \int \frac{d^4q}{(2\pi)^4} \xi(q) =  \int 
\frac{d^4q}{(2\pi)^4} \alpha_V(q) \xi(q).
\ee
Using the one-loop formula for the running of $\alpha_V$ from $q^\ast$ to
$q$, this gives
\be
\ln(q^\ast) = \frac{\int d^4 q \ln(q)\xi(q)}{\int d^4 q \xi(q)}. \label{eq:qast}
\ee

Other lattice schemes, e.g.\ the Schr\"odinger functional scheme are also used in recent determinations
of the strong coupling constant or alternatively $\lqcd$ from the lattice. For comparison to results from 
experiment for example one usually converts the lattice results to the $\overline{MS}$ scheme.
The first important scheme for the strong coupling constant in lattice calculations
is the bare lattice coupling $\alpha_b \equiv g^2/(2\pi)$ with $g$ given in 
Eq.~(\ref{eq:lataction}). Since perturbative calculation using this scheme usually suffer from 
large higher-order corrections from tadpole diagrams, it is more common to use 
'boosted' perturbation theory with $\alpha_s  = \alpha_b/u_0^4$. In Refs.~\cite{bowler2001,becirevic2001}
some renormalization constants calculated in the boosted scheme are employed.

We now give some recent results for the values of $\lqcd$ and $\alpha_s$ determined using the lattice.
The ALPHA collaboration obtains~\cite{dellamorte2005}
\ba
\Lambda_{QCD, N_f = 0}^{\overline{MS}} &  = & 238(19) \mbox{ MeV}; \nonumber \\
\Lambda_{QCD, N_f = 2}^{\overline{MS}} &  = & 245(16)(16) \mbox{ MeV}.
\ea
and the UKQCD-QCDSF collaboration~\cite{gockeler2005} 
\ba
\Lambda_{QCD, N_f = 0}^{\overline{MS}} &  = & 259(1)(20) \mbox{ MeV}; \nonumber \\
\Lambda_{QCD, N_f = 2}^{\overline{MS}} &  = & 261(17)(26) \mbox{ MeV},
\ea
where the first error is statistical and the second systematical. The UKQCD-QCDSF result
corresponds to a value of $\alpha_s$ at $5$ flavors of
\be
\alpha_{s, N_f = 5}^{\overline{MS}}(M_Z) = 0.112(1)(2).
\ee
That result can be compared with the world average including experimental 
measurements and lattice results as input~\cite{bethke2004} of
\be
\alpha_{s, N_f = 5}^{\overline{MS}}(M_Z) = 0.1182(27),
\ee
and a recent lattice calculation using $2+1$ dynamical quarks~\cite{mason2005}:
\be
\alpha_{s, N_f = 5}^{\overline{MS}}(M_Z) = 0.1170(12).
\ee
Recent lattice determinations of $\alpha_s$ and $\lqcd$ are reviewed in~\cite{rakow2004}.

In many lattice calculations one determines the lattice spacing from length scales from 
the potential.  For example one uses $r_0$, related to the interquark force 
\cite{sommer1994} with 
\be
r_0^2 \left.\frac{dV}{dr}\right|_{r=r_0} = 1.65 \label{eq:r0}.
\ee
It can be calculated on the lattice with high precision;  determinations
using quenched Wilson gauge fields are given in~\cite{guagnelli1998,edwards1998}.
However for the string tension $\sigma$ and $r_0$ it is not quite simple to assign experimental 
values.  
Relying on potential models to extract the physical result from experimental quarkonium level 
splittings, one finds values of $r_0 = 0.49-0.5$ fm. 
One can also determine $r_0$ from the lattice. With a two-flavor lattice calculation using the 
nucleon mass as input, Ref.~\cite{gockeler2005} quotes  $r_0 \sim 0.467$ fm. With $N_f = 2+1$ and 
using $\Upsilon$ $2S-1S$ and $1P-1S$ mass splittings to set the scale, Ref.~\cite{aubin2004} finds values 
of $r_0$ in the range $0.462-0.467$ fm.
Unless noted otherwise, we use an experimental value of $0.5$ fm when we use $r_0$ to convert 
lattice results to physical units. 
For the string tension $\sigma$, usually experimental values of $\sqrt{\sigma} = 
427$ or 440 MeV are assumed. 
The relation between the potential parameters $r_0$ and $\sqrt{\sigma}$ is relatively
independent of the gauge action and discretization errors. A dependence on the flavor
number can be observed e.g.\ in the results for $r_0\sqrt{\sigma}$ (see 
 Table~\ref{tab:r0sig}). 
Ref.~\cite{bernard2001b} finds indications for string
breaking in two-flavor staggered QCD. 
A recent study of string breaking in two-flavor QCD with 
a sea quark mass around the strange mass and at zero temperature~\cite{bali2005} finds
clear indications for string breaking at a distance of $r \simeq 1.2$ fm, which 
corresponds to roughly 6 GeV$^{-1}$. 
One should note that in the CP-PACS result shown in
Fig.~\ref{fig:rot_inv} there is no indication of a flattening at this distance. 
Indications for string breaking in two-flavor QCD at finite
temperature, however  below the deconfinement phase transition have been
found by Ref.~\cite{bornyakov2004}. 

\begin{table}[htb]
\begin{center}
\begin{tabular}{|ccc|}
\hline
Ref. & $\beta$ & $r_0 \sqrt{\sigma}$ \\
\hline
\multicolumn{1}{|c}{} & 
\multicolumn{2}{c|}{Wilson, $N_f = 0$} \\
\protect\cite{edwards1998} & 5.7 & 1.160(15) \\
\protect\cite{edwards1998} & 6.0 & 1.175(5) \\
\protect\cite{ukqcd1992,luscher1994} & 6.5 & 1.199(25) \\
\hline
\multicolumn{1}{|c}{} & 
\multicolumn{2}{c|}{1-loop SI, $N_f = 0$} \\
\hline
\protect\cite{bernard2001} & & 1.160-164 \\
\hline
\multicolumn{1}{|c}{} & 
\multicolumn{2}{c|}{RG, $N_f = 0$} \\
\hline
\protect\cite{cppacs2002hadr} & 2.187 & 1.158(18) \\ 
\protect\cite{cppacs2002hadr} & 2.575 & 1.158(8) \\ 
\hline
\multicolumn{1}{|c}{} & 
\multicolumn{2}{c|}{RG, $N_f = 2$} \\
\hline
\protect\cite{cppacs2002hadr} & 1.80 & 1.143(24) \\ 
\protect\cite{cppacs2002hadr} & 1.95 & 1.142(16) \\ 
\protect\cite{cppacs2002hadr} & 2.10 & 1.129(5) \\ 
\protect\cite{cppacs2002hadr} & 2.20 & 1.144(5) \\ 
\hline
\multicolumn{1}{|c}{} & 
\multicolumn{2}{c|}{1-loop SI, $N_f = 2+1$} \\
\hline
\protect\cite{bernard2001} & & 1.114(4) \\
\hline
\end{tabular}
\end{center}
\vspace{-0.2cm}
\caption{$r_0 \sqrt{\sigma}$ for Wilson, renormalization group improved and 1-loop Symanzik 
improved (SI) gauge actions. The $N_f = 2$ values for the RG action are from the data sets with the
smallest sea quark masses.}
\label{tab:r0sig}
\end{table}
\subsection{Setting the Scale \label{sec:scale}}
Masses and decay constants coming out of a simulation are at first dimensionless numbers in units of 
the lattice spacing.
The value of the lattice spacing is  determined by calculating a suitable quantity $aM$ on the lattice and
adjusting the corresponding dimensionful quantity $M$ to its physical value. 
If the calculation is free of
systematic errors such as lattice spacing, finite volume and quenching effects, using any quantity should
give the same result. In practical calculations all of those errors can occur. Then, `suitable' means
that systematic errors of the quantity used to set the scale and the quantity
that is supposed to be calculated cancel as well as possible. 

Length scales from the static potential often used in lattice calculations are the Sommer parameter
$r_0$ and the string tension, discussed in the next section.
Hadronic quantities used to set the scale are $m_\rho$, the nucleon mass, the decay constants $f_\pi$ and 
$f_K$, and charmonium and bottomonium level splittings.
\begin{figure}[thb]
\begin{center}
\centerline{
\epsfysize=4.9cm \epsfbox{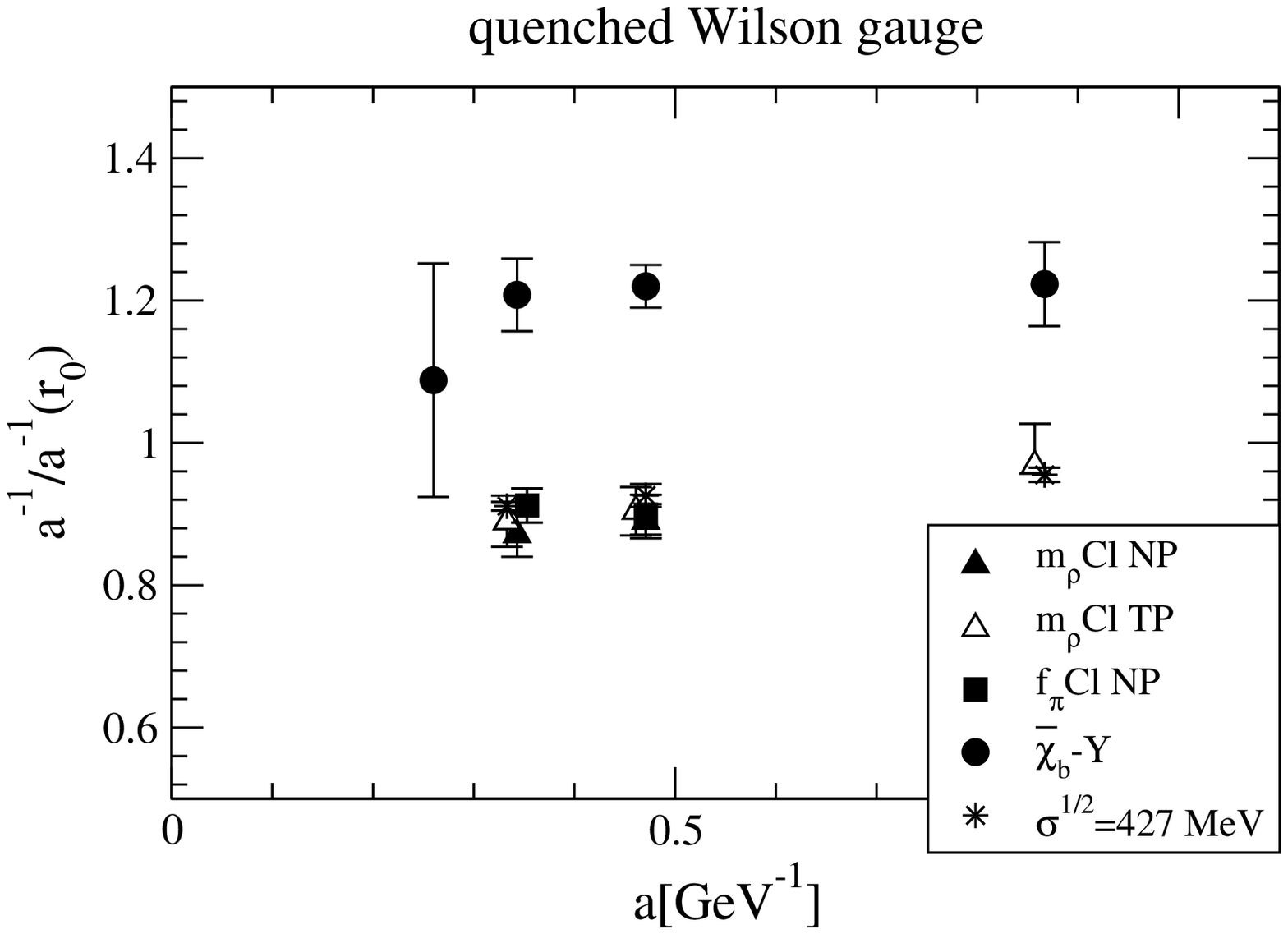}
\epsfysize=4.9cm \epsfbox{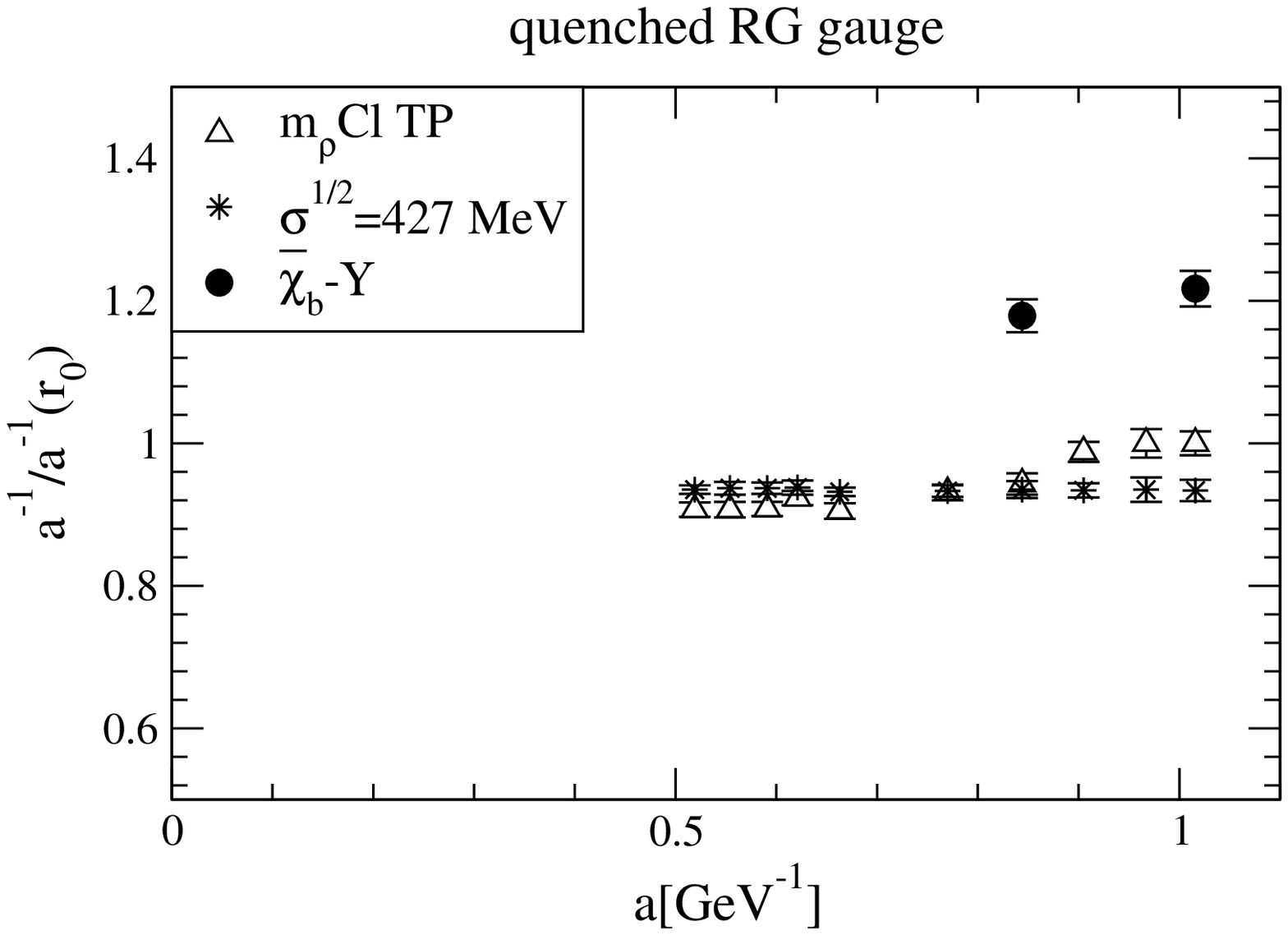}
}
\end{center}
\vspace{-0.5cm}
\caption{Ratio of inverse lattice spacings from various physical quantities to the inverse lattice
spacing from $r_0$ on quenched lattices. Results are from 
\protect\cite{guagnelli1998,cppacs2002hadr,jlqcd2003hadr,davies1997,manke2000,hein2000,bowler2001}.
The overbar in the legend denotes an average over spin-orientations.}
\label{fig:aquenched} 
\end{figure}

\begin{figure}[htb]
\vspace{0.1cm}
\begin{center}
\epsfysize=5cm \epsfbox{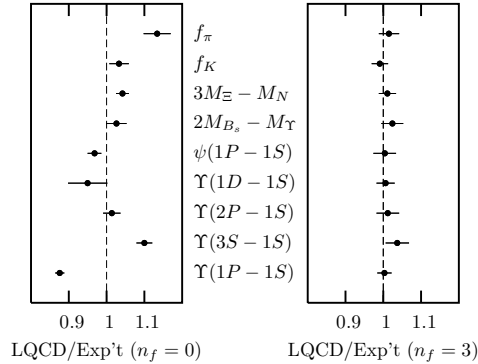}
\end{center}
\vspace{-0.5cm}
\caption{Comparison of lattice with experimental results from~\protect\cite{davies2004}, 
using zero (left) and $N_f = 
2_{\mathrm{light}} +1_{\mathrm{strange}}$ flavors of dynamical quarks(right). $a$ is determined
using the $\Upsilon^\prime-\Upsilon$ mass splitting.}
\label{fig:lat_milc} 
\end{figure}

As mentioned in the discussion of the light hadron spectrum, one expects a systematic difference
between the quenched and the real world. For example, in the quenched approximation the coupling runs 
differently, and the string between quark and antiquark does not break.

This is in particular of interest for heavy-light mesons since their physics contains several
rather different momentum scales.

In Fig.~\ref{fig:aquenched} we show examples for the discrepancy between lattice spacings 
determined from different physical quantities in the quenched approximation.  In the quenched 
approximation the $\rho$ does not decay, there is however a systematic uncertainty in the chiral 
extrapolation. The  $m_\rho$ 
values in these plots  result in some cases from simple linear fits in the quark mass, which 
might turn out as not an appropriate parameterization for very light quark masses (see 
Fig.~\ref{fig:mrho_quenched} in Section \ref{sec:llm}). 
To date many calculations with $b$ and light quarks do not reach
light quark masses much below the strange, which gives rise to a systematic uncertainty which 
has to be taken into account.

From HQET, typical momentum scales in $B$  mesons are expected to be of similar magnitude as 
in light-light mesons, namely $O(\lqcd)$. The use of $m_\rho$, $f_\pi$, and $f_K$ to set the
scale in heavy-light meson calculations is motivated by the expectation that quenching errors 
would cancel optimally. In calculations of the heavy-light meson spectrum one indeed finds 
good agreement with experiment, if $m_\rho$ is used to set the scale. 
Using $r_0$, one tends to find a smaller lattice spacing than e.g.\ using $m_\rho$. However,
in recent quenched calculations with chiral and chirally improved
fermions the ground state spectrum of light hadrons agrees
well with experiment already in the quenched approximation, depending on the procedures how the 
lattice spacing is determined and the physical quark masses are 
fixed~\cite{gattr2004,hasenfratz2004}. Their preferred
quantity to set the scale is $r_0$, but if their calculations are confirmed it would 
imply a rather good consistency of $a$ values determined from $r_0$, $m_N$ and $m_\rho$.

With two flavors of sea quarks, usually a better agreement between scales is found than 
demonstrated in Fig.~\ref{fig:aquenched}.
Using Wilson gauge fields and two flavors of $O(a)$ improved clover sea quarks, Ref.~\cite{jlqcd2003} 
quotes an agreement of scales  from $m_\rho$,  $f_K$ and $r_0 = 0.5$ fm. With renormalization 
group-improved glue and tadpole-improved clover quarks, lattice spacings from $m_\rho$ and
$r_0$ agree in the chiral limit of sea quarks except for very coarse lattice 
spacings~\cite{cppacs2002hadr}. Agreement of $a$ from $m_\rho$ and $r_0$ is also quoted 
in~\cite{yaoki2004}. 
However, in the two flavor calculations of~\cite{davies1997,collins1999} (Wilson gauge fields, 
staggered sea and clover valence quarks) and of~\cite{manke2000,cppacs2002hadr} 
(renormalization-group-improved gauge fields and tadpole-improved clover sea and valence quarks) at 
$a \sim 0.5$ GeV$^{-1}$, there is
a $\sim 20\%$ discrepancy between lattice spacings from $m_\rho$ and $\chi_b-\Upsilon$ mass splittings 
calculated using NRQCD. 

With two flavors of light and one flavor of strange dynamical quarks, using a 1-loop Symanzik $O(a^2)$ 
improved gauge action and a tree-level tadpole $O(a^2)$ improved staggered sea quark action,
Ref.~\cite{davies2004} finds agreement of a variety of physical quantities with experiment 
(see Fig.~\ref{fig:lat_milc}).  A recent calculation with $2+1$ flavors using $O(a)$-improved 
clover quarks finds that the continuum extrapolated strange meson masses agree with 
experiment~\cite{t.ishikawa2005}. 
\section{EFFECTIVE FIELD THEORIES \label{sec:eff}}
In many situations one can obtain quite a good description of physical
phenomena if only interactions determined by energy scales which are low
with respect to a cutoff scale are considered dynamical, and the interactions at higher energies are 
contained in the coupling constants of the low energy theory. 

If the low energy interactions are weak, they can be described using perturbative methods.
 An example for this is chiral perturbation theory which is an expansion
in terms of masses and momenta of the lightest particles  in
QCD, namely the pions, which can be thought of as would-be Goldstone
bosons of the spontaneously broken chiral symmetry of QCD. Certain constraints on 
the interactions can be derived from chiral symmetry. For example,
the pion self-interactions vanish in the limit of vanishing pion mass and momenta.

In hadrons with heavy quarks, mass splittings are often $O(\mbox{few}× \lqcd)$, i.e.\ much smaller 
than the mass of the heavy quark itself. Effective theories, Heavy Quark Effective Theory (HQET)
and Nonrelativistic QCD (NRQCD),  were developed to describe the 
interaction of soft gluons in heavy mesons with momentum scales much smaller than the meson mass. The 
heavy mass acts as a cutoff. The low-energy degrees of freedom are gluons
and light quarks interacting via the QCD gauge coupling as in QCD, and modified heavy quark fields.
The large momentum scale is given by the heavy quark mass, and low energy physics is again 
an expansion in terms of a small momentum scale. In NRQCD, this is a power of the 
heavy quark velocity in the meson, whereas in HQET it is the heavy quark recoil due to the interaction with 
soft gluons. Since the low-energy physics is non-perturbative, the expectation values of the corrections 
have to be calculated non-perturbatively, for example on the lattice. The effect of the high energy 
interactions modifies the coupling constants of these terms. 
Heavy meson or baryon chiral perturbation theory 
describes the dynamics of the low-energy interactions of a  heavy hadron with a pion in an expansion
around the limits of infinite hadron mass and small momenta, using similar modified fields for the heavy degrees
of freedom as HQET.

Constraints on the precision of a lattice calculation with heavy and light quarks result from the finite 
volume of the box and from the finite lattice spacing. 
On one hand is the particle mass the inverse of the correlation
length $\xi$. If discretization artifacts are supposed to be small, $\xi$ should  be large 
compared to the lattice spacing. 
As discussed in the previous section, discretization effects are in conventional lattice formalisms 
proportional to powers of the quark mass and momenta in lattice units. Using effective theories on the lattice it is 
 possible 
to re-formulate physics in terms of the energy splittings of hadrons, where energy scales
and the associated discretization errors are smaller.
On the other hand the volume has to be much larger that 
the wave functions of the particles to be described, or their Compton wavelength, such that finite
volume effects are small. 

In the following table we show schematically
the various relevant scales  in a lattice simulation of hadrons with a $b$ and a $u$ or $d$ quark together 
with the boundaries achievable on present lattices with dynamical quarks:

\vspace{0.5cm}

\begin{center}
\begin{tabular}{|c|ccc|}
\hline
Nature & &    $m_\pi = 140$ MeV  &    \\
       & &    $m_b \sim 5$ GeV   &      \\
       & &    $\Delta E(B^{\ast\ast}-B) \sim 400$ MeV  &                \\
\hline
Ideal & 
$
\begin{array}{c}
             \mbox{finite size } \\
(L^{-1}) \\
\end{array} 
\ll$ &
$
\begin{array}{c}
\mbox{physical quantities} \\
(\mbox{masses} \propto 1/\xi \\
\mbox{and energy splittings}) \\
\end{array}
$
& $ \ll
\begin{array}{c}
\mbox{cutoff }  \\
(a^{-1}) \\
\end{array}
$ \\
\hline
Lattice &
            {$L^{-1}\leq 100$ MeV} &     10 MeV $\leq m_q  \leq 0.5-1$ GeV   
& {$a^{-1}=2-4$ GeV}  \\
Reality  &            ($L=2-3$ fm) &  &                  \\
\hline
\end{tabular}
\end{center}

\vspace{0.5cm}

\subsection{Chiral Perturbation Theory $(\chi PT)$}
The chiral symmetry of massless QCD is assumed to be
spontaneously broken. As a  consequence one would expect massless Goldstone
bosons in the spectrum. One can identify the physical pions with the Goldstone bosons if 
small quark mass terms are included in the QCD Lagrangian  resulting in a small explicit breaking of 
chiral symmetry and a finite but small pion mass $m_\pi$. 
Typical momenta of low-energy interactions of pions will be $O(m_\pi)$, which is the momentum scale
that dominates also the integrals over the pion loops.

If hadronic interactions at low energies are to be understood within this scenario,
it is of interest to  be able to describe the interactions of the
Goldstone bosons of spontaneously broken chiral symmetry, within a field theoretical 
framework. 

The Goldstone boson couples to the symmetry current with the pion-to-vacuum matrix element given by
\be
\langle 0 | A_\mu(x) | \pi(p) \rangle = -ip_\mu f_\pi e^{-ip\cdot x},
\ee
where $A_\mu$ is the axial vector current. A relativistic normalization is assumed.
The zeroth component of an axial 
vector current is an interpolating operator for pseudoscalar states in general.
For Goldstone bosons the vacuum matrix element of $\partial_\mu A^\mu$
vanishes, since the Goldstone boson mass 
is zero. The matrix element of $\partial_\mu A^\mu$ between two hadron 
states is proportional to the pion momentum squared, and therefore also  has a vanishing zero-momentum limit 
if the symmetry is spontaneously broken. In the case of explicit symmetry breaking it is 
proportional to the mass-squared of the on-shell pion. 

The idea is therefore to construct  a Lagrangian with the required symmetries
which  can be used  to describe the low-energy interactions in terms of hadron fields and
hadron currents. One can formulate the low-energy interactions in terms of an 
expansion in powers of the pion mass and momenta $p =  O(m_\pi)$, the so-called $p$ 
expansion of chiral perturbation theory~\cite{gasser1984,gasser1985,weinberg1996}. Beyond tree level, 
quantum corrections generate new effective interaction terms with new coupling constants.
To make a perturbative calculation in the framework of the $p$ expansion meaningful, it should be
possible to determine how each of the perturbative loop diagrams contributes to the various powers of the
 $p$ expansion. In the systematic expansion of Ref.~\cite{weinberg1979},
$n$-loop corrections to amplitudes  are  $O(p^{2n})$ with respect to the tree-level amplitude.
The lower order terms which are generated by each diagram can be mapped onto redefinitions of couplings of
already existing terms in the Lagrangian, while higher order terms predict corrections to the momentum 
dependence of the amplitude under consideration.

To give meaning to the new interactions and to be able to assign the correct order to them a careful 
choice of the renormalization procedure is necessary.

The resulting formalism can be applied to low-energy hadronic interactions, for example pion-pion and
pion-nucleon scattering and many others which are of interest for determinations of
parameters of the Standard Model, e.g.\ weak interaction parameters.  It also gives predictions for the expansion of
hadron masses around the zero quark mass limit, which can be  used in the analysis of lattice
calculations to extrapolate  hadron masses from simulations at larger quark 
mass to the physical light ($u,d$) quark mass.

The quark condensate occurs within the chiral perturbation theory formalism
as a parameter which incorporates the dynamics of chiral symmetry breaking 
and can be determined e.g.\ using $\chi PT$ and comparison to experiment or lattice QCD.

Beside the chiral $SU(2)$ invariance, the quark Dirac Lagrangian is symmetric under 
transformations of global  $SU(2)$ vector transformations.
The left-and right-handed components of the field $q$ therefore have an
invariance under the independent transformations 
\be
q_L  \to  \exp[i\epsilon_L{\scriptsize {\frac{1}{2}}}(1-\gamma_5)t^a] q_L, \;
q_R  \to  \exp[i\epsilon_R{\scriptsize {\frac{1}{2}}}(1+\gamma_5)t^a] q_R,
\ee
and the corresponding transformations for $\overline{q}_L$ and $\overline{q}_R$.
If chiral symmetry is explicitly broken,  the
Lagrangian is only invariant under simultaneous $(\epsilon_L = \epsilon_R)$
transformations of left-and right-handed quarks. 

We begin with considering an effective Lagrangian for the pions coupling to $u$ and $d$
quarks which should be chirally symmetric up to quark mass terms which act as external 
sources for the pions. The pions are massless except for non-vanishing quark masses which
provide the couplings to the chiral condensate.
It is common to use a nonlinear $\sigma$ model to represent the pions: 
\be
U = \exp\left(\frac{i}{f_\pi} \vec{\tau}\vec{\pi}\right) \in SU(2), \label{eq:pion}
\ee
where the $\pi^i$ are the pion fields, and the $\tau^i$ are Pauli matrices.
$f_\pi$ is the pion decay constant. $U$
transforms according to the $(2,2)$ representation of $SU(2)_L \times SU(2)_R$:
\be
U(x) \to L U(x) R^\dagger,
\ee
where $R \in SU(2)_R$ and $L \in SU(2)_L$. 
The expansion of the pion Lagrangian begins at $O(p^2)$, i.e.\ quadratic in the momenta and
linear  in the quark mass:
\be
{\cal L}_\pi^{(2)} = \frac{1}{4}f_\pi^2\Tr\left[\partial_\mu U^\dagger
\partial^\mu U +\chi^\dagger U + \chi U^\dagger)\right].
\ee
$\chi = 2B_0 {\cal M}$, where ${\cal M}$ is the quark mass matrix. If
$u$ and $d$ quarks have masses $m_u$ and $m_d$, the mass matrix is:
\be
{\cal M} = \left(
\begin{array}{cc}
m_u &  0 \\
0   & m_d \\
\end{array}
\right),
\ee
$B_0$ is proportional to the chiral condensate: $B_0 = -\langle \overline{q} q \rangle/f_\pi^2$.
Keeping only quadratic terms in the pion fields one obtains a free Klein Gordon equation for
canonically normalized fields $\pi^i$ with mass $\mps^2 = B_0\Tr{\cal M}/2$, i.e.
\be
\mps^2 = 2 \frac{|\langle \overline{q}q\rangle |}{f_\pi^2}m_q, \label{eq:gmor}
\ee
the Gell-Mann-Oakes-Renner (GMOR) relation.
The relation has been verified by comparing with experimental results to hold to a high 
accuracy~\cite{leutwyler2004}. The result is taken as confirmation that the chiral
condensate is the leading order parameter of spontaneous chiral symmetry breaking.
In lattice calculations,  $\mps^2$ is used as a measure of the quark mass.
For heavier quark masses,  the GMOR relation is modified as can be seen from
lattice results~\cite{meinulf2004a,aubin2004a}.

At the next order, the pion Lagrangian contains many additional terms~\cite{gasser1984} with 
couplings which have to be determined from comparison with experiment or lattice QCD predictions.
If the couplings to external fields are not taken into account, the $O(p^4)$ terms 
Lagrangian are~\cite{leutwyler2002}:
\ba
{\cal L}_\pi^{(4)} &=& \frac{1}{4}l_1\left(\Tr[\partial_\mu U^\dagger
\partial^\mu U \right])^2 + \frac{1}{4}l_2\Tr[\partial_\mu U^\dagger\partial_\nu U]
\Tr[\partial^\mu U^\dagger\partial^\nu U] \nonumber \\
& +& \frac{1}{4}(l_3+l_4)\left(\Tr[\chi^\dagger U + U \chi^\dagger]\right)^2
 + \frac{1}{4}l_4\Tr[\partial_\mu U^\dagger \partial_\mu U]\Tr[\chi^\dagger U + \chi 
U^\dagger] \nonumber \\
& +&  \frac{l_7}{4}\left[\left(\Tr[\chi^\dagger U - U \chi^\dagger]\right)^2 
+\left(\Tr[\chi^\dagger U - U \chi^\dagger]\right)^2
-2\Tr[\chi^\dagger U\chi^\dagger U + U^\dagger \chi U^\dagger \chi]\right]
\ea

Extension of the formalism  to quenched QCD, i.e.\ without dynamical sea quarks, 
is discussed e.g.\ in~\cite{golterman1997}) and references therein, 
and extension to partially quenched QCD with different valence and sea quark 
masses is discussed in~\cite{sharpe1997,golterman1998}.

If the pions are light compared to the inverse lattice extent, but the lattice size 
is not too small,
one can study fluctuations around the zero modes within the
so-called $\epsilon$ expansion of $\chi PT$~\cite{epsilon1987}. 
\subsubsection{$\chi PT$ for relativistic baryons}
The nucleon (and other hadrons such as the $\rho$) remain massive in the chiral limit, and
their mass cannot be directly used as a small expansion parameter.
However typical momentum scales for the interaction with pions can still be of the order of 
the pion mass. Here, we employ $\chi PT$ to predict self-energy corrections to the nucleon mass
at finite pion mass.

The Lagrangian description of low-energy nucleon-pion interactions in chiral perturbation theory
 at lowest order $(O(p^1))$ is
\be
{\cal L}^{(1)} =  \overline{\Psi}\left(i\gamma_\mu D^\mu - m_0\right)\Psi + 
\frac{1}{2} g_A\overline{\Psi} \gamma_\mu \gamma_5 u^\mu \Psi,
\ee
where $\Psi$ is the Dirac spinor of the nucleon.

$g_A$ is the nucleon axial charge, related to the nucleon-pion coupling $G_{\pi N}$ and the pion 
decay constant through the Goldberger-Treiman relation
\be
G_{\pi N} = \frac{2m_N g_A}{f_\pi}. \label{eq:gb}
\ee
Although the nucleon is massive, the Lagrangian is
invariant under the chiral transformations
\ba
\Psi(x) &\to & V(x)\Psi(x), \nonumber \\
\overline{\Psi}(x) &\to & \overline{\Psi}(x)V^\dagger(x). \nonumber \\
\ea
The transformation matrix $V$ is a product of $L, R$ and the pion fields $U(x)$:
\be
V(x) = R\left[R^\dagger LU(x)\right]^{1/2} \left[U(x)^{1/2}\right]^\dagger
     = L\left[L^\dagger RU^\dagger(x)\right]^{1/2} U(x)^{1/2}.
\ee
with $u^2 = U$, the covariant derivative 
\be
D_\mu = \partial_\mu + \frac{1}{2}[u^\dagger,\partial_\mu u],
\ee 
and the pion current $u_\mu = iu^\dagger \partial_\mu U u^\dagger$. 
$m_0$ the nucleon mass in the chiral limit. 

The $O(p^2)$ Lagrangian is given by
\ba
{\cal L}^{(2)} &=&  c_1\mathrm{Tr}(\chi_+)\overline{\Psi}\Psi
+ \frac{c_2}{4m_0^2}\mathrm{Tr}( u_\mu u_\nu)\left( \overline{\Psi}
D^\mu D^\nu \Psi + h.c.\right) \nonumber \\
&+& \frac{c_3}{2}\mathrm{Tr}( u_\mu u^\mu ) \overline{\Psi}\Psi
-\frac{c_4}{4}\overline{\Psi} \gamma^\mu \gamma^\nu [u_\mu,u_\nu]\Psi,
\label{eq:l2}
\ea
with $\chi_+ = u^\dagger \chi u^\dagger + u \chi^\dagger u$.

In $\chi PT$, the quark mass dependence of the mass and the finite size effect are determined by the 
mass renormalization of the nucleon. The mass dependence
in infinite volume has been used to calculate the chiral extrapolation formula, and the finite size
effect is given by the difference of the self energy in finite and infinite volume.
The renormalized mass is determined by the pole of the nucleon propagator calculated to a certain 
order in the $p$ expansion with the external momentum set on-shell,
\be
 m = \Sigma(q\!\!\slash=m), \label{eq:onsh}
\ee
where $q$ is the momentum of the external nucleon line, and $p = O(q\!\!\slash -m)$. 
In practical calculations,
one will substitute for $m$ its expansion in powers of $p$ to the necessary order 
in Eq.~(\ref{eq:onsh}).
Extrapolation of lattice nucleon masses to the chiral limit and the calculation of 
the effect of the lattice size on nucleon masses in QCD with $N_f = 2$
is discussed in Section~\ref{sec:FS}.

In quenched QCD, chiral perturbation theory is modified because quark-antiquark pairs cannot
be dynamically generated. Additional terms which are non-analytic (logarithmic) in the quark mass
arise in quenched chiral perturbation theory. The modification for the nucleon mass
is given after Eq.~(\ref{HBmass}). Modifications also occur in partially quenched QCD, i.e.\
if the sea and valence quark masses are not equal. 
Partially quenched chiral perturbation theory is reviewed in~\cite{sharpe2000}. 
\subsubsection{Heavy baryon $\chi PT$}
Since the nucleon mass is large, of the order of the cutoff scale of chiral perturbation 
theory, it is non-trivial to retain the power counting of the effective theory in
calculations beyond tree-level within the relativistic formalism~\cite{relativistic}. 
In Heavy Baryon Chiral Perturbation Theory~\cite{jenkins1991}  ($HB\chi PT$), 
an infinitely heavy field is used for the baryon, and  finite
mass corrections can be included within 
an expansion in inverse powers of $M_B$, where $M_B$ is the baryon mass. 

If the baryon mass is considered to be very large,
the interactions introduce only small fluctuations $k$ of the baryon momentum, and one can write:
\be
P = M_B v + k,
\ee 
Corrections can be included in a $1/M_B$ expansion.
The baryon Dirac fields are projected onto fields $B_v(x)$ which are projected
to constant velocity:
\be
B_v(x) = \exp(iM_{B}v\cdot x)P_+(v) B(x).
\ee
$v$ is the velocity of the heavy nucleon. In the rest frame, the temporal component is one and
the spatial components are zero: $v = (1,0,0,0)$. 
$P_+(v) = \frac{1}{2}(1+\slash{\!\!v})$ projects onto  the forward moving part of the nucleon field:
\ba
P_+ B_v & = & B_v; \\
v\!\!\slash B_v & = & B_v.
\ea
Accordingly $P_-(v) = \frac{1}{2}(1-\slash{\!\!v})$ projects onto
the backwards moving part of the nucleon field. In the baryon rest frame,
the operators $P_{±}$ project onto the upper two or lower two components 
of the Dirac spinor.

At $O(p^1)$ the nucleon Lagrangian is given by:
\begin{eqnarray}
{\cal L}^{(1)}_{N_v} &=& \bar{N}_v\left[iv\cdot D + g_A S\cdot u\right]N_v.
\end{eqnarray}
with the definitions for the axial pion current $u_\mu$ and the covariant derivative
$D_\mu = \partial_\mu + \Gamma_\mu$ as before.
The axial vector coupling of the relativistic Lagrangian is substituted by the coupling to
the Pauli-Lubanski spin vector:
\be
 S_\mu = \frac{i}{2}\gamma_5 \sigma_{\mu\nu}v^\nu.
\ee

The tree-level Lagrangian at $O(p^2)$, 
\begin{eqnarray}
{\cal L}_{N_v}^{(2)}&=&\bar{N}\left\{c_1 Tr(\chi_+)+\ldots\right\}N_v
\end{eqnarray}
includes the low-energy constant, $c_1$, which has to be determined from experiment.
The leading pion mass dependence is quadratic.
In principle, $c_1$ can also be fixed from the lattice, however we will see later that
low order $\chi PT$ does not fit lattice data at the present large quark masses well.

A similar formalism exists for vector mesons~\cite{jenkins1995}.
\subsubsection{Lattice heavy baryon $\chi PT$}
There are various ways of regularizing $\chi PT$, e.g.\ with dimensional regularization or
a cutoff scheme. One can also construct a lattice version of $\chi PT$ 
for heavy baryons~\cite{ouimet2001} (a modification for $SU(2)$ and the 
representation~(\ref{eq:pion}) for the pion fields is given in~\cite{alikhan2002}).
The inverse of the free pion propagator is then:
\be
D^{-1}(q) = \left[\mps^2 + \frac{4}{a^2}\sum_\mu\sin^2\frac{aq_\mu}{2}
\right]. \label{eq:pi}
\ee
The lattice baryon propagator in the $HB \chi PT$ is 
\ba
S_B^{-1}(q) &=& \left[\frac{1}{a}(1- e^{-iaq_4})\right],
\ea

Lattice regularization for chiral perturbation theory for pions is discussed 
in~\cite{shushpanov1999}.
Use of a lattice formalism to calculate pion finite volume effects 
is described in~\cite{borasoy2005}.

At finite lattice spacing, the quark Lagrangian has a different symmetry structure from 
continuum QCD. Unless Wilson quarks are  $O(a)$ improved, 
chiral symmetry is also broken by terms $O(a)$ at finite lattice spacing, and these terms have 
to be included in the chiral Lagrangian. The leading terms have been derived for the unquenched
theory in
\cite{sharpe1998,rupak2002,aoki2003a,baer2004a}. For a review on chiral perturbation 
theory for
Wilson-like quarks on the lattice see e.g.\ Ref.~\cite{baer2004}. The chiral symmetry group of
staggered fermions is also reduced due to $O(a^2)$ corrections at finite lattice spacing. 
A chiral Lagrangian for staggered fermions was constructed in~\cite{lee1999}.

%
\subsection{Heavy Quark Effective Theories}
In effective theories for heavy quarks, one makes use of the fact that although the $b$ quark mass 
for example is around 5 GeV, the mass splittings of heavy hadrons are only of 
$O(\mbox{few 100 MeV})$, and that the dynamics of a very
heavy quark in a hadron can be described as a small correction to the infinite
quark mass limit. Using an effective
theory where momentum scales $> M$ are integrated out, a $1/M$ expansion for various 
quantities can be derived.

\begin{figure}[thb]
\begin{center}
\centerline{
\epsfysize=6cm \epsfbox{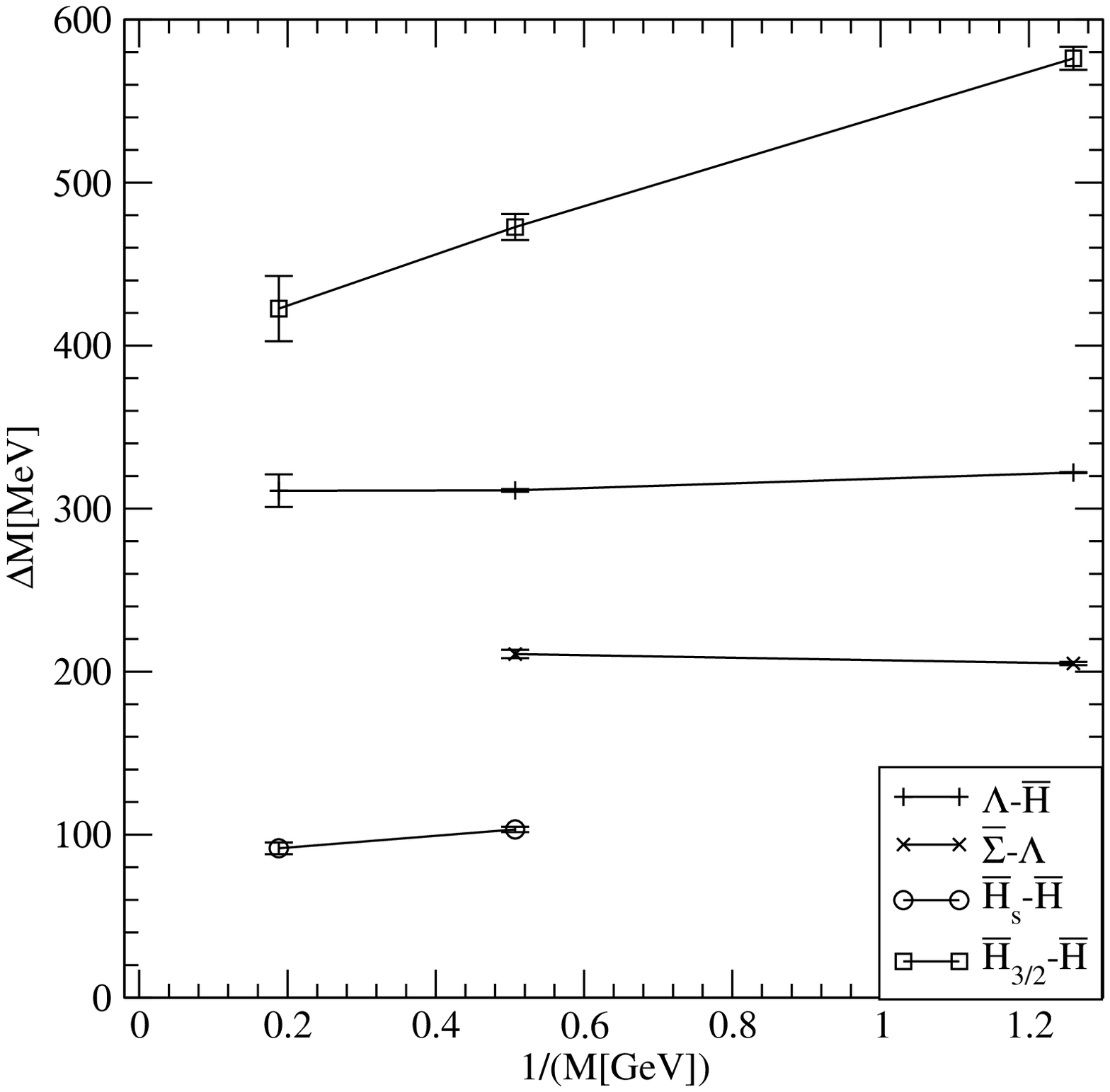}
\epsfysize=5.9cm \epsfbox{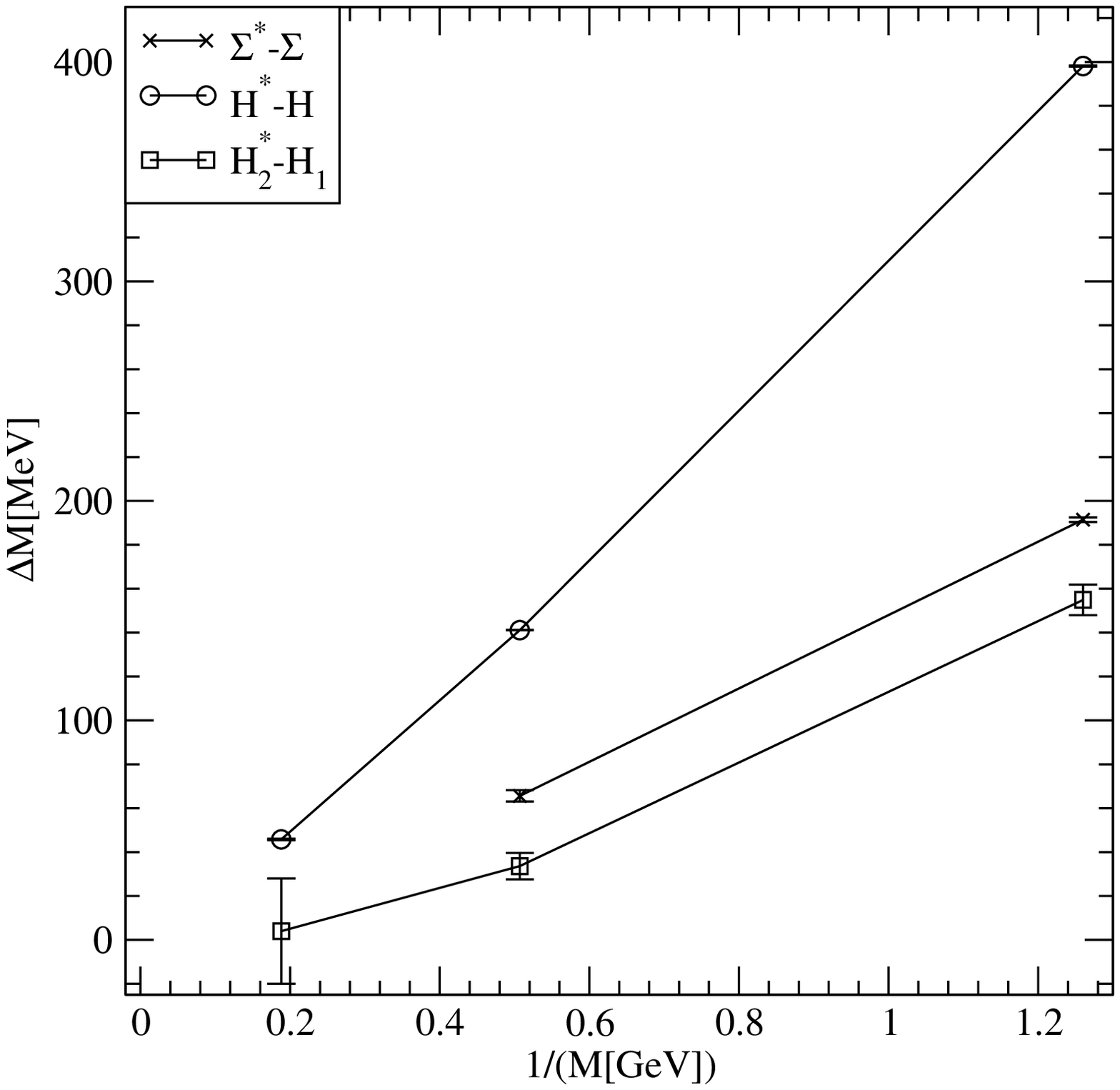}
}
\end{center}
\caption{
Mass splittings of heavy-light hadrons. On the left, splittings of states averaged over heavy quark
spin orientations. On the right, hyperfine splittings between states of different heavy quark spin
orientation.
}
\label{fig:spin_indep} 
\end{figure}

In Fig.~\ref{fig:spin_indep} on the left we show mass splittings of heavy-light hadrons
averaged over spin-orientations.
$H$ stands for a generic heavy-light meson where the
heavy quark is one of  $s,c,b$, and $H_s$ for a meson with a $c$ or $b$ quark and a strange quark.
Overbars denote averages over spin orientations. $\overline{H}_{3/2}$ denotes the $P$ wave meson doublet with light 
quark angular momentum $3/2$ averaged over the heavy quark spin orientations. For an explanation of the
$P$ wave mesons and of the baryons see Eqs.~(\ref{eq:pdoublets}) and (\ref{eq:barydoublets}).
They are largely independent of the heavy quark mass and have a finite limit in the limit of infinite
mass. 
Experimental results on spin splittings of heavy-light hadrons are shown in Fig.~\ref{fig:spin_indep}
on the right. Circles show ground state $S$ wave
hyperfine splittings, squares $j_l = 3/2$ $P$ wave hyperfine splittings and 
crosses ground state $S$ wave $j_l = 1$ baryon hyperfine splittings, where $j_l$ is
the light quark angular momentum in the quark model. For an explanation of the $P$ wave mesons and
of the baryon states see Eqs.~(\ref{eq:pdoublets}) and (\ref{eq:barydoublets}).

One observes that the spin splittings behave very roughly proportional to the inverse heavy meson mass
and appear to vanish in the limit of infinite mass.
\subsubsection{HQET}
A quark with infinite mass $M_Q$ does not experience recoil from interactions with gluons and 
moves at constant velocity with momentum $M_Qv$, 
i.e.\ it is static in the rest frame of the hadron.  The spinor components of the heavy quark decouple,
which gives rise to a spin symmetry, and the quark can be regarded  as a simple color source. 
Energy splittings are
determined by the interactions of gluons with momenta $O(\lqcd)$. Since in $B$ and $D$ mesons
the energy splittings are much smaller than the heavy quark mass, one expects to be able to
describe them to a certain degree of accuracy within an effective theory where energy scales above the 
heavy quark mass are integrated out. 

At finite heavy quark mass, exchange of gluons with momentum $O(\lqcd)$ 
exchange leads to a recoil momentum $k$ of the heavy quark with $k \sim O(\lqcd)$ which is small 
compared to the momentum of an on-shell heavy quark which is 
non-relativistically given by $M_Qv$. Recoil and spin flip of the heavy quark give rise to 
corrections $O(\lqcd)/M_Q$ to the expectation values of the physical quantities, which are
described by the interactions of higher-dimensional operators consisting of quark and gluon
fields. The coefficients of the operators have to be determined by matching to full QCD.
Since energy scales larger than the heavy quark mass
are $O(\mbox{several GeV})$, the physics they correspond to, which is 
integrated out in the effective theory, is to a large extent perturbative. 

For a quark of mass $M_Q$ moving at 
velocity $v$, one can introduce a velocity projected heavy quark field $Q_v$
which is  related to the heavy quark Dirac field $Q$ through the unitary 
transformation
\be
Q = e^{-iM_Qvx}Q_v.
\ee
We calculate the tree-level Lagrangians for $Q_v$ at zeroth and first 
order in $1/M_Q$ according to the description of \cite{neubert1994}.
Undoing the projection, the equation of motion for $Q_v$ is 
\be
\left[iD\!\!\!\!\slash + M_Q(v\!\!\!\slash-1)\right]Q_v = 0.
\label{eq:1}
\ee
Using the projection operators introduced in the $HB\chi PT$ section, 
one finds the forward $(+)$ or backwards $(-)$ moving fields
\be
Q_v^\pm = \frac{1 \pm v\!\!\!\slash}{2}Q_v.
\ee
Multiplying Eq.~(\ref{eq:1}) with $P^+$ and using
$a\!\!\!\slash b\!\!\!\slash = 2a\cdot b-b\!\!\!\slash a\!\!\!\slash$
one finds
\be
i v\cdot DQ_v^+  = -P^+ iD\!\!\!\!\slash Q_v^-.
\ee
An analogous equation can be obtained by multiplying Eq.~(\ref{eq:1}) with $P^-$:
\be
iv\cdot DQ_v^-+ 2M_QQ_v^- = P^- iD\!\!\!\!\slash Q_v^+.
\label{eq:2}
\ee
If one assigns $O(M_Q^0)$  to $Q_v^+$  and $O(1/M_Q)$ to $Q_v^-$, one finds
for the equation of motion at zeroth order in $1/M_Q$:
\ba
iv\cdot D Q_v &=& 0.
\ea
For the zeroth order Lagrangian in the $1/M_Q$ expansion one then obtains
\ba
{\cal L}^{(0)} &=& \bar{Q}_v iv\cdot D Q_v.
\ea
From this one can read off the free propagator for the $Q_v$ fields,
\be
\tilde{G}^{(2)}(k) = \frac{i}{v\cdot k+i\epsilon},
\ee
and the tree-level vertex
\be
igt^a v^\mu.
\ee
In the following we derive the $1/M_Q$ corrections to the Lagrangian at tree level.
Re-ordering Eq.~(\ref{eq:2}), one finds
\ba
Q_v^- = \frac{1}{2M_Q} P^- iD\!\!\!\!\slash Q_v^+ &-& i\frac {v\cdot D}{2M_Q}
Q_v^-
\ea
The second term is $O(1/M_Q^2)$ relative to the first term.
Substituting this into Eq.~(\ref{eq:1}) one obtains
\be
iv\cdot DQ_v^+ = -P^+ iD\!\!\!\!\slash \frac{1}{2M_Q}P^-iD\!\!\!\!\slash Q_v^+.
\ee
This can be simplified further. Using
\be
\frac{1+ v\!\!\!\slash}{2}D\!\!\!\!\slash \frac{1+ v\!\!\!\slash}{2}
D\!\!\!\!\slash \frac{1+ v\!\!\!\slash}{2}
= \frac{1+ v\!\!\!\slash}{2}\left[D\!\!\!\!\slash^2-2(D\cdot v)^2
\right]\frac{1+ v\!\!\!\slash}{2}. 
\ee
and  $G_{\mu\nu} = \frac{i}{g}[D_\mu,D_\nu]$, one finds
\be
D\!\!\!\!\slash^2 = \frac{1}{2} g\sigma^{\mu\nu} G_{\mu\nu}+D^2,
\ee
where $\sigma^{\mu\nu} = \frac{i}{2}[\gamma^\mu,\gamma^\nu]$.
The $1/M$ Lagrangian is therefore given by 
\be
{\cal L}^{(1)} = \frac{1}{2M_Q}\bar{Q}_v\left[D^2 - (v\cdot D)^2
+ \frac{1}{2}g\sigma^{\mu\nu} G_{\mu\nu}\right]Q_v. \label{eq:106}
\ee
In the $B$ meson rest frame the spatial components of the velocity are zero,
i.e.\ $ D^2 - (v\cdot D)^2 = -\vec{D}^2$.
Then the two contributions to Eq.~(\ref{eq:106}) are recognizable as the 
\begin{itemize}
\item kinetic energy operator
\be
 O_{\mathrm{kin}} \equiv -\frac{1}{2M_Q} \bar{Q}_v\vec{D}^2 Q_v
\ee
\item 
spin-magnetic energy operator
\be 
O_{\mathrm{mag}} = -\frac{g}{M_Q}\bar{Q}_v\vec{S}\cdot\vec{B}Q_v,
\ee
\end{itemize}
with the spin operator $S^k =  {\scriptsize\frac{1}{2}} \gamma_5
\gamma^0\gamma^k$ and the magnetic field $B^k = - {\scriptsize\frac{1}{2}}
\epsilon^{kji}F^{ji}$.
In the Heavy Quark Effective Theory (HQET), the mass of a heavy-light 
hadron $H$ therefore consists of the contributions
\be
M_H = M_Q + \overline{\Lambda} - \frac{1}{2M_Q}\left[ \frac{\langle H|\overline{Q} 
\vec{D}^2 Q|H\rangle}{2M_H} + \frac{\langle H|\overline{Q} 
\vec{\sigma}\cdot\vec{B}Q|H\rangle }{2M_H}\right] + O(1/M_Q^2),
\label{eq:HQET}
\ee
where $\overline{\Lambda}$ the binding energy
of the meson for $M_Q \to \infty$, and the other two terms are the
expectation values of the heavy quark kinetic energy and the spin-colormagnetic
interaction energy respectively.

The HQET predictions for the mass dependence of decay form factors in the heavy 
quark effective theory 
are useful for the analysis of experimental and lattice results of form factors,
e.g. in fits of the mass dependence.
We summarize the discussion of the infinite mass limit of $f_B$~\cite{neubert1994}.
In QCD, the decay constant of a meson $H$ with a heavy and a light quark can be defined through  
$\langle 0|A_\mu(0)|H(p)\rangle = -ip_\mu f_H$~(Eq.~(\ref{eq:axial})). We further assume that the 
meson states are normalized relativistically, according to
\be
\langle H(p^\prime)|H(p)\rangle = 2 p^0(2\pi)^3\delta^{(3)}(\vec{p}-\vec{p}^\prime).
\ee
The decay matrix element at infinite mass in the HQET should be similarly related to the heavy quark
velocity
\be
\langle 0|\tilde{A}_\mu(0)|H(v)\rangle = -i\tilde{f}_H v_\mu,
\ee
where we have denoted the HQET quantities with $\tilde{\hphantom{O}}$.
Using a non-relativistic normalization for the HQET fields:
\be
\langle \tilde{H}(v^\prime)|\tilde{H}(v)\rangle = 
2 v^0(2\pi)^3\delta^{(3)}(\vec{p}-\vec{p}^\prime)
\ee
one obtains  the heavy quark limit of $f_H$:
\be
\tilde{f}_H = f_H\sqrt{M_H},
\ee
i.e.\ $f_H\sqrt{M_H}$ is $M_Q$-independent up to $1/M_Q$ corrections, and the tree level
mass dependence of $f_H\sqrt{M_H}$ is expected to go as
\be
f_H\sqrt{M_H} = A + \frac{B}{M_Q} + \frac{C}{M^2_Q} + (O(M^{-3}_Q)).
\label{eq:f1overM}
\ee
In lattice calculations, when matrix elements 
are studied at a range of heavy quark masses, Eq.~(\ref{eq:f1overM}) or a 
renormalized version (see Eq.~(\ref{eq:logs})) is used to fit the results as
a function of heavy quark or meson masses. 

Beyond tree level, renormalization effects have to be taken into account. We
attempt to describe some general features, following the discussion 
in~\cite{politzer1988}. Since we will discuss lattice results on $f_B$,
we are interested in the matching of the axial vector 
current.
In QCD with relativistic quarks, the axial current is protected by chiral symmetry and is therefore not renormalized. To calculate the QCD matrix element from
the matrix elements in the effective theory, one requires that the matrix 
elements in full QCD and the effective  theory match at an energy scale $\mu$.
This statement can be expressed by:
\be
A_\mu^{QCD} = C(\mu) A_\mu(\mu)^{HQET},
\ee
where $C$ is a Wilson coefficient, at leading order in HQET. In general
there can be several operators contributing to the right hand side which 
mix under renormalization,
and we will discuss this in the context of matching the axial vector current in lattice NRQCD to continuum 
QCD.

A typical momentum scale for the gluons exchanged in the interaction is $\mu = \lqcd$. The scale dependence of the 
coefficient $C$ is determined by a renormalization group equation:
\be
\left(\mu\frac{d}{d\mu} - \gamma_A\right) C(\mu) = 0,
\ee
where $\gamma_A$ is the anomalous dimension of the axial vector current.
The cutoff for the effective theory being the heavy quark mass implies the initial condition
\be
C(\mu=M_Q) = 1.
\ee
The Wilson coefficients $C$ contain logarithms of $\mu/m_Q$ which are large for typical choices of
$\mu$.
At leading log approximation the result for the scaling behavior of the coefficients is
\be
C(\mu) = \left(\frac{\alpha_s(M_Q)}{\alpha_s(\mu)}\right)^{-2/b_0}. \label{eq:c}
\ee
The large logarithms can be resummed if the coefficients are multiplied by the inverse of the
factor~(\ref{eq:c}).
The axial current matrix elements $f_1\sqrt{M_1}$ and $f_2\sqrt{M_2}$ of two heavy-light mesons 
with different heavy quarks of masses $m_1$ and $m_2$  are then related in HQET by
\be
\frac{f_1\sqrt{M_1}}{f_2\sqrt{M_2}} = \left(\frac{\alpha_s(M_1)}{\alpha_s(M_2)}\right)^{-2/b_0},
\label{eq:logs}
\ee
i.e.\ Eq.~(\ref{eq:f1overM}) is modified by logarithmic corrections.

Beyond tree-level, one expects a similar renormalization of the  operators in the HQET Lagrangian. 
For the spin-magnetic operator, 
\be 
O_{\mathrm{mag}} \to -c_B(\mu) \frac{g}{M_Q}\bar{Q}_v\vec{S}\cdot\vec{B}Q_v .
\ee
Scale invariance of physical mass splittings gives rise to a similar
renormalization group equation as for the decay constant. 
Using 
\be
c_B(\mu = M_Q) = 1,
\ee
the leading-log scaling behavior of the coefficient is~\cite{wise1993}
\be
c_B (\mu) = \left(\frac{\alpha_s(M_Q)}{\alpha_s(\mu)}\right)^{9/(3b_0)}.
\ee
The kinetic energy operator is not renormalized due to invariance under 
reparameterizations~\cite{luke1992}
\ba
v & \to & v + \frac{\epsilon}{M_Q} \nonumber \\
Q_v & \to & e^{i\epsilon\cdot x}\left(1 + \frac{\slash{\!\!\!\epsilon}}{2M_Q}\right).
\ea
\subsubsection{NRQCD}
Non-relativistic QCD (NRQCD) is an effective theory formulated for heavy quarks assuming that their 
dynamics is non-relativistic, with  correction terms which can be added within a systematic
expansion. For quarkonia the higher order interactions are arranged in a $v^2$
expansion, where $v$ is the heavy quark velocity (see e.g.~\cite{lepage1992}). NRQCD has been very
successful in studying quarkonia. For a review discussing effective field theory methods in studies
of quarkonia see~\cite{brambilla2004}.
One may say that lattice calculations of the $\Upsilon$ use to a large extent non-relativistic effective 
field theory, and it has been applied with very good success (see e.g. the results 
in~\cite{lidsey1994,manke1997,manke2000,gottlieb2004}). We now discuss NRQCD for heavy-light mesons.

In hadrons with heavy and light quarks it is an expansion in $v$ or $1/M_Q$, where $M_Q$ is 
the heavy quark mass, similar to HQET.
At infinite mass, the heavy quark is just a source of the color electromagnetic field, whereas
at finite $M_Q$, there is a recoil of the heavy quark due to the interaction with soft gluons with
typical momenta of $O(\lqcd)$.
One can argue that the heavy  quark 
momentum $P_Q$ and light quark momentum $p_q$ are equal due to momentum conservation in the rest
frame of the meson:
\ba
M_Q v \simeq P_Q &=& p_q \sim O(\Lambda_{QCD}). 
\ea
Therefore $v = p/M_Q \sim \Lambda_{QCD}/M_Q \sim 0.1$ in $B$ mesons and should be a useful 
expansion parameter to specify corrections to the $M_Q \to \infty$ (static) limit
for quantities which involve a small heavy quark momentum, such as the spectrum or 
leptonic decay constants.

Contributions at $O(1/M_Q)$ are the kinetic and the spin-colormagnetic energy of the heavy quark; 
at $O(1/M^2_Q)$ 
a heavy quark spin-orbit interaction and a Darwin term are included.  The 
tree-level terms in the $1/M_Q$ series can for example be obtained with 
Foldy-Wouthuysen transformations on the heavy Dirac spinor which disentangle the
interactions between the upper and lower two components of the Dirac spinor (see e.g.\
the discussion in \cite{bjorken1964}). 
The Dirac Hamiltonian contains spatial $\gamma$ matrices which are off-diagonal.
%
Using unitary transformations which act on the heavy quark Dirac fields $Q$ as
\be
Q \to  Q^\prime = e^{iS} Q \,, Q^\dagger \to  Q^{\dagger\prime} =  
Q^\dagger e^{-iS}.
\ee
Chosing $S$ to be proportional to $\vec{\gamma}\cdot \vec{D}/M_0$, one obtains decoupling
of the Pauli equation for the heavy quark and a corresponding equation for the antiquark
 up to higher orders in $1/M_0$. $M_0$ is the bare heavy quark mass and agrees
with the previous $M_Q$ at tree level.

The $O(1/M^2)$ Lagrangian for the heavy quark is finally given by
\be
{\cal L} = \psi^\dagger(D_t + H)\psi, \label{eq:NRQCD}
\ee
with the heavy quark Pauli spinor $\psi$ and the Hamiltonian 
\ba
H &=& -\frac{\vec{D}^2}{2M_0} - \frac{g_sc_4}{M_0} 
\vec{\sigma}\cdot\vec{B} \nonumber \\
&+& \frac{ig_sc_2}{8M_0^2}(\vec{D}\cdot\vec{E}-\vec{E}\cdot
\vec{D})  - \frac{g_sc_3}{8M_0^2}\vec{\sigma}\cdot(\vec{D} \times
\vec{E}-\vec{E}\times\vec{D}) - \frac{c_1(\vec{D}^2)^2}{8M_0^3}. \label{eq:NRQCD_cont}
\ea
$\vec{E}$ is the color-electric field with $E^{ai} = -F^{a,0i}$, where $a$ is a color and $i$ a space
index.
The last term is the first relativistic correction to the kinetic energy of the heavy quark, which is
usually included in calculations at $O(1/M_0^2)$. 
The coefficients of the various terms can be found with matching calculations to full QCD in the 
continuum.  

\section{THE LIGHT HADRON SPECTRUM  \label{sec:lresults}}
Light hadrons, e.g. the $\pi$, $\rho$, $p$ and $n$ and their excited states,
are of course among the most interesting states to be studied on the lattice. To understand the 
dynamics of the forces between quarks and to study fundamental properties of QCD such as confinement
nonperturbatively, one investigates the potential between heavy quarks. For a detailed discussion
see e.g.~\cite{bali2001}. Here the potential is discussed in the context of setting the lattice scale and 
to calculate the strong coupling constant used in perturbative calculations. Only a very brief 
introduction is given.
 
In this section we are interested in hadrons consisting of $u,d$ and $s$ quarks in the quark model.
Using quenched gauge fields, Wilson quarks and quark-model inspired lattice operators,
it was found already some time ago that the features of the experimental light hadron
spectrum are reproduced well by the lattice~\cite{butler1994,cppacs_q2003}. Progress in recent years
was made using improved lattice actions and/or unquenched gauge fields.
In this section we discuss in particular results on  light meson masses, meson decay constants and the light
baryon spectrum from~\cite{cppacs2002hadr} using improved quark and gluon actions. 
We give a brief outline of the method used in~\cite{cppacs2002hadr} as an example for a standard 
lattice calculation, and then discuss some aspects which could not have been discussed in detail in the
paper~\cite{cppacs2002hadr}, re-analyse the comparison of quenched and dynamical results, and compare to 
other new lattice calculations.
The chiral extrapolation and the finite size effect of the nucleon is discussed using numerical results 
from QCDSF, UKQCD, CP-PACS and JLQCD.

The lattice simulations discussed here have been  performed around the strange mass. To avoid
extrapolation uncertainties we compare not extrapolated, i.e.\ strange lattice hadron masses, with experiment,
and keep in mind that with using strange sea quark masses a systematic uncertainty may be included.
\subsection{Technical Details \label{sec:tech}}
We now discuss some technical details of the calculation described in Ref.~\cite{cppacs2002hadr}, as an 
example for a standard lattice simulation with two flavors. The explanation is kept close to the 
paper~\cite{cppacs2002hadr}.

The operators employed to project onto the mesons and bayons are constructed of quark fields such 
that they project onto the quantum numbers of the state under consideration. It is usual to construct
them in the same way as quark model wave functions to ensure a good overlap with the hadron state.

The baryons under consideration in this study are the ground state spin $1/2$ and spin $3/2$ baryons.
The operators used for the spin $1/2$ octet baryons are 
\be
O_\alpha = \epsilon^{abc}\left(q^{aT}_1C\gamma_5
q^{b}_2\right)q^c_{3,\alpha}, 
\ee
where $C = \gamma_4 \gamma_2$ is the charge conjugation matrix. The subscripts  $1,2,3$ denote the 
different quark flavors. The index $\alpha$ denotes the spinor
component of the quarks which are not summed over and spinor components of the baryon fields. 
$a,b,c$ are color indices.  The quark model wave function is
antisymmetric under interchange of color indices. The spin and flavor wave functions are symmetric under 
interchange of two quarks, but not totally symmetric. The operators 
for the octet baryons except for the $\Lambda$ are given by
\be
-\frac{1}{\sqrt{2}}\epsilon^{abc}([q_1^a q_3^b]q_2^c +[q_2^a q_3^b] q_1^c),
\ee
with $[q_1q_2] \equiv q_1q_2 - q_2q_1$, where the operator $C\gamma_5$ is not written out explicitly. 
For the $\Lambda$, the operator
\be
O_\alpha = -\frac{1}{\sqrt{6}}\epsilon^{abc}([q_1^a q_3^b]q^c_{2\alpha} - [q_2^a q_3^b] q^c_{1\alpha}
-2[q_1^a q_2^b]q^c_{3\alpha})
\ee
is used.
The spinor components of the baryon are given by the spinor components of the
third quark.
For the spin $3/2$ decuplet baryons, operators projecting onto specific values of the
three-component of the baryon spin depend on the spinor component of the diquark and of the
third quark:
\ba
O_{3/2} & = & \epsilon^{abc} \left(q^{aT}_1 C \gamma_+ q^{ b}_2\right) 
q^{c}_{3,1} \nonumber \\
O_{1/2} & = & \epsilon^{abc} \frac{1}{3}[\left(q^{aT}C \gamma_3 
q^b\right) q^c_{3,1} - [\left(q^{aT}_1C \gamma_+ 
q^b_2\right) q^c_{3,2}]  \nonumber \\
O_{-1/2} & = & \epsilon^{abc} \frac{1}{3}[\left(q^{aT}C \gamma_3
q^b_2\right) q^c_{3,2} - [\left(q^{aT}C \gamma_-
q^b_2\right) q^c_{3,1}]  \nonumber \\
O_{-3/2} & = & \epsilon^{abc} \left(q^{aT}C \gamma_- q^b_2\right)
q^c_{2,3},
\ea
where the first subscript denote the flavor and the second the Dirac spinor components. 
$\gamma_± = (\gamma_1 \mp i\gamma_2)/2$. 
The gauge fields used in this simulation are renormalization group-improved~\cite{iwasaki1985}.
Both sea and valence quarks are clover quarks with  the clover
coefficient $c_{SW}$ being tree-level mean-field improved using the plaquette calculated at one loop
perturbation theory~\cite{aoki99a}.
This choice agrees rather well with the one-loop result, as shown in Table~\ref{tab:csw}.
\begin{table}[htb]
\begin{center}
\begin{tabular}{|lcc|}
\hline
\multicolumn{1}{|c}{$\beta$} & 
\multicolumn{1}{c}{$c_{SW}^{(tad)}$} & 
\multicolumn{1}{c|}{$c_{SW}^{(1)}$} \\
\hline
1.8  & 1.60 & 1.69 \\
1.95 & 1.53 & 1.60 \\
2.1 & 1.47 & 1.53 \\
2.2 & 1.44 & 1.49 \\
\hline
\end{tabular}
\end{center}
\caption{Tadpole-improved and one-loop clover coefficient $c_{SW}$ for renormalization group improved 
glue with $N_f = 2$.}
\label{tab:csw}
\end{table}

The hadron masses are extracted using fits of the hadron correlation functions projected onto zero spatial 
momenta, as a function of Euclidean time (the idea is outlined in Sec.~\ref{sec:corr}). The masses are
obtained for the values of the simulated bare gauge couplings and quark masses and have to be interpolated 
or extrapolated to the physical quark masses, defined by the condition that certain hadron masses or mass
ratios take physical values. 

As mentioned before the pion mass squared is a rough measure for the quark mass, and the meson masses
from \cite{cppacs2002hadr} can indeed be described as functions of the pion mass squared.
It is found that vector meson masses can be fitted for a given sea and valence
quark mass combination $(\ksea, \kval{}_1, \kval{}_2)$ as functions of
the degenerate valence pion mass
\ba
\mu_i &=& m^2_{PS}(\ksea, \kval{}_i,\kval{}_i) \nonumber \\
\mu_{\mathrm v} &=& \frac{1}{2}\left(\mu_1+ \mu_2\right)
\ea
and the pion mass for degenerate valence and sea quark masses
\ba
\mu_{\mathrm s} &=& \mps^2\left(\ksea,\ksea,\ksea \right),
\ea
including up to quadratic terms in the quark masses:
\be
m = A + B_\mathrm{s} \mu_{\mathrm s} + B_{\mathrm v} 
\mu_{\mathrm v}+ C_{\mathrm s} \mu_{\mathrm s}^2 + 
C_{\mathrm v} \mu_{\mathrm v}^2 + C_{\mathrm{sv}} \mu_{\mathrm s}
\mu_{\mathrm v}. \label{eq:mvfit}
\ee
\begin{figure}[thb]
\begin{center}
\centerline{
\epsfysize=6cm \epsfbox{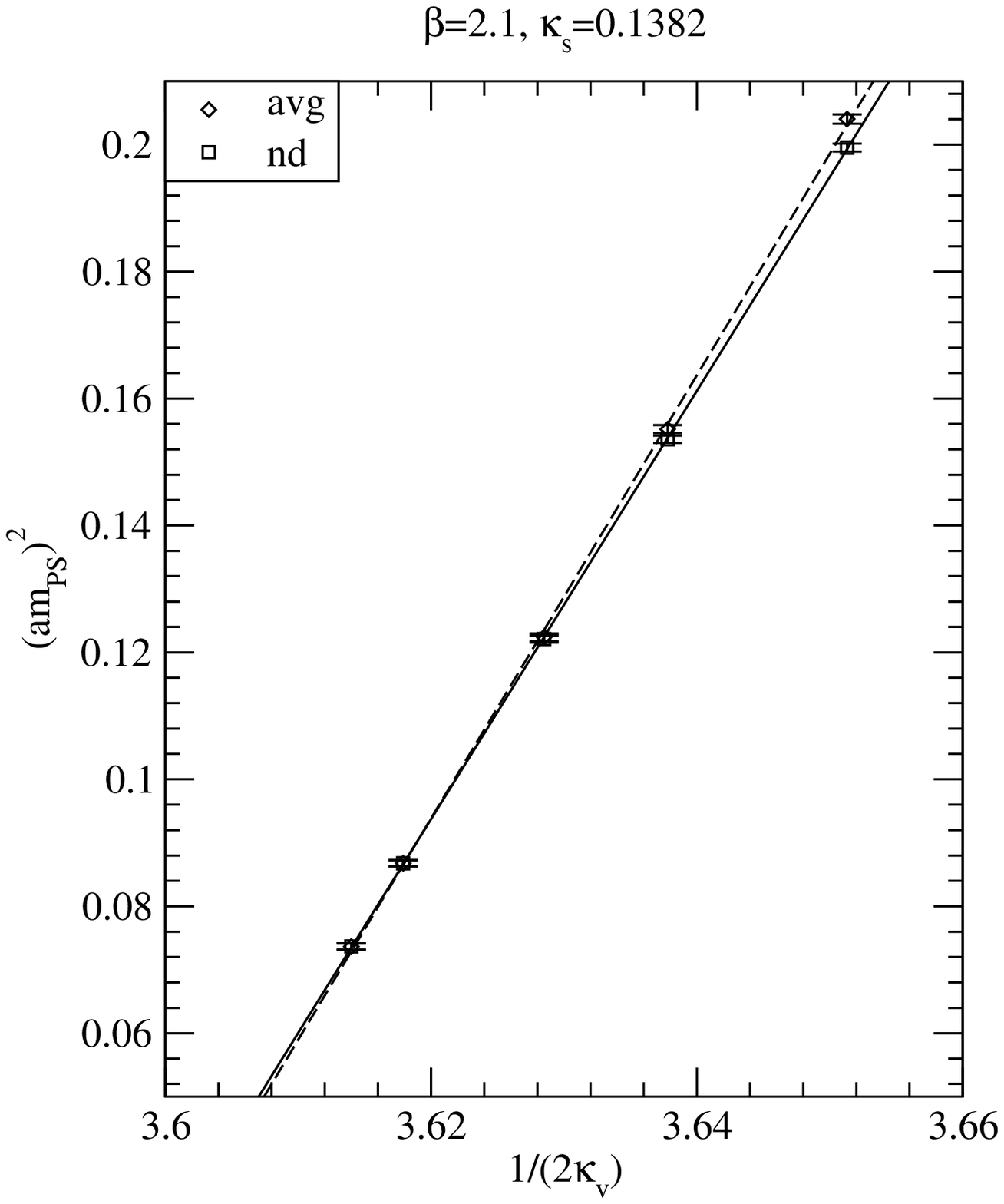}
\epsfysize=6cm \epsfbox{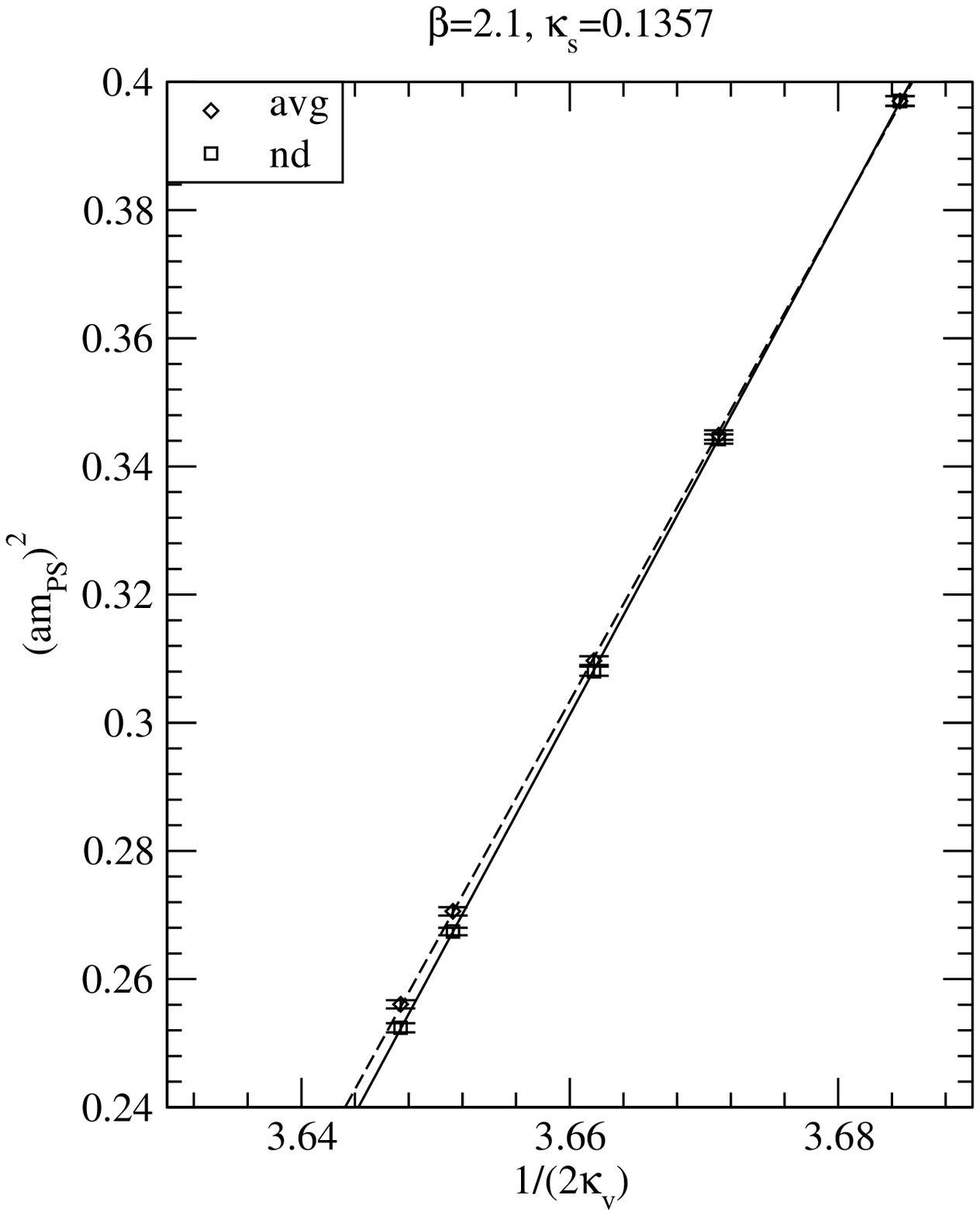}
}
\end{center}
\vspace{-0.5cm}
\caption{Squared pion mass as a function of the averaged valence hopping parameter $\kappa_v$, for the lightest
($\kappa_s=0.1382$) and the heaviest ($\kappa_s=0.1357$) sea quark at $\beta=2.1$. Squares denote the 
actual pion for a given $\kappa$ combination, and diamonds the average squared mass from degenerate valence
quarks. Dashed lines show linear fits to the averaged, solid lines to the non-degenerate data.}
\label{fig:mq_2.1} 
\end{figure}

The decuplet baryons are also fitted to the functional form given in  Eq.~(\ref{eq:mvfit}), where 
$\mu_{\mathrm v}$ corresponds to the valence quark mass averaged over all three quarks.

The octet baryon masses are described as functions of $\mu_{\mathrm v1}$ and  $\mu_{\mathrm v2}$
(corresponding to the masses of both valence quark flavors) and  the sea quark mass, including up to
quadratic terms in $\mu_i$. 
One should note that the mass of the actual pion with valence quark parameters $\kval{}_1$ and
$\kval{}_2$ agrees with the averaged mass of
two degenerate pions with the respective mass parameters $\kval{}_1$ and $\kval{}_2$  only if 
the quark masses are similar. In Fig.~\ref{fig:mq_2.1} we show the averaged and real pion masses in lattice units 
as functions of the VWI quark mass shifted by the critical mass.

The lattice spacing and the quark masses can now be fixed to the physical point.
For the calculation of quark masses and spectrum in~\cite{cppacs2002hadr}, the lattice spacing is determined
from $m_\rho$, and the  determination of the lattice spacing and the determination of the $\kappa$ values 
corresponding to the physical $u,d$ and $s$ quark masses are related. The procedure used in 
Ref.~\cite{cppacs2002hadr} is described in the following. 
First the pion and rho masses in lattice units are extrapolated to the physical ratio $m_\pi/m_\rho$ using a diagonal 
extrapolation (setting  $\mu_{val} = \mu_{sea}$ in  Eq.~(\ref{eq:mvfit}). This gives the pion mass in lattice units,
which is in turn substituted in Eq.~(\ref{eq:mvfit}) to determine $a m_\rho$, from which one obtains the lattice spacing $a$.
$\kappa_l$, the $\kappa$ value corresponding to the light ($u,d$) quark mass is calculated using 
a polynomial fit of $m_{PS}^2$ in terms of the VWI quark masses Eq.~(\ref{eq:vwi})
and setting $m_{PS}^2(\kappa_l;\kappa_l,\kappa_l) = m_\pi^2$. The $\kappa$ value corresponding to the strange quark 
mass ($\kappa_s$) is determined using strange meson masses. One method is to use the $K$ meson and to fix the 
ratio $m_{PS}^2(\kappa_l;\kappa_l,\kappa_s)/m_\pi^2$ to the physical value $m_K^2/m_\pi^2$.
Then one can define a fictitious strange-strange pseudoscalar meson, the $\eta_{ss}$, from the relation 
$m^2_{\eta_{ss}} = m_{PS}^2(\kappa_l;\kappa_s,\kappa_s)$.  
$m^2(\eta_{ss})$ can then be used to parameterize the strange quark physical 
point in the fits of the other hadrons in terms of the $\mu$'s. The strange quark mass is alternatively determined
using the $\phi$, an $\overline{s}s$ vector meson. 

Parameterizing the $m_\phi$ using Eq.~(\ref{eq:mvfit}) with $\mu_{sea}$ set to 
$m_\pi$ and fixing the ratio $m_\phi/m_\rho$ to the physical value one obtains a
result for $\mu_{val}=\eta_{ss}$ which is used to fix the strange point for baryon spectrum. Using the
$\phi$ to determine the strange quark mass, the $K$ meson mass is a prediction, determined by
the equation $m_K^2 = m_{PS}^2(\kappa_l;\kappa_l,\kappa_s)$.

The continuum results are obtained using linear extrapolations in $a$.
\subsection{Light Meson Masses \label{sec:llm}}
\subsubsection{Quenched results}
The quenched approximation is only an approximation, and some physics is expected to be different 
in the quenched world  from real QCD.

\begin{figure}[thb]
\vspace{0.1cm}
\begin{center}
\epsfysize=6cm \epsfbox{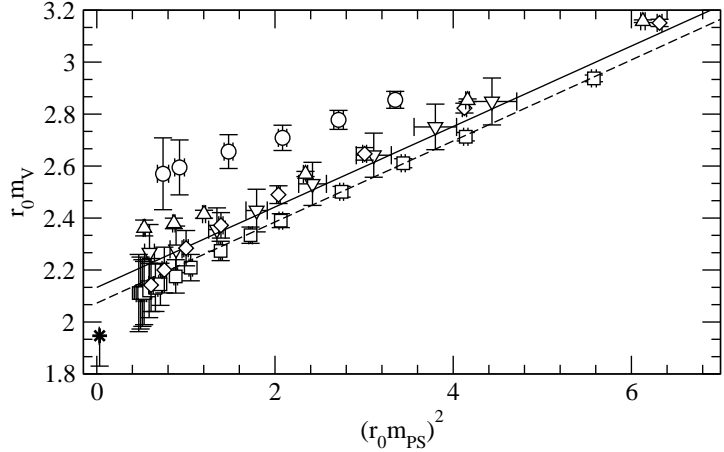}
\end{center}
\caption{Quenched vector meson masses as a function of $\mps^2$ in units of $r_0$.
The lines denote linear fits to results from~\protect\cite{cppacs2002hadr} at
the coarsest (dashed) and finest (solid) lattice spacings. Triangles up use overlap quarks
from~\protect\cite{galletly2005}, diamonds fixed point quarks
from~\protect\cite{hasenfratz2004}, circles overlap quarks with coarse Wilson gauge fields
from~\protect\cite{bieten2004}, 
squares use chirally improved quarks from \cite{gattr2004} and triangles down use FLIC quarks
from~\protect\cite{zanotti2004}. The experimental value is denoted by a star, with the error bar depicting 
the variation if $r_0 = 0.47$ fm instead of 0.5 fm is used.}
\label{fig:mrho_quenched} 
\end{figure}

Since the $\rho$ mass is an important result of QCD and CP-PACS's favorite quantity to set the lattice scale, 
we first compare the quenched vector meson mass to experiment, using $r_0$ to set the scale. Plotting
the chiral extrapolation of the $\rho$ mass from the coarsest and finest lattice of \cite{cppacs2002hadr} in
Fig.~\ref{fig:mrho_quenched} one sees that the result does not agree with experiment. To investigate whether 
this may be due to discretization errors and whether the behavior changes qualitatively if yet smaller 
quark masses are used, we compare 
to a set of recent quenched results using improved gauge field and fermion actions, which are expected to
scale well, at small quark masses. Refs.~\cite{galletly2005,bieten2004,hasenfratz2004} use 
chiral lattice fermions. We compare the results using chirally improved fermions 
of~\cite{gattr2004}, and results  using clover fermions with a fattened-link Wilson term
which also enables simulations at small quark masses~\cite{zanotti2004}.
We find an agreement of the CP-PACS lines with most of the new lattice results at larger quark masses. For smaller
quark masses masses both overlap results seem to extrapolate to higher values of $m_\rho$. 
A possible reason for the difference of the results from overlap 
fermions is that \cite{bieten2004} uses unimproved Wilson glue on coarse lattices and 
\cite{galletly2005} the L\"uscher-Weisz gauge action~\cite{luscher1985} on fine lattices. The plot
however indicates that the overlap quark results might extrapolate to a relatively larger value.
The fixed point and 
chirally improved fermions \cite{hasenfratz2004,gattr2004} appear to extrapolate 
towards the experimental result. Using a chiral extrapolation guided by the functional form predicted by
quenched chiral perturbation theory they find quenched $\rho$ and $N$ masses in agreement with experiment.
\subsubsection{Unquenched results}
\begin{figure}[thb]
\begin{center}
\epsfysize=6cm \epsfbox{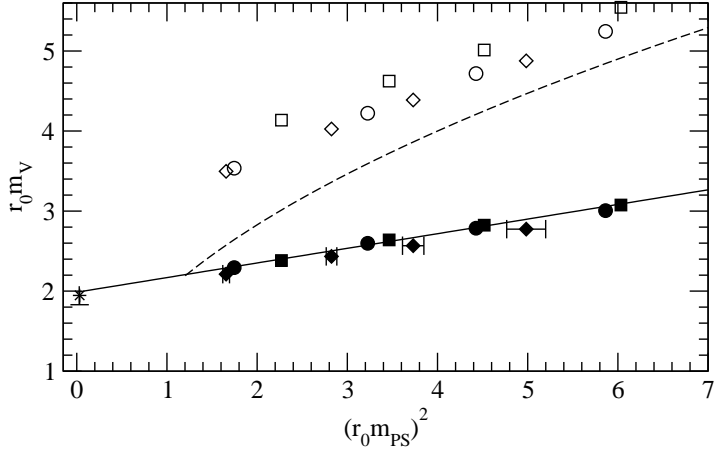}
\end{center}
\caption{The $\rho$ meson mass from 
\protect\cite{cppacs2002hadr}. Filled symbols denote the simulation results. Errors in the $y$
direction are roughly of the size of the symbols. 
Open points indicate the corresponding thresholds given by $2r_0\sqrt{m_{PS}^2 + (2\pi/L)^2}$.
Diamonds stand for $\beta = 1.95$, circles for $\beta = 2.1$ and squares for
$\beta = 2.2$. The dashed curve indicates $r_0m_V = 2r_0m_{PS}$.
The solid  line indicates a linear fit to the data at  $\beta = 2.20$. The star denotes the experimental
value. The error bar gives the variation if $r_0 = 0.47$ fm instead of 0.5 fm is used.}
\label{fig:r0mv_cppacs} 
\end{figure}
In Figure~\ref{fig:r0mv_cppacs} the vector meson mass is shown as a function of the pion mass at 
finite lattice spacing, for equal sea and valence quark masses. The agreement of data from various
$\beta$ values is rather good, and a continuum extrapolation seems to be not necessary.
The result is rather linear in the quark mass.
A potential systematic error in this relation is due to the fact that the unquenched $\rho$ 
can undergo hadronic decays.
For the given set of masses, the decay thresholds of $2\sqrt{m_{PS}^2 + (2\pi/L)^2}$ are about 400 MeV 
larger than the vector mass.  
$2m_{PS}$ is also higher but for small quark mass close to the vector masses. 

\begin{figure}[thb]
\vspace{0.1cm}
\begin{center}
\epsfysize=6cm \epsfbox{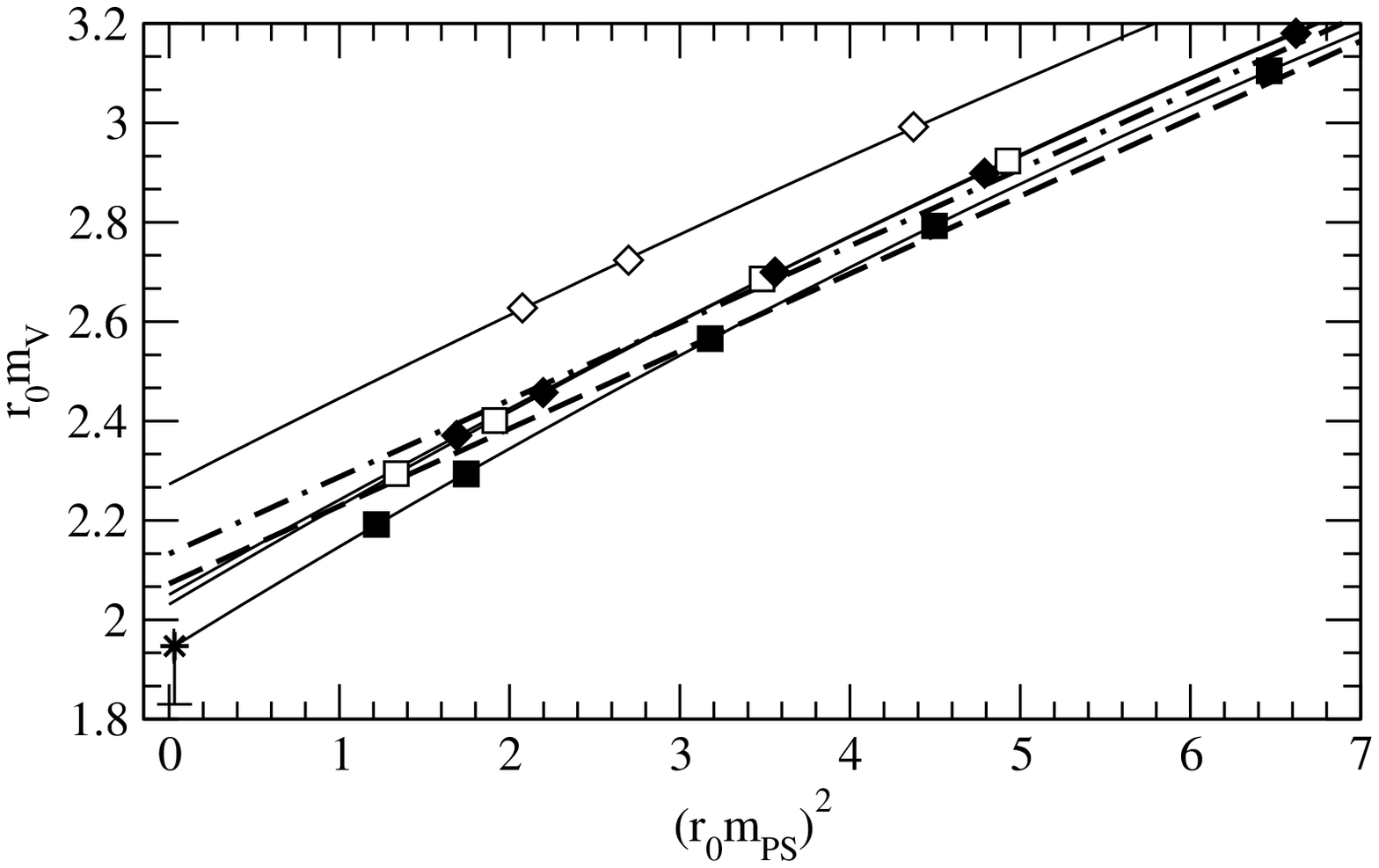}
\end{center}
\caption{Partially quenched vector meson mass as a function of $(\mps)^2$ at $\beta = 2.1$. Diamonds 
denote $\ksea = 0.1367$, squares  $\ksea = 0.1382$. Filled 
symbols correspond to the $r_0$ values with equal sea and valence quark masses. Open symbols correspond to chirally 
extrapolated values of $r_0$. Error bars are smaller than the symbols. The star denotes the experimental value. 
Fat dot-dashed and dashed lines denote the quenched fits on the finest and coarsest lattice
respectively. Thin dashed lines denote quadratic fits to the various data sets.}
\label{fig:mv_PQ} 
\end{figure}
We have seen that the $\rho$ mass agrees with the experimental value if the lattice scale is set using
$r_0$ in unquenched QCD. It is of interest to study also extrapolations in the valence quark mass at
finite fixed sea quark masses. In Fig.~\ref{fig:mv_PQ} we compare the chiral extrapolation in the valence
quark on a relatively fine lattice for two different values of the sea quark mass to the quenched 
extrapolation. Again, the scale is set using $r_0$.  As in the previous plots, we can use the $r_0$ values 
determined at the respective  sea quark masses, and find agreement with experiment already at finite sea quark mass.
In a mass-independent
renormalization scheme, the lattice spacing depends only on the coupling, not on the quark 
mass~\cite{massren} up to  discretization errors,  
and we compare the meson masses also with chirally extrapolated $r_0$ values,
$r_0^{(\chi)}$. Those were determined in \cite{cppacs2002hadr} assuming a linear quark mass dependence:
\be
\frac{1}{r_0}(\mu_{\mathrm{sea}}) = \frac{1}{r_0^{(\chi)}} + C \mu_{\mathrm{s}}.
\ee
We find that partially quenched results using the chirally extrapolated $r_0$ into the comparison in
Fig.~\ref{fig:mv_PQ} also move towards the experimental value, while the agreement 
is less obvious at finite sea quark mass.
%
\subsection{Light Meson Decay Constants}
Lattice results for light decay constants are important tests of QCD 
as a theory of light hadron physics and also a calibration of the lattice method for calculations of
hadron matrix elements. Since they are likely to be less sensitive to systematic errors, 
it is of interest to calculate  ratios of light and heavy-light decay constants.
We discuss the results of~\cite{cppacs2002hadr}.

The  decay constant of a pseudoscalar meson $P$ of mass $M$ in Euclidean space-time can be defined as:
\be
\langle 0 | A_4(0)| P(\vec{p}=0) \rangle = f_{P} M.
\ee

For the chiral extrapolation of the matrix elements, the same fitting form as for the vector 
meson masses is used,
Eq.~(\ref{eq:mvfit}). The quenched decay constants can be described by a linear function
in $\mu_{\mathrm v}$ only.

In~\cite{cppacs2002hadr} the decay constant was calculated with  one-loop 
tadpole-improved  renormalization constants. The axial vector current could then be
expressed by the relation:
\be
A_\mu^R(x) = 2\kappa u_0 Z_A\left(1 + b_Am/u_0\right) \left(A_\mu(x) - c_A\partial_\mu
P(x) \right),
\ee
with one-loop values  for $Z_A, b_A$ and $ c_A$. 
The effect of the improvement term proportional to $c_A$ on the decay constant
is  a few MeV.
\subsubsection{Quenched results}
The quenched results for $f_\pi$ and $f_K$ appear to have reasonably small discretization
errors at lattice spacings up to around $a \sim 0.15$ fm. 
The plateau occurs however at a roughly $20\%$ higher value than the quenched result of 
\cite{cppacs_q2003} using Wilson quarks.

Using the result from the finest lattice, one can quote for the quenched light meson 
decay constants
\ba
f_\pi &=& 145(5) \;\mbox{MeV}, \nonumber \\
f_K   &=& 162(4) \;\mbox{MeV, $K$ input,} \nonumber \\
   & & 166(4) \;\mbox{MeV, $\phi$ input,}
\ea
i.e.
\ba
\frac{f_K}{f_\pi} & = & 1.12(5) \;\mbox{, $K$ input} \nonumber \\
                  & = & 1.15(4) \;\mbox{, $\phi$ input} 
\ea
Errors are statistical and extrapolation errors only.
$f_K$ is close to experiment, while $f_\pi$ is $\sim 10\%$ higher than experiment. 
\subsubsection{Unquenched results}
Scaling violations in the unquenched decay constants are large, i.e.\ 50\% 
between $0.2 \geq a \geq 0.1$ fm. To avoid systematic errors from the continuum
extrapolation, we quote results from the second finest unquenched lattice at $\beta = 2.1$, 
since at $\beta = 2.2$ the error bars are very large.
\ba
f_\pi &=& 127(6) \;\mbox{MeV} \nonumber \\
f_K   &=& 149(6) \;\mbox{MeV, $K$ input}, \nonumber \\
f_K   &=& 152(6) \;\mbox{MeV, $\phi$ input},
\ea
where only statistical errors are quoted.
The results are to be compared to the experimental values of $132$ MeV for $f_\pi$ and
$160$ MeV for $f_K$.

In the ratio of vector and pseudoscalar meson decay constants, $f_\rho/f_\pi$, 
discretization errors largely cancel, and the continuum
extrapolation of the ratio is in agreement with experiment, however within large error bars.
\subsection{Light Baryons}
\subsubsection{Quenched results}
\begin{figure}[htb]
\begin{center}
\centerline{
\epsfysize=6.2cm \epsfbox{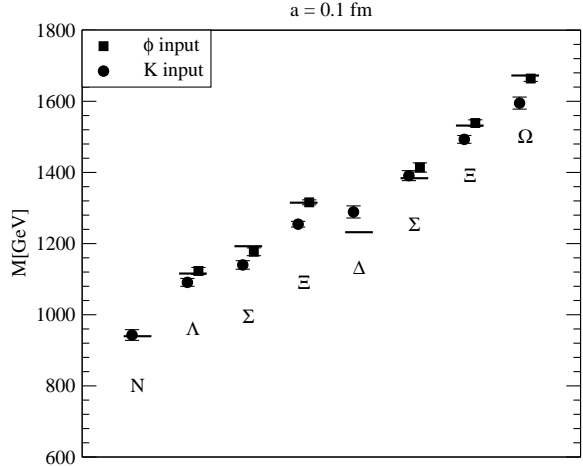}
}
\end{center}
\caption{Quenched light baryon spectrum from CP-PACS from lattices with $a \sim 0.1$ 
fm~\protect\cite{cppacs2002hadr}.}
\label{fig:bary_cppacs} 
\end{figure}
First we would like to compare the quenched baryon masses from \cite{cppacs2002hadr} to
other recent quenched calculations and experiment. The values were extracted using the
method described in Sec.~\ref{sec:tech}.  
In the quenched calculation, the decuplet baryon masses were extrapolated using just a linear fit in the
pion mass, while for the octet states a quadratic term was included.
The results were continuum extrapolated using a linear fit in $a$. The quenched spectrum from this 
procedure is found to agree with the quenched continuum extrapolated results using Wilson gauge and
fermion actions, see Fig.~\ref{fig:bary_cppacs}. The lattice spacing has been set with
$m_\rho$ in both cases. The features of the experimental spectrum are reproduced, while strange baryons
with $K$ input and with $\phi$ input may differ among each other and with experiment by several $\sigma$.
Before trying to make definite statements of whether this is a quenching effect, we should consider possible 
other systematic errors. 
It is likely that  discretization artifacts will affect also any unquenched calculation with the same action
at similar quark masses and lattice spacings, and we therefore try to give some more thoughts to the question.

\begin{figure}[thb]
\begin{center}
\vspace{-0.3cm}
\epsfysize=6cm \epsfbox{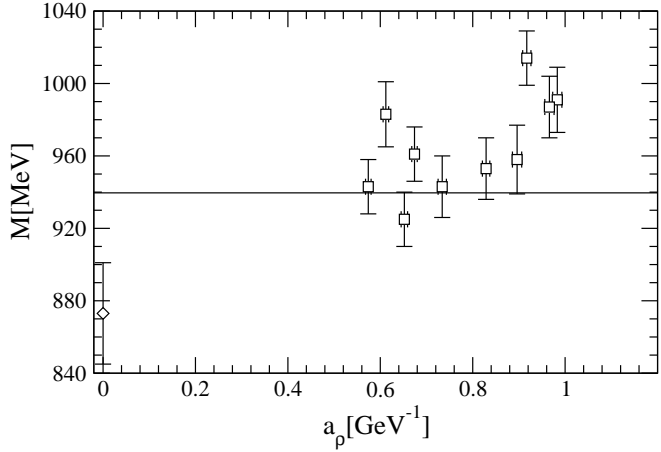}
\end{center}
\caption{Quenched nucleon masses from CP-PACS \protect\cite{cppacs2002hadr} as function of the
lattice spacing (squares) with the continuum extrapolated value (diamond) on the left. The line
denotes the experimental value for the nucleon mass.}
\label{fig:mN_quenched_vs_a}
\end{figure}

In the present case, the quark action is tree-level tadpole improved. The dominant
discretization effects should be $O(a^2)$ and $O(\alpha_s a)$, i.e.\ in principle there are
linear and quadratic contributions in $a$. In Fig.~\ref{fig:mN_quenched_vs_a} we show the nucleon mass as
a function of the lattice spacing.

We find that the data  from the finer lattices scale well, i.e.\ they are independent of the
lattice spacing, and a linear continuum extrapolation including all
data points does not appear more reliable than using the result from the finest
lattice as continuum estimate. 
We plot the quenched light baryon spectrum using the results from the finest quenched lattice at
$\beta = 2.575$
in Fig.~\ref{fig:bary_cppacs}. The result changes by several $\sigma$, and in most cases, 
agreement with experiment is better for the results which are not continuum extrapolated.

\begin{figure}[htb]
\vspace{0.1cm}
\begin{center}
\epsfysize=7cm \epsfbox{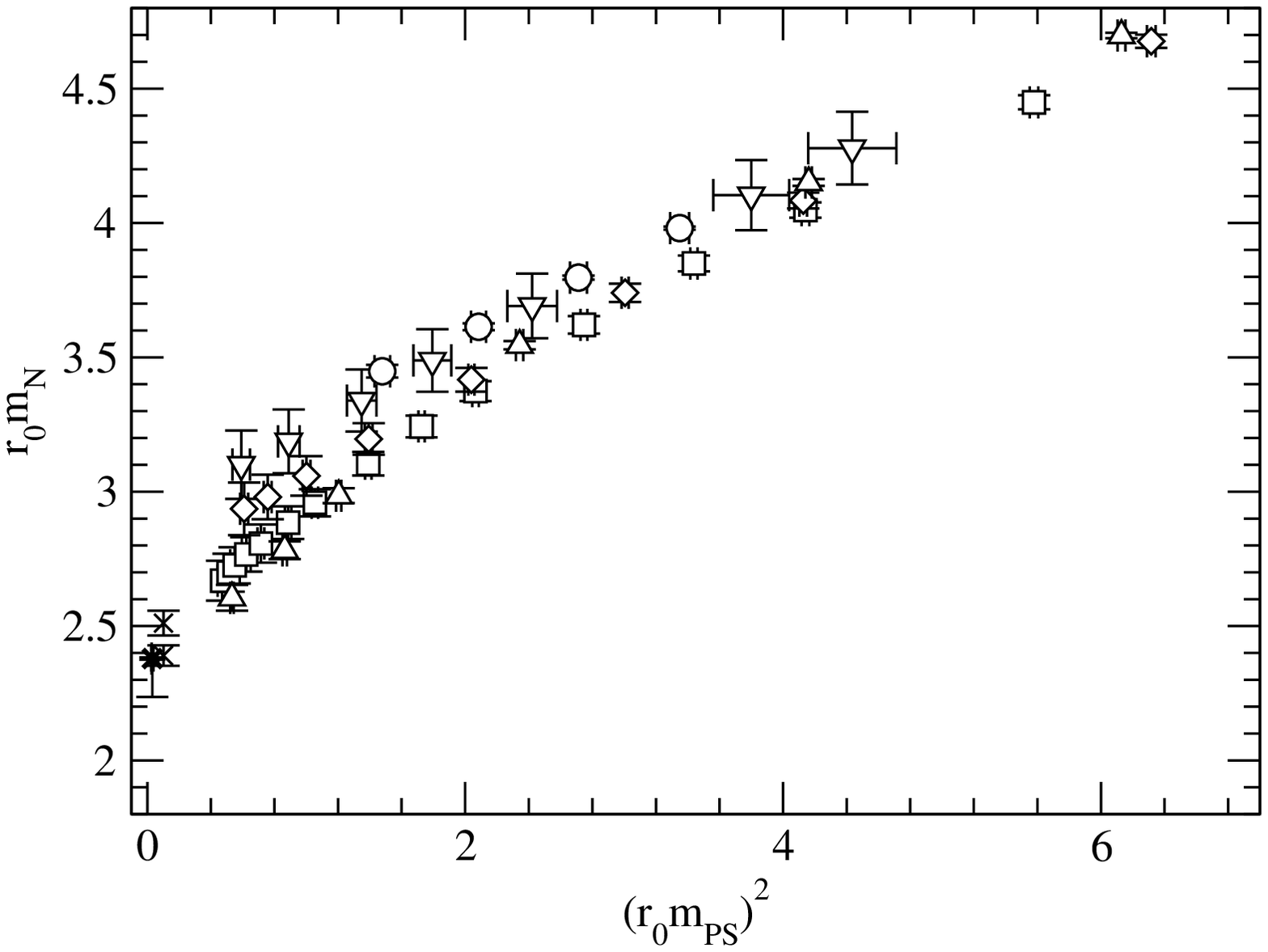}
\end{center}
\caption{Quenched nucleon mass as a function of $\mps^2$ in units of $r_0$.
The chirally extrapolated results from \protect\cite{cppacs2002hadr} from the coarsest and 
finest quenched lattices are denoted by crosses (slightly shifted to the right for clarity), 
triangles use overlap fermions from~\protect\cite{galletly2005},
diamonds fixed point fermions from~\protect\cite{hasenfratz2004}, 
 circles overlap fermions with coarse Wilson gauge fields from~\protect\cite{bieten2004},
 squares chirally improved fermions from\protect\cite{gattr2004} and triangles down FLIC fermions 
from~\protect\cite{zanotti2004}. The star denotes the experimental value, with the error bar giving the
variation if $r_0 = 0.47$ fm instead of 0.5 fm is used.}
\label{fig:mN_quenched} 
\end{figure}

Chiral extrapolation uncertainty could be another source of error. 
The extrapolated nucleon masses from CP-PACS using naive polynomial extrapolations
agree quite well with experiment. The results at finite quark masses are not available, therefore
we show only the chirally extrapolated values from the finest and the coarsest quenched lattice in 
Fig.~\ref{fig:mN_quenched}.  The
fixed point~\cite{hasenfratz2004} 
The overlap results from QCDSF~\cite{galletly2005} and the chirally improved quarks~\cite{gattr2004} 
 seem to extrapolate to the experimental value.
For the fixed point results plotted here~\cite{hasenfratz2004} it is not so clear whether the experimental value 
is reached, however their results from coarser lattices and smaller quark masses do extrapolate to the
experimental value~\cite{gattr2004}.  For the overlap results from~\cite{bieten2004} and the FLIC
fermions we have used here~\cite{zanotti2004}, the extrapolation is again not obvious from the data on this plot. 
\subsubsection{Unquenched results}
\begin{figure}[htb]
\begin{center}
\epsfysize=5cm \epsfbox{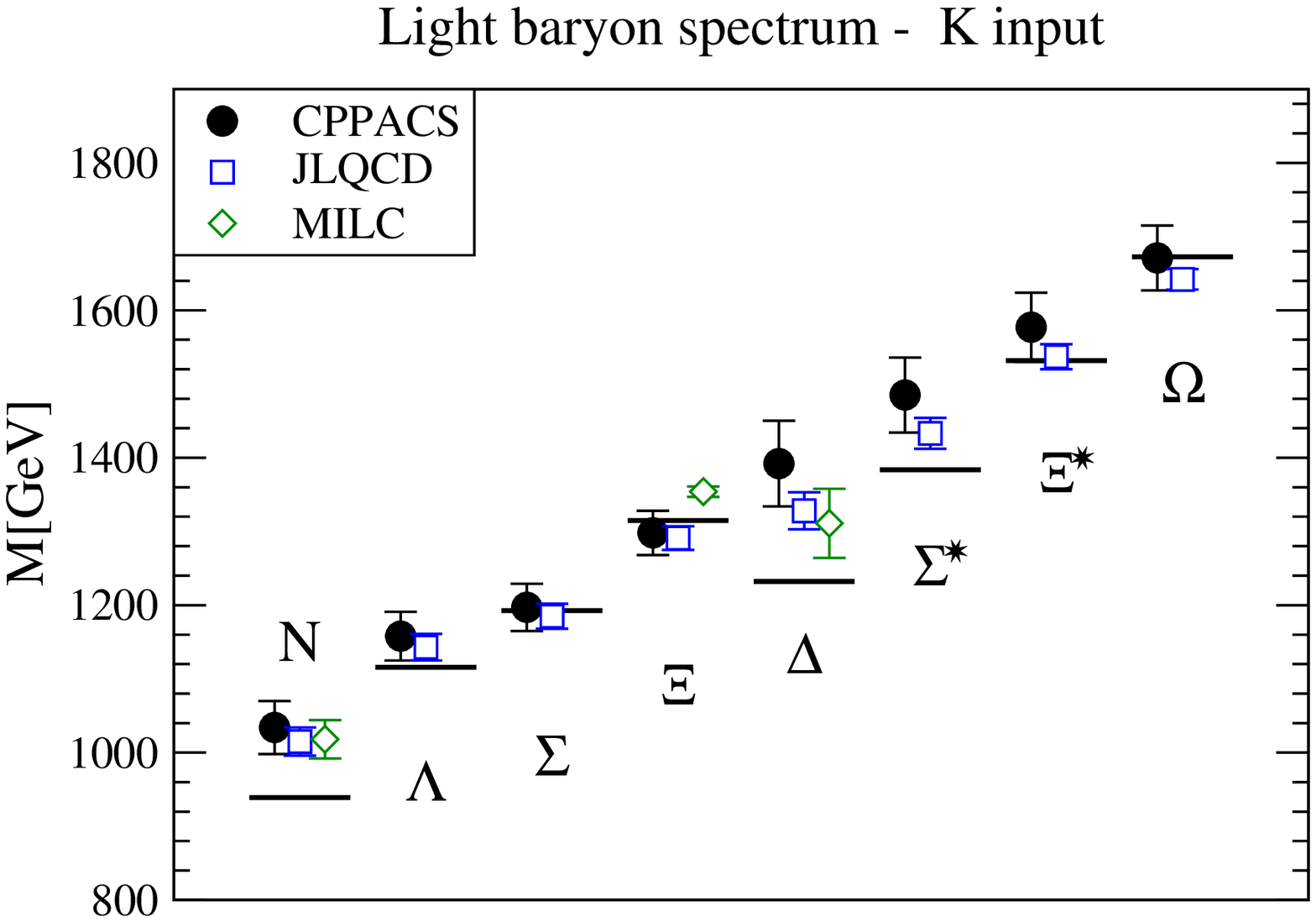}
\end{center}
\caption{Unquenched light baryon spectrum from recent unquenched lattice calculations.}
\label{fig:bary_unquenched} 
\end{figure}

Ref.~\cite{jlqcd2003hadr} works at a fixed lattice spacing and thus cannot 
perform a continuum extrapolation. For the chiral extrapolation, a similar formula
as in~\cite{cppacs2002hadr} is used.
In the baryon mass calculation of Ref.~\cite{aubin2004} with two light and one 
strange sea quarks, the continuum extrapolation is performed first, and then a
chiral extrapolation is done.

In Fig.~\ref{fig:bary_unquenched} we compare the the baryon spectrum from 
the recent unquenched simulations of~\cite{cppacs2002hadr,jlqcd2003hadr,aubin2004}.
Ref.~\cite{jlqcd2003hadr} assigns additional systematic errors of up to
25 MeV from the  chiral extrapolation uncertainty and the determination of $r_0$.
We do not add them to Table~\ref{tab:light_split} and just note that
their additional error bars would point towards the opposite direction to the 
experimental value. 
We find discrepancies with experiment of up to $\sim 2\sigma$ for the nucleon and $\Lambda$, 
where improvement with respect to the quenched spectrum is not visible. 
Chiral extrapolation uncertainty might be relevant. An attempt to use chiral 
perturbation theory to extrapolate the nucleon mass is discussed in the next section.
\begin{table}[htb]
\begin{center}
\begin{tabular}{|l|l|l|l|l|l|}
\hline
                     & scale    & $\Lambda-N$[MeV]  & $\Delta-N$[MeV] & $\Sigma^\ast-\Sigma$[MeV]
& $\Xi^\ast - \Xi$[MeV] \\
\hline
\multicolumn{6}{|c|}{$N_f = 0$} \\
\hline
\protect\cite{cppacs2002hadr}  & $m_\rho$ & 
$151(34)(^{28}_{0})$ & 346(41) &  $252(33)(^{0}_{10})$ & $247(26)(^{0}_{7})$ \\
\protect\cite{jlqcd2003hadr}   & $m_\rho$ & 
 $116(33)(^{26}_0)$   & $314(37)$ & $252(30)(^0_{10})$ & $250(23)(^0_{11})$ \\
\hline
\multicolumn{6}{|c|}{$N_f = 2$} \\
\hline
\protect\cite{cppacs2002hadr}  & $m_\rho$ & $124(49)(^2_0)$ & 358(68) & 
$288(60)(^0_2)$ & $279(56)(^2_0)$ \\
\protect\cite{jlqcd2003hadr}   & $m_\rho$ & $128(26)(^{16}_0)$ & $313(31)$ &
$248(27)(^{+0}_{-5})$ & $246(23)(^0_5)$ \\
\hline
\multicolumn{6}{|c|}{$N_f =  2_{\mathrm{light}} +1_{\mathrm{strange}}$} \\
\hline
\protect\cite{aubin2004}  & $\chi_b-\Upsilon$ &  & 293(54) &  &  \\
\hline
\multicolumn{6}{|c|}{Model calculations} \\
\hline
\protect\cite{capstick1986}     &    & 155  & 270 &  180(15)  & 200(15)   \\
\hline
\multicolumn{6}{|c|}{Experiment} \\
\hline
       &   & 177 &  294 & 191 & 215 \\
\hline
\end{tabular}
\end{center}
\vspace{-0.2cm}
\caption{Light baryon mass splittings. The first error is statistical,
the second is the difference from fixing
the strange quark mass using the $K$ or $\phi$ meson where applicable.
The quantity used to fix the lattice scale is shown on the left.}
\label{tab:light_split}
\end{table}
In Table~\ref{tab:light_split} we give results for light baryon mass splittings corresponding to the
results shown in Fig.~\ref{fig:bary_cppacs}.   For the splittings, the agreement with experiment 
is at the $1-2\sigma$ level. A recent calculation \cite{aubin2004} with 
$N_f = 2_{{\mathrm{light}}}+ 1_{{\mathrm{strange}}}$ finds a $\Delta-N$ splitting which agrees well 
with experiment. The mass ratio $m_N/m_\rho$ has also been calculated using 
domain-wall fermions with $N_f = 0$ and $N_f = 2$~\cite{yaoki2004}. 
Their quenched results seem to extrapolate to a value in agreement with 
experiment. Their unquenched value is in quite good agreement with experiment.

These lattice studies do not take into account that in the unquenched theory
the $\Delta$ can decay strongly; it is an $N \pi$ resonance with a Breit-Wigner width of
$115-125$ MeV~\cite{pdg} and one would expect a shift in the mass.

To avoid the uncertainties due to extrapolation in the comparison beween theory and
experiment, we now consider relations between strange hadron masses in two-flavor QCD.

\begin{figure}[htb]
\vspace{0.1cm}
\begin{center}
\epsfysize=6cm \epsfbox{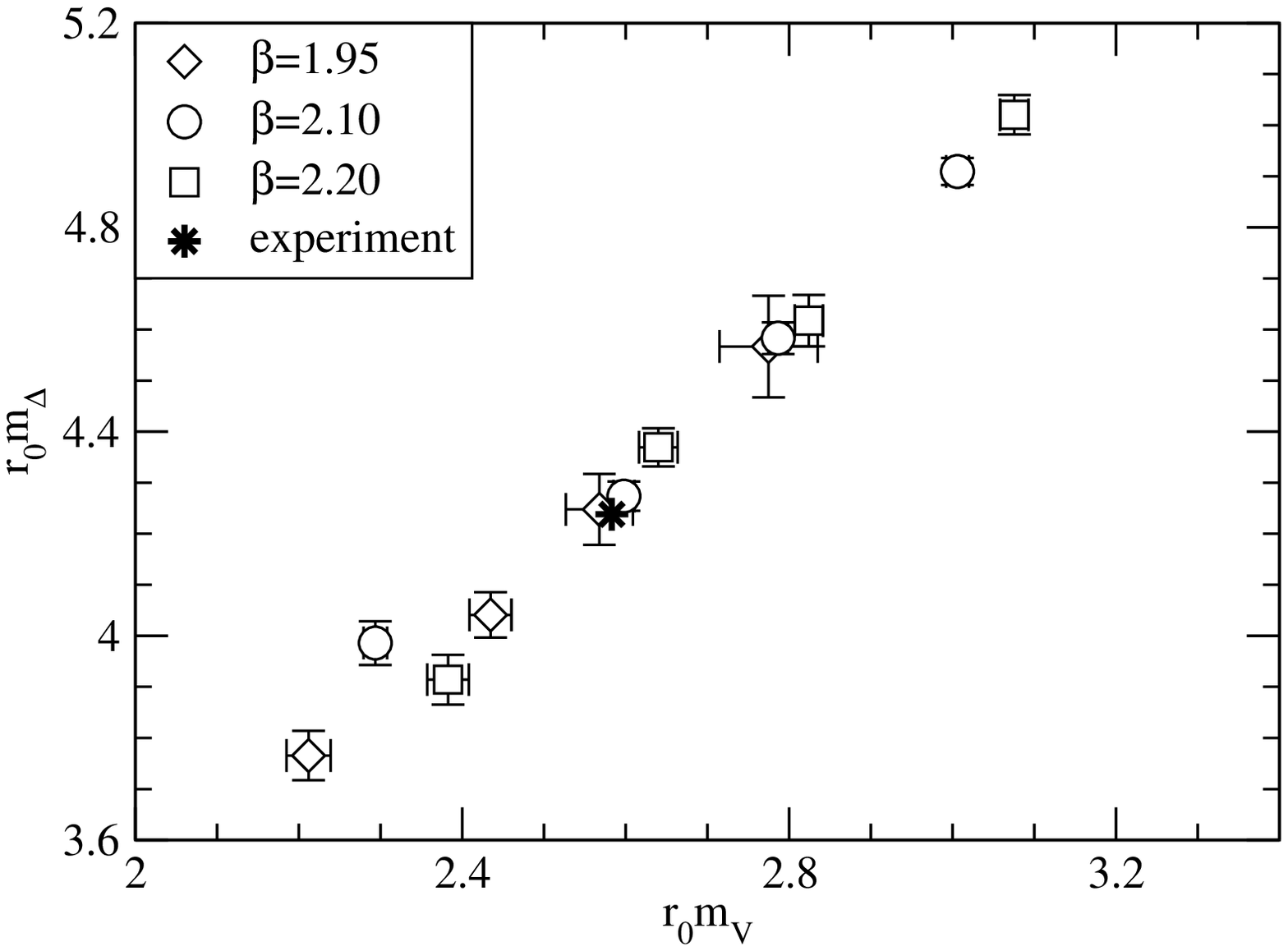}
\end{center}
\caption{Unquenched decuplet baryon mass as a function of the vector meson mass corresponding to the same
gauge coupling and quark mass. Different symbols correspond to different values of the inverse 
bare gauge coupling. Valence and sea quark masses are equal. The experimental point
corresponds to the $\phi$ meson and the $\Omega$ baryon, with $r_0= 0.5$ fm. }
\label{fig:omega} 
\end{figure}

In Figure~\ref{fig:omega} we plot the decuplet baryon mass from three different $\beta$ values
versus the vector meson mass in units of $r_0$. Sea and valence quark masses are degenerate.
The physical point corresponds to the position of the $\phi$ meson and the $\Omega^-$ baryon mass
and agrees with the lattice result. Mixing  between the $\phi$ and the $\omega$ has been ignored
at this point.

\begin{figure}[htb]
\begin{center}
\centerline{
\epsfysize=6cm \epsfbox{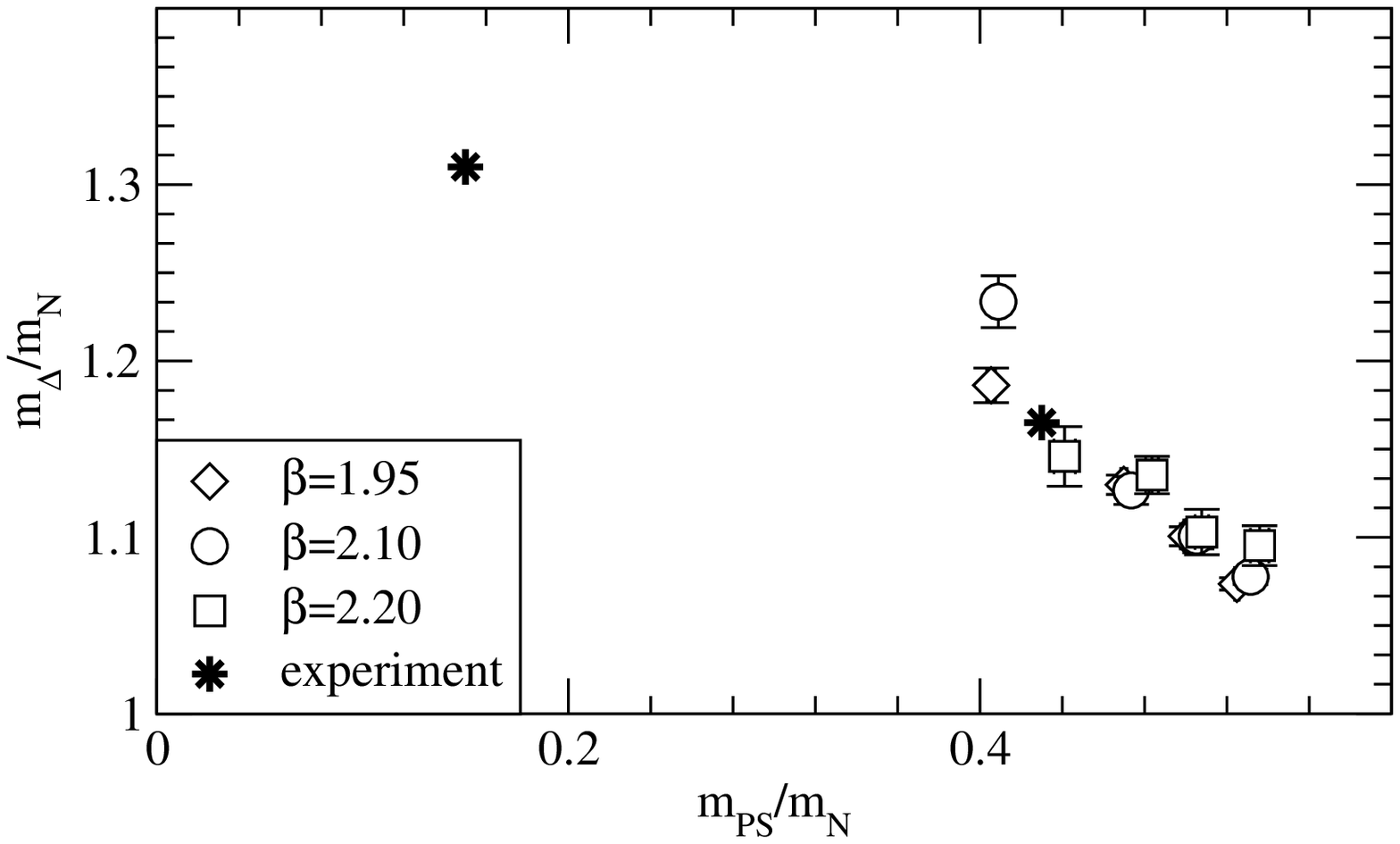}
}
\end{center}
\caption{Unquenched mass ratio of decuplet and octet baryons.  Different symbols correspond to different 
values of the inverse bare gauge coupling. Sea and valence quark masses are equal.
The experimental points respectively correspond to the physical light quark mass and quark mass averaged over the $ssl$ 
quark triplet, with $r_0= 0.5$ fm. }
\label{fig:berlin} 
\end{figure}
In Fig.~\ref{fig:berlin} we show the decuplet/octet mass ratio as a function of the
pion/octet baryon mass ratio for three $\beta$ values from CP-PACS. The experimental
points correspond to quark masses averaged over all three valence quarks. Sea and 
valence quark masses of the lattice data in the plot are taken to be equal.
The `experimental' $ssu$ pseudoscalar meson mass is determined by interpolation using
the physical values $m_K^2$ and $m_\pi^2$. 
Although the plot has to be interpreted with some caution since the sea quark mass is not
extrapolated to the physical point,  the lattice octet baryon masses cannot be well described as a 
function of the average quark mass, and the baryons can be affected by
strong decays, we nevertheless find the agreement with experiment remarkable.
\subsection{Nucleon Mass: Finite Size Effects and Chiral Extrapolation
 \label{sec:FS}}
In this section we discuss chiral extrapolation and calculation of the finite
size effect of the nucleon mass using chiral perturbation theory.

\begin{figure}[thb]
\begin{center}
\centerline{
\epsfysize=1.3cm \epsfbox{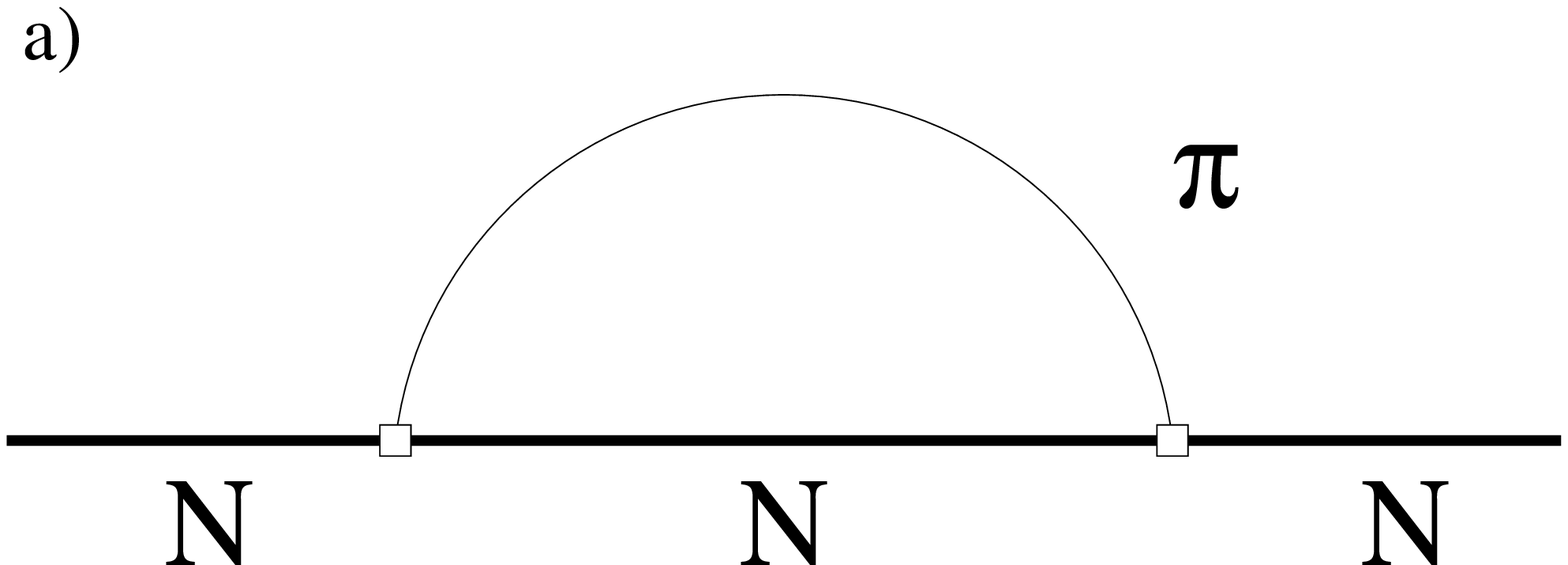}
\epsfysize=1.5cm \epsfbox{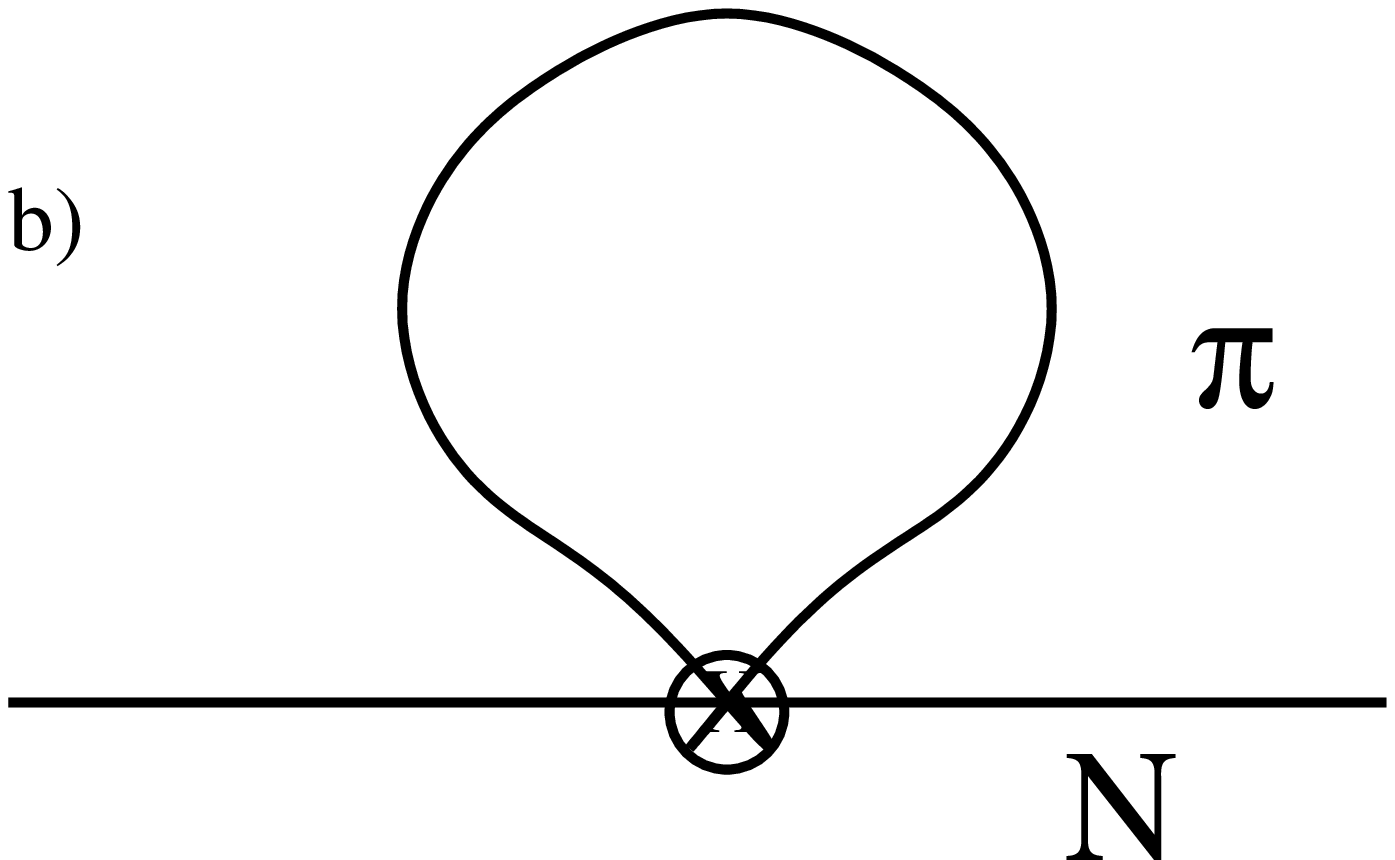}
\begin{minipage}[b]{2cm}\epsfysize=1.3cm \epsfbox{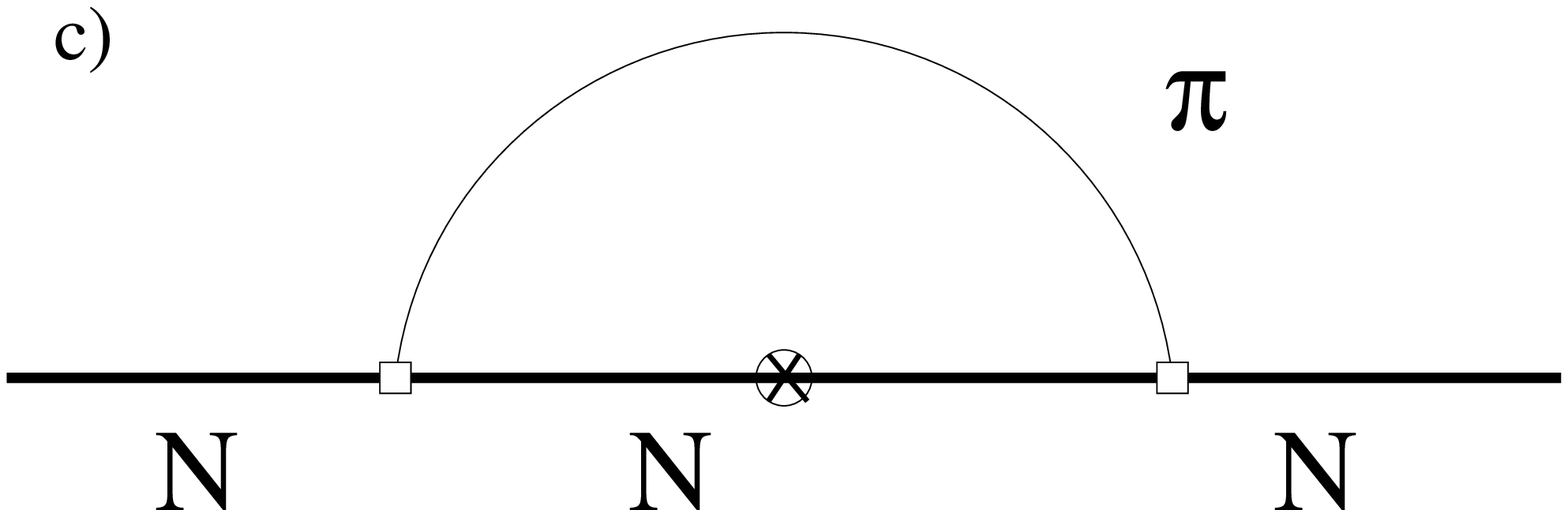}
\end{minipage}\vspace{-0.3cm}
}
\end{center}
\caption{Nucleon self-energy diagrams in $\chi PT$ at $O(p^4)$.}
\label{fig:pNN2}
\end{figure}

We begin with the Heavy Baryon $\chi PT$. The self energy of the nucleon propagator is considered
on-shell, i.e.\ with external momentum zero.
The $O(p^3)$ correction to the quadratic dependence on $\mps$ from $HB\chi PT$ 
is represented by the loop diagram in  Fig.~\ref{fig:pNN2} on the left. 
Evaluating the diagram one finds for the mass at $O(p^3)$:
\begin{eqnarray}
m^{(3)}&=&m_0-4\,c_1\mps^2-\frac{3\,(g_A)^2}{32\pi F_\pi^2}\,\mps^3+
{\cal O}(p^4)\;.\label{HBmass}
\end{eqnarray}
In quenched chiral perturbation theory, 
the $\mps^3$ contributions are modified, and new terms proportional to $\mps$ and $\mps^2\ln \mps$
are introduced~\cite{sharpe1996}. 

\begin{figure}[t]
\begin{center}
\centerline{
\epsfysize=5cm \epsfbox{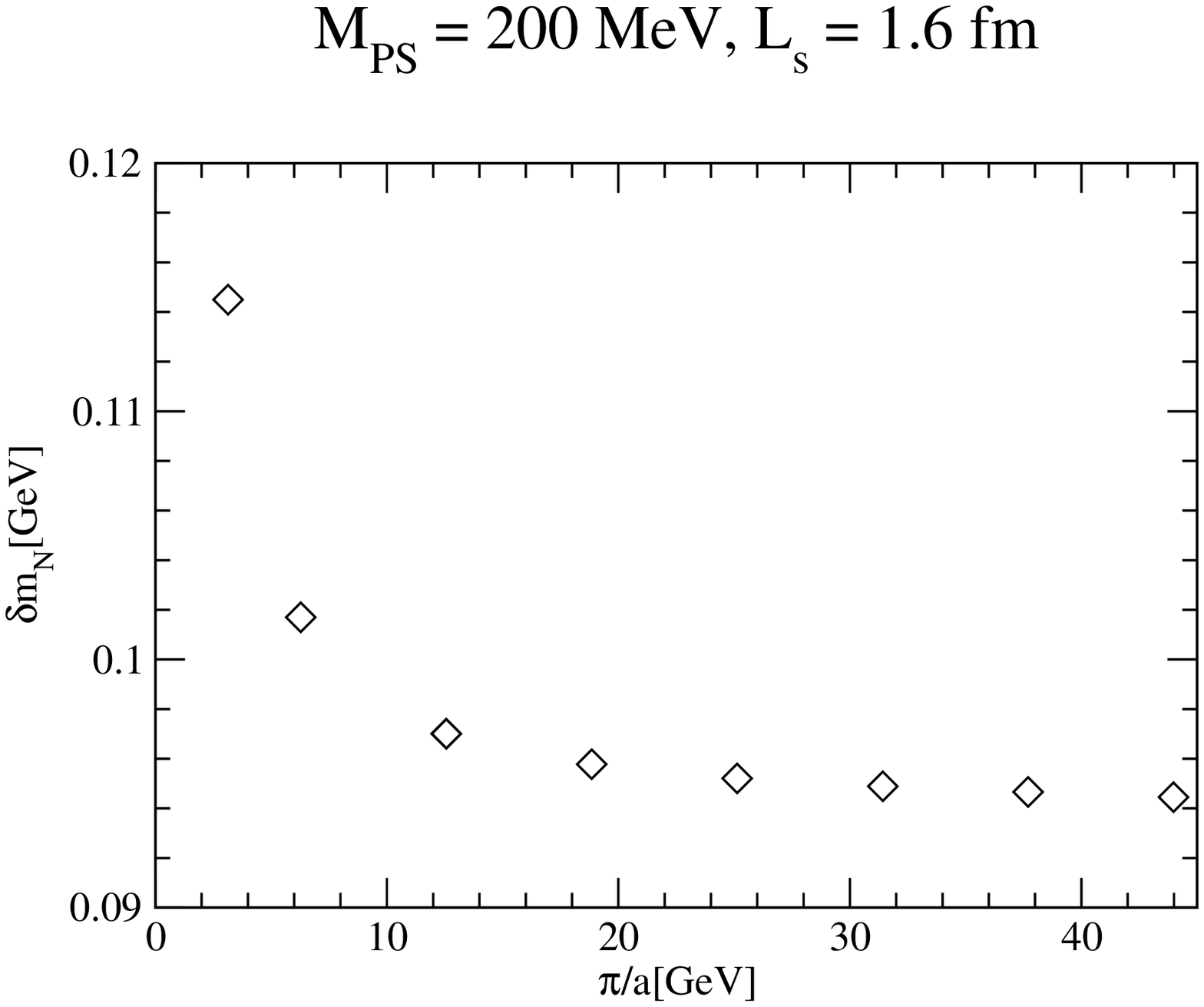}
\epsfysize=5cm \epsfbox{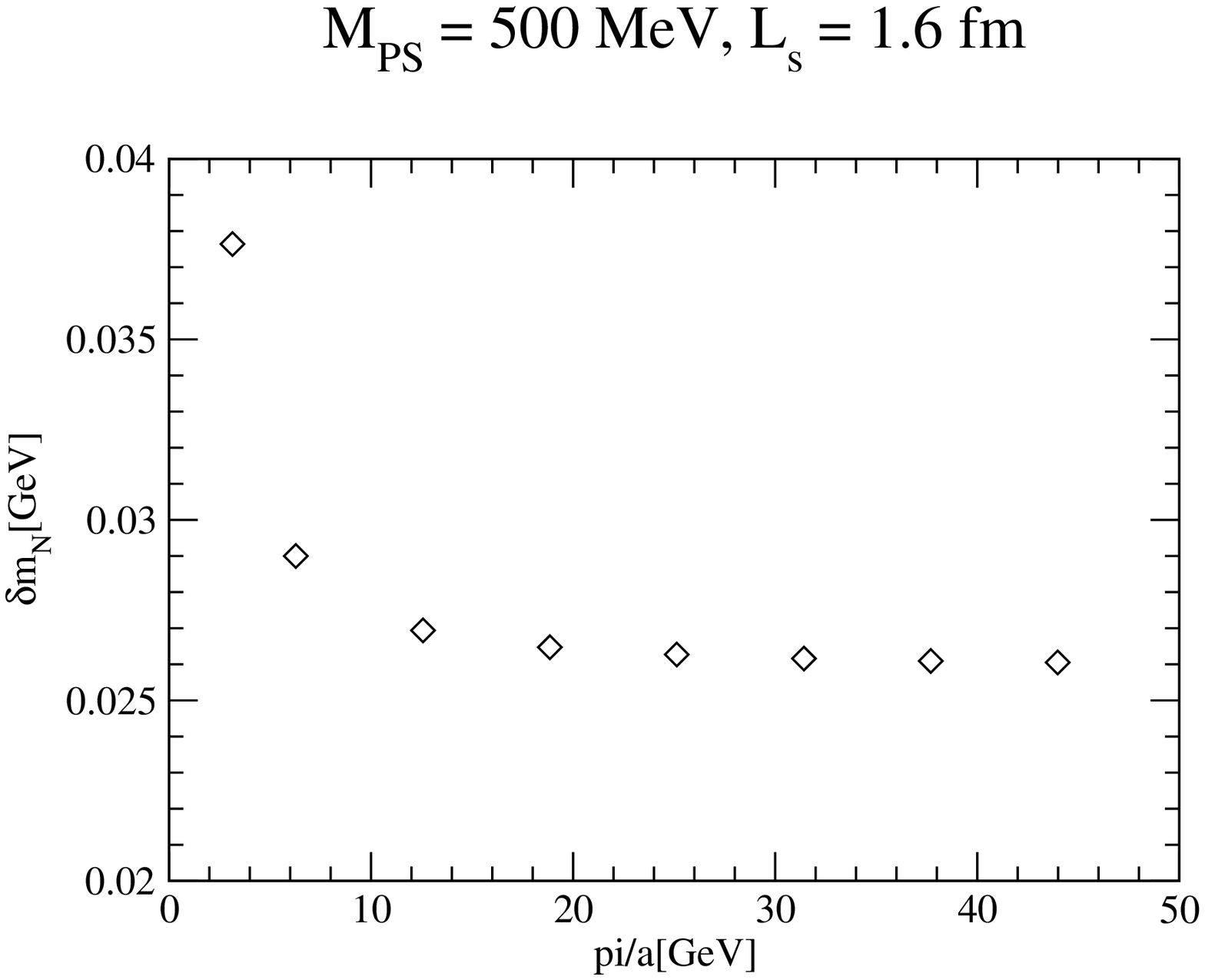}
}
\caption{Finite size effects on the nucleon mass from lattice heavy baryon chiral perturbation theory
at $O(p^3)$ as a function of the inverse lattice spacing.}
\end{center}
\label{fig:FVlattice}
\end{figure}

One can also attempt to calculate the finite size effect using lattice $\chi PT$. The lattice
provides a way to regularize integrals, if one use substitutes the continuum $HB \chi PT$
actions for the nucleons and pions with lattice ones.
In Fig.~\ref{fig:FVlattice} we study the dependence of the
finite volume effect on the lattice cutoff for various pion masses. For $\mps$ around
500 MeV, one finds a  10\% difference between a lattice cutoff of $a^{-1}=2$ GeV 
and the continuum limit $a \to 0$, whereas the difference is almost 50\% if
a lattice with $a^{-1} = 1$ GeV is assumed. For $\mps = 200$ MeV, the differences are
slightly less than 10\% and 20\% respectively.

$HB \chi PT$ at $O(p^3)$ yields a pion mass dependence rather different from the nucleon mass lattice
data~\cite{alikhan2003,alikhan2002 }. Moreover, finite size effects predicted by $HB \chi PT$ are found to be clearly 
too small with respect to the finite size effects of the lattice data. The contribution of high pion
momenta in the internal  loops is of particular interest since on a lattice of $1.6$ fm extent the lowest
mode satisfying periodic boundary conditions has a momentum of $\sim 0.8$ GeV, which  implies 
that there is only one mode within the domain where $\chi PT$ is theoretically expected to be valid.
Using lattice regularized
chiral perturbation theory one sees an indication that the contributions from pion modes at high momenta 
can indeed be sizable, in particular for larger pion masses. An example is plotted in 
Fig.~\ref{fig:FVlattice}. 
Including a $\Delta$-nucleon coupling, $HB\chi PT$ 
predicts roughly  $50-60\%$ of the actual finite size effect for pion masses around 500 MeV~\cite{alikhan2002}.

Also, since for the unquenched clover lattice results, quark masses are rather large such
that $\mps/m_N \sim 0.5$, the non-relativistic approximation does not appear particularly justified.
The calculation was therefore repeated with a relativistic formalism.

In the relativistic formalism at $O(p^3)$, the nucleon mass can be written as~\cite{procura2004}:
\be
m = m_0 - 4c_1\mps^2 + m_a + e_1 \mps^4.
\ee
$m_a$ is the one-loop contribution to the mass renormalization. 
The corresponding Feynman diagram is shown in Fig.~\ref{fig:pNN2} on the left.
The term proportional to $\mps^2$ comes from the tree-level Lagrangian  ${\cal L}^{(2)}$,
and the term proportional to $e_1$ is necessary as a counterterm to regularize 
the loop integral, although it contributes to the Lagrangian formally only at $O(p^4)$. 
Using the infrared regularization scheme for baryons explained in~\cite{becher1999}, 
\begin{eqnarray}
 m_a & = & 
-i\frac{3g_A^2m_0\mps^2}{2F^2}\int_0^\infty dx \int\frac{d^4p}{(2\pi)^4}
\left[p^2-m_0^2x^2 - \mps^2(1-x)+i\epsilon\right]^{-2}  
\end{eqnarray}
in Minkowski space. To summarize the idea of infrared regularization one might say that
the integral is decomposed into an infrared finite and an infrared singular part; the infrared
singular part is retained in the calculation which is relevant to the calculation of the low
energy structure of the theory. The Feynman parameter then runs from 0 to 
$\infty$ instead from 0 to 1. Calculation of the integral yields the following expression for 
the nucleon mass at $O(p^3)$:
\begin{eqnarray} 
m_N &=& m_0 - 4c_1 \mps^2 
+ \left[e_1(\lambda) + \frac{3g_A^2}{64\pi^2F^2m_0}
\left(1-2\ln\frac{m_{PS}}{\lambda}\right)\right]m_{PS}^4 \nonumber \\
&- & \frac{3g_A^2}{16\pi^2F^2}m_{PS}^3\sqrt{1-\frac{m_{PS}^2}{4m_0^2}} 
\left[\frac{\pi}{2} + \arctan \frac{m_{PS}^2}{\sqrt{4\mps^2 m_0^2 - \mps^4}}\right].
\end{eqnarray}

At $O(p^4)$ there is an additional tadpole diagram with an insertion from the 
$c_1$ term in ${\cal L}^{(2)}$ (see Eq.~\ref{eq:l2} and in Fig.~\ref{fig:pNN2},
middle), and the nucleon mass then reads:
\be
m_N = m_0 - 4c_1\mps^2 + m_a + m_b  + e_1 \mps^4. \label{eq:chiextrap}
\ee
The tadpole contribution, $m_b$, is given by the integral
\begin{eqnarray} 
m_b &=& - \mathrm i \frac{3}{f_\pi^2} \int \! \frac{\mathrm d^4 k}{(2 \pi)^4} 
 \frac{2 c_1 \mps^2 - (c_2 + c_3) (k^0)^2 + c_3 \vec{k}^{\,2}}
     { \mps^2 - k^2 - \mathrm i \epsilon}.
\end{eqnarray}
For this diagram, infrared and dimensional regularization give the same result.
The diagram on the left of Fig.~\ref{fig:pNN2} also receives a correction 
due to a third diagram with a $c_1$ insertion to the nucleon propagator (the right diagram
in  Fig.~\ref{fig:pNN2}). 
If the nucleon mass is redefined using $m_0 \to m_0 - 4c_1\mps^2$, the contribution of
the additional graph is absent.

The $\arctan$ can be expanded in powers of $\mps$, with the first terms being:
\begin{eqnarray}
 m_N & = &\Black{ m_0 - 4 c_1 \mps^2}-\frac{3 g_A^2}{32 \pi f_\pi^2} \mps^3 
  \nonumber \\
& & +\left[\Black{e_1^r(\lambda)}-\frac{3}{64 \pi^2 f_\pi^2}
    \left( \frac{g_A^2}{m_0} - \Black{\frac{c_2}{2}} \right) 
  - \frac{3}{32 \pi^2 f_\pi^2}
       \left( \frac{g_A^2}{m_0} - \Black{8c_1 + c_2 + 4 c_3} \right)
   \ln{\frac{\mps}{\lambda}} \right] \mps^4
\nonumber \\ & {} & {}
 + \frac{3 g_A^2}{256 \pi f_\pi^2 m_0^2}\mps^5 + O(\mps^6) \,.
\end{eqnarray}
If chiral perturbation theory is a valid low-energy description of QCD, this formula
should be predict the relation between the pion and nucleon mass if the quark mass
is small.

\begin{figure}[thb]
\begin{center}
\epsfysize=6cm \epsfbox{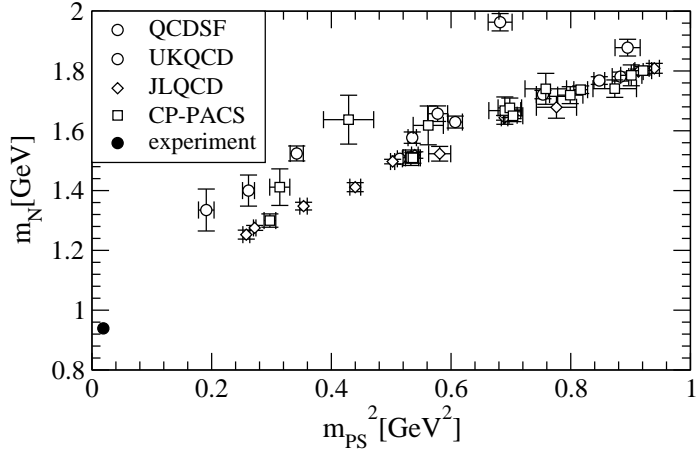}
\end{center}
\caption{Nucleon mass from QCDSF and UKQCD~\protect\cite{meinulf2004}, 
CP-PACS~\protect\cite{cppacs2002hadr} and JLQCD~\protect\cite{aoki2002}.}
\label{fig:allmNdata}
\end{figure}

A collection of lattice nucleon masses for $N_f = 2$ from clover-improved Wilson quarks is
plotted in Fig.~\ref{fig:allmNdata}. The QCDSF and UKQCD collaboration~\cite{meinulf2004} uses
non-perturbatively $O(a)$ improved clover quarks and Wilson gluons.
The UKQCD analysis is discussed in~\cite{allton2002}. The JLQCD collaboration uses 
the same quark and gluon actions~\cite{jlqcd2003}, and the CP-PACS collaboration 
uses tadpole-improved clover quarks and RG-improved gluons~\cite{cppacs2002hadr}.

\begin{figure}[thb]
\begin{center}
\epsfysize=6cm \epsfbox{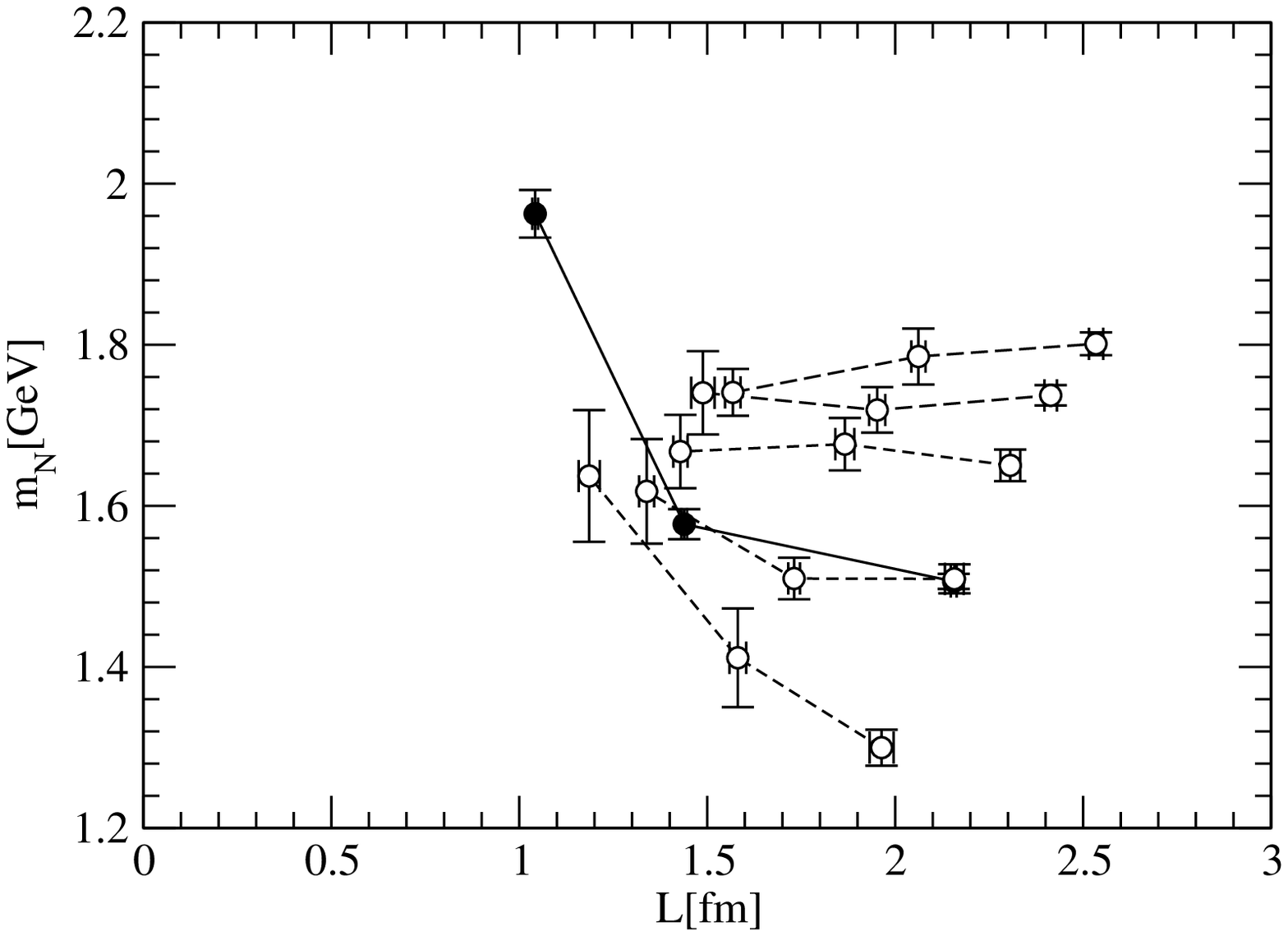}
\end{center}
\caption{Finite size effect of the nucleon mass for JLQCD~\protect\cite{aoki2002}
(open circles) and QCDSF/UKQCD data (filled circles)~\protect\cite{meinulf2004}. Lines connect 
sets belonging to the same $\beta, \kappa$ values.}
\label{fig:FVdata}
\end{figure}

The lattice spacing is set using $r_0 = 0.5$ fm
at each finite value of the sea quark mass. The data are not on a universal curve. If we
compare results at various values of $\beta$, it appears that
discretization effects in the relation between pion and nucleon masses are small.
In Fig.~\ref{fig:FVdata} we plot results from QCDSF and JLQCD  as a function of the spatial lattice 
size at fixed values of simulation parameters $\beta$ and $m_{sea}$. For smaller pion masses, 
one finds an increase of the mass on smaller lattices. 
If only data on larger lattices are considered (here $\mps L > 5$), they do lie on a universal 
curve and we use the results with pion masses $< 800$ MeV to fit to the chiral perturbation 
theory formula. 
\begin{figure}[htb]
\begin{center}
\centerline{
\epsfysize=5.4cm \epsfbox{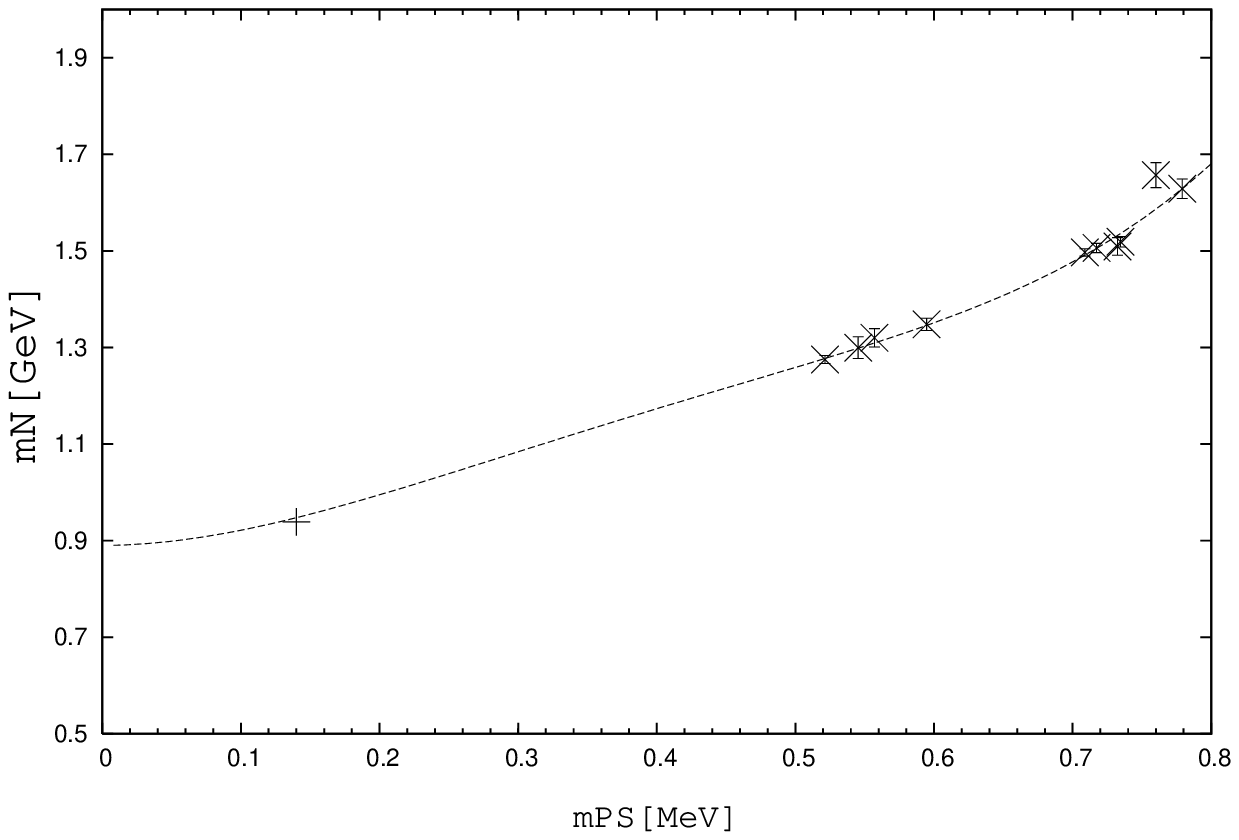}
}
\end{center}
\caption{Chiral extrapolation of nucleon masses on large lattices
in two-flavor QCD. The crosses denote the lattice results. The dashed line denotes relativistic  $\chi PT$ 
at $O(p^4)$. The plus denotes the experimental value.}
\label{fig:mNfit} 
\end{figure}

Setting $c_2 = 3.2$ GeV$^{-1}$ which is in reasonable agreement with the experimental
values and $c_3 = -3.4$ GeV$^{-1}$, the other parameters can be determined by a fit
to the lattice results from large lattices and  are 
\ba
  m_0 & = & 0.89(6) \mbox{ GeV}, \\ 
  c_1 & = & -0.93(5) \mbox{ GeV}^{-1}, \\
  e_1 & = & 2.8(4) \mbox{ GeV}^{-3}, \\
\ea
with $g_A$ = 1.267, $\lambda = 1$ GeV.

As seen in Fig~\ref{fig:mNfit}, the formula with these parameter choices is in good agreement with the 
data, and the chiral extrapolation is in agreement with experiment.  
Fitting the lattice results to the $O(p^3)$ formula one finds relatively similar values for the fitting
parameters~\cite{alikhan2003}, although the coefficient of the additional term is numerically not very small: 
$(8 c_1 + c_2 + 4 c_3) \sim 18-25$ and $3/(32\pi^2f_\pi^2) \simeq 1$.

It should be noted that with a
simple three parameter fit allowing for constant and quadratic and cubic or quartic contributions from 
the pion mass, one can also obtain a reasonable description of the lattice data with
$\chi^2/dof \simeq 2$. However then the nucleon mass extrapolates to a chiral limit of $\sim 1.2$ GeV.

In Ref.~\cite{vbernard2004}, a good description of nucleon data up to pion masses $\sim 600$ MeV could 
be achieved using a non-relativistic formalism at $O(p^4)$.
Octet baryon masses in QCD with $2_{light}+1_{strange}$ flavors of sea quarks have been calculated in 
Ref.~\cite{frink2004}  at $O(p^4)$ using relativistic heavy baryons with infrared regularization.
The octet baryon mass has been calculated for 3 flavors at $O(p^3)$ in Ref.~\cite{chen2002}, where
partially quenched heavy baryon $\chi PT$ has been first formulated.
Partially quenched results have been presented for the decuplet baryon mass 
at $O(p^4)$ for 3 flavors in Ref.~\cite{tiburzi2005a}, for 2 and 3 flavors in
Ref.~\cite{tiburzi2005b} and for 3 flavors at $O(p^4)$ in Ref.~\cite{wloud2005}. In 2-flavor partially
quenched $\chi PT$ the nucleon mass was calculated by Ref.~\cite{bean2002}, and the nucleon and decuplet masses
to $O(p^4)$ by Ref.~\cite{tiburzi2005c}.

The corrections to the $\rho$ meson mass  within chiral perturbation theory using 
infrared regularization are described in~\cite{meissner2004}.

In the following we attempt to describe the finite volume effect using $\chi PT$ in the $p$ expansion.
The temporal extent of the lattice is assumed to be infinite.

On a finite lattice, the integral in the self energy correction is substituted by a sum over
the allowed modes. The difference between sum and integral is not ultraviolet divergent and
can be brought into a closed form according to~\cite{hasenfratz1990}
\begin{eqnarray}
 & &  
\frac{\Gamma (r)}{L^3} \sum_{\vec{k}} \left( \vec{k}\,^2 + M^2 \right)^{-r}
- \frac{\Gamma (r)}{(2 \pi)^3} \int \! \mathrm d^3 k 
 \left( \vec{k}\,^2 + M^2 \right)^{-r} 
\nonumber \\ & & \quad {}
 = \frac{1}{4 \pi^{3/2}} \sum_{\vec{n}\neq 0} \, 
  \left( \frac{L^2 |\vec{n}|^2}{4 M^2} \right)^{r/2 - 3/4} 
  K_{r-3/2} (\sqrt{L^2 |\vec{n}|^2 M^2}) \,, \nonumber
\label{basic}
\end{eqnarray}
where $K_\nu (x)$ is a modified Bessel function.

\begin{figure}[htb]
\begin{center}
\centerline{
\epsfysize=6cm \epsfbox{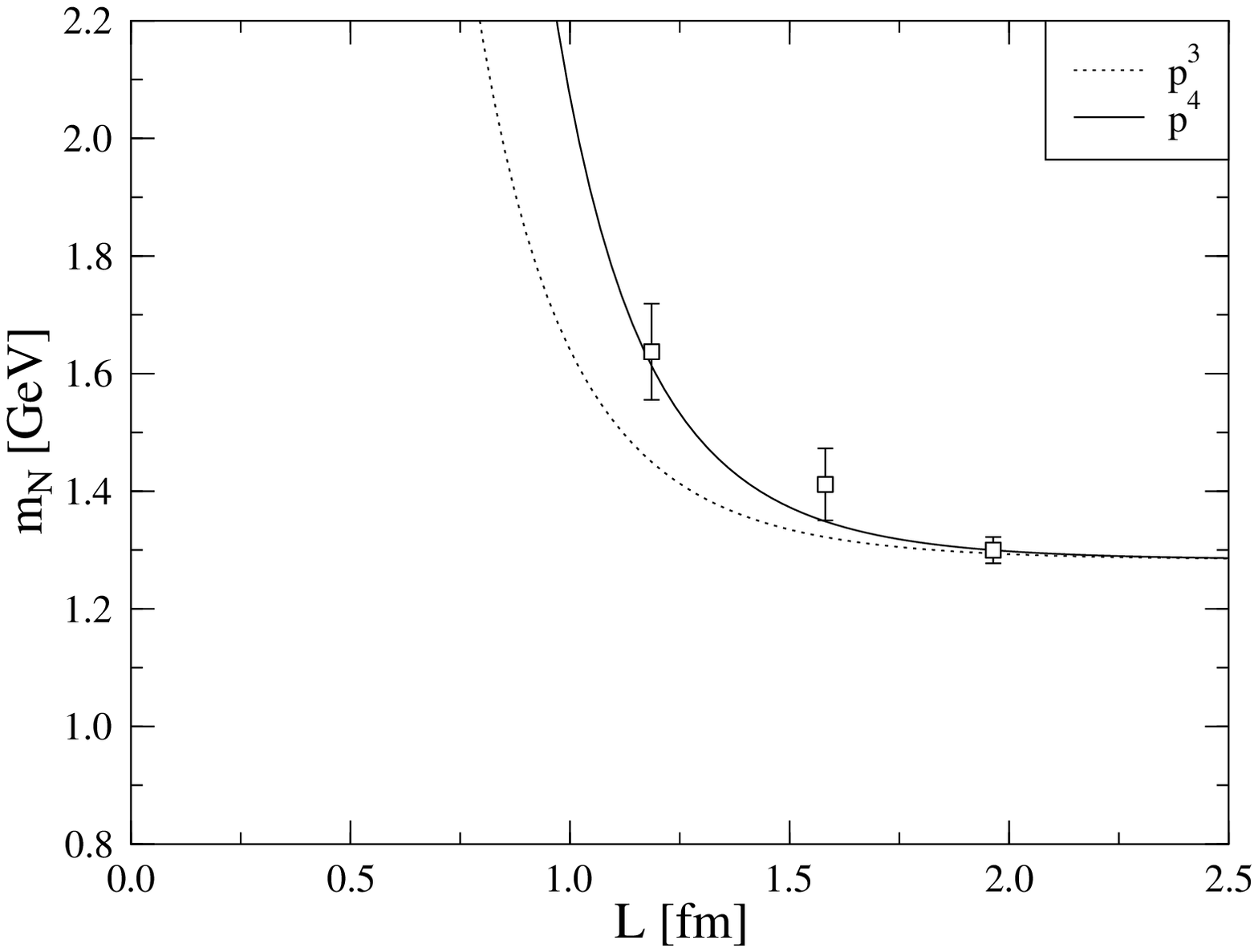}
}
\end{center}
\vspace{-0.7cm}
\caption{Volume dependence of the nucleon mass relativistic $\chi PT$ compared with
$N_f = 2$ lattice results, from Ref.~\protect\cite{meinulf2004}. Solid line: $O(p^4)$ result, 
dashed line: only $O(p^3)$ terms.}
\label{fig:FS}
\end{figure}

The contribution of diagram $a$ to the finite size effect is
\be
\delta m_a(L) =  \frac{3 g_A^2 m_0 m_{PS}^2}{16 \pi^2 f_\pi^2}
 \int_0^\infty \! \mathrm d x \,
 \sum_{\vec{n}\neq 0} \, 
  K_0 \left( L |\vec{n}| \sqrt{m_0^2 x^2 + \mps^2 (1-x)} \right) \,.
\label{Deltaa}
\ee
The contribution of diagram $b$ is 
\be
\delta m_b(L) = \frac{3\mps^4}{4\pi^2f_\pi^2} \sum_{\vec{n} \neq 0}
\left[ (2c_1 - c_3)\frac{K_1(L|\vec{n}|\mps)}{L|\vec{n}|\mps} + 
c_2\frac{K_2(L|\vec{n}|\mps)}{(L|\vec{n}|\mps)^2}\right].
\ee
The $\chi PT$ result at $O(p^4)$ indeed finds 
a good agreement with the finite size effects of the lattice results.
An example for a pion mass around 550 MeV is given in Fig.~\ref{fig:FS}. 
Agreement is also found for larger pion masses,  $\mps \sim 700$ MeV.
$\chi PT$ is considered to be an effective field theory valid for momentum scales $ < 1$ GeV.
Using a cutoff regularization, Ref.~\cite{vbernard2004} calculates a validity bound of 
$\chi PT$ for pion masses less than 600 MeV. Therefore further verification of the 
nucleon mass formula in comparison to lattice results would be very important, e.g.\  
by comparison with simulation data at smaller quark masses once 
they will be available. It should be mentioned
that the result is not very sensitive to the exact experimental value of $r_0$.

Good agreement of lattice nucleon data at relatively high masses with chiral perturbation 
theory is also achieved if a finite nucleon radius is taken into account using
a dipole form factor~\cite{awthomas2002}.

Finite volume effects on $f_K$ and $B_K$ are discussed in \cite{becirevic2004a}
in dynamical, partially quenched and quenched QCD. Finite volume effects
on heavy-light meson matrix elements are discussed in~\cite{arndt2004}.
In the next section we discuss formalisms which we used to simulate $b$ quarks on the lattice.
\section{HEAVY QUARKS ON THE LATTICE\label{sec:heavy}}
HQET predicts $1/M$ expansions for various mass splittings, decay constants and heavy
meson decay form factors, where $M$ denotes the heavy quark mass. The expansions are useful for 
guiding extrapolations or interpolations of 
experimental or lattice results as a function of the heavy quark or meson mass.

Moreover, in  simulations of massive quarks using a relativistic action,
discretization effects of the form $O((aM)^n)$ where $n$ is a positive number,
can occur. This effect can be avoided by simulating the heavy quarks  on the
lattice using Heavy Quark Effective Theory  or Non-relativistic QCD.
\subsection{Lattice HQET \label{sec:lhqet}}
The HQET Lagrangian for the static quark can be put on the lattice 
directly~\cite{eichten1990}. In
the rest frame of the heavy  meson, the static  tree-level lattice Lagrangian
is given by:
\be
\psi^\dagger \nabla_4^\ast \psi = 0.
\ee
Beyond tree-level there is an additive mass renormalization.
Corrections to the infinite mass limit break the spin symmetry by terms
$O(1/M_Q)$. Relativistic corrections are incorporated in a systematic
expansion in terms $O(\lqcd/M_Q)$.

The formalism enabled many studies of $B$ meson matrix elements. For one of the
first very extensive ones see~\cite{duncan1995}. Recent work using 
non-perturbative renormalization is discussed in \cite{heitger2004,heitger2004a}.

\subsection{Lattice NRQCD\label{sec:NRQCD}}
In NRQCD on the lattice one simulates a discretized version of the action corresponding to
Eq.~(\ref{eq:NRQCD}) which includes the terms to the desired order in  $1/M$.
One solves a discretized Pauli equation with a covariant coupling to the color gauge field
and relativistic corrections. In this section we summarize some general features of the
calculation of masses and decay constants in NRQCD on the lattice, in particular based on
the discussions in Refs.~\cite{lepage1992,morningstar93,morningstar98}.

\subsubsection{NRQCD lattice action\label{sec:latNRQCD}}
Analogous to the Dirac fermion action, the Hamiltonian~(\ref{eq:NRQCD_cont}) can be put 
on the lattice by substituting the derivatives by lattice derivatives.
The dominant discretization error in the spatial derivative of the heavy quark action
is $O(a^2p_i^2) \sim \lqcd^2 a^2$, and $O(a M v^2/2)$ and the dominant discretization error in the 
temporal derivative should be proportional to the kinetic energy: $E_{\mathrm{kin}} a
= \vec{p}\,^2/(2M) a = \lqcd^2 a/(2M)$. Since in typical NRQCD calculations $a \sim M^{-1}$,
the discretization corrections are introduced for consistency in 
calculations which are supposed to be correct at $O(1/M^2)$.
Discretization errors in the hyperfine interaction are of higher order in the expansion in 
$1/M$ and $a$.  

The lattice Lagrangian density used for Refs.~\cite{alikhan1998,alikhan2000} is
\ba
  L & = & 
  \psi_t^{\dagger} 
    \psi_t  \nonumber  \\
& -& \psi_t^{\dagger} \left( 1-\frac{a\delta H}{2} \right)_t
    \left( 1-\frac{aH_0}{2n} \right)_t^n 
    U^\dagger_4
    \left( 1-\frac{aH_0}{2n} \right)_{t-1}^n 
    \left( 1-\frac{a\delta H}{2} \right)_{t-1}
    \psi_{t-1}
  \label{eq:evol},
\ea
where the Hamiltonian is $H = H_0 + \delta H$. The contributions of the
$1/M^2$ terms to the low-lying spectrum and $f_B$ is small, therefore also
some recent calculations include only the $1/M$ terms, i.e.\ the kinetic energy and the 
spin-magnetic operator, (e.g.~\cite{cppacs2001NR}), and using $M_0$ for the bare lattice
mass, the kinetic energy operator becomes
\begin{eqnarray}
  H_0 & \equiv & - \frac{\delsq}{2M_0},
\end{eqnarray}
with the second lattice derivative given by
\be
\delsq \equiv \sum_i \nabla_i \nabla^\ast_i=\sum_i \nabla_i^\ast \nabla_i.
\ee
$M_0$ is the bare lattice heavy quark mass.
\begin{eqnarray}
\delta H 
&=& - \frac{g}{2\Mbz}\,\sigmav\cdot\Bv \nl
& & + \frac{ig}{8(\Mbz)^2}\left(\delv\cdot\Ev - \Ev\cdot\delv\right)
 - \frac{g}{8(\Mbz)^2} \sigmav\cdot(\delv×\Ev - \Ev×\delv)\nl
& & - \frac{(\delsq)^2}{8(\Mbz)^3} 
  + \frac{a^2\delfour}{24\Mbz}  - \frac{a(\delsq)^2}{16n(\Mbz)^2} \;.
\label{deltaH}
\end{eqnarray}
The parameter $n$ is introduced to avoid instabilities due to large momenta 
contributing to the kinetic energy operator $H_0$ which 
cause an increase of the correlation function in time. For $p_i ~\sim
1/a$, the instability can be avoided if $n > 3/(2Ma)$.
In all previous calculations, the coefficients are set to their tree-level value (one) with mean field 
improved gauge links. 
The last two terms in Eq.~(\ref{deltaH}) are corrections for the leading discretization errors
of the kinetic energy operator and the temporal derivative respectively. The fourth derivative 
in the kinetic energy correction is defined as
\be
\delfour \equiv \sum_i (\nabla_i \nabla^\ast_i)^2.
\ee

One can use this action to simulate $b$ quarks directly on the lattice,
since $O((aM)^n)$ discretization errors do not occur.
\subsubsection{Meson mass in NRQCD}
The exponential decay rate of the correlation function in NRQCD or HQET,
denoted by $E_{\mathrm{sim}}$, 
is the unrenormalized meson binding 
energy whose contributions should after renormalization be related to the
energies appearing in the $1/M$ expansion of the HQET:
the binding energy of the light quark due to the static color source, the heavy quark
kinetic energy, the spin-colormagnetic energy and so on. 

Also in NRQCD, the inverse propagator at one loop, $S(p)$,  is related to the
tree-level propagator $S^{(0)}(p)$ by the formula~\cite{morningstar93}
\ba
S^{-1}(p) &=& S^{(0) -1}(p) - \Sigma(p),
\ea
where $\Sigma(p)$ is the heavy quark self-energy.
By matching the heavy quark dispersion relation to the continuum~\cite{morningstar93}, 
one can find the one-loop on-shell mass and wave function renormalization constants.
$C_m M_0$ is identified with the pole mass of the heavy quark, $M_Q$.
At one loop, $C_m = 1 + O(\alpha_s)$.
The quark mass receives an additive renormalization which starts at $O(g^2)$
and has the form 
\ba
E_0 &=& -\log(1 + \alpha_s A).
\ea
This cancels the leading $1/a$ correction in the exponential decay rate 
of the heavy quark propagator.
The wave function renormalization also starts at $O(1)$, 
$C_Q = 1 + O(\alpha_s)$.


The mass $M_{HL}$ 
of a heavy-light hadron with an NRQCD heavy quark is related to the 
exponential decay rate of the hadron propagator by
\be
M_{HL} = M_{\mathrm{pert}} \equiv E_{\mathrm{sim}} + \Delta, \label{eq:hl}
\ee
where $\Delta$ contains the renormalized heavy quark mass and zero point energy
\be
\Delta = C_m M_0 + E_0,
\ee
with $C_m M_0$ being the above-mentioned heavy quark pole mass.

In hadrons with several heavy quarks or antiquarks of bare mass $M_0$, the mass renormalization should
contribute in the same way for each heavy quark, i.e.\ the quarkonium mass $M_{HH}$ can be
calculated using:
\be
M_{HH} = E_{\mathrm{sim}, HH} + 2 \Delta. \label{eq:hh}
\ee

The meson mass can be determined in simulations by considering the momentum 
dependence of the meson binding energy. Formulae (\ref{eq:hl}) and (\ref{eq:hh})
should also hold for the mass of excited states and of meson states at finite  
momentum. Then, the energy difference between zero and finite momentum 
should be described by a dispersion relation:
\be
E_{\mathrm{sim}}(p^2) - E_{\mathrm{sim}}(0) = \sqrt{M^2_{\mathrm{kin}} + \vec{p}^2} - 
M_{\mathrm{kin}}. \label{eq:disp}
\ee
$M_{\mathrm{kin}}$ is the kinetic mass of the heavy meson.

Mass splittings are simply given by the difference of the binding energies of the
states. 

If states are close together, statistical fluctuations of
the individual correlators may be larger than the energy difference. If the 
overlap with excited states is neglected, one can however extract the energy 
difference from a fit of the ratio of correlators to an exponential 
decay~\cite{lidsey1994}
\be
\frac{C_1(t)}{C_2(t)} \propto \exp(-\Delta E t), \label{eq:ratiofit}
\ee
where $\Delta E$ is the energy splitting of the ground states belonging to the correlation 
functions $C_1$ and $C_2$.
%
\subsubsection{Currents in NRQCD \label{sec:cNRQCD}}
If matrix elements of operators
are to be calculated in this formalism, the $1/M$ corrections to the operators have to be 
taken into account as well. 
To determine for example $f_B$ from the lattice, it is necessary to calculate the 
renormalization factors to match the unrenormalized lattice matrix element of the axial vector
current to the corresponding matrix element in continuum QCD. 
The matching is discussed e.g.\ in~\cite{morningstar98,morningstar99}, and we briefly repeat
some features of interest here.
The bare current operators at $O(1)$ in the $1/M$ expansion are the same as in full
QCD. For the time component of the axial vector current this means
\ba
A_4^{(0)} &=& \overline{q}(x) \gamma_\mu\gamma_5 Q(x),
\ea
where the heavy quark spinor $Q$ is projected onto the upper two components:
\be
Q(x) = \left( \begin{array}{c} 
                    \psi(x) \\ 0 \\
                  \end{array} \right).
\ee
The tree-level corrections occuring at higher order can be derived applying a 
Foldy-Wouthuysen transformation on the Dirac heavy quark spinor. At $O(1/M)$ this is
\ba
A_4^{(1)} &=& -\frac{1}{2M_0} \overline{q}(x) \gamma_\mu\gamma_5 \vec{\gamma}\cdot \vec{\nabla} 
Q(x).
\ea
Beyond tree-level, all operators with the appropriate symmetries can contribute. 
Some of them are related to each other by equations of motion. For practical 
calculations it is sufficient to consider a linearly independent set of operators.
At $O(1/M)$, there is only one additional operator and one can choose:
\ba
A_4^{(2)} &=& -\frac{1}{2M_0} \overline{q}(x) \vec{\gamma}\cdot \stackrel{\gets}{\nabla}\gamma_0
\gamma_\mu\gamma_5  Q(x).
\ea

The corrections at $O(1/M^2)$ are given by
\ba
A_4^{(3)} &=& \frac{1}{8M_0^2} \overline{q}(x)  \gamma_\mu\gamma_5 \vec{\nabla}^2 Q(x), \nonumber \\
A_4^{(4)} &=& \frac{g}{8M_0} \overline{q}(x) \gamma_\mu\gamma_5 \vec{\Sigma}\cdot \vec{B} Q(x), \nonumber \\
A_4^{(5)} &=& -\frac{2ig}{8M_0} \overline{q}(x) \vec{\alpha}\cdot \vec{E} Q(x), 
\ea
with $\vec{\alpha} = \gamma_0\vec{\gamma}$, $\vec{\Sigma } = \mathrm{diag}(\vec{\sigma},
\vec{\sigma})$. 
They have also been calculated on the lattice~\cite{alikhan2000,ishikawa2000,collins2001}, but without
inclusion of the matching renormalization constants.

In~\cite{morningstar98}, 
the continuum matrix element is first matched to the matrix element in the continuum effective
theory which is an expansion in terms of the inverse heavy quark pole mass. Each of the continuum matrix 
elements then equals the renormalized lattice matrix element at the same order:
\be
\langle A_4^{QCD} \rangle = \sum_i \eta_i \langle A^{(i)}_{4,\mathrm{ren}}\rangle 
= \sum_i \eta_i \sum_j Z_{ij}^{-1} \langle A^{(j)}_{4}\rangle ,
\ee
where the lattice renormalization factors have the structure
\be
Z_{ij} = \delta_{ij} + \alpha_s\left(\frac{1}{2}(C_Q+C_q)\delta_{ij} + C_m\delta_{ij}\delta_{1j}
+\zeta_{ij}\right).
\ee
Because of the mixing of lattice operators, the renormalization constants depend on
two indices $i,j$.
Expanding the renormalization factors in terms of their lattice and continuum contributions, 
one finds~\cite{sgo1997}
\ba
\langle A_{4, QCD} \rangle &=& \left[1 + \alpha_s \rho_0\right]
\langle A^{(0)}_4\rangle + \left[1 + \alpha_s \rho_1\right]\langle A^{(1)}_4\rangle
+ \alpha_s \rho_2 \langle A^{(2)}_4\rangle,
\ea
with the one-loop coefficients 
\ba
\rho_0 &=& B_0 - \zeta_{00} - \zeta_{10}
- \scriptsize{\frac 12}(C_Q + C_q), \nonumber \\
\rho_1 &=& B_1 - \zeta_{01} - \zeta_{11} - \scriptsize{\frac 12}(C_Q 
+ C_m + C_q),  \nonumber \\
\rho_2 &=&  B_2 - \zeta_{02} - \zeta_{12} - \scriptsize{\frac 12}(C_Q 
+ C_q).
\ea
$B_i$ denote the continuum matching factors for HQET to QCD, $\zeta_{ij}$ the lattice 
vertex renormalizations (which contribute to mixing of matrix elements for $i\neq j$), 
$C_m$ the heavy  quark mass renormalization and $C_Q$ and $C_q$ the 
heavy and light quark wave function renormalizations. For tadpole-improved calculations,
additional contributions to the renormalization factors have to be included~\cite{morningstar98}. 

The 
continuum matrix element is independent of the lattice spacing. Since NRQCD results are always 
at finite lattice spacing, this is only true up to lattice spacing corrections. However one can
derive an approximate renormalization group equation~\cite{morningstar98}:
\be
a\frac{d}{da} \sum_kC^k(\alpha_s, aM_0) \langle A_4^{(k)}\rangle = 0 + O(a^2),
\ee
with $C^k(\alpha_s,aM_0) = 1 + \alpha_s \rho_k$.
One can derive the scaling behavior of the coefficient functions from this:
\be
\left(a\frac{d}{da} \delta^{ij} - \gamma^{ij}\right)C^j = 0.
\ee
$\gamma$ is the anomalous dimension matrix of the current operators. Since the coefficients are functions of
the lattice scale and $\alpha_s$, and $\alpha_s$ is also a function of the the lattice scale, the renormalization
group equation  is given by
\be
\left(\left[a\frac{\partial}{\partial a} - \beta\frac{d}{d\alpha_s}\right]\delta^{ij} - 
(\gamma^{T})^{ij}\right)C^j = 0,
\ee
where $\gamma$ is the anomalous dimension matrix of the axial vector current in NRQCD, and the 
$\beta$ function  determines the running of $\alpha_s$ with the matching scale, which is 
inversely proportional to $a$:
\be
\beta = - a \frac{d}{da}\alpha_s(a^{-1}).
\ee
In the leading-log approximation \cite{morningstar98}, $\gamma$ is found to be:
\be
(\gamma^{ij}) = \mathrm{diag}
\left(-\frac{\alpha_s}{\pi},-\frac{\alpha_s}{\pi},-\frac{\alpha_s}{\pi}\right)
\ee 
Therefore there is a simple scaling relationship of the Wilson coefficients 
as a function of the lattice spacing or the inverse heavy quark mass 
\be
\frac{C^i(a_2M_0^b)}{C^i(a_1M_0^b)} = \left(\frac{\alpha_s(1/a_2)}{\alpha_s(1/a_1)}\right)^{2/b_0},
\label{eq:coeff}
\ee
$i = 1,2$, where the mass on the left-hand side is the bare $b$ quark mass.
\subsubsection{Systematic errors}
Inclusion of higher dimensional operators in the action and currents leads to 
non-renormalizability, and unrenormalized lattice results may contain divergences 
going as powers in $1/a$. A continuum extrapolation is therefore not possible. Since results
are obtained at finite $a$, they also contain $O(a^n)$ discretization errors which have
to be accounted for in the error estimate. 
To be reliable it is necessary to compare simulations at various values of the lattice spacing. 
One can attempt to make a theoretical estimate of the error size at each lattice spacing, and 
continuum estimates of physical quantities can be taken from simulations at 
parameters which minimize discretization effects as well as $1/a$ effects. The lattice with the
minimal error can be used for the continuum estimate.
The renormalization group arguments of the previous paragraph apply only if a scaling region
of lattice spacings can be found where the result remains rather constant as a function of $a$.

To compare lattice results  at a fixed heavy quark mass from a range of lattice spacings,
the leading lattice spacing dependence of the coefficients is given by Eq.~(\ref{eq:coeff}) in a 
scaling window of the resummed, renormalized currents as a function of $a$. In the calculations  of
\cite{ishikawa2000,collins2001,cppacs2001NR} it is found that the scaling of the 
currents resummed using Eq.~(\ref{eq:coeff}) is indeed flatter.
\subsection{Relativistic Heavy Quark Actions \label{sec:fnal}}
Another method to avoid large discretization errors used in many lattice calculations
is to simulate heavy quarks at smaller masses (e.g. around the charm mass) and extrapolate
to the $b$. If the simulated masses are much smaller than the $b$ mass,
this might give rise to substantial uncertainties which might however be resolved if the extrapolation
is substituted by an interpolation to an $O(a)$ improved static value. 
Static current matrix elements may contain sizable 
discretization errors as has been observed for  the axial current in Refs.~\cite{sgo1997,morningstar98}.
Renormalization constants for (partly) non-perturbative $O(a)$ improvement have been calculated recently 
by Ref.~\cite{kurth2001}.
If the simulated heavy mass is not very small in lattice units, one might worry about discretization 
errors also if a mass-independent renormalization scheme is implemented non-perturbatively.

Ref.~\cite{elkhadra1997} formulates an improvement program for on-shell quantities in heavy quark 
physics, here briefly denoted as Fermilab formalism or FNAL. An important result of their work is the 
realization that a non-relativistic re-interpretation of the 
Wilson or clover action leads to smaller discretization effects in simulations of very heavy quarks. 
Amplitudes are polynomials in $ap$, where $p$ is the three-momentum, while they are bounded functions
of the heavy quark mass in lattice units.
Similar to NRQCD, the rest mass of the heavy quark does not affect physical quantities. 
Relevant are the mass calculated from the dispersion relation and mass splittings. 
An important difference to NRQCD is that the continuum limit is well-defined.
We briefly present some of the important formulae for massive Wilson and clover quarks 
from their work, see e.g.~\cite{elkhadra1997,kronfeld2000} and~\cite{aoki1998}.       

Expanding the energy-momentum dependence of a particle around spatial momentum zero
one finds
\be
E^2 = M_1^2(1 + \left.\sum_i\frac{\partial^2 E}{\partial p_i^2}\right|_{p_i=0}
\vec{p}^2 + ...)
\ee
where equal indices are summed over. One can define the kinetic mass, $M_2$ by the
relation
\be
M_2^{-1} = \left.\sum_i\frac{\partial^2 E}{\partial p_i^2}\right|_{\vec{p}=0},
\ee
$M_2 = M_1$ for a relativistic particle, but at finite lattice spacing the relativistic
dispersion relation is modified.

The tree-level relation between the quark mass parameter in the Lagrangian and $M_1$ 
can be derived from the 
energy-momentum relation of a free Wilson fermion. If the bare mass is denoted as $m$ 
and the Wilson parameter is set to $r = 1$, this is given by
\be
\cosh a E = 1 + \frac{1}{2}\frac{(am + \frac{1}{2}a^2\vec{P}^2)^2 + a^2\vec{S}^2}{1+am+ 
\frac{1}{2}a^2\vec{P}^2}
\ee
with $S_i = \frac{1}{a}\sin ap_i$ and $P_i = \frac{2}{a}\sin(ap_i/2)$.
The energy at zero spatial momentum equals 
\be
M_1  = \frac{1}{a} \ln(1 + a m). 
\ee
$M_1$, the pole mass,  determines the exponential decay of the quark propagator. 

$M_2$ is given by:
\ba
\frac{1}{aM_2} &=& \frac{2}{am(2+am)} + \frac{1}{1+am}, \;\mbox{or} 
\nonumber \\
aM_2 &=& \exp(aM_1)\frac{\sinh(aM_1)}{1+\sinh(aM_1)},
\ea
with  $M_1=M_2$ only for small $am$. 

The Wilson and clover Lagrangians already contain the 
interaction terms of the heavy quark Lagrangian at $O(1/M)$.
The temporal derivatives of the Wilson term, combined with the $D_4\gamma_4$ part 
of the Dirac kinetic term, project onto the upper or lower two components of the 
Dirac field, see e.g.~\cite{kronfeld2000}. The temporal term is therefore the temporal 
derivative of the heavy quark spinor in the HQET formalism.
The spatial derivatives of the  Wilson term contribute to 
the kinetic term of the heavy quark Lagrangian. 
The clover term can be split up into a color-electric and a color-magnetic
contribution,
\be
S_{SW} = \frac{i}{2} c_{SW}\kappa \left(\sum_{x,ijk} \overline{\psi}_x \epsilon_{ijk}
\sigma_{ij}B_k \psi_x + \sum_{x,i} \overline{\psi}_x \sigma_{0i} E_i \psi_x\right),
\ee
and therefore contains the spin-colormagnetic interaction term of the HQET at 
$O(1/M)$. The heavy quark limit of the clover action indeed contains the kinetic and 
the colormagnetic terms of the HQET Lagrangian proportional to $1/M_2$ with coefficients 
of $O(1)$~\cite{elkhadra1997}. 
The $\vec D\cdot \vec\gamma$ term of the naive Lagrangian also contributes to the kinetic and
the spin-magnetic operators, however with a factor proportional to
$\lqcd/M$ where $M$ is the heavy quark mass relative to the contribution from the Wilson 
term~\cite{elkhadra1997}.

The residue of the pole of the free quark propagator is mass dependent:
\be
Z_2 = e^{-aM_1} = \left(1 + am\right)^{-1}. \label{eq:z2}
\ee
The canonically normalized field at non-vanishing quark mass is therefore given by
\be
Q_x = e^{aM_1/2} q_x,
\ee
In the hopping parameter representation the canonical normalization for the
massive free quark is given by 
\be
Q_x = {\sqrt{1-6\kappa}}a^{-3/2}\psi_x, \label{eq:field_norm}
\ee
which approaches the usual norm for Wilson fermions~Eq.(\ref{eq:contnorm}) at
vanishing quark mass.
To remove the tree-level $\vec{p}/M$ mass dependence of the heavy quark field 
$Q$, an additional application of the spatial derivative is necessary.
The complete tree-level field redefinition at $O(1/M)$ is then
\be
Q^\prime_x =  e^{aM_1/2}\left(1 + ad_1\vec{\gamma}\cdot \vec{D}\right) q_x,
\ee
with 
\be
d_1 = \frac{1+am}{am(2+am)} - \frac{1}{2aM_2}, \label{eq:d1}
\ee
i.e.\  for $am = 1$, $d_1 = 1/12$, while for $am \to 0$, $d_1 \to 0$.

The Hamiltonian of heavy Wilson fermions can then be written as an expansion in terms 
of the inverse quark mass:
\be
H = \overline{Q}_x \left(M_1 - \frac{\vec{D}^2}{2M_2} - g\frac{\vec{\sigma}
\cdot\vec{B}}{2M_B} + O(1/M^2)\right) Q_x.
\ee
where the pole mass $M_1$ just acts as an additive zero point of the heavy quark mass.
The different masses parameterize the departure of the lattice action
from the Foldy-Wouthuysen transformed Dirac action. As mentioned before,
$M_B = M_2$ for clover fermions at tree level.
In general, the coefficient of the spin-magnetic term is related to the pole mass and 
kinetic mass through $M_B/M_2 = 1+\sinh aM_1$.

Ref.~\cite{aoki2002} discusses a program to improve heavy quarks with $am \simeq 1$ 
on-shell to $O(a)$. It is assumed that
the discretization errors are of the form $f_n(am) (a\lqcd)^n$ with $f_n(am)$ 
analytic functions of $am$. The $f_0$  and $f_1$ terms are removed in their procedure.
In~\cite{aoki2004}, results are given for tree-level and one-loop $O(a)$ 
improvement.

Anisotropic lattices with a finer lattice spacing in the time direction are also used in 
simulations of heavy quarks, to reduce discretization errors from 
the quark mass at least at tree level and to allow for more points in the time direction to be 
used in the fits. 
Different hopping parameters and clover coefficients corresponding to time- and spacelike 
links are needed, which leads to the necessity of tuning many parameters if one wishes to
improve the actions. Programs for simulations of heavy quarks can  e.g.\ be found in 
Refs.~\cite{matsufuru2004,foley2004}.  Discussions of earlier work are given in 
Refs.~\cite{klassen1999} and \cite{liao2002a,liao2002b}. 

For reviews about heavy quark lattice formalisms see e.g.~\cite{bernard2001a,kronfeld2003}. 
\section{THE $b$ HADRON SPECTRUM \label{sec:hresults}}
The success of NRQCD in calculations of the quarkonium spectrum (for the $\Upsilon$,  
see e.g.~\cite{davies1994,lidsey1994,davies1995} and 
the subsequent studies~\cite{manke1997,davies1998,manke2000,gottlieb2004}; for charmonium
see e.g.~\cite{davies1995b}),  motivated us to calculate
the spectrum of hadrons with $b$ and light quarks using NRQCD at $O(1/M)$ with $a\sim 0.1$ fm. 
First calculations 
of the ground state hyperfine structure, a $P$ wave excitation and the $\Lambda_b$ baryon obtained
a good signal for the lattice data and  agreement with the experimental spectrum~\cite{alikhan1996,collins1996}.
This was followed by a calculation at $O(1/M^2)$ at  $a\sim 0.1$ fm (published in Ref.~\cite{alikhan2000}
in 2000), which obtained good signals for many spectral states and resolved several mass splittings for the first 
time. Also preliminary results on the $B_c$ spectrum were obtained~\cite{alikhan1998b}.
The results were supported by a study of the $B$ and $D$ meson spectrum at larger and smaller lattice 
spacings~\cite{hein2000}. 
From a further calculation of the $N_f = 2$ and quenched $B$ spectrum on a range of lattices using a
renormalization group improved gauge action only preliminary results were obtained~\cite{cppacs2000}.

In this section we discuss our results on the $b$-light spectrum, in particular 
Refs.~\cite{alikhan2000,cppacs2000}.

\begin{figure}[htb]
\begin{center}
\setlength{\unitlength}{.020in}
\begin{picture}(130,100)(30,500)
\put(23,500){\line(0,1){100}}
\multiput(21,500)(0,50){3}{\line(1,0){4}}
\multiput(22,500)(0,10){11}{\line(1,0){2}}
\put(20,500){\makebox(0,0)[r]{{\normalsize5.0}}}
\put(20,550){\makebox(0,0)[r]{{\normalsize5.5}}}
\put(20,600){\makebox(0,0)[r]{{\normalsize 6.0}}}
\put(20,570){\makebox(0,0)[r]{{\normalsize GeV}}}
\put(23,500){\line(1,0){160}}
     \put(28,510){\makebox(0,0)[t]{{\normalsize $B$}}}
     \put(29,529.5){\circle{3}}
     \put(29,529.5){\line(0,1){0.4}}
     \put(29,529.5){\line(0,-1){0.4}}
     \multiput(23,527.9)(3,0){4}{\line(1,0){2}}
     \put(28,589.8){\circle{3}}
     \put(28,589.8){\line(0,1){8.5}}
     \put(28,589.8){\line(0,-1){8.5}}
     \put(38,595){\makebox(0,0)[t]{{\normalsize $2S$}}}
%
     \put(45,510){\makebox(0,0)[t]{{\normalsize $B^{*}$}}}
     \put(46,531.9){\circle{3}}
     \put(46,531.9){\line(0,1){0.1}}
     \put(46,531.9){\line(0,-1){0.1}}
     \multiput(40,532.6)(3,0){4}{\line(1,0){2}}
     \multiput(40,532.4)(3,0){4}{\line(1,0){2}}
%
     \put(60,510){\makebox(0,0)[t]{{\normalsize $B_s$}}}
     \put(60,538.4){\circle{3}}
     \put(60,538.4){\line(0,1){3.0}}
     \put(60,538.4){\line(0,-1){1.1}}
     \multiput(55,537.1)(3,0){4}{\line(1,0){2}}
     \multiput(55,536.7)(3,0){4}{\line(1,0){2}}
     \put(60,594.5){\circle{3}}
     \put(60,594.5){\line(0,1){6.1}}
     \put(60,594.5){\line(0,-1){5.3}}
     \put(72,601){\makebox(0,0)[t]{{\normalsize $2S$}}}
%
     \put(75,510){\makebox(0,0)[t]{{\normalsize $B^{*}_s$}}}
     \put(76,541.2){\circle{3}}
     \put(76,541.2){\line(0,1){3.4}}
     \put(76,541.2){\line(0,-1){1.4}}
     \multiput(70,541.9)(3,0){4}{\line(1,0){2}}
     \multiput(70,541.3)(3,0){4}{\line(1,0){2}}
%
     \put(100,510){\makebox(0,0)[t]{{\normalsize $B^{\ast\ast}$}}}
     \put(88,567.0){\circle{3}}
     \put(88,567.0){\line(0,1){3.7}}
     \put(88,567.0){\line(0,-1){3.7}}
     \put(94,563.0){\makebox(0,0)[t]{{\normalsize $B^\ast_0$}}}
     \put(102,582.2){\circle{3}}
     \put(102,582.2){\line(0,1){4.5}}
     \put(102,582.2){\line(0,-1){4.5}}
     \put(106,589.2){\makebox(0,0){{\normalsize $B^*_2$}}}
     \multiput(84,570.6)(3,0){8}{\line(1,0){2}}
     \multiput(84,569.0)(3,0){8}{\line(1,0){2}}
     \multiput(86,5681)(3,0){8}{\line(1,0){1}}
     \multiput(86,5732)(3,0){8}{\line(1,0){1}}
     \multiput(84,575.2)(3,0){3}{\line(1,0){0.8}} 
     \multiput(84,571.2)(3,0){3}{\line(1,0){0.8}}
     \multiput(100,575.2)(3,0){3}{\line(1,0){0.8}} 
     \multiput(100,572.4)(3,0){3}{\line(1,0){0.8}}
%
     \put(132,510){\makebox(0,0)[t]{{\normalsize $B_s^{\ast\ast}$}}}
     \put(122,574.2){\circle{3}}
     \put(122,574.2){\line(0,1){4.2}}
     \put(122,574.2){\line(0,-1){2.7}}
     \put(128,569.0){\makebox(0,0)[t]{{\normalsize $B^*_{s0}$}}}
     \put(146,593.5){\makebox(0,0){{\normalsize $B^*_{s2}$}}}
     \put(136,587.3){\circle{3}}
     \put(136,587.3){\line(0,1){3.7}}
     \put(136,587.3){\line(0,-1){2.6}}
     \multiput(119,586.8)(3,0){8}{\line(1,0){2}}
     \multiput(119,583.8)(3,0){8}{\line(1,0){2}}
     \multiput(132,584.7)(3,0){4}{\line(1,0){0.8}}
     \multiput(132,585.7)(3,0){4}{\line(1,0){0.8}}
\end{picture}
\begin{picture}(130,150)(30,500)
  \put(15,500){\line(1,0){175}}
     \put(30,510){\makebox(0,0)[t]{{\normalsize $\Lambda_b$}}}
     \put(30,567.9){\circle{3}}
     \put(30,567.9){\line(0,-1){7.1}}
     \put(30,567.9){\line(0,1){7.1}}
     \multiput(25,563.3)(3,0){4}{\line(1,0){2}}
     \multiput(25,561.5)(3,0){4}{\line(1,0){2}}
%
     \put(50,510){\makebox(0,0)[t]{{\normalsize $\Sigma_b$}}}
     \put(50,588.7){\circle{3}}
     \put(50,588.7){\line(0,-1){4.6}}
     \put(50,588.7){\line(0,1){4.6}}
     \put(70,510){\makebox(0,0)[t]{{\normalsize $\Sigma_b^*$}}}
     \put(70,590.9){\circle{3}}
     \put(70,590.9){\line(0,1){4.7}}
     \put(70,590.9){\line(0,-1){4.7}}
     \put(90,510){\makebox(0,0)[t]{{\normalsize $\Xi_b$}}}
     \put(90,578.5){\circle{3}}
     \put(90,578.5){\line(0,-1){5.4}}
     \put(90,578.5){\line(0,1){6.9}}
     \put(110,510){\makebox(0,0)[t]{{\normalsize $\Xi'_b$}}}
     \put(110,596.2){\circle{3}}
     \put(110,596.2){\line(0,-1){4.0}}
     \put(110,596.2){\line(0,1){6.4}}
     \put(130,510){\makebox(0,0)[t]{{\normalsize $\Xi_b^*$}}}
     \put(130,598.2){\circle{3}}
     \put(130,598.2){\line(0,1){6.4}}
     \put(130,598.2){\line(0,-1){3.9}}
     \put(150,510){\makebox(0,0)[t]{{\normalsize $\Omega_b$}}}
     \put(150,604.8){\circle{3}}
     \put(150,604.8){\line(0,-1){3.3}}
     \put(150,604.8){\line(0,1){6.7}}
%
%
     \put(170,510){\makebox(0,0)[t]{{\normalsize $\Omega_b^*$}}}
     \put(170,606.9){\circle{3}}
     \put(170,606.9){\line(0,1){6.9}}
     \put(170,606.9){\line(0,-1){3.4}}
%
\end{picture}
\end{center}
\caption{The $B$ meson spectrum from quenched lattice 
NRQCD~\protect\cite{alikhan2000}. Long dashed lines denote the experimental error 
bounds  from
the Particle Data Book~\cite{pdg}, dotted lines experimental error bounds from
DELPHI~\protect\cite{delphi2004}. 
We use the collective notation $B^{\ast\ast}$ for the $P$ states.}
\label{fig:Bspec}
\end{figure}
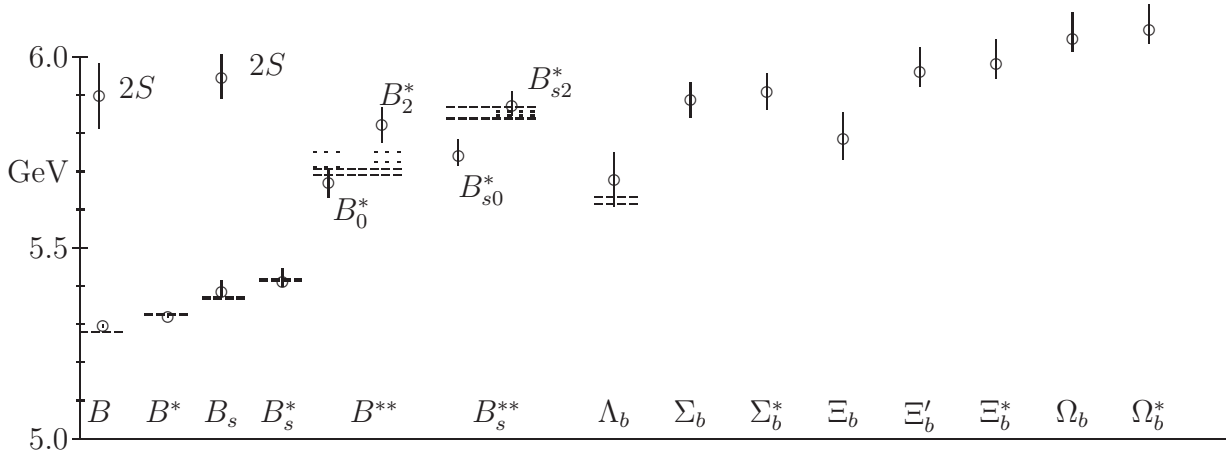

To demonstrate that a lattice calculation using NRQCD can reproduce the $b$-light hadron spectrum, we first show
the results of~\cite{alikhan2000} compared with the available experimental data in Fig.~\ref{fig:Bspec}.
At the same time, predictions for a number of ground state baryons can be made.
In the next subsections we discuss the results in more detail in comparison with other lattice and with 
potential model calculations.

In this section we discuss results on the spectrum of hadrons with $b$ quarks
from \cite{alikhan2000,cppacs2000} and compare with \cite{collins1999,hein2000}. 

\begin{table}[thb]
\begin{center}
\setlength{\tabcolsep}{0.25cm}
\begin{tabular}{|ll|l|}
\hline
\multicolumn{1}{|c|}{$J^P$}&\multicolumn{1}{|c|}{$(^{2S+1}L_J)$}&\multicolumn{1}{c|}{operator} \\
\hline
$0^-$   & $(^1S_0)$     & $\bar u \;Q$ \\
$1^-$   & $(^3S_1)$     & $\bar u \;\vec \sigma \;Q$ \\
$1^+$   & $(^1P_1)$     & $\bar u \;\vec \Delta \;Q$ \\
$0^+$   & $(^3P_0)$     & $\bar u \;\vec \sigma \cdot \vec\Delta \;Q$ \\
$1^+$   & $(^3P_1)$     & $\bar u \;\vec \sigma × \vec\Delta \;Q$ \\
$2^-$   & $(^3P_2)(T)$  & $\bar u \;(\sigma_i \Delta_{j} + \sigma_j \Delta_{i}) \;Q$ \\
$2^-$   & $(^3P_2)(E)$  & $\bar u \;(\sigma_i \Delta_i - \sigma_j \Delta_j) \;Q$ \\
\hline
$\Lambda_Q$  & ($s_z = +1/2$) & $\bar u^c d \; Q_\uparrow$ \\
$\Sigma_Q$   & ($s_z = +1/2$) & $(\bar u^c \sigma_z d \; Q_\uparrow 
                                - \sqrt{2} \bar u^c \sigma_+ d \; Q_\downarrow)
                              /\sqrt{3}$ \\
$\Sigma^*_Q$ & ($s_z = +3/2$) & $\bar u^c \sigma_+ d \; Q_\uparrow$ \\
$\Sigma^*_Q$ & ($s_z = +1/2$) & $(\sqrt{2} \bar u^c \sigma_z d \; Q_\uparrow 
                                +  \bar u^c \sigma_+ d \; Q_\downarrow)
                              /\sqrt{3}$ \\
\hline
\end{tabular}
\end{center}
\caption{ Ground state $S$ and $P$ wave heavy-light meson  and $S$ wave baryon operators
used in~\protect\cite{alikhan2000}. 
}
\label{tab:ops}
\end{table}
The spin operators used to project onto different hadron states in \cite{alikhan2000,cppacs2000}
are listed in Table~\ref{tab:ops}. Operators from two 
representations of the cubic group have been used, denoted as  $T$ and $E$, which correspond to the 
continuum quantum numbers $^3P_2$ in the $^SL_J$  notation used for heavy-heavy mesons. 
The baryon operators are constructed using only 
the upper two components of the quark spinors coupled with  Clebsch-Gordon coefficients. 
The operators projecting onto the maximal $z$ component of the spin are shown in the Table, the other
components are defined in an analogous way. For naming conventions of the $P$ wave heavy-light mesons
see Eq.~(\ref{eq:pdoublets}) and of the heavy-light-light baryons Eq.~(\ref{eq:barydoublets}).
The same operators are used for the heavy-heavy-light baryons. For their naming conventions see 
Eq.~(\ref{eq:hhlstates}).

The lattice scale has been set using $m_\rho$.
In \cite{alikhan2000}, the $b$ quark mass is fixed using the spin-averaged $\overline B(1S)$,
in \cite{cppacs2000}, both the $B$ and $B_s$ masses were used for consistency.
More simulation details can be found in the respective papers. 
\subsection{$S$ Wave States}
The splitting between the $B_s$ and $B_d$ mesons is expected to be dominated by 
the strange and light quark mass difference. The spin-averaged splitting depends in
addition only on the difference of heavy quark kinetic energies in the
cloud of gluons and relativistic light or strange quarks. 
\begin{figure}[thb]
\begin{center}
\centerline{
\epsfysize=4.9cm \epsfbox{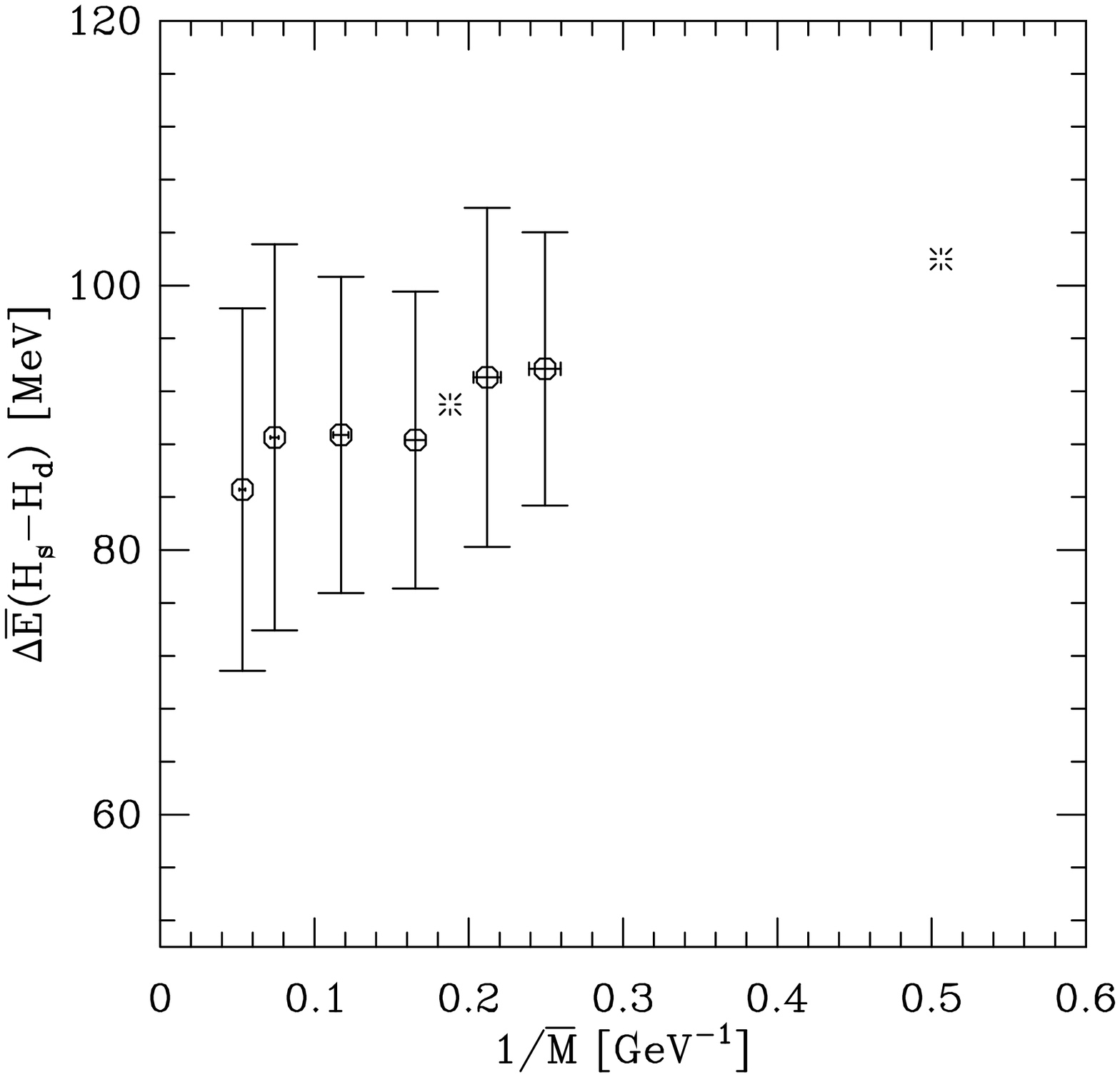}
\epsfysize=4.9cm \epsfbox{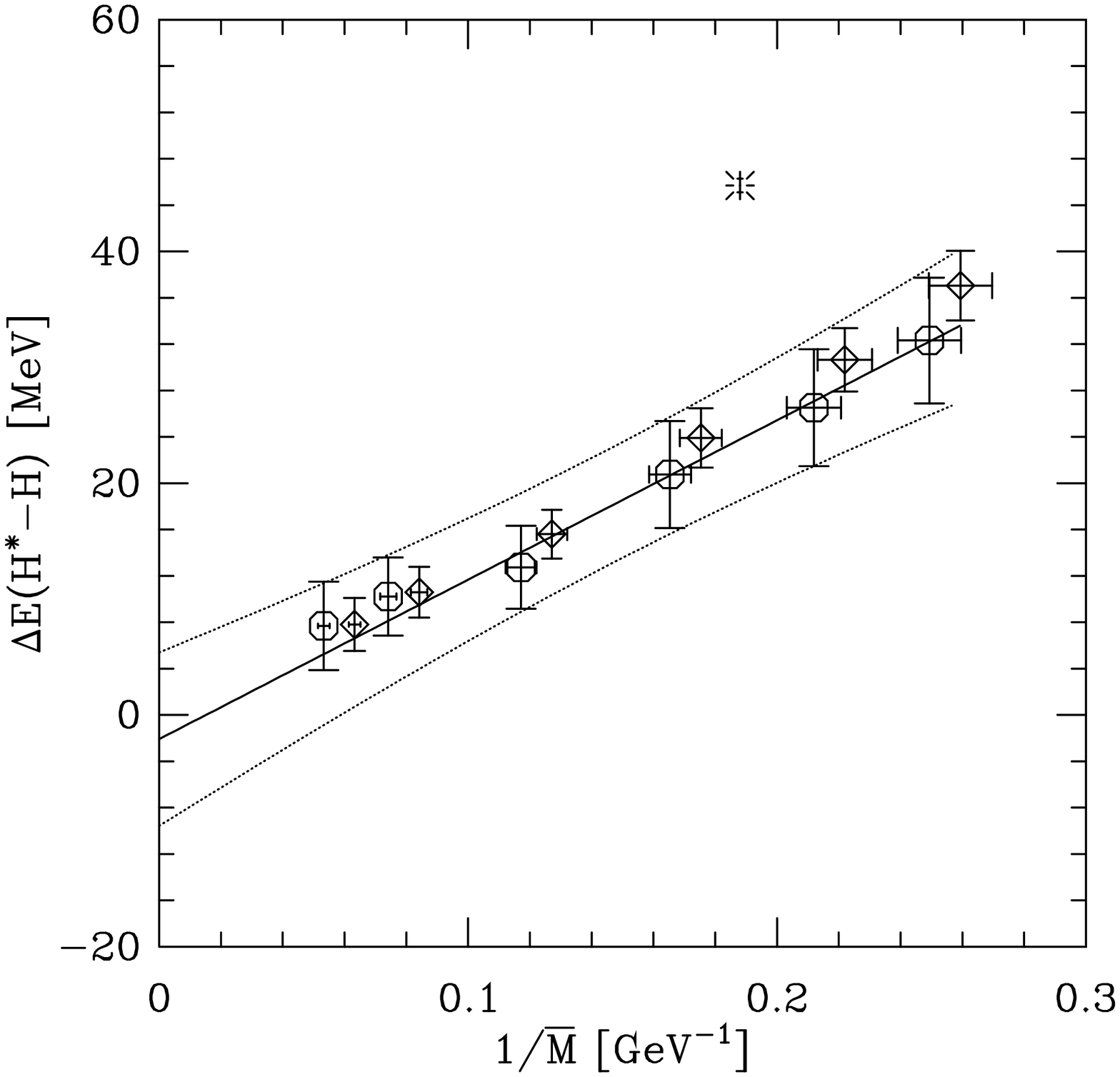}
}
\end{center}
\vspace{-0.5cm}
\caption{ $S$ wave meson splittings from~\protect\cite{alikhan2000}. On the left, the 
$\overline{H}_s-\overline{H}$, on the right the $H^\ast-H$ hyperfine splitting. 
Circles denote results for mesons with a light quark 
($H = H_{u,d}$), diamonds for mesons with a strange quark ($H_s$). The solid line denotes a linear fit to the 
($H$) mesons with the errors denoted by dashed lines. The experimental value for the $B$ (and on the right also
for the $D$) meson is given by a burst.}
\label{fig:BsBd}
\end{figure}
In Fig.~\ref{fig:BsBd} on the left, the spin-averaged energy difference from \cite{alikhan2000} is plotted
for various heavy meson masses.
The dependence is relatively flat, which is also found in experiment.
The experimental 
$\overline{D}_s-\overline{D}$ splitting is roughly 10\% higher than the experimental
$\overline{B}_s-\overline{B}$ splitting.

In HQET and NRQCD, the hyperfine splitting is generated by the spin-magnetic interactions.
The leading mass dependence should therefore  be $\sim 1/M_Q$ or $\sim 1/\overline{M}$
where $\overline{M}$ denotes the spin-averaged meson mass.
In Fig.~\ref{fig:BsBd} on the right we show that results indeed are consistent with this 
expectation for strange as well as for chirally extrapolated light quark masses. 

Results are shown in Table~\ref{tab:mes_splittings}. We find the results to be $30-50\%$ lower
compared to the experimental values of $45.8(4)$ MeV and $47.0(2.6)$ MeV~\cite{pdg} for the
$B$ and $B_s$ splitting respectively. An error source could e.g.\ be radiative corrections to the 
hyperfine coefficient, since the spin-magnetic coefficient in the calculations discussed here is only 
set to its tree-level tadpole-improved value.
Results from other quenched lattice NRQCD calculations of the $B^\ast-B$ and $B^\ast_s-B_s$ 
splittings fixing the scale with $m_\rho$ (e.g.\cite{hein2000,cppacs2000,mathur2002,ishikawa2000}), 
and $r_0=0.5$ fm~\cite{wing2003.stag} are  around $25-35$ MeV. 
Refs.~\cite{hein2000,alikhan2000}, and \cite{lewis2000,ishikawa2000} 
do not resolve a lattice spacing dependence of the hyperfine splitting.
Using relativistic $O(a)$ improved heavy quarks, Ref.~\cite{bowler2001} obtains a splitting 
around $10-20$ MeV with an error of 10 MeV. 
The quenched calculation of~\cite{mackenzie1998} gives $\Delta E(B^\ast_s-B_s) \simeq 40(10)$ MeV,
but using a different quantity to set the scale, i.e.\ the charmonium $1P-1S$ splitting.

\begin{figure}[thb]
\begin{center}
\epsfysize=5cm \epsfbox{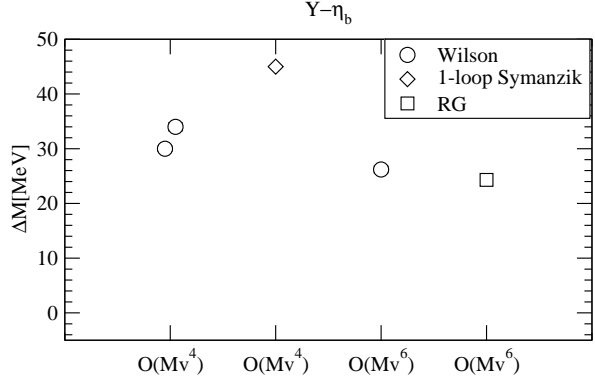}
\end{center}
\vspace{-0.5cm}
\caption{
Quenched $\Upsilon-\eta_b$ hyperfine  splitting from NRQCD
at $O(Mv^4)$ and $O(Mv^6)$. From left to right, the points are from 
Refs.~\protect\cite{davies1998} at $a \simeq 0.08$ and $0.06$ fm, 
and~\protect\cite{gottlieb2004,manke1997,manke2000} (from left to right).
Different symbols denote different gauge actions. 
The scale is set using the $\Upsilon'-\Upsilon$ splitting except
for~\protect\cite{manke1997,manke2000}
which prefers to set the scale using the $\overline{\chi}_b-\Upsilon$ splitting to set the scale but find the ratio 
$\Delta m(\Upsilon'-\Upsilon)/\Delta m(\overline{\chi}_b-\Upsilon)$ within errors in 
agreement with the experimental value already in the quenched approximation.
}
\label{fig:ups}
\end{figure}

In the CP-PACS preliminary result  of \cite{cppacs2000} we do not find a significant unquenching
effect. The unquenched result is with $33$ MeV still much lower than experiment. The sea quark mass 
is around $3\times$ the strange quark and the lattice is quite coarse, and perhaps the 
observed absence of unquenching is due to a large systematic error.
An NRQCD calculation with three dynamical flavors~\cite{wing2002.stag} and staggered light quarks
does find the $B^\ast_s-B_s$ splitting to increase from $25(5)$ to  $42.5(3.7)$ MeV,
in agreement with the experimental value. An appropriate choice of gauge action might improve
quantities which could be sensitive to discretization and perturbative effects. E.g.\ 
for quarkonia, there is a $\sim 50\%$ variation of the quenched $\Upsilon-\eta_b$ 
splitting between simulations using only different gauge actions, as shown in 
Fig.~\ref{fig:ups}. Using relativistic quarks on quenched anisotropic lattices, 
Ref.~\cite{liao2002a} finds a $^3S_1-{^1S_0}$ splitting around $50-60$ MeV which is in agreement with
the potential model result of \cite{ebert2000} of 60 MeV
and close to the NRQCD result given in~\cite{gottlieb2004}.
There is one experimental candidate for the $\eta_b$, which is roughly 160 MeV lighter than the $\Upsilon$.
It is quoted to have around $20$ MeV statistical and  $20$ MeV systematic 
errors~\cite{pdg}. For a detailed review
of  theoretical and experimental information on quarkonium physics see~\cite{brambilla2005}.
\subsection{Orbitally Excited $B$ Mesons}
The spectrum  of $P$ wave $B$ mesons is also e.g.\ of interest for the experimental study
of $CP$ violation in $B(\overline{B})$ experiments. Resonance 
structures will be noticeable in the decay $B^\ast \to B^0 \pi^±$.

The fact that in the limit $M \to \infty$ the heavy quark is a static
source of a color field in which the light quark moves relativistically, 
motivates to treat heavy-light mesons similar to the hydrogen
atom. The energy levels can be characterized by the light quark angular and radial 
quantum numbers. The light quark angular momentum, denoted as $j_l$, is then coupled
with the heavy quark spin to the  angular momentum of the meson, $J$. This can be
sketched as follows:
\be
\begin{array}{ccc}
j_l = \frac{3}{2}   & \begin{array} {c}
          {}_\nearrow \\
          {}^\searrow \\
          \end{array}
          & \begin{array}{cc}
           J = 2: & B_2^\ast \\ 
           J = 1: & B_1 \\      
\end{array} \\
j_l = \frac{1}{2}   & \begin{array} {c}
          {}_\nearrow \\
          {}^\searrow \\
          \end{array}
          & \begin{array}{cc}
           J = 1: & B^\prime_1 \\ 
           J = 0: & B^\ast_0 \\      
\end{array} \\
\end{array}
\label{eq:pdoublets}
\ee
Experimental knowledge of the $P$ state level structure is still sparse. The Particle Data
Group~\cite{pdg} lists two candidates.
One is a relatively broad resonance, the $B_J^\ast(5732)$, which is believed to come from 
several narrow and broad $P$ wave states, and the relatively narrow 
$B_{sJ}^\ast(5850)$ which can be interpreted as stemming from
excited $B_s$ states. Recent experimental results from  D0~\cite{ziem2004}, who measures the
$B_1$ mass and for the first time finds separate signals for the $B_1$ and the $B_2^\ast$,
and from DELPHI~\cite{delphi2004} are given in Table~\ref{tab:p_exp}.
\begin{table}[thb]
\begin{center}
\begin{tabular}{|l|l|l|}
\hline
\multicolumn{1}{|c}{Ref.} &
\multicolumn{1}{|c|}{state} &
\multicolumn{1}{c|}{$M$ or $\Delta E$[MeV]} \\
\hline
DELPHI~\cite{delphi2004} & $B_1$ &  $5732(20)$ \\
                         & $B_2^\ast$    & $5738(14)$ \\
                         & $B_{s2}^\ast$ & $5852(5)$  \\
D0~\cite{ziem2004}       & $B_1$         & $5724(4)(7)$ \\
                         & $B_2^\ast-B_1$& $23.6(7.7)(3.9)$\\
Particle Data Book~\cite{pdg}  & $B_J^\ast(5732)$ & $5698(8)$ \\
                               & $B_{sJ}^\ast(5850)$ & $5853(15)$ \\
\hline
\end{tabular}
\end{center}
\caption{Experimental measurements for $P$ wave $B$ meson masses.}
\label{tab:p_exp}
\end{table}
Heavy-light mesons can be characterized by the quantum numbers $J^P$. The lattice operators
corresponding to the $^1P_1$ and $^3P_1$ states in the $LS$ coupling scheme 
both onto the $J^P = 1^+$ states.
Since in the correlators of $^3P_1 \leftrightarrow {^1P_1}$ errors are very large, 
we cannot disentangle the splitting between the $^1P$ states. 
If one assumes that the energy difference of the $^1P$ states is very small
(potential model calculations give e.g.\ $\sim 40$ MeV~\cite{ebert1998}), 
we can get a reasonable estimate of the spin-averaged $J=1$ energy~\cite{bhat1996} with
the present lattice data. Since the spin-average of the measured $^3P$ energies is 
close to the energies in the $^1P_1$ channel, the $^1P_1$ might be used as an estimate
for the spin-average of the $P$ states.
\begin{table}[thb]
\begin{center}
\begin{tabular}{|clccccc|}
\hline
Ref. & NRQCD & $\overline{B}_s-\overline{B}_d$ &  $B^\ast - B$ &  $\overline{P}-\overline{S}$ & 
$B_2^\ast-B$ & $B_2^\ast-B_0^\ast$ \\
\hline
\hline
\multicolumn{7}{|c|}{$N_f = 0$} \\
\hline
\hline
\multicolumn{7}{|c|}{light}\\
\hline
\protect\cite{alikhan2000} & $1/M^2$ & $90(9)(^{21}_{03})$ & $24(5)(^{2}_{3})$ & $457(31)(^{24}_{35})$ &
$526(45)(^{27}_{35})$ & $155(32)(^{09}_{13})$ \\
\protect\cite{hein2000}  & $1/M^2$ & $84.1(20)(^{214}_{9})$ & $29.5(15)(^{47}_{17})$ &  &
 &  \\
\protect\cite{hein2000}  & $1/M^2$ & $109(26)(^{11}_{5})$ &  &  &
 &  \\
\protect\cite{cppacs2000} &   $1/M$ & & 32(3) & & $426(17)$  & \\
\hline
\multicolumn{7}{|c|}{strange} \\
\hline
\protect\cite{alikhan2000} &  $1/M^2$ &  & $27(3)(^{2}_{3})$ & 
$428(27)(^{27}_{41})$   &  $493(30)(^{25}_{33})$ & $136(23)(^{10}_{14})$ \\
\protect\cite{hein2000} &  $1/M^2$ &  & $28.3(10)(^{34}_{16})$ & $385(70)(^{19}_{4})$ &
 & $41(94)(^{18}_{11})$ \\
\protect\cite{hein2000} &  $1/M^2$ &  & $27.3(20)(^{28}_{23})$ & $395(45)(^{17}_{15})$ &
 & $179(65)(^{7}_{7})$ \\
\hline
\hline
\multicolumn{7}{|c|}{$N_f = 2$} \\
\hline
\hline
\multicolumn{7}{|c|}{light}\\
\hline
\protect\cite{collins1999} &  $1/M$ &$103(8)(^{0}_{12})$ &  &  &   & \\
\protect\cite{cppacs2000} &  $1/M$ & & 33(2) & & $461(14)$  & \\
\hline
\end{tabular}
\end{center}
\caption{Summary of own results on the $b$ meson spectrum. The light quark content and
the flavor number are denoted in the headers. 
The errors are taken over from the
papers, adding the systematic errors in quadrature where applicable.}
\label{tab:mes_splittings}
\end{table}

The spin-averaged $P-S$ splittings from our calculations are given in Table~\ref{tab:mes_splittings}. 
Refs.~\cite{alikhan1996} and~\cite{collins1999} give results for $^1P_1-S$ splittings which are larger
than 500 MeV. The former use a very small set of configurations. The
latter might have increased finite volume effects because they are unquenched.
Our lattice simulations (see Table~\ref{tab:mes_splittings}) are in general in agreement with the preliminary experimental
results, although statistical and systematical uncertainties are both large. We attempt to give an estimate of
discretization and perturbative errors in Section~\ref{sec:otherspec}.

An important result of~\cite{alikhan2000} is to resolve the $B_2^\ast-B_0^\ast$ energy splitting on the lattice
for the first time. In the scaling study of  \cite{alikhan2000,hein2000} 
(see Table~\ref{tab:mes_splittings}), there are indications of discretization errors in the $P$ states. 

A source of systematic error in our calculation~\cite{alikhan2000} are finite volume effects, although
the lattices are quenched, since the lattice extent is only around $1.6$ fm. 
Ref.~\cite{michael1998} finds that the Bethe-Salpeter wave function of a static-light $P$ wave meson 
decays to around $1/4$ of its nodal peak at a distance of $1.6\times r_0$, which might affect the
energy levels.  
CP-PACS~\cite{cppacs2000} find a lower quenched result for the $B_2^\ast$ mass. An advantage of
their calculation is that the volume is larger, on the other hand the discretization effects are
expected to be larger too, and the NRQCD action is only $1/M$ improved.

\begin{figure}[tbh]
\begin{center}
\centerline{
\epsfysize=5cm \epsfbox{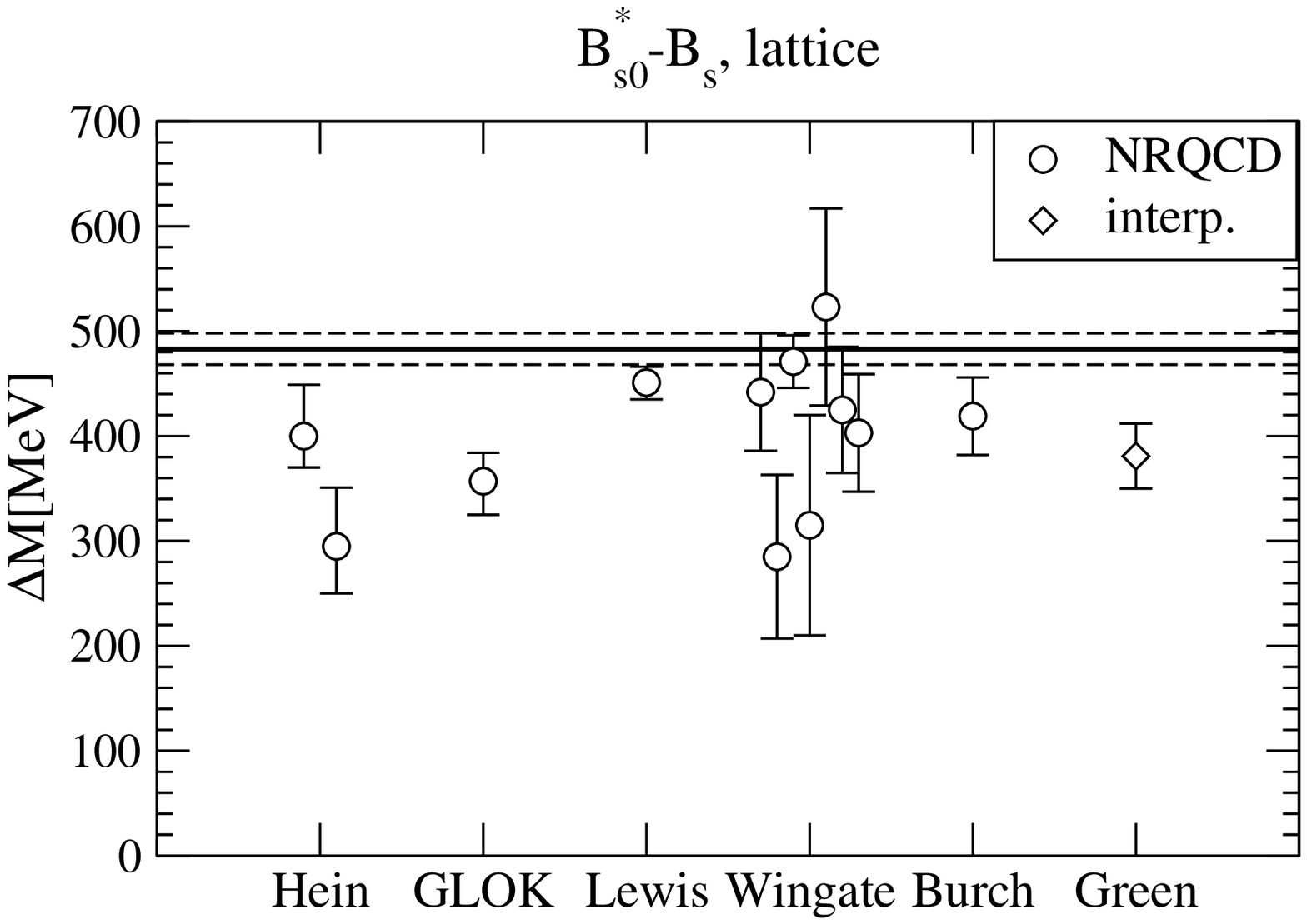}
\epsfysize=5cm \epsfbox{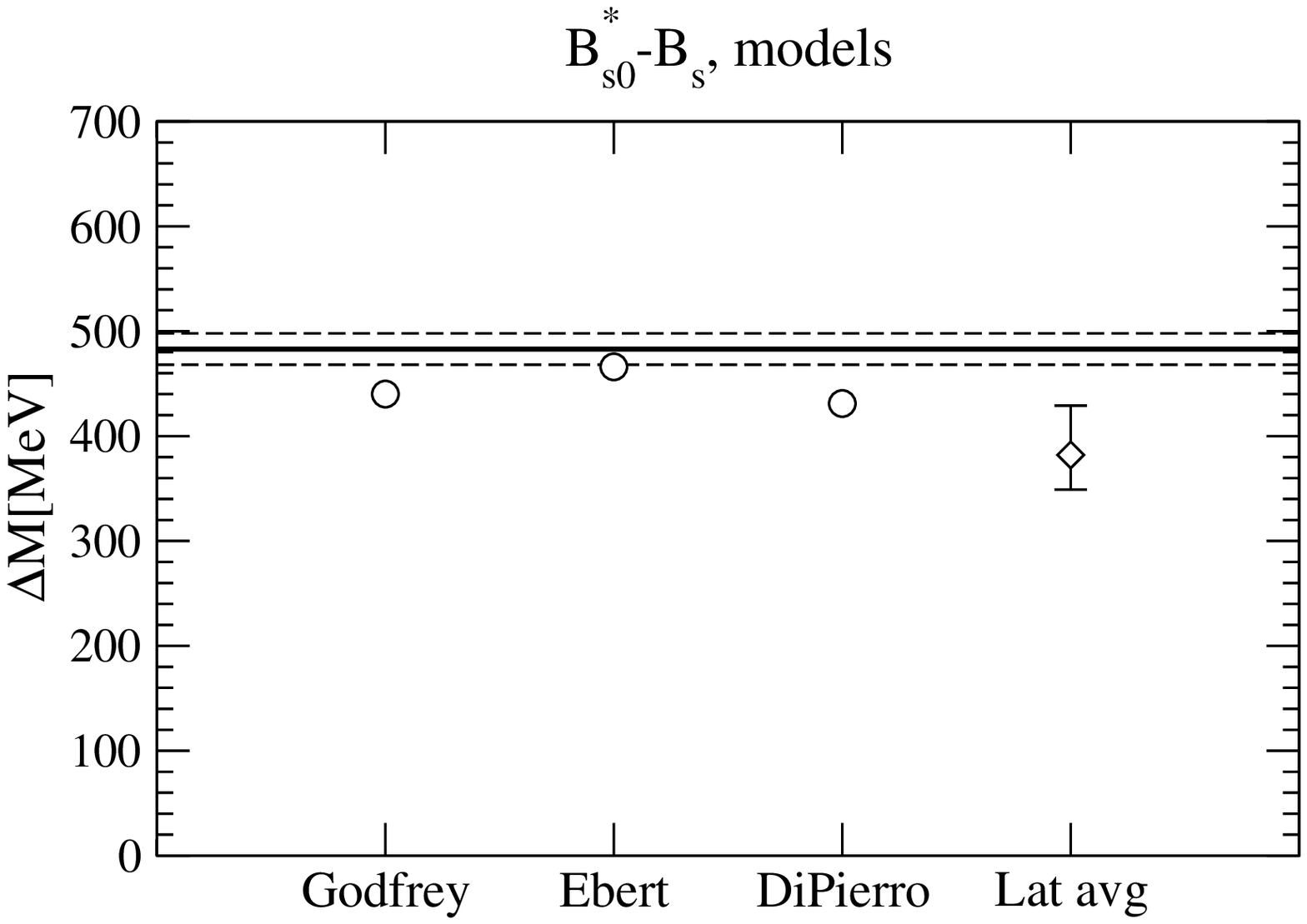}
}
\centerline{
\epsfysize=5cm \epsfbox{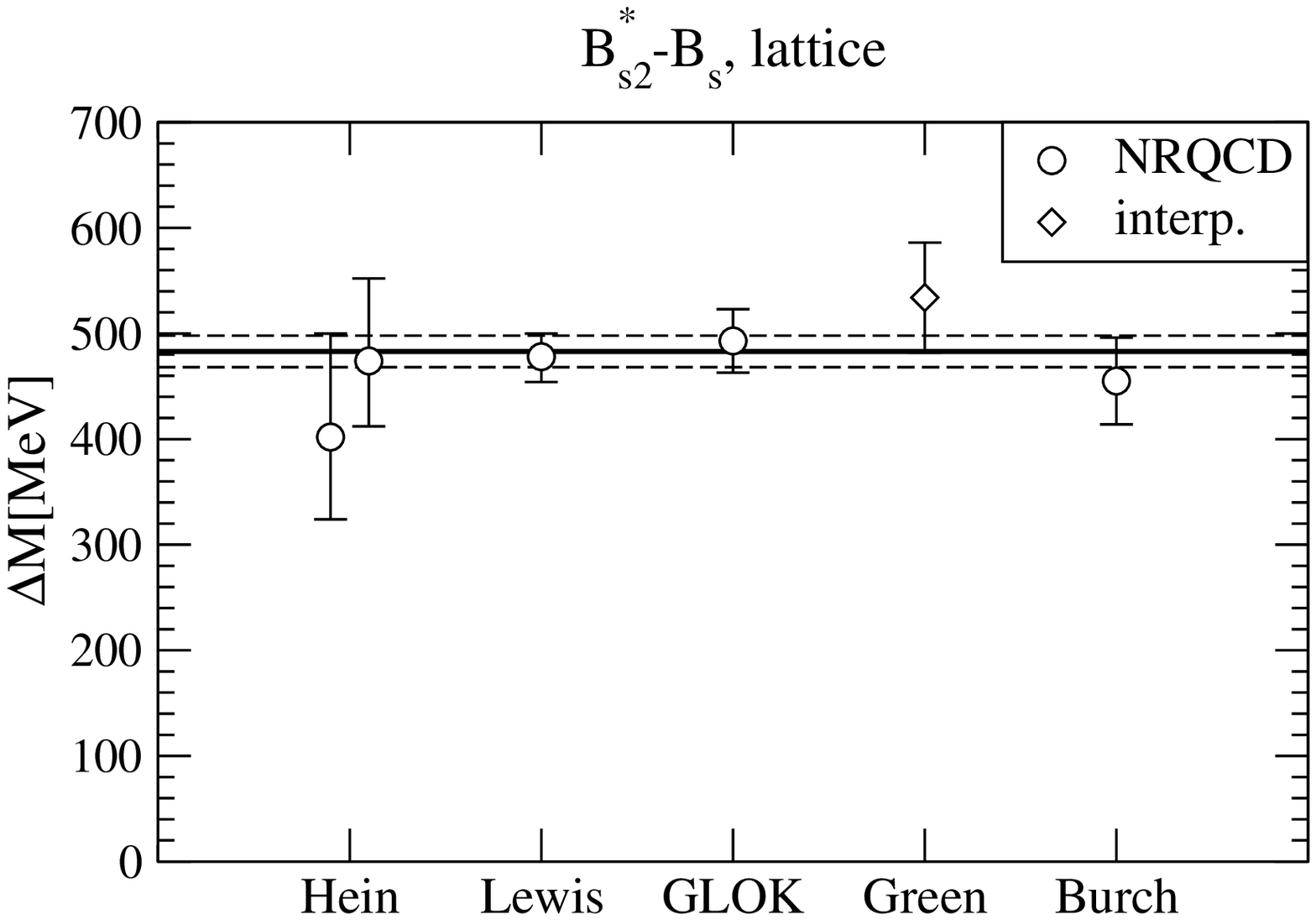}
\epsfysize=5cm \epsfbox{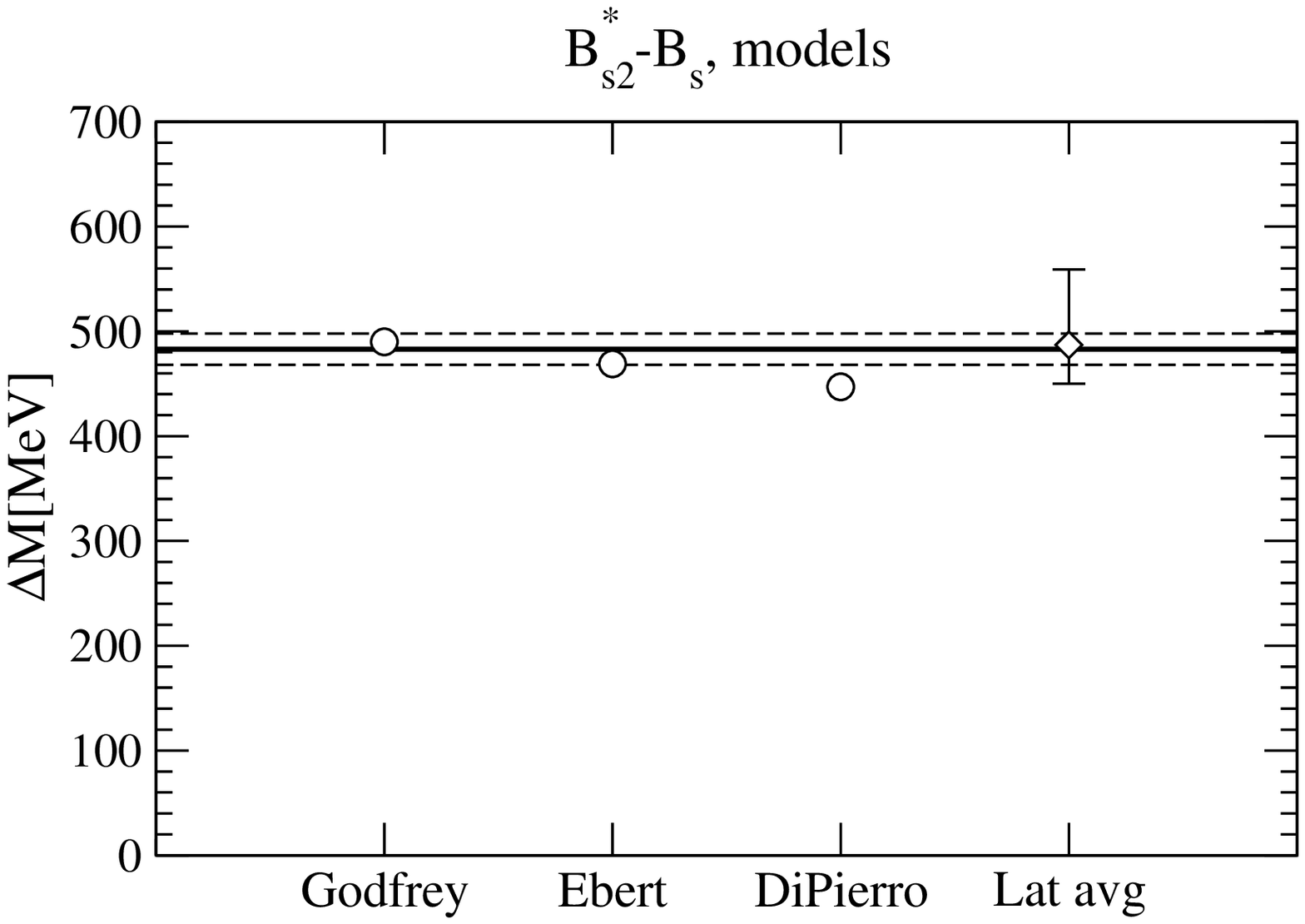}
}
\end{center}
\caption{Comparison of splittings between $P$ wave and the ground  state 
of $B_s$ mesons from 
Refs.~\protect\cite{hein2000,lewis2000,wing2003.stag,alikhan2000,burch2004,green2003}
(lattice) and Refs.~\protect\cite{godfrey1985,ebert1998,dipierro2001} (models).
The lines denote the experimental value of the narrow $B_{sJ}^\ast(5850)$ resonance 
believed to come from orbitally excited $B_s$ mesons~\protect\cite{pdg}. }
\label{fig:Bsstarstar-Bs}
 \end{figure}
A comparison of results from the lattice, model calculations and experiment for 
$P-S$ splittings of the $B_s$ meson is given in Fig.~\ref{fig:Bsstarstar-Bs}.

There are only few dynamical simulations so far. In 
unquenched calculations, the $j_l = 1/2$ lattice states can decay into or mix with
$S$ wave $B \pi$ or $B^\ast \pi$ systems if the mass of the $P$ wave states is 
above or near the threshold. This would result in a shift of the meson mass. 
Decays of the other two $P$ states occur at higher angular momentum and are therefore suppressed.

\begin{figure}[tbh]
\begin{center}
\centerline{
\epsfysize=5cm \epsfbox{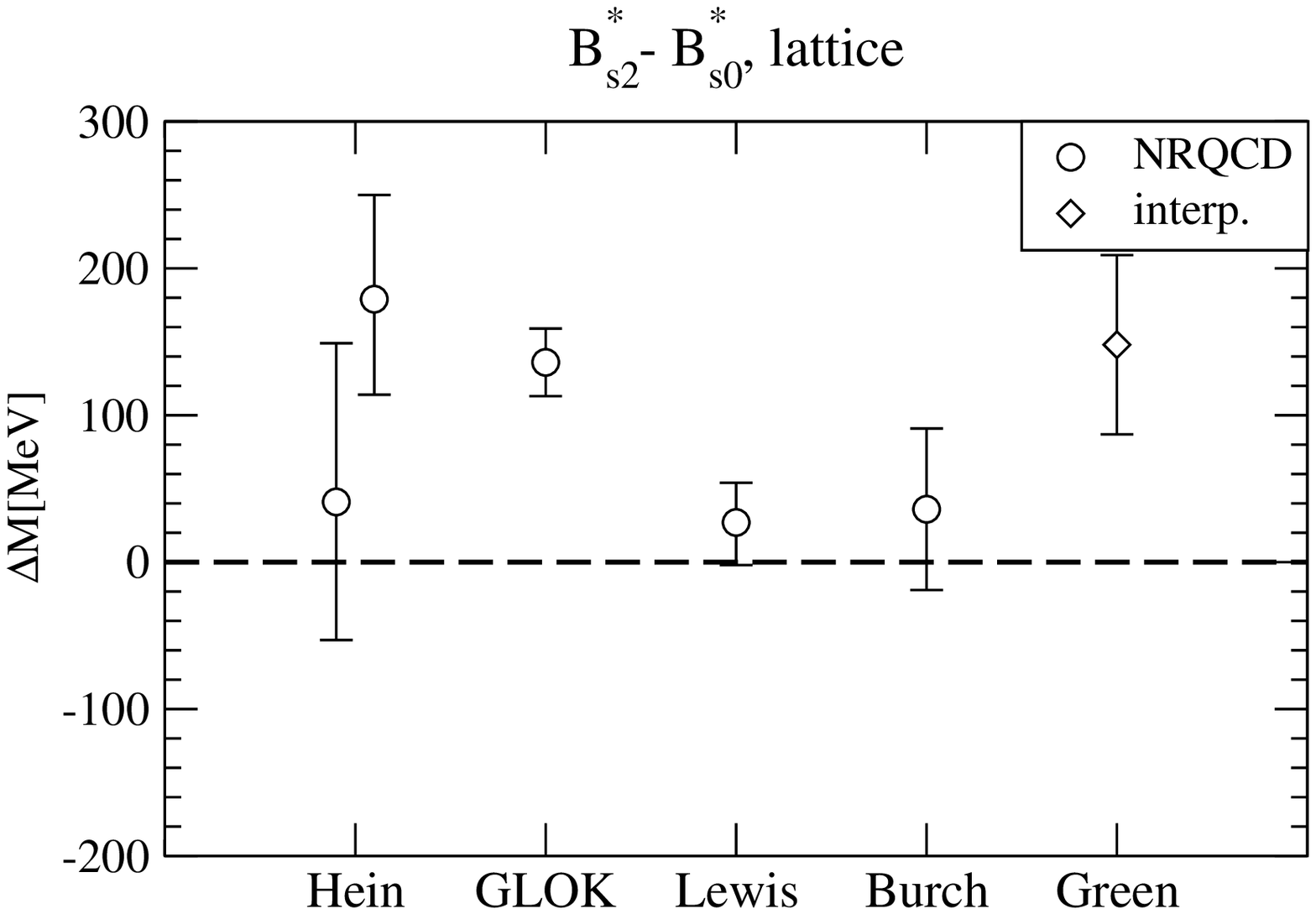}
\epsfysize=5cm \epsfbox{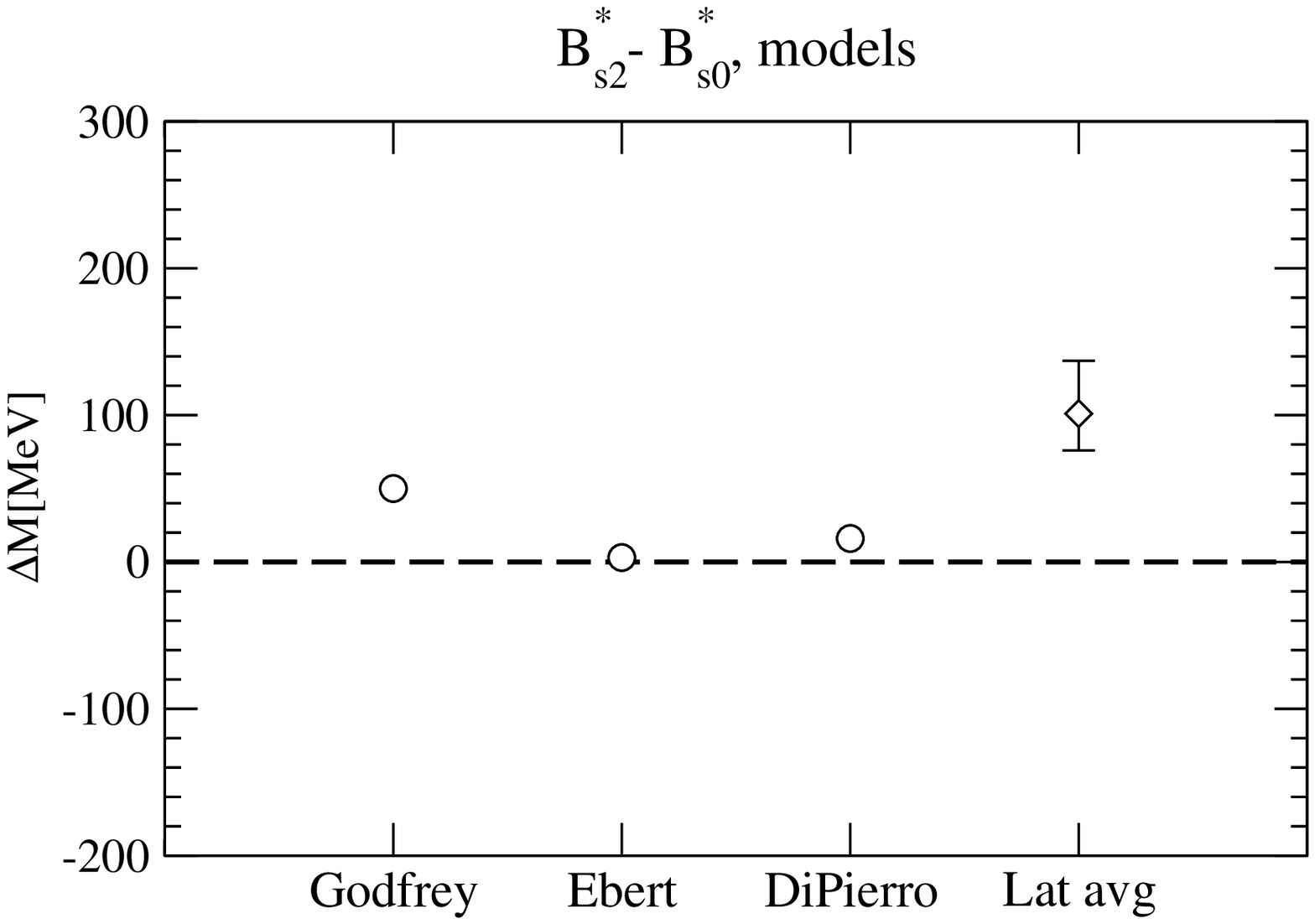}
}
\centerline{
\epsfysize=5cm \epsfbox{plots/B2star-B0star-comp.eps}
\epsfysize=5cm \epsfbox{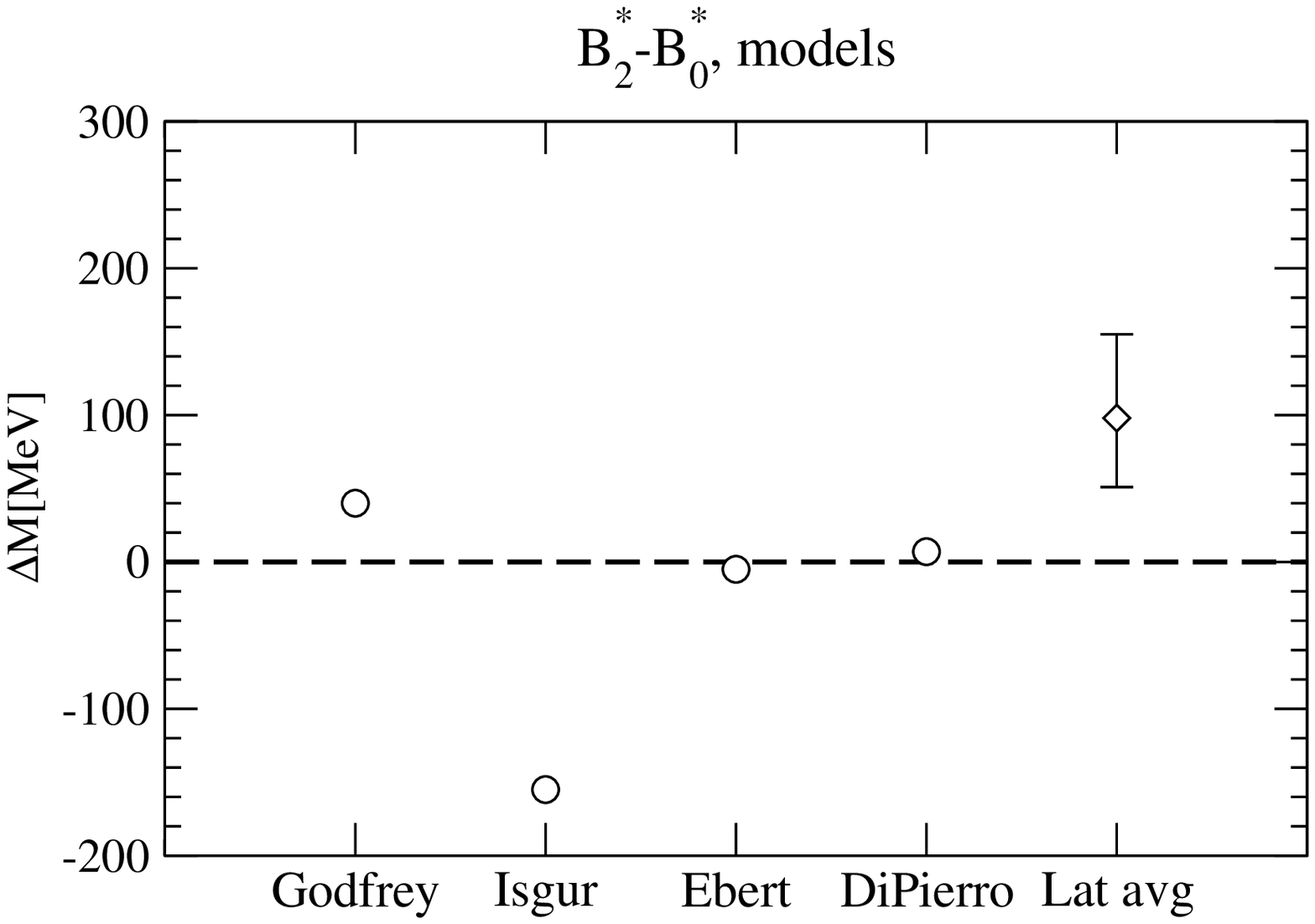}
}
\end{center}
\caption{Comparison of $P$ wave fine splittings of $B_s$ and $B$ mesons from 
Refs.~\protect\cite{hein2000,alikhan2000,lewis2000,burch2004,green2003}
(lattice) and Refs.~\protect\cite{godfrey1985,isgur1998,ebert1998,dipierro2001} (models).
The dashed line is to guide the eye.}
\label{fig:B2star-B0star}
\end{figure}
Among potential model calculations (e.g. \cite{godfrey1985,isgur1998,ebert1998,dipierro2001})
the sign of the $B_2^\ast-B_0^\ast$ mass difference is disputed, see Table~\ref{tab:psplit} and 
Fig.~\ref{fig:B2star-B0star}.
Some lattice calculations~\cite{lewis2000,burch2004} and \cite{hein2000} at $a \sim 0.2$ fm
find a small positive splitting which is within errors compatible with the potential models
predicting small negative splittings.
In Ref.~\cite{burch2004} the spin-chromomagnetic interaction term of the NRQCD action is not taken
into account in the calculation. From potential models one would  expect 
the $P$ wave hyperfine splitting to be close to $\sim 10-20$ MeV~\cite{ebert1998}.
Potential model calculations would predict \cite{ebert1998,ebert2005a} that including the hyperfine splitting 
rather increases the $B_2^\ast-B_0^\ast$ mass difference instead of decreasing it. 
In Refs.~\cite{alikhan2000,green2003}, and on the finer lattice of \cite{hein2000}, a positive splitting with 
$2 \sigma$ significance or more is found. 
Refs.~\cite{alikhan2000,hein2000,green2003} are in close agreement with each other, and
the static value of Ref.~\cite{green2003} is in agreement with the $M \to \infty$ extrapolation of 
\cite{alikhan2000}.
The present lattice average  of the $B_2^\ast-B_0^\ast$ splitting is positive with a central value around
100 MeV.
The chiral extrapolation of the splittings has still large errors, but since the
potential model predictions for the $(j_l = 3/2) - (j_l = 1/2)$ splitting depend on the ratio of
heavy quark masses, an inverted level ordering would already have been predicted in the 
static-strange system. 
\subsection{$b$ Baryons \label{sec:bbary}}
\begin{figure}[thb]
\begin{center}
\epsfysize=4.9cm \epsfbox{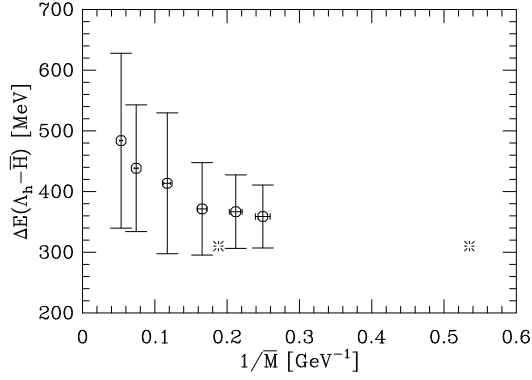}
\end{center}
\vspace{-0.5cm}
\caption{The $\Lambda_Q-\overline{H}$ splitting from quenched lattice NRQCD~\protect\cite{alikhan2000}.
The experimental values for the $\Lambda_b-\overline{B}$ and $\Lambda_c-\overline{D}$ splittings are
denoted by bursts.}
\label{fig:lambdab_glok}
\end{figure}
According to potential models, baryons with one $b$ and two light quarks 
can be described as a light diquark of spin zero or spin one which then couples to the heavy 
quark~\cite{ebert2005}. The resulting states are:
\be
\begin{array}{ccc}
j_l = 0 & \begin{array} {c}
            \to \\
            \end{array}
          & \begin{array}{cc}
             J = \frac{1}{2}: & \Lambda_b \\
            \end{array} \\
j_l = 1   & \begin{array} {c}
          {}_\nearrow \\
          {}^\searrow \\
          \end{array}
          & \begin{array}{cc}
           J = \frac{3}{2}: & \Sigma_b^\ast,\;\Xi_b^\ast,\; \Omega_b^\ast \\ 
           J = \frac{1}{2}: & \Sigma_b, \;\Xi_b,\; \Omega_b, \\      
\end{array} \\
\end{array}
\label{eq:barydoublets}
\ee
where the quark content is $(llb), (lsb), (ssb)$ from left to right.
Own results for the baryon-meson splittings are summarized in Table~\ref{tab:bary_sum}.
A summary of lattice and model results is given in Table~\ref{tab:bary}.

\begin{figure}[thb]
\begin{center}
\centerline{
\epsfysize=4.9cm \epsfbox{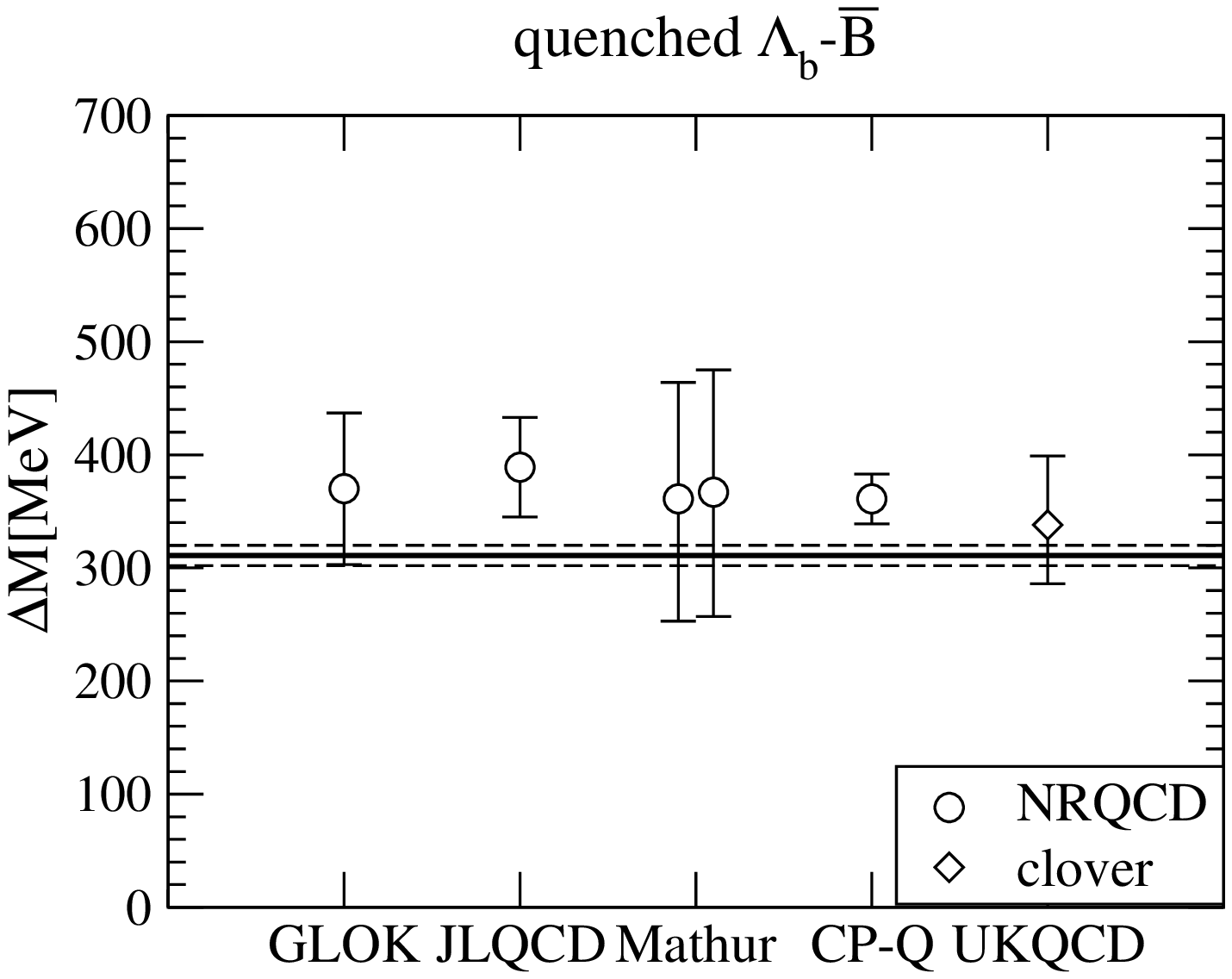}
\epsfysize=4.9cm \epsfbox{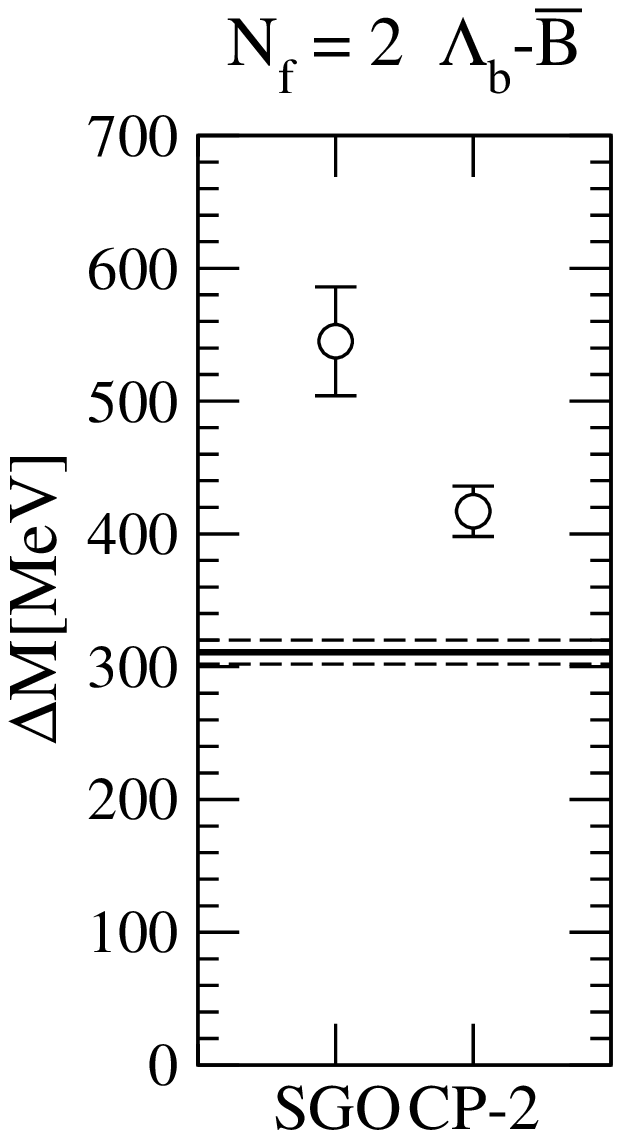}
\epsfysize=4.9cm \epsfbox{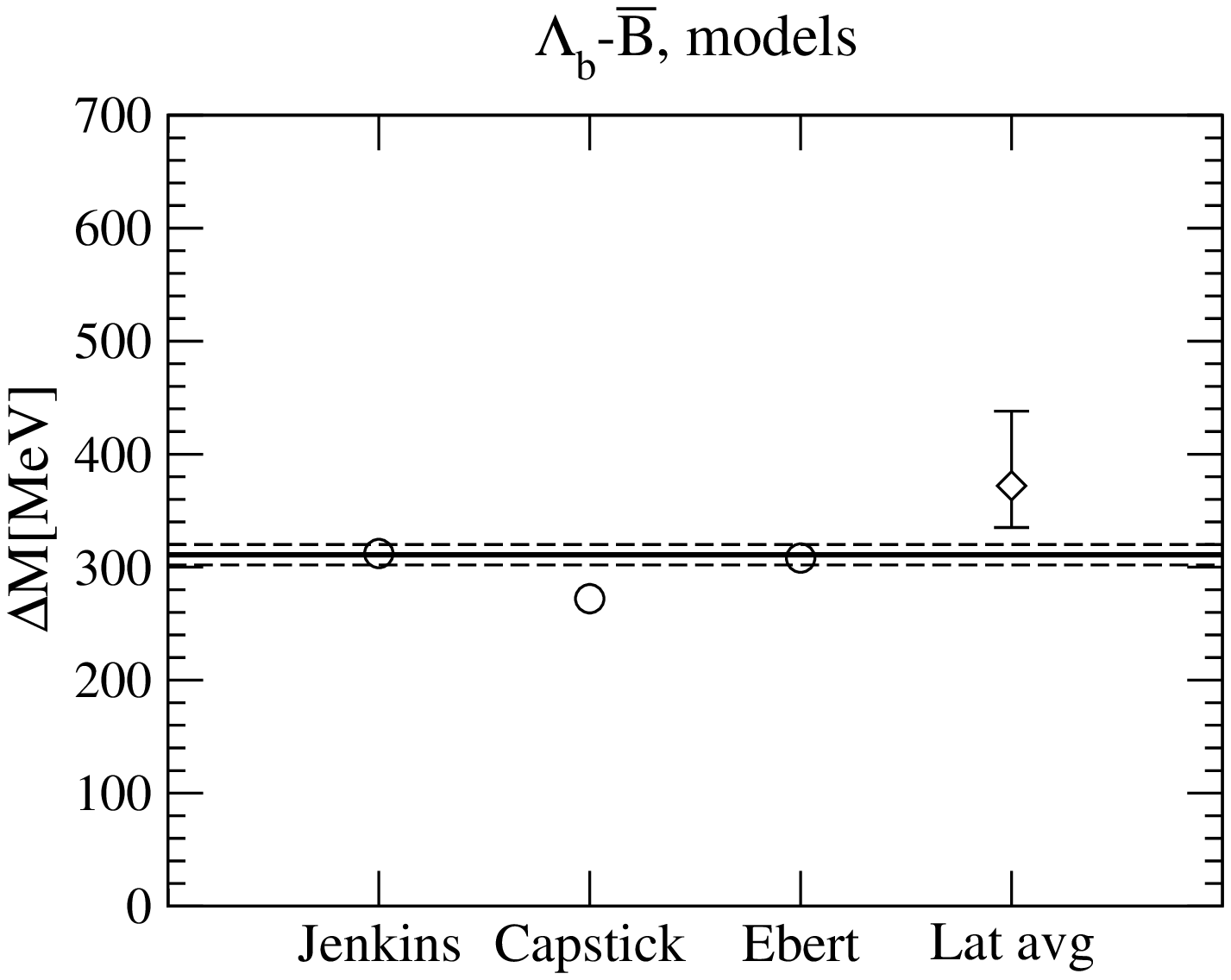}
}
\end{center}
\vspace{-0.5cm}
\caption{$\Lambda_b-\overline{B}$ splitting from  quenched (left) and
$N_f = 2$ (middle) lattices 
(Refs.~\protect\cite{collins1999,alikhan2000,cppacs2000,mathur2002,aoki2003,bowler1996}). 
The lattice average shown on the right is quenched. CP-Q and CP-2
denote quenched and unquenched results from \protect\cite{cppacs2000}, respectively. Model results are from 
Refs.~\protect\cite{capstick1986,jenkins1992}.}
\label{fig:lambdab-comp}
\end{figure}

The spin-averaged splitting $\Lambda_b-\overline{B}$ is plotted in Fig.~\ref{fig:lambdab_glok}.
Since in the quark model the light quarks in the $\Lambda_b$ are in a ground state with total spin 
zero, the expectation is that the splitting is dominated by the constituent mass of the 
additional light quark. The small slope of the splitting suggests that the heavy quark kinetic 
energy in the baryon is close to that in the meson, for experiment as well as for the lattice data. 
In Fig.~\ref{fig:lambdab-comp}, we compare quenched and $N_f = 2$ lattice results and results of model
calculations on the $\Lambda_b-\overline{B}$ splitting.
On quenched lattices, the result is $\sim 2 \sigma$ higher than the experimental value.
Preliminary results with two flavors of dynamical quarks around the 
strange quark mass~\cite{collins1999,cppacs2000} are even higher. On coarse lattices, we find
an increase by $\sim 15\%$ if $N_f$ is changed from zero to two dynamical
quarks of around  $3\times$ the strange quark mass~\cite{cppacs2000}.
Chiral extrapolation might be a source of uncertainty. Ref.~\cite{guo2003} estimates
the mass splittings of unquenched heavy baryons as a function of the light quark mass using
$\chi PT$ and predicts deviations from linear behavior.

\begin{figure}[thb]
\begin{center}
\epsfysize=4.9cm \epsfbox{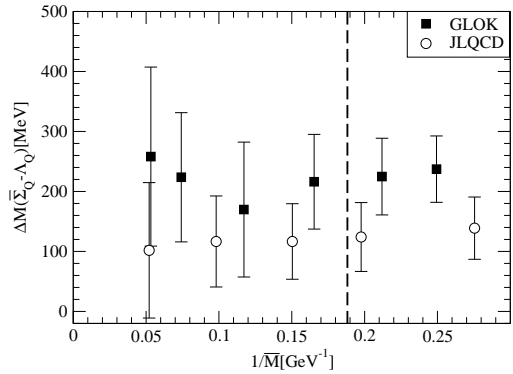}
\end{center}
\vspace{-0.5cm}
\caption{Mass dependence of the $\overline{\Sigma}_Q-\Lambda_Q$ splitting using quenched lattice 
NRQCD from ~\protect\cite{alikhan2000} and \protect\cite{aoki2003}. The spin-averaged meson mass
is denoted by $\overline{M}$. }
\label{fig:lambdasigma_glok}
\end{figure}
The splitting between the spin-averaged $j_l = 1$ and the $j_l = 0$ state is also
expected to be relatively heavy quark mass independent. This is indeed found in the lattice 
data shown in Fig.~\ref{fig:lambdasigma_glok}. 
\begin{figure}[thb]
\begin{center}
\epsfysize=4.9cm \epsfbox{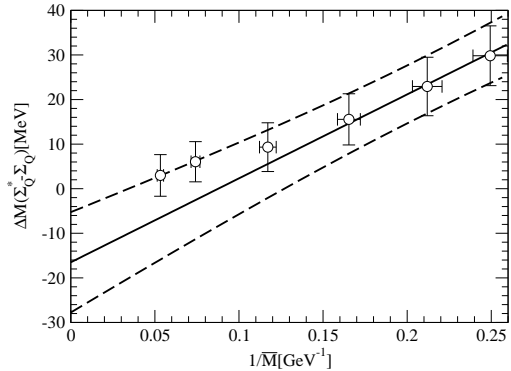}
\end{center}
\vspace{-0.5cm}
\caption{The $\Sigma_Q^\ast-\Sigma_Q$ splitting from quenched lattice NRQCD~\protect\cite{alikhan2000}.
The solid line denotes a fit to the three lightest meson masses. Dashed lines show the fitting errors.}
\label{fig:sigmastarsigma_glok}
\end{figure}
The light quark mass dependence of this splitting is very small. 
We now try to compare it with the results from other lattice calculations and 
results using a constituent quark model~\cite{capstick1986},
a Skyrme model~\cite{jenkins1992} where the
baryon is described as a bound state between a soliton and a heavy quark, and a quark model
with a relativistic description of the light quarks~\cite{ebert2005}.
Ref.~\cite{jenkins1992} predicts the relation between the heavy-light and light baryon splittings
to be $\Delta M\left(\overline{\Sigma}_Q - \Lambda_Q\right)/\Delta M\left(\Delta-N\right) = 2/3$,
for $Q = c,b$. For $Q=c$, the equality is experimentally well satisfied. 
The lattice results for the ratio with $Q=b$ still have large errors and vary between 0.5 and 1. 

\begin{figure}[thb]
\begin{center}
\centerline{
\epsfysize=4.5cm \epsfbox{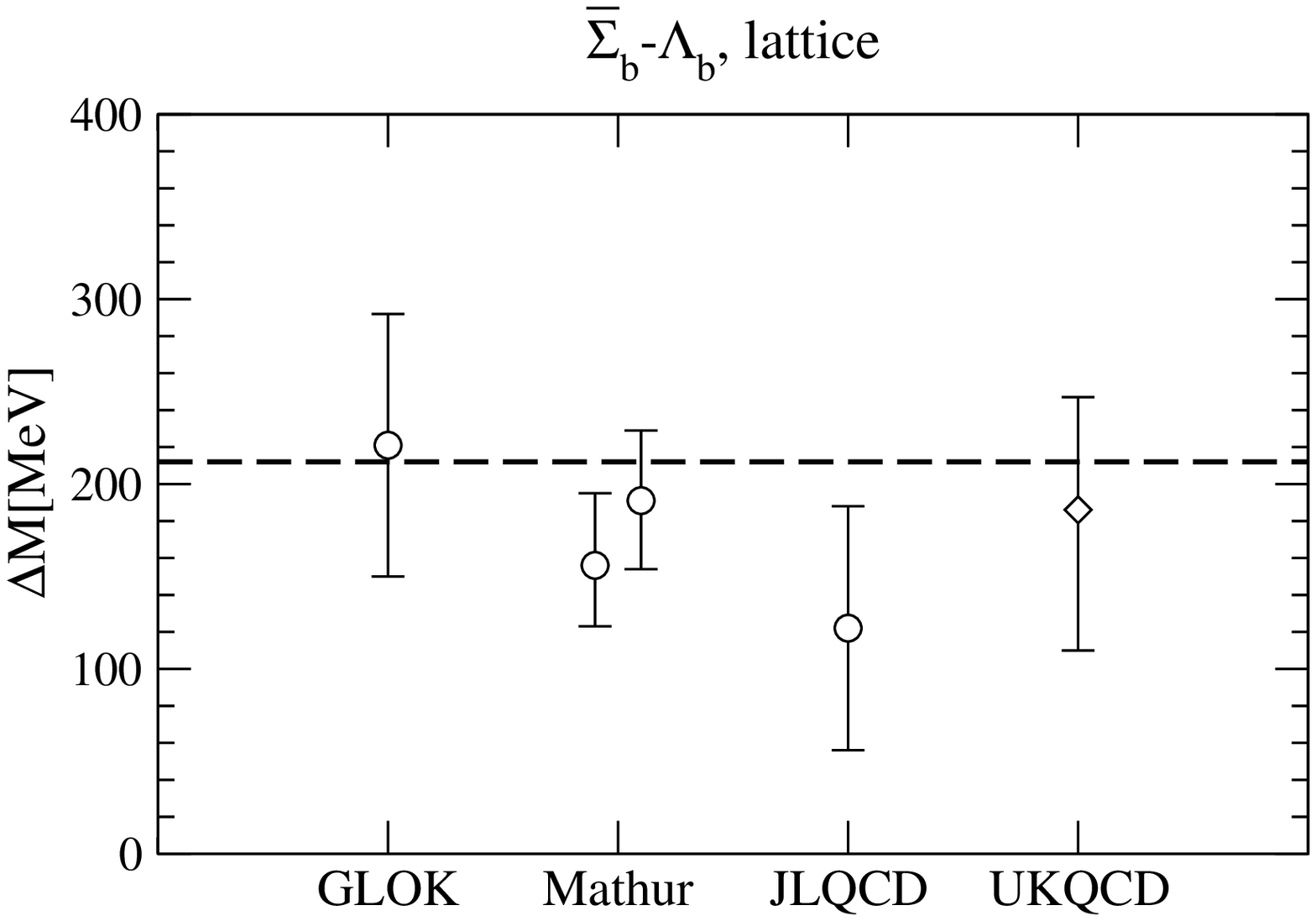}
\epsfysize=4.5cm \epsfbox{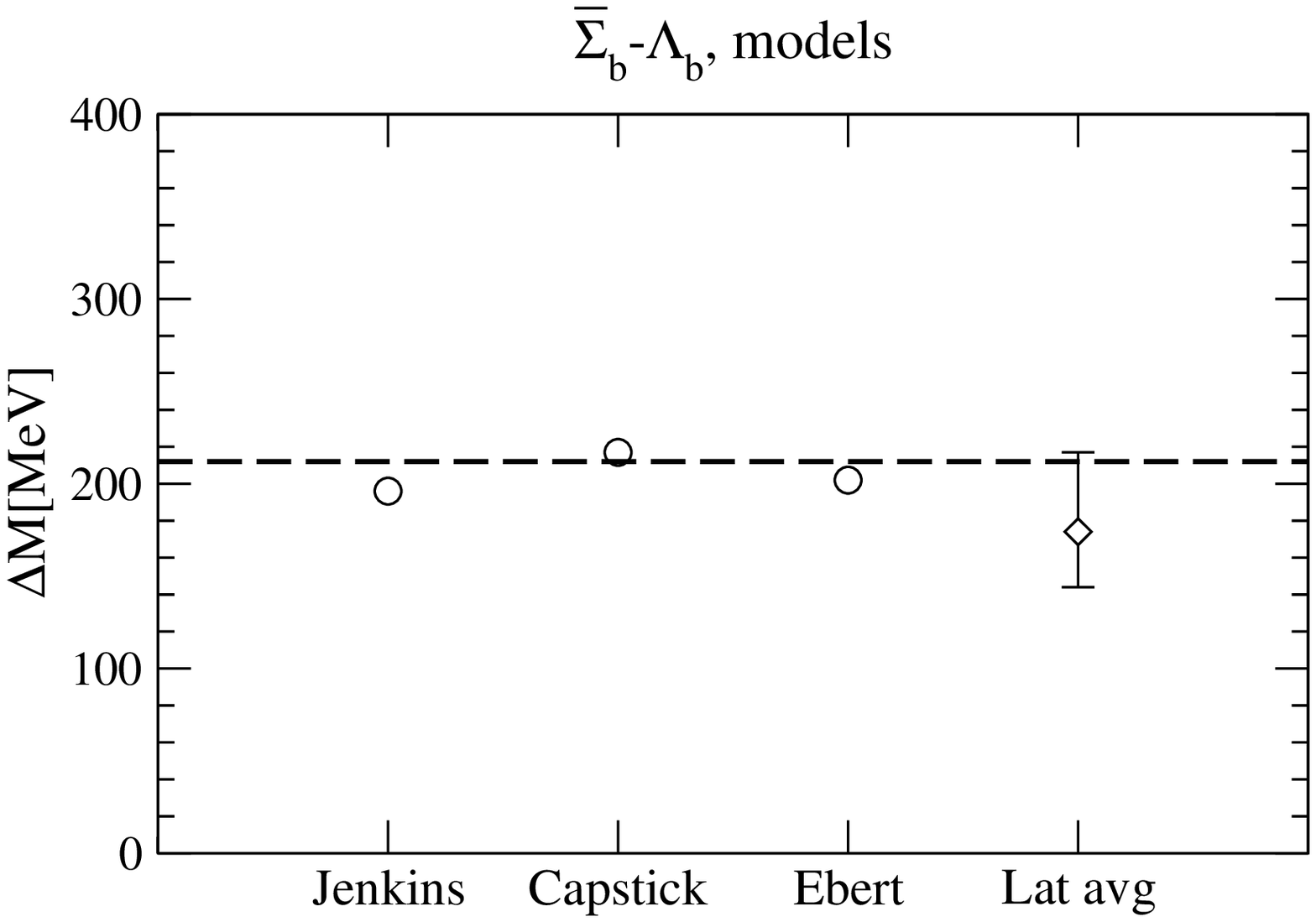}
}
\centerline{
\epsfysize=4.5cm \epsfbox{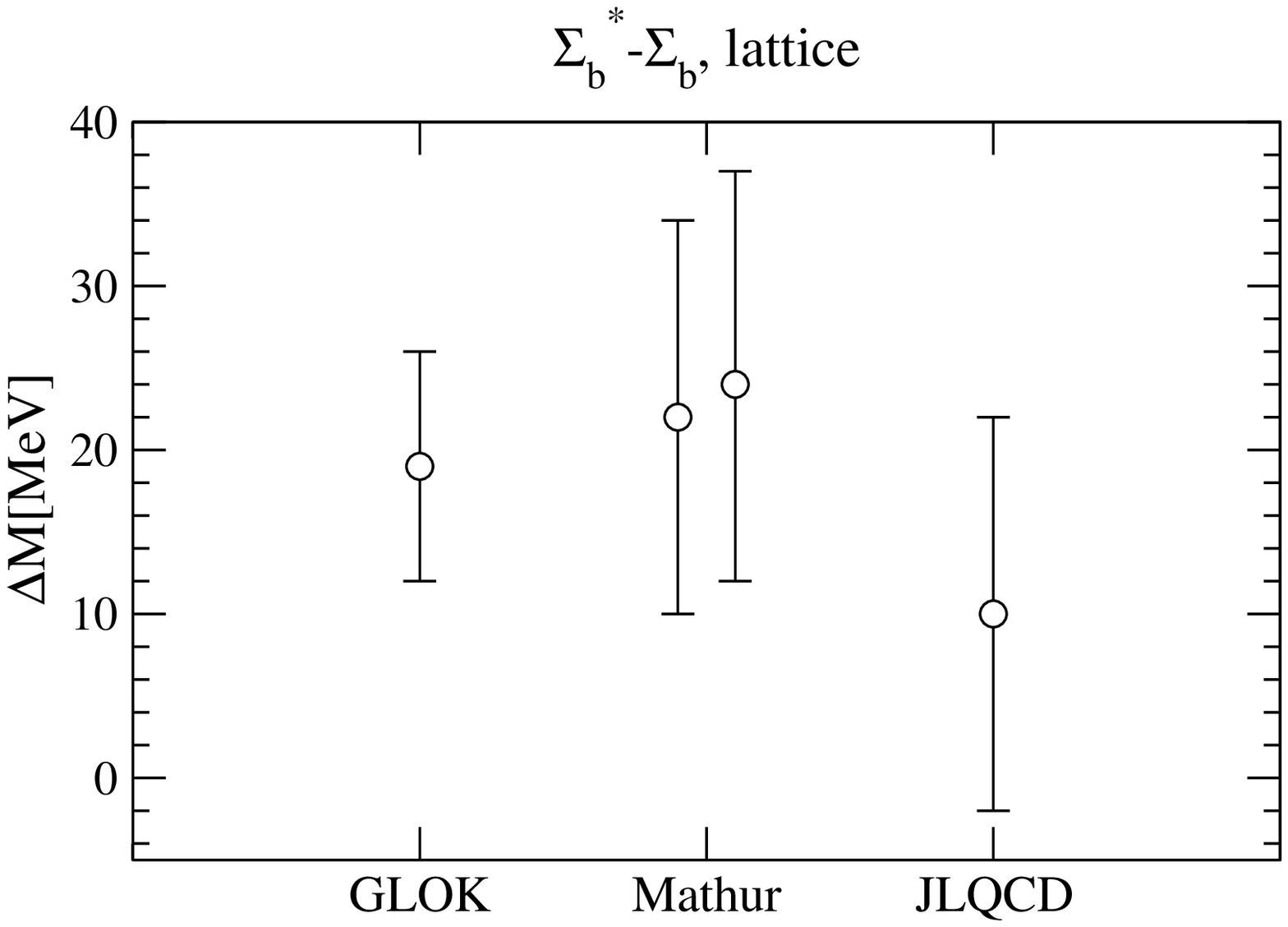}
\epsfysize=4.5cm \epsfbox{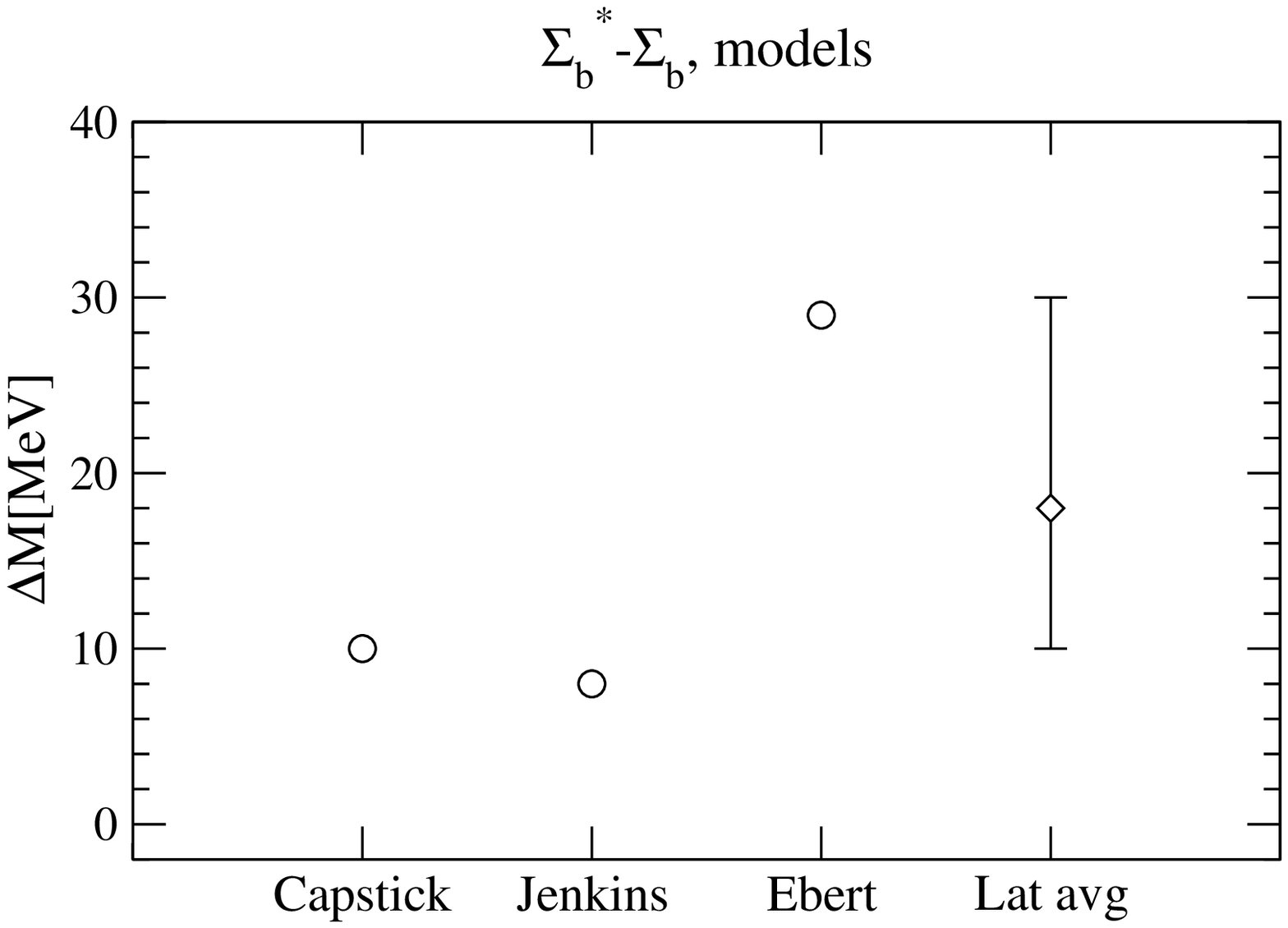}
}
\end{center}
\vspace{-0.5cm}
\caption{Above: $\overline{\Sigma}_b-\Lambda_b$ splitting from the lattice (left) and model calculations
(right). The dashed line shows the
experimental  value for the $\overline{\Sigma}_c-\Lambda_c$ splitting. Below: 
${\Sigma_b}\hspace{-0.1cm}^\ast-\Sigma_b$ splitting. Lattice data from 
Refs.~\protect\cite{alikhan2000,mathur2002,aoki2003,bowler1996}, model
results from Refs.~\protect\cite{capstick1986,jenkins1992}.}
\label{fig:sigmab-comp}
\end{figure}

In Fig.~\ref{fig:sigmab-comp} above we summarize lattice results for the spin-averaged 
$\overline{\Sigma}_b-\Lambda_b$  splitting. 
Where results from one collaboration at several lattices are plotted, the leftmost point is from
the coarsest lattice.

The expectation from HQET for the $\Sigma_Q$  hyperfine splitting (for baryons containing one
heavy quark $Q$) is that it is generated 
by the spin-chromomagnetic interaction (see Eq.~(\ref{eq:HQET})). It should therefore in first
approximation be linear in the inverse heavy quark or meson mass, up to logarithmic corrections. 
The data is compatible with this within errors, 
as shown in Fig.~\ref{fig:sigmastarsigma_glok}. 
On larger lattices and with $O(a)$-improved light quarks, Ref.~\cite{aoki2003} finds a similar behavior.
Their results for the $b$ meson and baryon splittings are in general agreement with ours~\cite{alikhan2000}.
Some trends in the data are slightly different. For example, their ratio 
$(B^\ast_s-B_s)/({\Omega_b}\hspace{-0.1cm}^\ast -\Omega_b)$ is closer to $3.5$ instead of our value of 
$\sim 1.5$, and their $\overline{\Sigma}_b-\Lambda_b$ splitting is smaller; however errors are
still large.

Assuming that the $\Sigma_b^\ast-\Sigma_b$ splitting is proportional to
$1/M_Q$ up to logarithmic corrections, one can rescale the experimental value of 
$\Delta M({\Sigma_c}\hspace{-0.1cm}^\ast- \Sigma_c) \simeq 67$ MeV by the ratio of $b$ and $c$ quark 
masses and finds 
$\Delta M({\Sigma_b}\hspace{-0.1cm}^\ast-\Sigma_b) \sim 20$ MeV, a value compatible with 
the lattice results.

\subsection{$bb$ Baryons \label{sec:bbbary}}
It is expected that the level structure of baryons containing two $b$ and one light 
quark can be understood as arising from a $b$ diquark color antitriplet  with size much smaller than 
$\lqcd$
which interacts with the light quark in a similar way as the heavy antiquark in a heavy-light meson.
\be
\begin{array}{ccc}
j_l = 0 & \begin{array} {c}
            \to \\
            \end{array}
          & \begin{array}{cc}
             J = \frac{1}{2}: & 
\Xi^{\prime 0}_{bb'}, \Xi^{\prime -}_{bb'}, 
\Omega^{\prime -}_{bb'}, \\
            \end{array} \\
j_l = 1   & \begin{array} {c}
          {}_\nearrow \\
          {}^\searrow \\
          \end{array}
          & \begin{array}{cc}
           J = \frac{3}{2}: & \Xi^{\ast 0}_{bb}, \Xi^{\ast -}_{bb}, 
\Omega^{\ast -}_{bb}, \\
           J = \frac{1}{2}: & \Xi^{ 0}_{bb}, \Xi^{ -}_{bb}, 
\Omega^{ -}_{bb} \\
\end{array} \\
\end{array}
\label{eq:hhlstates}
\ee
For two identical heavy quarks the $j_l = 0$ state does not occur because of the Pauli 
exclusion principle. The
three states for each $J$ value with $j_l = 1$ correspond to the quark content
$(bbu), (bbd)$ and $(bbs)$ respectively.
\begin{figure}[thb]
\begin{center}
\epsfysize=4.9cm \epsfbox{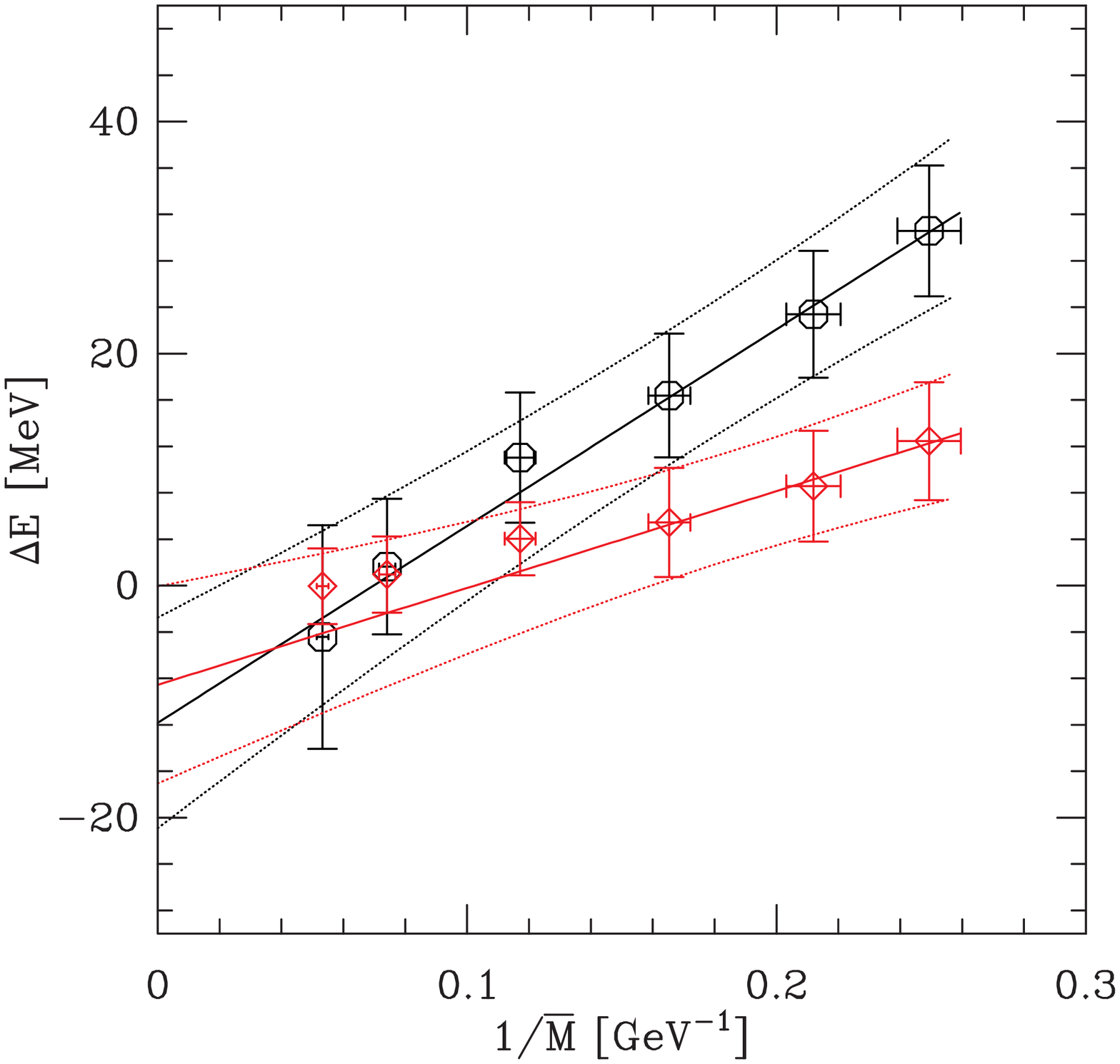}
\end{center}
\vspace{-0.5cm}
\caption{$\Xi_{QQ}^\ast-\Xi_{QQ}$ (circles) and 
$\Xi_{QQ}-\Xi_{QQ}^\prime$ (diamonds) splitting from quenched lattice NRQCD~\protect\cite{alikhan2000}.
The solid lines denote linear fits, }
\label{fig:sigmastarsigmabb_glok}
\end{figure}
The mass difference between the $J = 1/2$ and $J = 3/2$ states is a hyperfine splitting
depending on the heavy and light quark spin orientation.
Its leading behavior is expected to be linear in the inverse heavy quark or meson mass up to 
logarithmic corrections, which is also found  in the lattice data shown in 
Fig.~\ref{fig:sigmastarsigmabb_glok}.

The results are:

\vspace{0.5cm}

\begin{center}
\begin{tabular}{cl}
$\Xi_{bb}^\ast-\Xi_{bb}:$          &  $20(6)(^{+2}_{-3})$ MeV  \\
$\Omega_{bb}^\ast-\Omega_{bb}:$    &  $20(4)(^2_3)$       MeV  \\
$\Xi_{bb}:$                        &  $10314(46)(^{-10}_{+11})$ MeV \\
$\Omega_{bb}:$                     & $10365(40)(^{-11}_{+12})$  MeV\\
\end{tabular}
\end{center}

\vspace{0.5cm}

The lattice calculation of Ref.~\cite{mathur2002} quotes very similar values for the
hyperfine splittings of doubly heavy baryons, namely $22(6)(^4_3)$ MeV and $20(6)(^3_4)$
MeV with spatial lattice spacings of $a =1.1$GeV$^{-1}$  $a =0.9$GeV$^{-1}$ for 
$\Delta m(\Xi_{bb}^\ast-\Xi_{bb})$. For $\Delta m(\Omega_{bb}^\ast-\Omega_{bb})$ they
find $20(4)(^3_3)$ MeV and $19(4)(^3_3)$ MeV respectively.

The result for the $\Omega_{bb}$ mass is in agreement with the potential model 
calculation of \cite{ebert2002a}. The hyperfine splittings are slightly smaller
than the potential model results of $\sim 35$ MeV which might be due to a 
quenching effect in the lattice results. The potential model predicts
the $\Xi_{bb}$ mass to be around 100 MeV ($2 \sigma$) lower than the lattice result. 

Ref.~\cite{brambilla2005a} finds a value for the $\Xi_{bb}^\ast-\Xi_{bb}$ of 34(4) MeV within the
framework of potential NRQCD, in agreement 
with the potential model~\cite{ebert2002a}. The most plausible explanation for the difference 
is again that the lattice result for the splitting is slightly underestimated in the quenched approximation.
\subsection{The $B_c$ Spectrum}
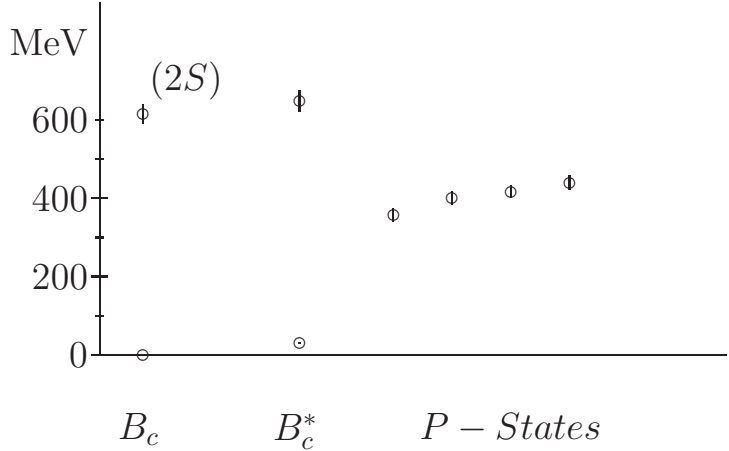
\begin{figure}[t]
\begin{center}
\setlength{\unitlength}{.0205in}
\begin{picture}(110,100)(30,100)

\put(15,100){\line(0,1){90}}
\multiput(13,100)(0,20){3}{\line(1,0){4}}
\multiput(14,100)(0,10){7}{\line(1,0){2}}
\put(12,100.0){\makebox(0,0)[r]{{\large 0}}}
\put(12,120.0){\makebox(0,0)[r]{{\large 200}}}
\put(12,140.0){\makebox(0,0)[r]{{\large 400}}}
\put(12,160.0){\makebox(0,0)[r]{{\large 600}}}
\put(12,180){\makebox(0,0)[r]{{\large MeV}}}
\put(15,100){\line(1,0){160}}

     \put(25,85){\makebox(0,0)[t]{{\large $B_c$}}}
     \put(26,100){\circle{2.9}}
     \put(26,161.6){\circle{2.9}}
     \put(26,161.6){\line(0,1){2.5}}
     \put(26,161.6){\line(0,-1){2.5}}
     \put(37,175){\makebox(0,0)[t]{{\large $(2S)$}}}

     \put(65,85){\makebox(0,0)[t]{{\large $B_c^{*}$}}}
     \put(66,103){\circle{2.9}}
     \put(66,103){\line(0,1){0.1}}
     \put(66,103){\line(0,-1){0.1}}
     \put(66,164.9){\circle{2.9}}
     \put(66,164.9){\line(0,1){2.6}}
     \put(66,164.9){\line(0,-1){2.6}}

     \put(120,85){\makebox(0,0)[t]{{\large $P-States$}}}
     \put(135,144){\circle{2.9}}
     \put(135,144){\line(0,1){1.8}}
     \put(135,144){\line(0,-1){1.8}}
     \put(90,135.7){\circle{2.9}}
     \put(90,135.7){\line(0,1){1.6}}
     \put(90,135.7){\line(0,-1){1.6}}
     \put(105,140.1){\circle{2.9}}
     \put(105,140.1){\line(0,1){1.7}}
     \put(105,140.1){\line(0,-1){1.7}}
     \put(120,141.7){\circle{2.9}}
     \put(120,141.7){\line(0,1){1.6}}
     \put(120,141.7){\line(0,-1){1.6}}

\end{picture}
\end{center}
\vspace{0.5cm}
\caption{$B_c$ level splittings~\cite{alikhan1998b}. The error bars are purely statistical.}
\label{fig:Bcspect}
\end{figure}
If the mass of the $u$ quark in the $B^+$ meson is increased towards the charm quark mass one
obtains the $B_c^+$ meson.
In Fig.~\ref{fig:Bcspect} we summarize results on the $B_c$ spectrum with clover
charm and NRQCD $b$ quarks~\cite{alikhan1998b}. The lattice scale is set using the charmonium
$\overline{P}-\overline{S}$ splitting from clover quarks~\cite{boylephd}. 
The $B_c$ mass can be calculated using the non-relativistic interpretation for the clover charm 
as well as for the NRQCD $b$ quarks and  relation~(\ref{eq:hh}), 
\be
M_{B_c} = E_{\mathrm{sim}} + \frac 1 2 \Delta_{NRQCD, HH} + \frac 1 2 \Delta_{\mathrm{clover},HH},
\ee
with the result $6.3(1)$ GeV, in agreement with the experimental value of $6.4(4)(1)$ GeV from 
\cite{pdg}, and a recent measurement from CDF~\cite{acosta2005} of 6287(5) MeV.

The quenched calculation of \cite{shanah1999}) uses the same lattice actions as our calculation, 
however at a finer
lattice with $a \simeq 0.06$ fm if the $\chi_b-\Upsilon$ splitting is used to set the scale. The bare quark mass
of their heavier quark is fixed to the $b$ while the charm quark mass is reached using an interpolation.
They find a $B_c$ ground state mass of 6.386(9)(99) GeV, where the first error is statistical and the
second is an estimate of uncertainties due to quenching and estimation of $M_{\eta_b}$, added in quadrature. 
Determining the charmed $\kappa$ value by fixing the ratio $m_{D_s}/(\Delta m(\Upsilon'-\Upsilon))$ to the 
experimental value and interpolating the ground state energy to the charmed $\kappa$ value, 
we obtain an estimate of the energy splittings to the ground state from their data.
Their $P-S$ mass splittings are  in general larger than the lattice ones by $\sim 100$ MeV (or $20-25\%$).
In particular, they find $\Delta m(B_{c0}^\ast-B_c) \sim 500$ MeV compared to our value of $\sim 360$ MeV. 
The difference can to a large extent be explained by the fact that in the quenched approximation 
the lattice spacings from the $\Upsilon$ are larger 
than the ones from charmonium. At $\beta = 6.0$, the difference between the value used in our calculation and the
$\chi_b-\Upsilon$ splitting is $\sim 20\%$.
The $B_{c2}^\ast-B_{c0}^\ast$ splitting  is found to be around 80 MeV in both calculations.
An earlier calculation of~\cite{davies1996} uses NRQCD for $b$ and $c$ quarks on a coarser lattice. 
The bare $b$ and $c$ quark masses are fixed using $\overline{b}b$ and $\overline{c}c$
quarkonium splittings respectively, using in each determination the different values for the lattice scale 
from mass splittings of the respective quarkonium. They fix the lattice spacing by assuming that the experimental value
for the $B_c$ $\overline{P}-\overline{S}$ splitting agrees with the one for the $\Upsilon$ and obtain a value
similar to the lattice spacing from the $\Upsilon'-\Upsilon$ splitting. 
Their results for the $P-S$ mass splittings are between
ours and the results from \cite{shanah1999}, while their $P$ state splittings tend to be smaller, e.g.\
$\Delta m(B_{c2}^\ast-B_{c0}^\ast)\sim 50$ MeV.
A recent simulation using three $N_f = 2+1$ lattices ranging from $0.1 \leq a \leq 0.2$ fm
with NRQCD $b$ and Fermilab $c$ valence quarks finds a $B_c$ mass in agreement with the previous 
results: $m_{B_c} = 6.304(12)(^{18}_0)$ GeV, where the first error is a quadratic sum of statistical and
systematical errors, and the second is an estimate of the heavy quark discretization uncertainty.
The results for the ground state masses are therefore not very sensitive to quenching effects and support the 
idea that heavy quark energies are physically relevant. Once
the quark mass has been fixed, e.g.\ by matching the dispersion relation of one meson to the physical value, the 
mass of another meson containing the same heavy quark can be predicted reasonably well.
\subsection{Comparison to other work\label{sec:otherspec}}
In this Section we give technical details of the other lattice calculations occuring in the
comparisons in the previous sections and describe how the lattice averages presented in the 
previous sections were determined.

Results on $P$ wave states and baryons from other authors  are summarized in the 
Tables~\ref{tab:p2},~\ref{tab:psplit} and~\ref{tab:bary}. 
Masses averaged over spin-orientations (spin-averaged) are denoted by an 
overbar. The first error on the individual lattice results includes statistical errors and uncertainties
fixing the masses to the physical values, the second, where applicable, is a chiral extrapolation uncertainty.
Comparison plots of the results, showing the statistical errors only, were already shown in 
Sec.~\ref{sec:hresults}. 

For comparison to experiment and to model calculations 
one would like to have an estimate for the `lattice average'. In
\cite{alikhan2004} it is attempted to calculate the error using a similar method as for the
averages of experimental data~\cite{pdg}. The assumption is that sets of  lattice calculations using
similar methods have the same systematic error, 
or that  the systematic error of sets of  lattice results can be divided into a common 
part and a rest which is treated as independent. The common part is
taken to be of the order of the error of the calculation with the smallest uncertainty.
The error on the average is rescaled by $r = \sqrt{\chi^2/(N-1)}$ if $r > 1$, where $N$ is the number of 
results. 

Among the states we could resolve, we found in the previous subsection that the spectrum changes only little 
if the NRQCD action is changed from $1/M$ to $1/M^2$. There have been calculations of the effect of higher 
NRQCD corrections on ground state $B$ mesons.
Ref.~\cite{lewis2000} quotes a 6\% increase of the $B^\ast-B$ splitting
when the action is changed from $1/M$ to $1/M^2$, and, depending on the exact discretization of the
NRQCD action, a $0-4$\% decrease when the action is changed from $1/M^2$ to $1/M^3$.
For the ${B_0}\hspace{-0.1cm}^\ast-B$ splitting, Ref.~\cite{lewis2000} finds no change between $1/M$ and
$1/M^2$ and a $0-6$\% increase from $1/M^2$ to $1/M^3$.

Most of the calculations use NRQCD, except for~\cite{bowler1996} which uses heavy clover quarks and
\cite{green2003} which simulates $B_s$ mesons in the static approximation
and interpolates between the static and experimental $D_s$ mesons.
Discretization 
errors with non-perturbatively $O(a)$ improved clover light quarks (finer lattice of~\cite{hein2000}
and~\cite{aoki2003}) are $O(a^2\lqcd^2)$, whereas the tadpole-improved light
clover action has $O(a^2\lqcd^2)$ and $O(\alpha_s a\lqcd)$ errors (coarser lattice of~\cite{hein2000}
and~\cite{alikhan2000,cppacs2000}). Refs.~\cite{lewis2000,mathur2002} use $O(a^2)$ 
tree-level tadpole-improved clover light actions respectively. 
Ref.~\cite{wing2003.stag} uses staggered light quarks. 

The NRQCD action has errors $O(\alpha_s\lqcd/M)$ from  corrections to the spin-magnetic coefficient.
Errors on spin splittings are treated as being dominated by an error on the spin-magnetic coefficient of
$O(\alpha_s) \sim 20-30\%$.

It is obvious that this way of estimating the systematic error is rather schematic.
Using this method one would expect that the  discretization
errors of the light hadron spectrum with staggered quarks ($O(a^2)$) are rather smaller than those of 
tadpole-improved Wilson quarks $O(a^2)$ and $O(\alpha_s a)$ at the same lattice spacing.
In calculations of the light spectrum, discretization errors of staggered quarks appear not so small. 
However this is the usual method to estimate the systematic error in lattice calculations of weak matrix 
elements using a non-relativistic effective theory.

The scale is determined using $m_\rho$ except in Ref.~\cite{cppacs2000} which 
sets the scale with $\sqrt{\sigma} = 427$ 
MeV. Refs.~\cite{green2003} and ~\cite{burch2004} use $r_0$ with physical values of  0.525  and 0.5 
fm respectively, and~\cite{wing2003.stag} uses $r_0 = 0.5$ fm and 
quarkonia at and around the charm~\cite{alford1998}. All other calculations from the set discussed here
use $m_\rho$.

The variation due to the $10\%$ ambiguity between using $a$ from $m_\rho$ and
$a$ from $r_0 = 0.5\mbox{ fm}$ in the quenched case, and asymmetric chiral 
extrapolations where applicable, are included in the second error on the averages.
The $\chi_b-\Upsilon$ mass difference is not included in the estimate of the scale variation
since it gives values for spin-independent mass splittings which are much higher than experiment. 
For example, Ref.~\cite{collins1999} quotes an increase of 
the $B_s-B$ and the $\Lambda_b-B$ splittings from 98 and 560 MeV by $\sim 20\%$ if the scale is 
set with the $\chi_b-\Upsilon$ splitting instead with $m_\rho$. The experimental values are $90$ and $345$ MeV
respectively~\cite{pdg}.

For the error estimates we use  nominal values of $\lqcd = 400$ MeV,  $M = 5$ GeV and
$\alpha_s = \alpha_V(1/a)$, 
where $\alpha_V(q^\ast)$ is  defined in the potential scheme~\cite{lepage1993,hornbostel2003} at the scale
$q^\ast$ of Eq.~(\ref{eq:qast}). 
Since the lattice results have rather varying central values we do not calculate the error in percent of the
individual lattice splittings but of the experimental splittings or nominal estimates thereof.
We use guessed values where no experimental value is available: 400 MeV for $B_0^\ast-B$ and $B_{s0}^\ast-B_s$,
500 MeV for $B_2^\ast-B$ and $B_{s2}^\ast-B_s$ and 100 MeV for $B_2^\ast-B_0^\ast$. 

\begin{table}[thb]
\begin{center}
\begin{tabular}{|l|ll|}
\hline
\multicolumn{1}{|c}{Ref.} &
\multicolumn{1}{c}{$\Delta M(B  )$[MeV]} &
\multicolumn{1}{c|}{$\Delta M(B_s)$[MeV]} \\
\hline
\hline
 \multicolumn{3}{|c|}{$\bf{B_0^\ast-B}$} \\
\hline
\multicolumn{3}{|c|}{\bf{Lattice}} \\
\hline
\protect\cite{hein2000}, $a \sim 1.1$GeV$^{-1}$  
                               &                         & $400(30)(^{19}_0)$ \\
\protect\cite{hein2000}, $a \sim 2.6$GeV$^{-1}$  
                                 &                         & $295(45)(^{11}_0)$ \\
\protect\cite{alikhan2000}      &  $374(37)$         &  $357(27)(^0_5)$  \\
\protect\cite{lewis2000}        &$475(^{19}_{20})$ &  $451(^{15}_{16})$ \\ 
\protect\cite{wing2003.stag}, 1l  &                   &  $442(56)$ \\
\protect\cite{wing2003.stag}, 1l  &                   &  $471(25)$ \\
\protect\cite{wing2003.stag}, 1l  &                   &  $315(105)$ \\
\protect\cite{wing2003.stag}, 1l  &                   &  $425(60)$ \\
\protect\cite{wing2003.stag}, Asq  &                   &  $285(78)$ \\
\protect\cite{wing2003.stag}, Asq  &                   &  $523(94)$ \\
\protect\cite{wing2003.stag}, Asq  &                   &  $403(56)$ \\
\protect\cite{green2003}        &                   &  $386(31)$ \\
\protect\cite{burch2004}        &  $408(67)$           &   $419(37)$ \\
average                         &  $402(41)(^{33}_7)$&  $382(23)(^{27}_{13})$      \\
\hline
\multicolumn{3}{|c|}{\bf{Model calculations}} \\
\hline
\protect\cite{godfrey1985}      & 450                     & 440        \\
\protect\cite{ebert1998}        & 453                     & 466      \\
\protect\cite{dipierro2001}     & 427                     & 431   \\
\hline                        
\hline
 \multicolumn{3}{|c|}{$\bf{B_2^\ast-B}$} \\
 \hline
 \hline
\multicolumn{3}{|c|}{Lattice} \\
\hline
\protect\cite{hein2000}, $a \sim 1.1$GeV$^{-1}$ 
                                &                         &$402(78)(^{20}_0)$ \\
\protect\cite{hein2000}, $a \sim 2.6$GeV$^{-1}$  
                               &                         & $474(62)(^{16}_0)$ \\
\protect\cite{alikhan2000}     &   $526(45)$             & $493(26)$   \\
\protect\cite{lewis2000}       &   $493(^{29}_{32})$     & $478(^{22}_{24})$   \\
\protect\cite{cppacs2000}      &   $426(17)$             &    \\
\protect\cite{green2003}       &                         & $534(52)$   \\
\protect\cite{burch2004}        &  $440(77)$             &  $455(41)$ \\
average                        &   $498(49)(^{42}_8)$    & $487(31)(^{44}_8)$ \\
\hline
\multicolumn{3}{|c|}{\bf{Model calculations}} \\
\hline
\protect\cite{godfrey1985}     & 450                     & 490        \\
\protect\cite{ebert1998}       & 453                     & 469      \\
\protect\cite{dipierro2001}    & 435                     & 447  \\
\hline                        
\end{tabular}
\end{center}
\caption{$B_2^\ast-B$ splittings. Only statistical errors and, where quoted by the authors, 
errors due to chiral extrapolation and fixing the $b$ quark mass are shown. 
}
\label{tab:p2}
\end{table}
\begin{table}[thb]
\begin{center}
\begin{tabular}{|l|ll|}
\hline
\multicolumn{1}{|c}{Ref.} &
\multicolumn{1}{c}{$\Delta M(B  )$[MeV]} &
\multicolumn{1}{c|}{$\Delta M(B_s)$[MeV]} \\
\hline
\hline
\multicolumn{3}{|c|}{\bf{Lattice}} \\
\hline
\protect\cite{hein2000}, $a \sim 1.1$GeV$^{-1}$  
                                &                         & $41(94)(^{14}_{0})$  \\
\protect\cite{hein2000}, $a \sim 2.6$GeV$^{-1}$  
                           &                         & $179(65)(^{6}_{0})$  \\
\protect\cite{alikhan2000} &   $155(32) $             & $136(23)$   \\
\protect\cite{lewis2000}   &   $18(^{36}_{38})$     & $27(^{27}_{29})$   \\
\protect\cite{green2003}   &                         & $148(61)$   \\
\protect\cite{burch2004}   &  $32(87)$               & $36(55)$     \\
average              &        $98(47)(^{10}_0)$            & $101(25)(^{11}_0)$ \\
\hline
\multicolumn{3}{|c|}{\bf{Model calculations}} \\
\hline
\protect\cite{godfrey1985}      & 40                      & 50        \\
\protect\cite{isgur1998}        & $-155$                  &          \\
\protect\cite{ebert1998}        & $-5$                    & 3      \\
\protect\cite{dipierro2001}     & 7                       & 16  \\
\hline
\end{tabular}
\end{center}
\caption{$P$ state fine structure of $B$ mesons. Errors include statistical uncertainties and
uncertainties due to chiral extrapolation and in determination of the $b$ quark mass.}
\label{tab:psplit}
\end{table}
The lattice results for $B_0^\ast-B$ splittings are summarized in Table~\ref{tab:p2}, for
$B_2^\ast-B$ splittings in Table~\ref{tab:p2} and for $B_2^\ast-B_0^\ast$ splittings in
Table~\ref{tab:psplit}.

\begin{table}[thb]
\begin{center}
\begin{tabular}{|lcccc|}
\hline
\multicolumn{5}{|c|}{$N_f  = 0 $}  \\
\hline
\hline
\multicolumn{5}{|c|}{$llb$}  \\
\hline
\multicolumn{2}{|c}{Ref.}  &   $\Lambda_b -\overline{B}$ & $\overline{\Sigma}_b-\Lambda_b$ &
$\Sigma_b^\ast-\Sigma_b$            \\
\hline
SGO & \protect\cite{alikhan1996}     &    410(10)                  &             &  \\
GLOK & \protect\cite{alikhan2000}    & $370(67(^{14}_{20})$ &     $221(71)(^{12}_{16})$   &
$19(7)(^2_3)$ \\
CP-PACS & \protect\cite{alikhan2000} & 361(22) &                   &     \\
\hline
\multicolumn{5}{|c|}{$lsb$}  \\
\hline
\multicolumn{2}{|c}{Ref.} &  
\multicolumn{1}{c}{$\Xi_b - \overline{B}_s$}  &  
\multicolumn{1}{c}{$\overline{\Xi'}_b-\Xi_b$}          & 
\multicolumn{1}{c|}{$\Xi_b^\ast-\Xi'_b$}   \\
GLOK & \protect\cite{alikhan2000} &  $392(50)(^9_{15})(^{15}_0)$   & $186(51)(^{+13}_{-17})(^{+0}_{-10})$ &
$19(5)(^2_3)$                   \\
\hline
\multicolumn{5}{|c|}{$ssb$}  \\
\hline
\multicolumn{2}{|c}{Ref.} &  
\multicolumn{1}{c}{$\Omega_b - \overline{B}_s$}    & 
\multicolumn{1}{c}{ }    & 
\multicolumn{1}{c|}{$\Omega_b^\ast-\Omega_b$}  \\
GLOK  & \protect\cite{alikhan2000} &  $392(50)(^{+9}_{-15})(^{+15}_{-0})$  &  $18(4)(^2_3)$ &  \\
\hline
\hline
\multicolumn{5}{|c|}{$N_f  = 2 $}  \\
\hline
\hline
\multicolumn{5}{|c|}{$llb$}  \\
\hline
\multicolumn{2}{|c}{Ref.}  &   $\Lambda_b -\overline{B}$ & $\overline{\Sigma}_b-\Lambda_b$ &
$\Sigma_b^\ast-\Sigma_b$            \\
\hline
SGO & \protect\cite{collins1999}     &    545(40)(22)      &             &  \\
CP-PACS & \protect\cite{alikhan2000} & 417(19) &                   &     \\
\hline
\end{tabular}
\end{center}
\caption{Own results on $b$ baryon masses and mass splittings for various light quark flavor
combinations. The first error is statistical. Where applicable, 
the second error comes from the variation of $a^{-1}$ between 1.8 and 2 GeV, and the third from fixing
the strange quark mass using the $K^\ast$ or $\Phi$ instead of $K$.}
\label{tab:bary_sum}
\end{table}

\begin{table}[thb]
\begin{center}
\begin{tabular}{|l|l|l|}
\hline
\hline
\multicolumn{1}{|c}{Ref.} &
\multicolumn{1}{c}{$\Lambda_b-\overline{B}$[MeV]} &
\multicolumn{1}{c|}{$\overline{\Sigma}_b-\Lambda_b$[MeV]} \\
\hline
\hline
\multicolumn{3}{|c|}{{\bf Lattice}} \\
\hline
\protect\cite{bowler1996}                   &  $338(^{61}_{52})$    & $186(^{61}_{76})$  \\
\protect\cite{alikhan2000}                  &  $370(67)$              & 221(71)   \\ 
\protect\cite{mathur2002}, $a_s=1.1$GeV$^{-1}$
                                            &  $361(^{103}_{108})$  & $156(^{39}_{33})$  \\
\protect\cite{mathur2002}, $a_s=0.9$GeV$^{-1}$
                                            &  $367(^{108}_{110})$  & $191(^{38}_{37})$  \\
\protect\cite{aoki2003}                     &  $389(44)$            & 122(65)            \\
\protect\cite{cppacs2000}, quenched         &  $361(22)$                       &  \\
\protect\cite{collins1999}, $N_f=2$         &  $545(40)(22)$         &  \\
\protect\cite{cppacs2000}, $N_f=2$          &  $417(19)$                       &  \\
quenched average                            &  $372(33)(^{33}_5)$         &  $174(28)(^{15}_2)$  \\
\hline
\multicolumn{3}{|c|}{{\bf Models}} \\
\hline
\protect\cite{jenkins1992}                   & 312 & 196     \\ 
\protect\cite{capstick1986}            &     & 217     \\
\protect\cite{ebert1998,ebert2005}        308    &     & 202     \\
\protect\cite{capstick1986}$c$ quark   &     &  212     \\
\hline
\multicolumn{3}{|c|}{{\bf Experiment}} \\
\hline
\protect\cite{pdg}                       & 311(10) &               \\
\hline
\hline
\multicolumn{1}{|c}{Ref.} &
\multicolumn{1}{c}{$\Sigma_b^\ast-\Sigma_b$[MeV]} &
\multicolumn{1}{c|}{$\Omega_b^\ast-\Omega_b$[MeV]} \\
\hline
\hline
\multicolumn{3}{|c|}{{\bf Lattice}} \\
\hline
\protect\cite{alikhan2000}        &    $19(7) $           & $18(4)$   \\
\protect\cite{mathur2002}, $a_s=1.1$GeV$^{-1}$  
                                   & $22(12)$           &  $18(^9_8)$ \\
\protect\cite{mathur2002}, $a_s=0.9$GeV$^{-1}$  
                                   & $24(^{13}_{12})$           & $20(9)$  \\
\protect\cite{aoki2003}            & 10(12)            & 7(4)    \\
average                            & $18(8)(^4_0)$     & $13(7)(^3_0)$  \\
\hline
\multicolumn{3}{|c|}{{\bf Models}} \\
\hline
\protect\cite{capstick1986}           &  10  &                 \\
\protect\cite{jenkins1992}   & 8           &     \\
\protect\cite{ebert2005}   & 29          &  23   \\
\hline
\end{tabular}
\end{center}
\caption{$b$ baryons. Only statistical errors and, where applicable, systematical errors due to scale 
setting except for the quenched scale ambiguity and fitting are shown.}
\label{tab:bary}
\end{table}
In Table~\ref{tab:bary} we summarize results for 
the spin-independent splittings $\Lambda_b-\overline{B}$ and 
$\overline{\Sigma_b}-\Lambda_b$ with $M(\overline{\Sigma}_b) = [2M(\Sigma_b) + 
4M({\Sigma_b}\hspace{-0.1cm}^\ast)]/6$ and $M(\overline{B}) = [3M(B^\ast)+M(B)]/4$, and the 
${\Sigma_b}\hspace{-0.1cm}^\ast-\Sigma_b$ 
hyperfine splitting.
\section{$f_B$ WITH NON-RELATIVISTIC HEAVY QUARKS \label{sec:fB}}
Since $B$ meson decay constants are  experimentally undetermined, it is important to
calculate them on the lattice. Their lattice calculation also serves as the calibration of the
lattice methods for the calculation of more complicated matrix elements. 

To implement the heavy quark directly at the $b$ quark mass on the lattice, we used
 NRQCD. The first result was an unquenched ($N_f=2$) study of the tree-level 
contributions to the pseudoscalar and vector decay constants at 
$O(1/M)$~\cite{collins1997} (1997). 
Then (1997) we published a quenched calculation at a lattice spacing 
$a \sim 0.1$ fm including most of the renormalization factors at 
$O(\alpha_s/M)$~\cite{sgo1997}, accompanied by a calculation using two flavors which was
published in 1999~\cite{collins1999}. To estimate the
effect of the truncation of the $1/M$ expansion, and including a very broad range of states in
the spectrum, we performed a quenched calculation which included $1/M^2$ terms at 
tree-level~\cite{alikhan1998} (1998). Renormalization constants correct to 
$O(\alpha_s/M)$ were included.
Since our previous unquenched lattice calculation \cite{collins1999} used only one lattice 
spacing, a small physical volume and a different sea and valence quark action, the major 
remaining systematic uncertainty appeared to be quenching. We then performed a calculation 
comparing two and zero flavors, with equal sea and valence quark actions, at a range of lattice 
spacings. NRQCD and Fermilab heavy quarks were simulated in parallel to minimize remaining
systematic uncertainties from the heavy quark action (see Refs.~\cite{cppacs2001R,cppacs2001NR}
of 2001).

In this section, we discuss the calculations of
Refs.~\cite{sgo1997,alikhan1998,cppacs2001NR,cppacs2001R}, 
including  systematic effects, and 
compare to other lattice results and results using different methods.
A typical analysis procedure is discussed in the context 
of the unquenched simulations~\cite{cppacs2001NR,cppacs2001R}.
\subsection{Quenched calculations at an intermediate lattice spacing}
\subsubsection{NRQCD at $O(1/M)$}
Matching of the current matrix element was performed at one loop perturbation theory,
using the strong coupling constant $\alpha_s$  in the potential scheme~\cite{lepage1993}. 
It is plausible that the matching scale $q^\ast$ is in the region of the lattice cut-off 
$1/a \leq q^\ast \leq \pi/a$. A value of $aq^\ast = 2.18$  for the matching of 
static-light axial currents has been calculated by Ref.~\cite{hernandez1994},
while the newer calculation of Ref.~\cite{degrand2002}, which retains 
parts of the momentum integral which were set to zero in the calculation of 
 finds $aq^\ast = 1.43$.
Renormalization constants for Refs.~\cite{sgo1997} and \cite{alikhan1998} were calculated 
for $q^\ast = 1/a$ and $q^\ast = \pi/a$. 
%

Numerical results presented here are from the calculation with tadpole-improved clover light
quarks (Run B) of~\cite{sgo1997}.
\begin{table}[thb]
\begin{center}
\begin{tabular}{|cclllll|}
\hline
$aM_0$ &  $a\Delta_{HL}$ & $a\Esim(\kappa_s)$ & $aM_{PS}(\kappa_s)$ & $M_{PS}(\kappa_s)$[GeV] 
& $aq^\ast= 1$ & $aq^\ast = \pi$ \\
1.71   &  1.73(10)       & 0.513(3)           & 2.24(10)            & 4.46(22)         
& 0.169(4) & 0.172(4)  \\
2.0    &  2.02(9)        & 0.515(4)           & 2.54(9)             & 5.05(21)          
& 0.169(3) & 0.174(3)  \\
4.0    &  4.07(5)        & 0.528(4)           & 5.00(5)             & 9.95(22)          
& 0.169(3) & 0.187(4)  \\
8.0    &  7.63(16)       & 0.527(6)           & 8.16(16)            & 16.24(45)         
& 0.167(5) & 0.195(5)  \\
\hline
\end{tabular}
\end{center}
\caption{Parameters and results for strange heavy-light mesons from~\protect\cite{sgo1997}. 
The last two columns show renormalized decay matrix elements $a^{3/2}f\sqrt{M}$ for 
matching scales $q^\ast$.}
\label{tab:masses_sgo}
\end{table}
Since the mass renormalization constant $C_m$ was not included in the renormalization of the
matrix element $J^{(1)}$ in~\cite{sgo1997}, we list $f\sqrt{M}$ as a function of the
heavy quark masses  in Table~\ref{tab:masses_sgo} including the renormalization. Along with this
we summarize the bare heavy quark masses in lattice units, the
mass shift $\Delta_{HL}$ and the strange meson masses in lattice and physical units. 
$m_\rho$ was used to set the scale. The values for the shifts  are calculated from the
quarkonia shifts $\Delta_{HH}$ with a $1/M^2$ NRQCD action. Values for $aM_0 = 1.71, 2.0$ 
are taken from~\cite{lidsey1994} and were newly calculated for 
$aM_0 = 4.0$. For $aM_0 = 8.0$, perturbation theory was used.
To determine the errors on the meson masses in Table~\ref{tab:masses_sgo}, the individual errors 
were added in quadrature.

$aM_0 = 2.0$ can be used as a reasonable estimate of the bare $b$ quark mass.
A linear fit of the meson mass to the form $\frac{M_{PS}}{M_0} = A + \frac{B}{M_0}$
for the three smaller mass values of Table~\ref{tab:masses_sgo}, using $m_{B_s} = 
5.37$ GeV as physical input, gives  $aM_0^b = 2.12$. 

Using the matrix elements at $aM_0 = 2.0$ and including the mass renormalization 
constant from \cite{morningstar98} which was not used in the matching calculation for
\cite{sgo1997}, we find for the decay constants:
\be
\begin{array}{ccccc}
q^\ast = 1/a: & f_B     & = & 182(28)(27)(16) & \mbox{MeV,} \\
              & f_{B_s} & = & 207(8)(30)(17)  & \mbox{MeV,} \\
q^\ast = \pi/a: & f_B     & = & 190(30)(29)(16) & \mbox{MeV,} \\
                & f_{B_s} & = & 216(8)(32)(17)  & \mbox{MeV,} \\
\end{array}
\ee
where $q^\ast$ denotes the scale of $\alpha_s$ used in the matching calculation.
We quote the same error bars as the $aM_0 = 2.0$ result of Ref.~\cite{sgo1997}. 
The first error denotes the  statistical and fitting error, the
second the uncertainty in the determination of $a$ and the
third, the error due to perturbation theory and higher orders in the $1/M$ expansion.
\subsubsection{NRQCD at $O(1/M^2)$}
For the results from \cite{alikhan1998}, the current operators were included up to
$O(1/M^2)$ while the matching renormalization was calculated only to $O(\alpha_s/M)$. The 
decay constants are found to be:
\be
\begin{array}{cccc}
f_{B} & = & 147(11)(^{+8}_{-12})(9)(6)  & \mbox{MeV,} \\
f_{B_s} & = & 175(8)(^{+7}_{-10})(11)(7)(^{+7}_{-0})  & \mbox{MeV}. \\
\end{array}
\ee
The first error is statistical, the second reflects the variation if the lattice spacing
is changed between $a^{-1} = 1.8$ and $2.0$ GeV. The third error gives the estimate of higher order 
perturbative corrections (estimated from the variation of the one-loop result using $q^\ast = 1/a$ and
$\pi/a$), and uncancelled $1/M^2$ corrections (estimated as $\alpha_s/(aM_0)^2$), and the fourth error
bar is an estimate of the $O(a^2\lqcd^2) = 4\%$ discretization errors.
\subsection{Investigation of Unquenching and Scaling}
\subsubsection{Technical details}
First we discuss a calculation of $f_B$ and $f_{B_s}$ using NRQCD at
$O(1/M)$ at various lattice spacings, with quenched and dynamical ($N_f = 2$)
renormalization group-improved~\cite{iwasaki1985} gauge fields at high statistics.
The current and mass renormalization constants were calculated by Ishikawa {\it et al.}
\cite{ishikawa2000a}.
$B$ decay constants have been calculated with dynamical configurations at two $\beta$ values 
corresponding to lattice spacings of respectively $a^{-1}=1.3$ and $1.8$ GeV in the
chiral limit of the sea quark mass and at four values of the sea quark mass each to
study an extrapolation to physical $u,d$ sea quark masses. In parallel, 
quenched configurations at 4 $\beta$ values are studied at a comparable range of lattice spacings.
using five values of heavy and five values of light valence quark masses 
for reliable interpolations and extrapolations to the physical quark masses.
\begin{figure}[thb]
\begin{center}
\centerline{
\epsfysize=7cm \epsfbox{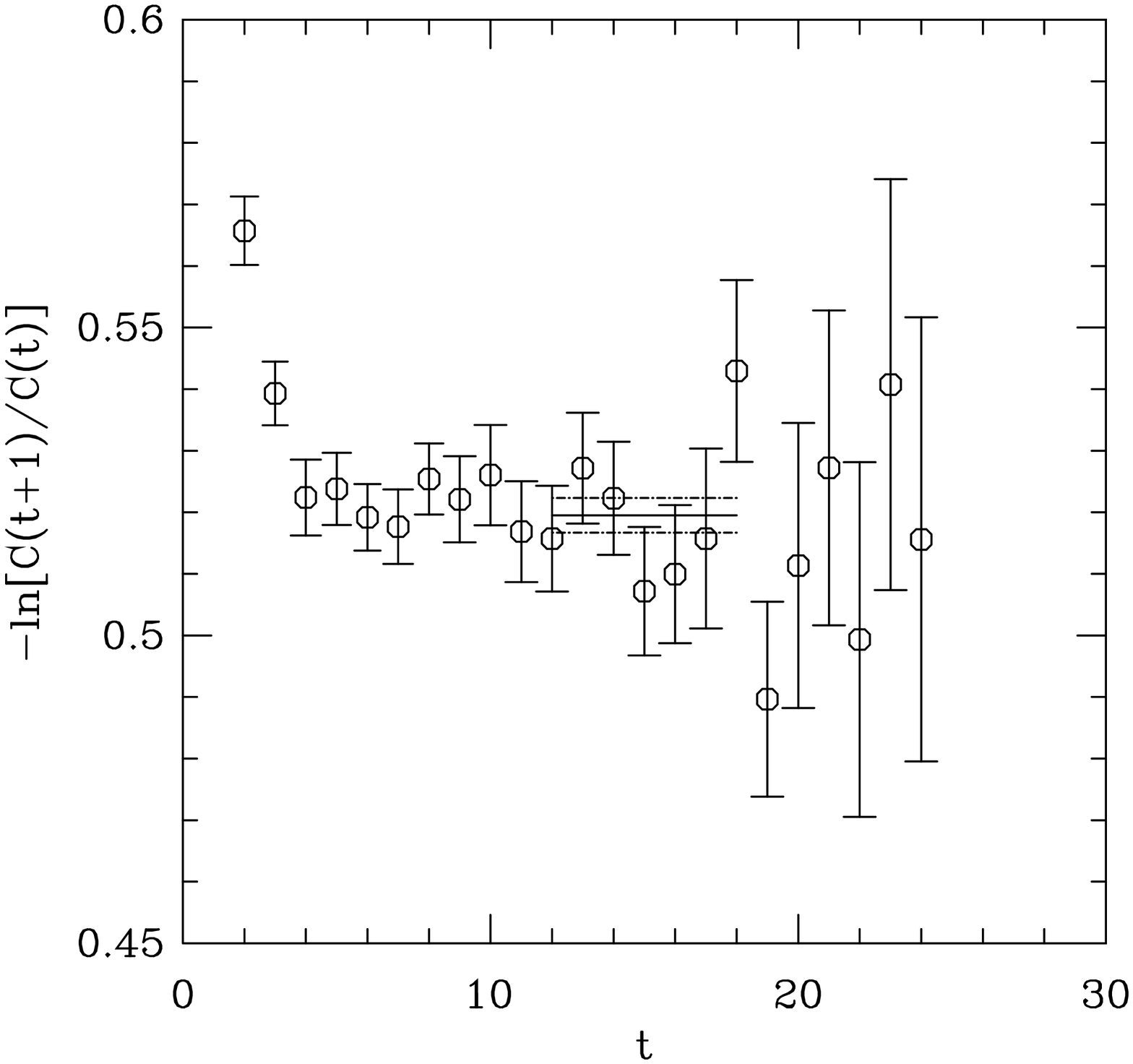}
\epsfysize=7cm \epsfbox{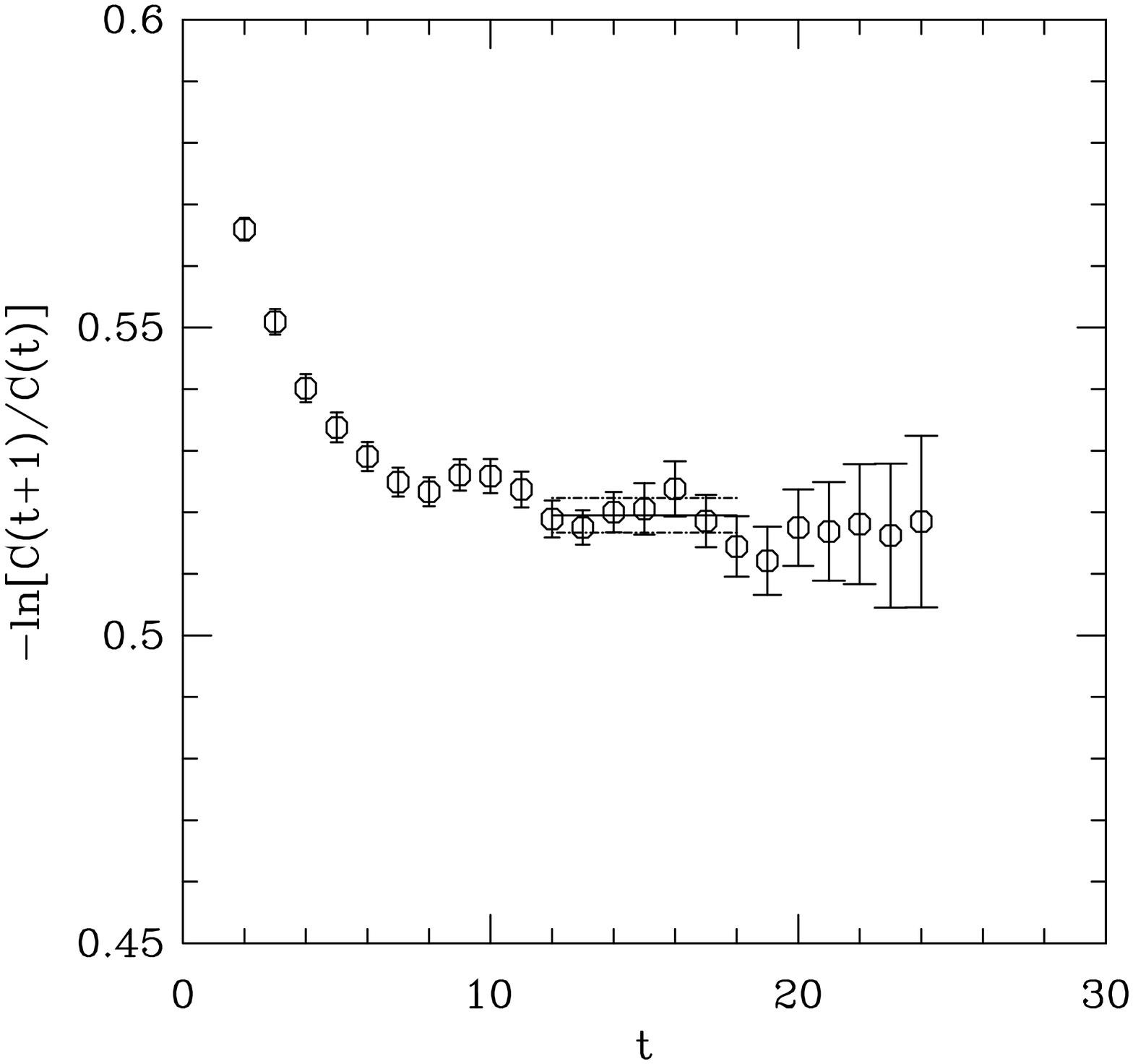}
}
\vspace{-0.5cm}
\end{center}
\caption{'Effective mass' of heavy-light pseudoscalar-pseudoscalar correlation functions on a dynamical lattice
at the smallest value of the gauge coupling, the lightest sea quark mass, the valence heavy quark slightly 
heavier than the $b$, and the lighter valence quark mass close to the strange quark mass~\protect\cite{cppacs2001NR}.
On the left, the source and the sink are smeared, on the right, a smeared source and a local sink
are used.}
\label{fig:effmass} 
\end{figure}
Correlations functions are smeared at the source. 
The contributions to the bare decay constants are related to the matrix elements of the local 
lattice operators $A_4^{(i)}$ which contribute to the time component of the axial vector current by
\be
a^{3/2} f^{(i)}_B\sqrt{M_B} = \frac{1}{\sqrt{M_B}} \langle 0|A_4^{(i)}(0)|B(\vec{p}=0) \rangle,
\ee
where $M_B$ is the $B$ meson mass. Assuming a relativistic normalization of the meson states $|B(p)\rangle$, 
one can define a local amplitude by
\be
Z_L^{(i)} \equiv \frac{1}{\sqrt{2M_B}} \langle 0|A_4^{(i)}(0)|B(\vec{p}=0) \rangle
\Rightarrow a^{3/2} f^{(i)}_B\sqrt{M_B} = \sqrt{2} Z_L^{(i)}.
\ee
The normalization of the pseudoscalar operator used in the simulation for the $B$ meson, 
which is in general smeared and denoted as $P$, is taken into account by measuring the pseudoscalar density 
\be
Z_S \equiv \frac{1}{\sqrt{2M_B}} \langle 0|P(0) |B(\vec{p}=0) \rangle.
\ee
The following correlation function of the current $A_4^{(i)}$  is  calculated on the lattice:
\be
\langle 0|A_4^{(i)}(t,\vec{p})|P(\vec{p}=0) \rangle = Z_{L}^{(i)} Z_S e^{-\Esim t},
\label{eq:matrixel}
\ee
inserting a complete set of mass eigenstates and assuming that the decay is dominated by the smallest
mass at large times. The exponent is the same for the various $A_4^{(i)}$. 
The pseudoscalar density $Z_S$ can be extracted from the correlator of
the smeared meson operator at source and sink:
\be
\langle 0|P(t,\vec{p}) |P(\vec{p}=0) \rangle = Z_S^2 e^{-\Esim t}.
\ee

Masses and amplitudes for the current matrix elements are extracted from single
exponential fits to smeared-smeared and smeared-local correlation functions. Note that for NRQCD
heavy quarks the correlator $\langle A_4^{(0)} | P\rangle$ is identical to  $\langle P | P\rangle$
since spinors where only the upper components are nonzero do not distinguish between $\gamma_0$ 
and the unit matrix. Plots of typical heavy-light correlation functions are shown in 
Fig.~\ref{fig:effmass}.

\begin{figure}[t]
\begin{center}
\epsfysize=5cm \epsfbox{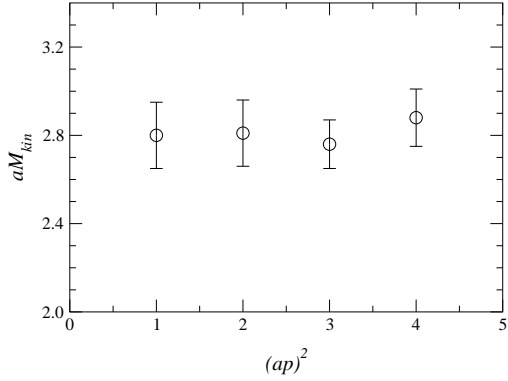}
\end{center}
\vspace{-0.5cm}
\caption{Kinetic meson masses determined from (\protect\ref{eq:disp}) as a function
of momentum, from \protect\cite{cppacs2001NR}.}
\label{fig:Ekin} 
\end{figure}

The bare heavy quark mass corresponding to the $b$ meson mass is determined from
the difference of energies at finite and zero spatial momentum, Eq.~(\ref{eq:disp}), with
$|\vec{p}| = 2\pi/L$, where $L$ is the spatial lattice extent. Since the energy 
difference is small, the states are expected to be strongly correlated, and the
splitting was extracted using a  one exponential fit to the ratio of finite and zero 
momentum correlators~(Eq.~(\ref{eq:ratiofit})). That the result is rather
independent of the momentum can be seen in Fig.~\ref{fig:Ekin} for the finest
quenched lattice.

\begin{figure}[thb]
\begin{center}
\centerline{
\epsfysize=6cm \epsfbox{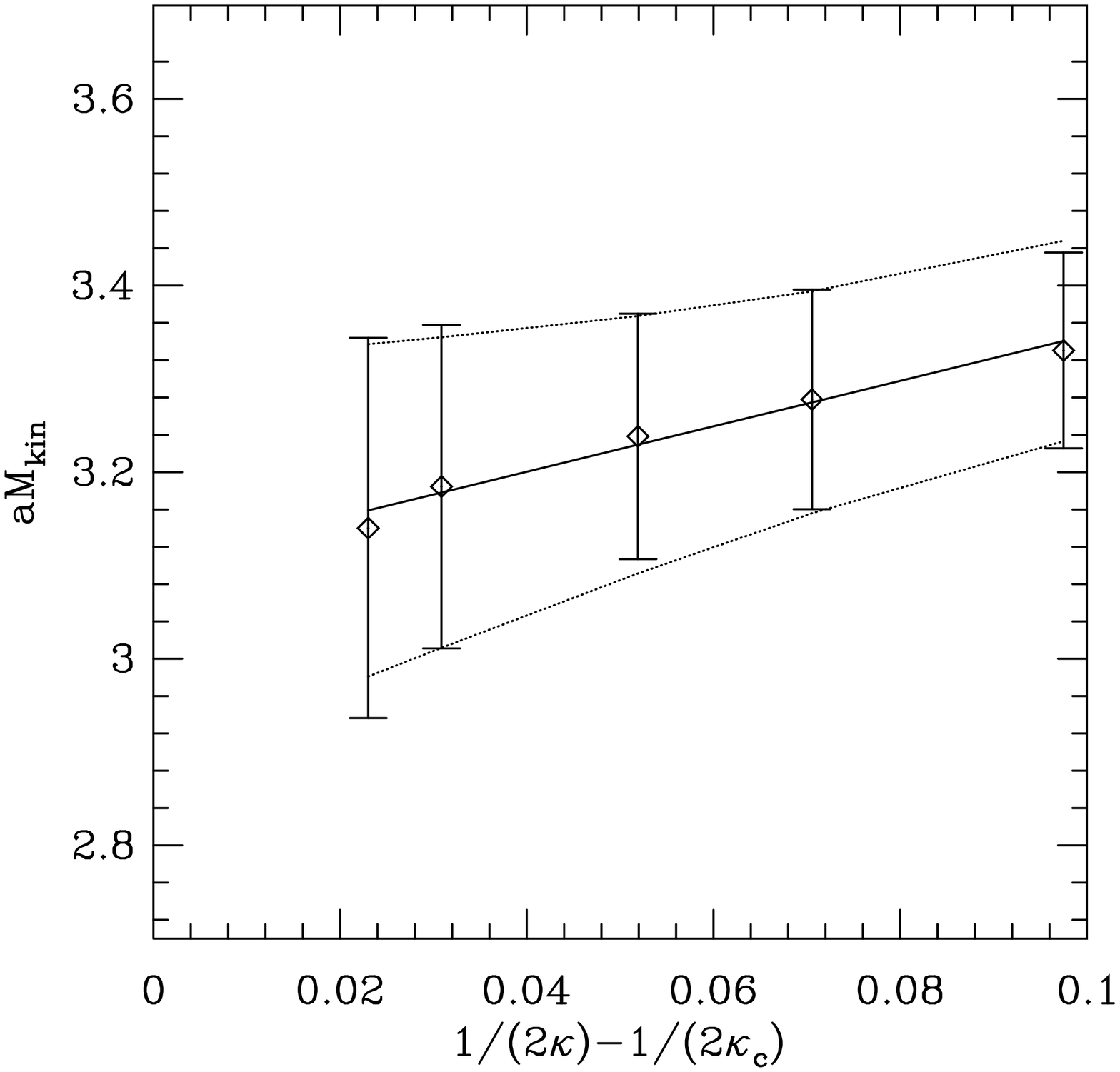}
\epsfysize=6cm \epsfbox{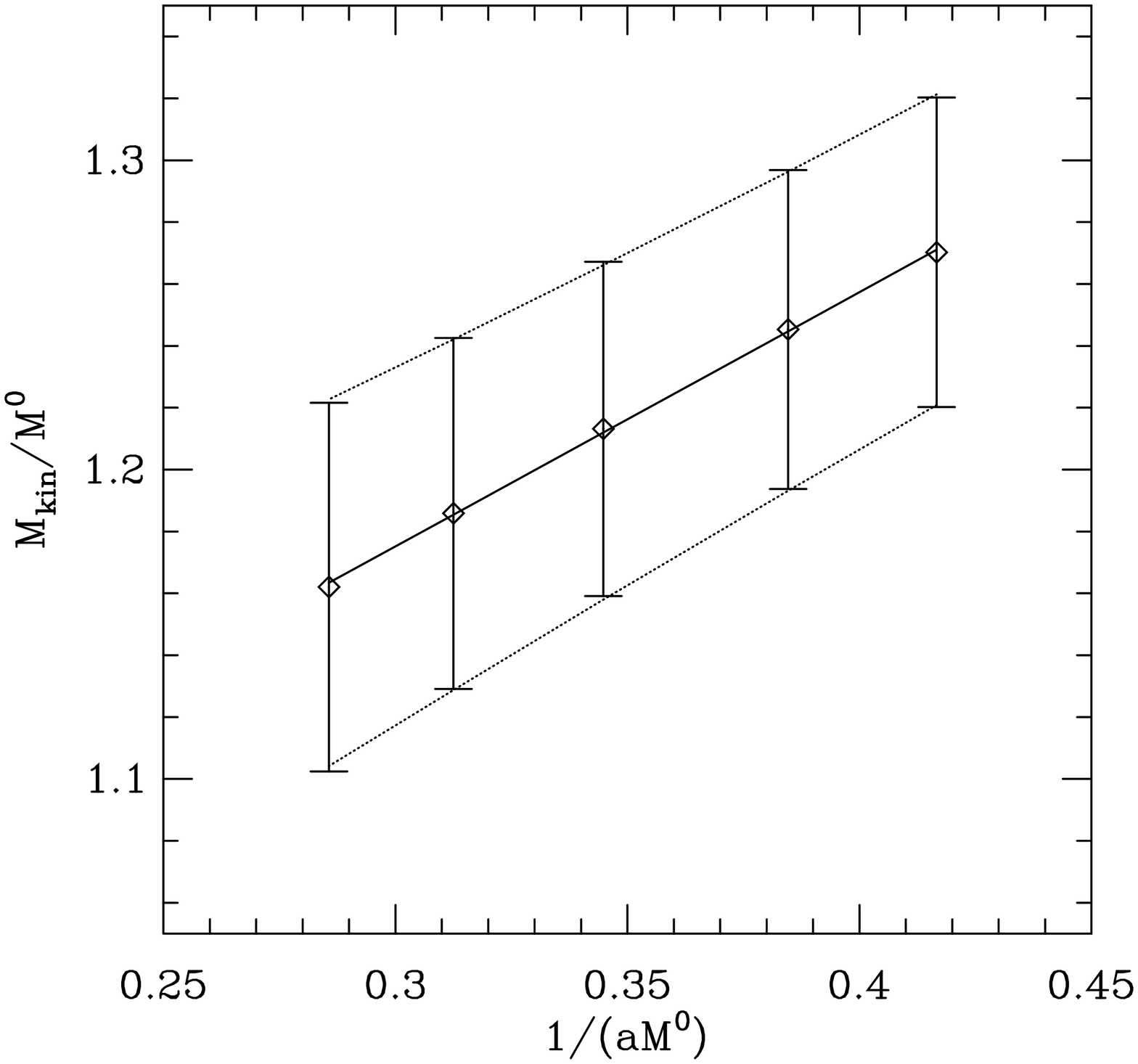}
}
\vspace{-0.5cm}
\end{center}
\caption{Dynamical meson masses from~\protect\cite{cppacs2001NR} at the smaller value of the
gauge coupling ($\beta=2.1$) and a sea quark mass around the strange. On the left, fit of the meson mass 
as a function of the VWI light valence quark mass. The heavy quark mass is close to the $b$ mass.
On the right, fit of the heavy-strange meson mass as a function of the heavy quark mass to
determine the $b$ quark mass.}
\label{fig:mmeson} 
\end{figure}
The meson masses are fitted as a linear function in the VWI light quark mass to
find the physical points corresponding to the light and strange quark. The result
is fitted as a linear function in $M_{PS}/M_0$. The heavy quark mass 
parameter corresponding to the $b$ quark is determined by setting the heavy-strange
meson mass to the $B_s$ mass. Setting the heavy-light meson mass to the physical
$B$ meson mass gives the same result, however with larger error.
Fits of the meson mass as function of the light and
heavy quark mass are shown in Fig.~\ref{fig:mmeson}.

\begin{figure}[thb]
\begin{center}
\centerline{
\epsfysize=5cm \epsfbox{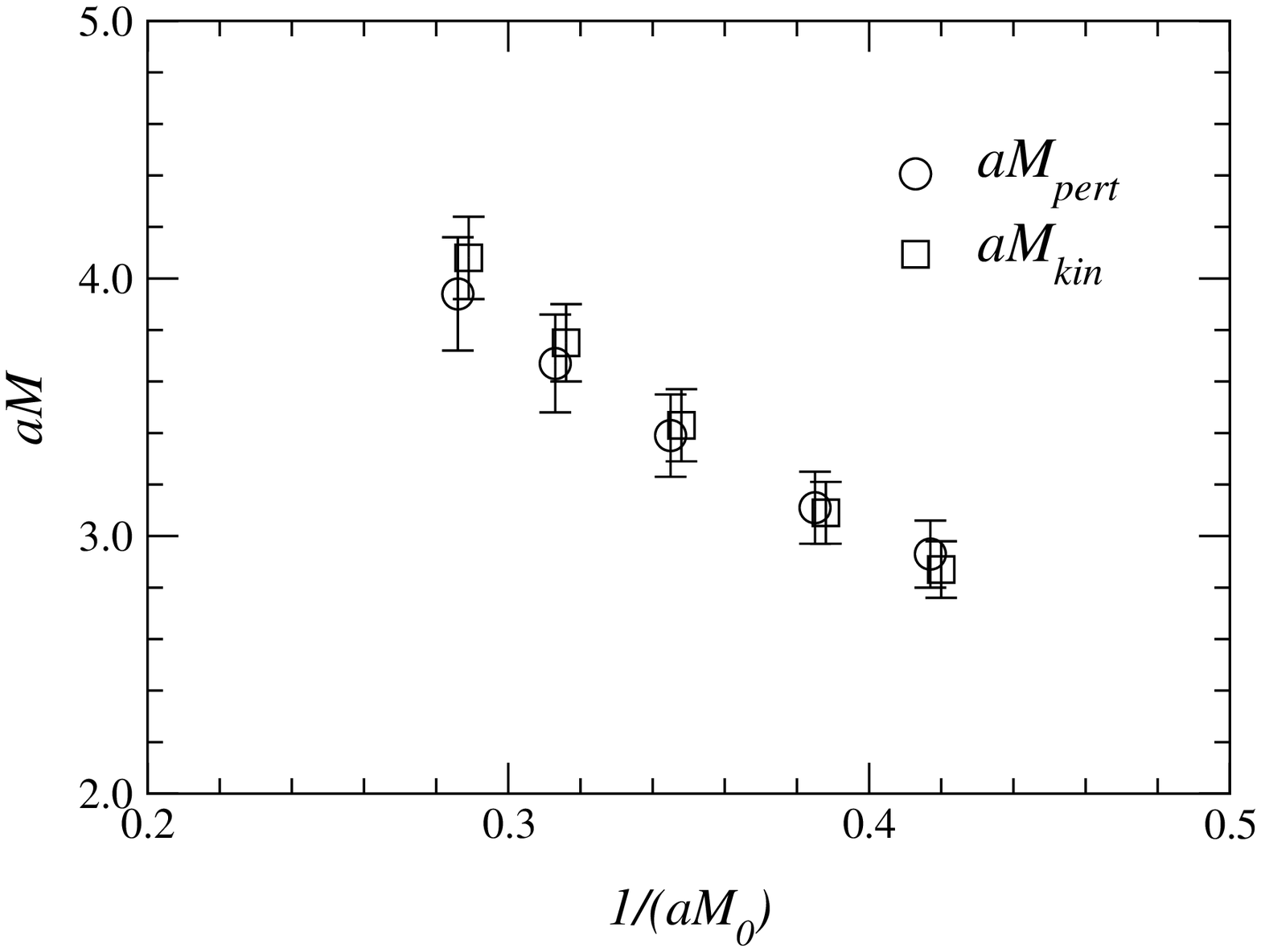}
}
\end{center}
\caption{Dynamical meson masses determined using the dispersion relation~\protect\ref{eq:disp} ($M_{kin}$) 
compared to the masses using one-loop values for the mass shift $\Delta$ ($M_{pert}$) 
at the smaller gauge coupling, the lightest sea quark mass, and a light valence quark mass
around the strange.}
\label{fig:mdisppert} 
\end{figure}
In Fig.~\ref{fig:mdisppert} we show comparisons of the meson mass using 
the perturbative shift $\Delta$ with the kinetic mass. Usually the agreement is good;
the worst disagreements are still less than 2 combined $\sigma$ and 
occur for the quenched configurations at $\beta = 2.575$ (shown
in the Figure)  and for $N_f = 2$ at $\beta = 2.1,\kappa_{sea}=0.1357$. 
The dependence of the matrix elements on the heavy and light quark masses is 
analyzed in a partially quenched setup, i.e.\ setting the valence quark masses 
to their physical values (light, $s$ and $b$) at fixed value of the sea quark 
mass, and then extrapolating to zero sea quark mass.
\begin{figure}[thb]
\begin{center}
\centerline{
\epsfysize=5cm \epsfbox{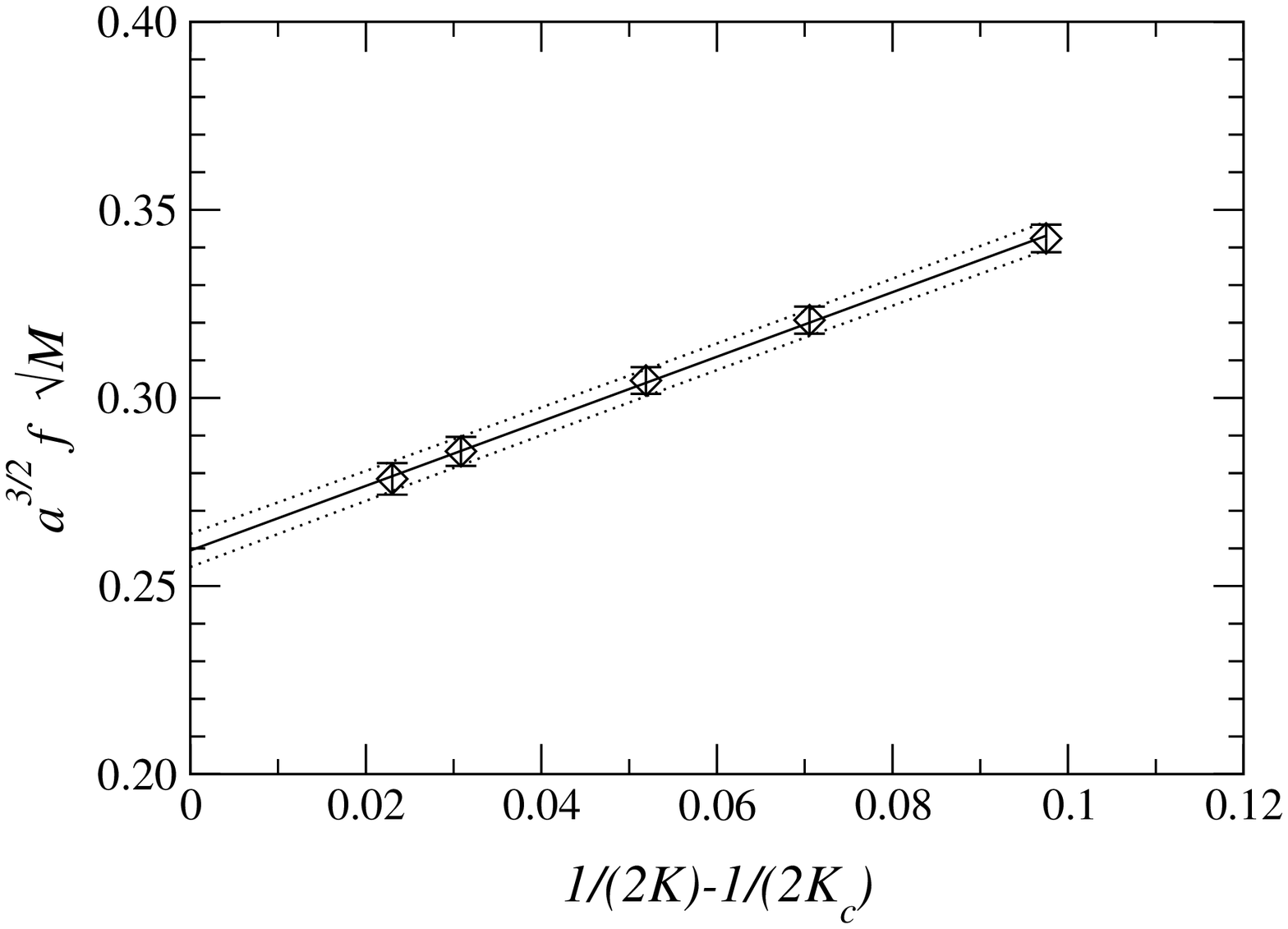}
\epsfysize=5cm \epsfbox{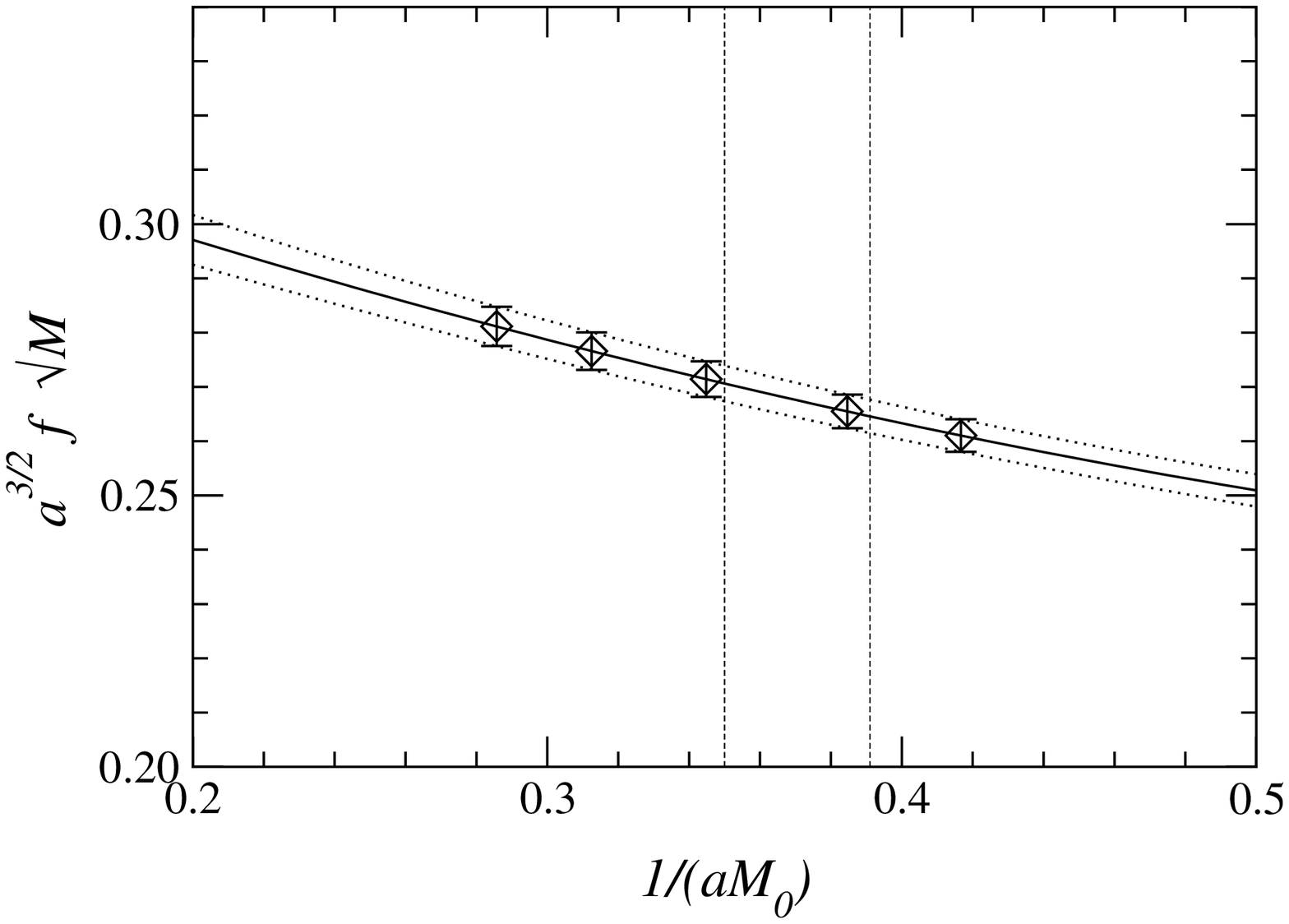}
}
\end{center}
\caption{Dynamical $b$ decay matrix elements from~\protect\cite{cppacs2001NR} at the
smaller gauge coupling $(\beta = 2.1)$ and a sea quark mass around the strange. On the left, fit of 
$f\sqrt{M}^{(0)}$ at a heavy quark mass close to the $b$ as a function of the VWI quark mass 
(left) and the renormalized matrix element $f\sqrt{M}$ with the light valence quark mass
interpolated to the strange (right). The $b$ quark mass corresponds to $1/(aM_0) = 0.36$.}
\label{fig:massextrap} 
\end{figure}
In Fig.~\ref{fig:massextrap} on the left 
we show the light quark mass dependence of the zeroth
order current at a typical set of parameter values. The light quark mass is
parameterized as $1/(2\kappa_{val}) - 1/(2\kappa_{c, val})$, where 
$\kappa_{c, val}$ denotes the point of zero valence quark mass at a given fixed
sea quark mass. The data can be described well by a linear behavior.

To correct the normalization of the light quark in the lattice operators for
$A_4^{(i)}$, we multiplied
Eq.~(\ref{eq:matrixel}) with $R=\sqrt{1-3\kappa/(4\kappa_c)}$, the tadpole improved equivalent of
Eq.~(\ref{eq:field_norm}) using $u_0 = 1/(8\kappa_c)$.
After interpolating the bare matrix elements $f\sqrt{M}^{(i)}, i=1,2$
to the strange  or extrapolating to the light $u,d$ quark mass, the renormalized
matrix element is  interpolated as a 
quadratic function in $1/(aM_0)$ to the $b$ quark mass. An example is 
shown on the right in Fig.~\ref{fig:massextrap}. 

\begin{figure}[thb]
\begin{center}
\epsfysize=6cm \epsfbox{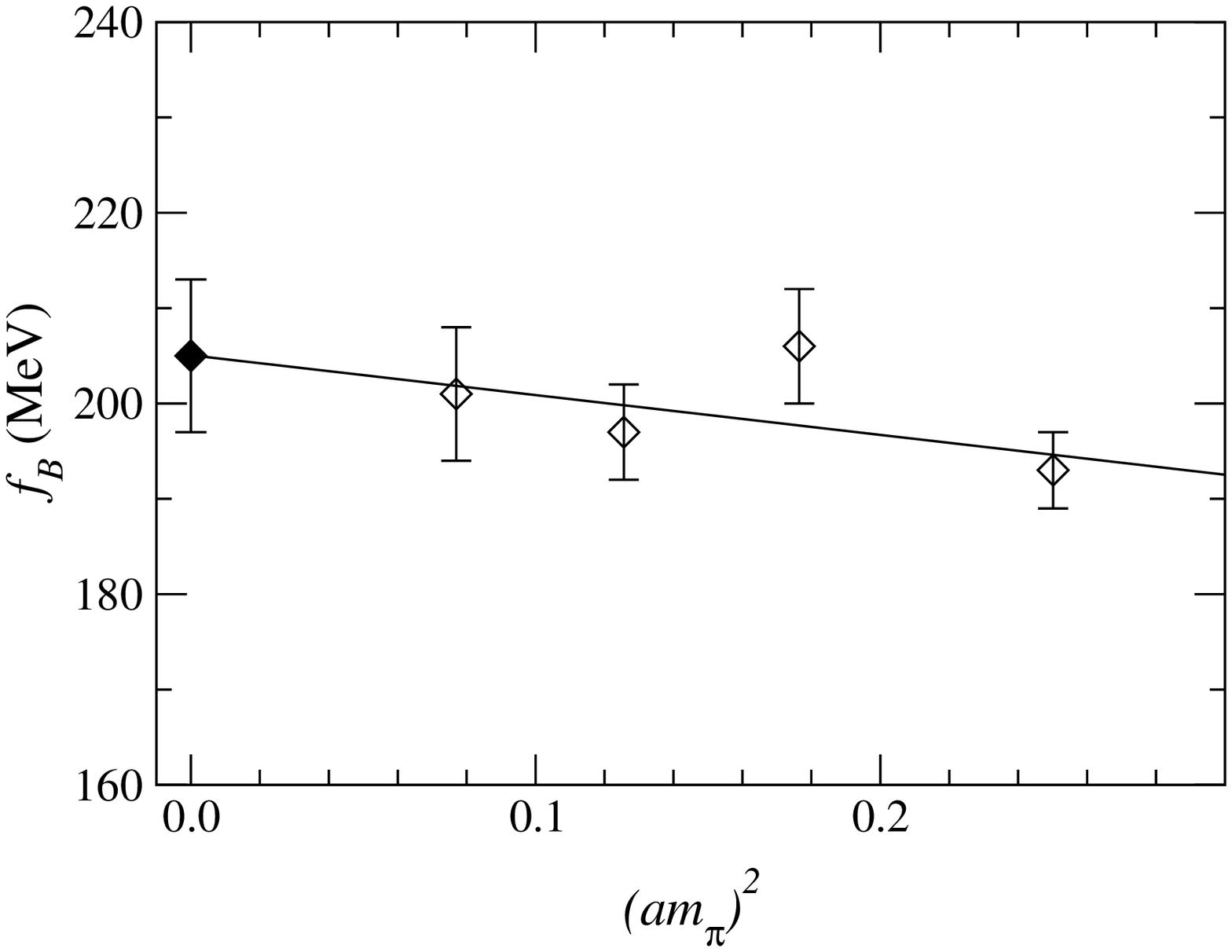}
\end{center}
\caption{Sea quark mass dependence of $f_B$ at $\beta = 2.1$, from \protect\cite{cppacs2001NR}.}
\label{fig:msea} 
\end{figure}

The dependence of the decay constant on the sea quark mass is mild. At $\beta = 1.95$,
no real dependence can be resolved, whereas at $\beta=2.1$ there appears to be a slight 
increase if the sea quark mass is taken to zero. A linear fit is shown in 
Fig.~\ref{fig:msea}.

In addition,  $f_B$ and  $f_D$  were 
calculated using clover heavy quarks with the Fermilab non-relativistic 
reinterpretation~\cite{elkhadra1997}.
Current operators at zeroth and first order in the $1/M$ expansion were used:
\ba
A^{(0)}_{4, rel}(x) &=& \overline{q}(x) \gamma_5\gamma_4 q_h(x), \\
A^{(1)}_{4, rel}(x) &=& \overline{q}(x) \gamma_5\gamma_4\gamma\cdot\vec{D} q_h(x).
\label{eq:currel}
\ea

Local and smeared
hadron operators were used at the source; operators at the sink were always local. 
Amplitudes and masses were extracted from two-exponential fits to 
axial vector-pseudoscalar and pseudoscalar-pseudoscalar correlators. 
%
Since the meson energy at finite momentum was affected by rather large statistical
errors, the mass of the meson was calculated by adding the tree-level difference of
the kinetic and pole heavy quark masses $M_2^Q$ and $M_1^Q$  to the meson pole mass, instead of
using the kinetic mass of the heavy meson:
\begin{equation}
  \label{eq:HQET_mass}
  aM_{2} = aM_1 + (aM_2^Q - aM_1^Q),
\end{equation}
where the masses with superscript $Q$ are quark masses and those without
superscript meson masses.

Resumming the logarithmic mass dependence of the matrix elements (see Eq.~(\ref{eq:logs}))
one can define the matrix element $\Phi$ which is, according to HQET,  in the heavy quark limit
constant up to power corrections in $1/M_Q$. It is normalized such that it agrees with
$f_B\sqrt{M_B}$ if the heavy quark mass is the $b$ mass. 
For a general heavy-light meson mass $M_P$ it has the form:
\be
\Phi(M_P) = (\alpha_s(M_P)/\alpha_s(M_B))^{2/b_0} f\sqrt{M_P},
\label{eq:leadinglog}
\ee
with $b_0 = 11 - \scriptsize{\frac 2 3} N_f$. The heavy and light quark mass dependence 
of the resummed matrix element $\Phi(M_P)$ was fitted as a polynomial in the light quark mass
$m_q$ and the inverse heavy meson mass:
\begin{equation}
  \label{eq:fit_form}
  a^{3/2} \Phi(M_P) =   
  A_0 + A_1 am_q +  A_2 (am_q)^2
  + \frac{1}{aM_P} \left[ B_0 + B_1 am_q \right]
  + \frac{1}{(aM_P)^2} C_0. \label{eq:phi_fit}
\end{equation}
\subsubsection{Results for $f_B$ and $f_{B_s}$}
\begin{table}[thb]
\begin{center}
\begin{tabular}{|c|l|l|}
\hline
\multicolumn{1}{|c}{} & 
\multicolumn{1}{c|}{$N_f = 0$} & 
\multicolumn{1}{c|}{$N_f = 2$} \\
\hline
\multicolumn{1}{|c}{} & 
\multicolumn{2}{c|}{NRQCD} \\
\hline
$f_B$[MeV] & 191(4)(27)        &  204(8)(29)\\
$f_{B_s}$[MeV]  & 220(4)(31)   &  242(9)(34)\\
\hline
\multicolumn{1}{|c}{} & 
\multicolumn{2}{c|}{relativistic} \\
\hline
$f_B $[MeV]    & 188(3)(26)               & 208(10)(29)  \\
$f_{B_s}$[MeV] & $220(2)(31)(^{+8}_{-0})$ & $250(10)(35)(^{+8}_{-0})$  \\
$f_D $[MeV]    & 218(2)(39)               & 225(14)(40)               \\
$f_{D_s}$[MeV] & $250(1)(45)(^{+6}_{-0)}$  & $267(13)(48)(^{10}_0)$ \\
\hline
\end{tabular}
\end{center}
\caption{Heavy-light decay constants from 
\protect\cite{cppacs2001NR,cppacs2001R}. The scale is set using $m_\rho$.}
\label{tab:fb_own}
\end{table}
The results for decay constants are listed in Table~\ref{tab:fb_own}.
The first error is statistical, the second $a$ and perturbation theory errors,
and the third error comes from setting the strange quark mass using $K$ or $\phi$
input. The agreement of the results from NRQCD and Fermilab heavy quarks is an
important result since one might expect that discretization and perturbative errors
are different between the two formalisms, although naively one would expect them to 
be of a similar magnitude.

In ratios of unquenched and quenched matrix elements calculated at the same
value of the lattice spacing one may expect that most of the discretization
and perturbative errors cancel. From the NRQCD calculation we find
\be
\begin{array}{cccccc}
  \frac{f_B^{N_f=2}}{f_B^{N_f=0}} 
  &=& 1.07(5); &
  \frac{f_{B_s}^{N_f=2}}{f_{B_s}^{N_f=0}} 
  &=& 1.10(5).\\
\end{array}
\ee
where only statistical errors are considered. If the scale is set with $m_\rho$,
we find a $\sim 10\%$ unquenching effect on the decay constants from the NRQCD calculation.

In the calculation with Fermilab heavy quarks we find
\be
\begin{array}{cccccc}
  \frac{f_B^{N_f=2}}{f_B^{N_f=0}} 
  &=& 1.11(6);  & \frac{f_{B_s}^{N_f=2}}{f_{B_s}^{N_f=0}}   &=& 1.14(5); \\
  \frac{f_D^{N_f=2}}{f_D^{N_f=0}} 
  &=& 1.03(6); &
  \frac{f_{D_s}^{N_f=2}}{f_{D_s}^{N_f=0}} 
  &=& 1.07(5), \\
\end{array}
\ee
Again, the scale is set with $m_\rho$.

With the Fermilab heavy quarks, the effect of setting the scale with $f_\pi$ has been 
studied. This leads to a better scaling of the decay constants  as function of 
$a$~\cite{cppacs2001R}. If one takes the central value at the smallest lattice 
spacing for the result, the unquenched value remains the same as with the $\rho$ scale,
whereas the quenched result for $f_B$ with $a$ from $f_\pi$ changes from 188 to 175 MeV.
\subsubsection{The flavor breaking ratio}
The extrapolation to the chiral limit could introduce a considerable uncertainty.
E.g.\ Ref.~\cite{jlqcd2003} find a decrease of
$f_B$ by almost 20 MeV if they use a chiral extrapolation supported by $\chi PT$ instead of a 
simple polynomial fit. We use linear extrapolations in the partially quenched NRQCD analysis 
and a linear and quadratic fit for the simultaneous
extrapolation in sea and light valence quark masses in the analysis with Fermilab heavy quarks.
The result is
\ba
f_{B_s}/f_B & = & 1.179(18)(23), \; N_f = 2, \nonumber \\
f_{B_s}/f_B & = & 1.150(09)(20), \; N_f = 0
\ea
from NRQCD, and 
\ba
f_{B_s}/f_B & = & 1.203(29)(28)(^{+38}_{-0})\; N_f = 2 \nonumber \\
f_{B_s}/f_B & = & 1.148(08)(20)(^{+39}_{-0})\; N_f = 0 \nonumber \\
f_{D_s}/f_D & = & 1.182(39)(25)(^{41}_0)\; N_f = 2 \nonumber \\
f_{D_s}/f_D & = & 1.138(05)(18)(^{29}_0)\; N_f = 0 
\ea
from heavy clover quarks. The first error is statistical, the second systematical
(using only discretization effects of the clover quarks since $O(\alpha^2_s)$ 
and $1/M$ errors should largely cancel in the ratios), and the third from the difference in setting
the strange quark mass using the $K$ or $\phi$ meson.

The flavor breaking ratio $f_{B_s}/f_B$ has with both heavy quark formalisms no
significant unquenching effects, setting the scale with $m_\rho$.
\subsection{Discussion}
For the decay constants and some of the mass results presented here, matching has  been performed at 
$O(\alpha_s)$.
In calculations of $1/M$ corrections in HQET and of higher twists in operator product
expansions  relatively large higher-loop corrections may be expected.
On the lattice, higher dimensional operators generate power divergences 
under renormalization which are difficult to cancel in perturbation theory~\cite{martinell96}.
One might worry how precisely $1/M$ corrections can be quantified using effective theories
on the lattice. The size of the higher order corrections in 
perturbation theory depends on the choice of renormalization scheme and scale used at one loop.
$\alpha_V$ at the scale $q^\ast$, described in Sec.~\ref{sec:pot}, should be a rather
optimized choice.

The $1/a$ terms affect the simulation energies as well as the matrix elements. 
For example we fit the matrix elements $f\sqrt{M}^{(0)}$ from 
Ref.~\cite{cppacs2001NR}\footnote{The unquenched
matrix elements in Tables XIV-XIX of Ref.~\protect\cite{cppacs2001NR} are mistakenly arranged in the
 order  opposite from the valence quark masses}, with the 
light quark mass interpolated to the strange with equal sea and valence quark masses,
according to 
\be
 f^{(i)}_P\sqrt{M_P} = A + \frac{B}{M_0} + \frac {C}{M_0^2} \label{eq:mfit}, 
\ee
where $M_0$ is the bare heavy quark mass.  
 which yields the fit parameters given in Table~\ref{tab:power}.
The $1/M^2$ terms are unphysical since the action is only $1/M$ improved, they are 
however also on finer lattices not very large. The lattice spacing dependence gives an idea of to what
extent the $1/a$ terms contribute to the matrix element. Whether the power contributions make out
a substantial part of the decay constant can only be seen if the renormalization constants are
included. 
\begin{table}[htb]
\begin{center}
\begin{tabular}{lr@{.}lr@{.}lr@{.}lr@{.}l}
\hline
\multicolumn{1}{c}{Matrix el.} &
\multicolumn{2}{c}{$\beta$} &
\multicolumn{2}{c}{$A_1$ [GeV$^{3/2}$]} &
\multicolumn{2}{c}{$A_2$ [GeV$^{5/2}$]} &
\multicolumn{2}{c}{$A_3$[GeV$^{7/2}$]} \\
\hline
$f^{(0)}_P\sqrt{M_P}$ & 1&95  & 1&236(3) & $-1$&977(26)  & 2&03(5) \\
             &    2&1   & 0&912(30) & $-0$&72(31)  & $-0$&08(79) \\
$f^{(1)}_P\sqrt{M_P}$ & 1&95  & $-0$&0134(21) & $-0$&402(17)  & 0&329(35) \\
             &    2&1  & $-0$&0086(19) & $-0$&422(19)  & 0&395(50) \\
\hline
\end{tabular}
\end{center}
\caption{Fit parameters of the unrenormalized matrix elements as a function of $1/M_0$
according to Eq.~(\ref{eq:mfit}).}
\label{tab:power}
\end{table}

In the calculation~\cite{alikhan1998} using $1/M^2$ NRQCD and clover quark actions and Wilson gluons at 
$a \sim 0.1$ fm, we find for the ratio of the unrenormalized $B_s$ matrix elements 
$f^{(1)}_{B_s}\sqrt{M_{B_s}}/f^{(0)}_{B_s}\sqrt{M_{B_s}} \sim 0.12$ and for 
$f^{(i)}_{B_s}\sqrt{M_{B_s}}/f^{(0)}_{B_s}\sqrt{M_{B_s}} \sim 0.02$ or less for $i = 3,4,5$.
With renormalization group improved gluons~\cite{cppacs2001NR}, using $1/M$ 
NRQCD and clover quark actions, the ratio of matrix elements has a similar value, 
$f^{(1)}_{B_s}\sqrt{M_{B_s}}/f^{(0)}_{B_s}\sqrt{M_{B_s}} \sim 0.1$.
\begin{figure}[thb]
\begin{center}
\centerline{
\epsfysize=9cm \epsfig{file=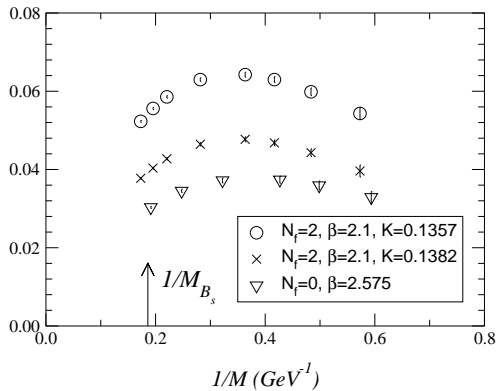, bb=50 10 550 600, width=7cm,clip=}
}
\end{center}
\vspace{-2cm}
\caption{
Ratio of decay constants with rotated and unrotated currents: 
$f_P^{\mathrm{rotated}}/f_P^{\mathrm{unrotated}}-1$
as a function of the inverse heavy meson mass.}
\label{fig:ratio_currents}
\end{figure}

To estimate the accuracy of the perturbative matching calculation it is of interest to  how well the 
cancellation of the $\alpha_s/(aM)$ contribution works at one loop. As discussed for NRQCD in more detail 
in~\cite{collins2001}, for $\alpha_s \sim 0.2-0.25$ and the perturbative coefficient
to the radiative correction of $A_4^{(0)}$ proportional to $1/(aM_0)$ around $-(0.3-0.4)$,
the physical $\lqcd/M$ contribution of the current correction is less than 50\% of
the bare current correction. However, there is no indication of very large higher order corrections.

The change from the bare to the 
one-loop decay constant is around $6-10\%$ in~\cite{sgo1997} using an $O(1/M)$ action and
$9-14\%$ in~\cite{alikhan1998} with an $O(1/M^2)$ action and Wilson gluons at $a^{-1} = 2$ GeV,
however less than 5\% with a $1/M$ NRQCD action and renormalization group-improved 
glue~\cite{cppacs2001NR}.

In contrast for relativistic heavy quarks~\cite{cppacs2001R}, the $1/M$ current operator
Eq.~(\ref{eq:currel}) contributes for $B$ as well as for $D$ mesons around $4-6\%$ to the decay matrix 
element, see Fig.~\ref{fig:ratio_currents}. 

\begin{figure}[thb]
\begin{center}
\centerline{
\epsfysize=6cm \epsfbox{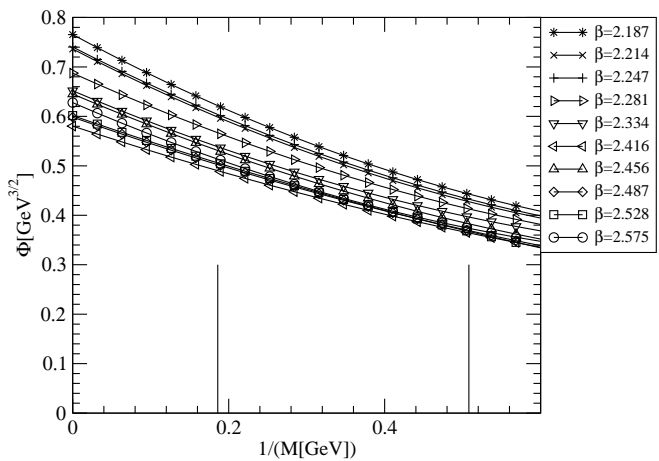}
}
\end{center}
\vspace{-0.5cm}
\caption{
Fit results for the heavy-strange matrix element $\Phi$ (Eq.~(\protect\ref{eq:phi_fit})), 
with the strange quark mass determined using the $K$ meson from \protect\cite{cppacs2001R}, 
as a function of the inverse heavy meson mass. Lines with different symbols describe fit 
functions from different lattices. The symbols aare just to giude the eye and do not correspond 
to actual data points. The two vertical lines indicate the $B_s$ and $D_s$ meson masses.}
\label{fig:hugh_quenched_fits}
\end{figure}
To see the  extrapolations to the static values and to be able to estimate the $1/M$ corrections
at a range of values of the heavy quark mass, we show in Fig.~\ref{fig:hugh_quenched_fits} 
the fits of the quenched FNAL results as a function of the heavy meson mass for all the quenched lattices. 
This will be compared to other quenched lattice results later.

\begin{figure}[thb]
\begin{center}
\centerline{
\epsfysize=5cm \epsfbox{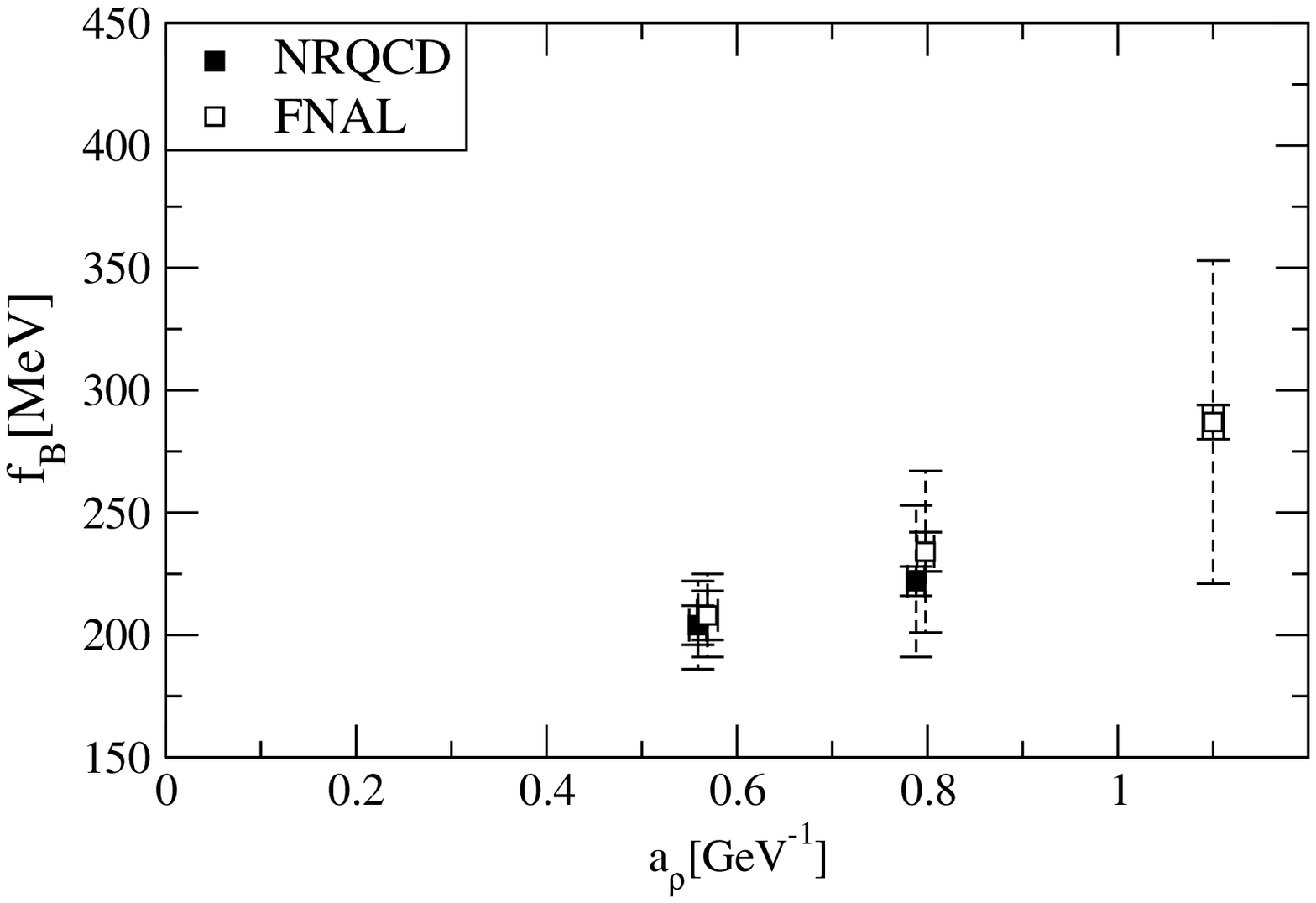}
\epsfysize=5cm \epsfbox{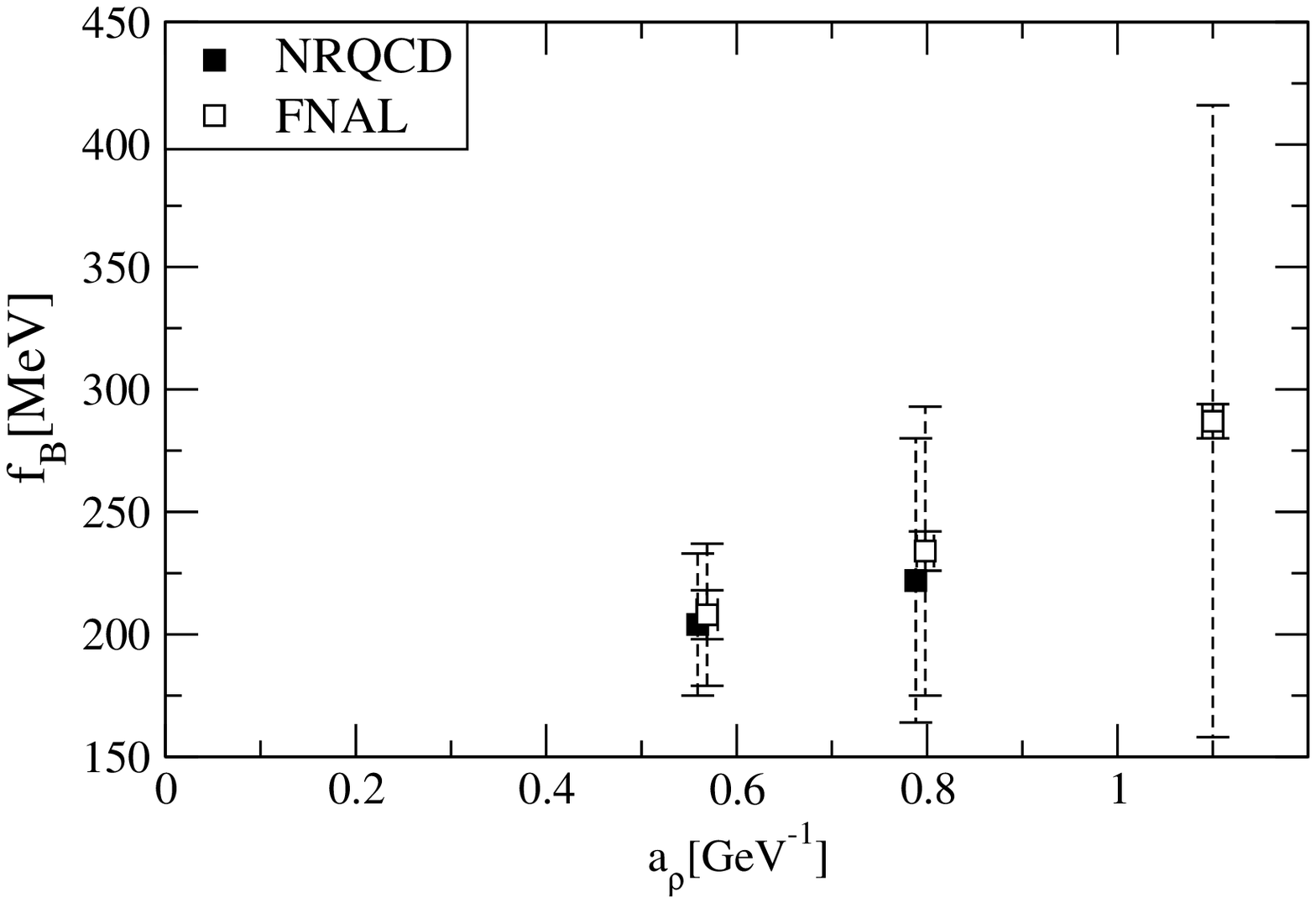}
}
\end{center}
\vspace{-0.5cm}
\caption{
$f_B$ as a function of $a$. Solid error bars: statistical errors, dashed
error bars: systematical errors using $\lqcd = 400$ MeV (left) and
$\lqcd = 600$ MeV (right).
}
\label{fig:fB_vs_a}
\end{figure}

\begin{figure}[thb]
\begin{center}
\centerline{
\epsfysize=6cm \epsfbox{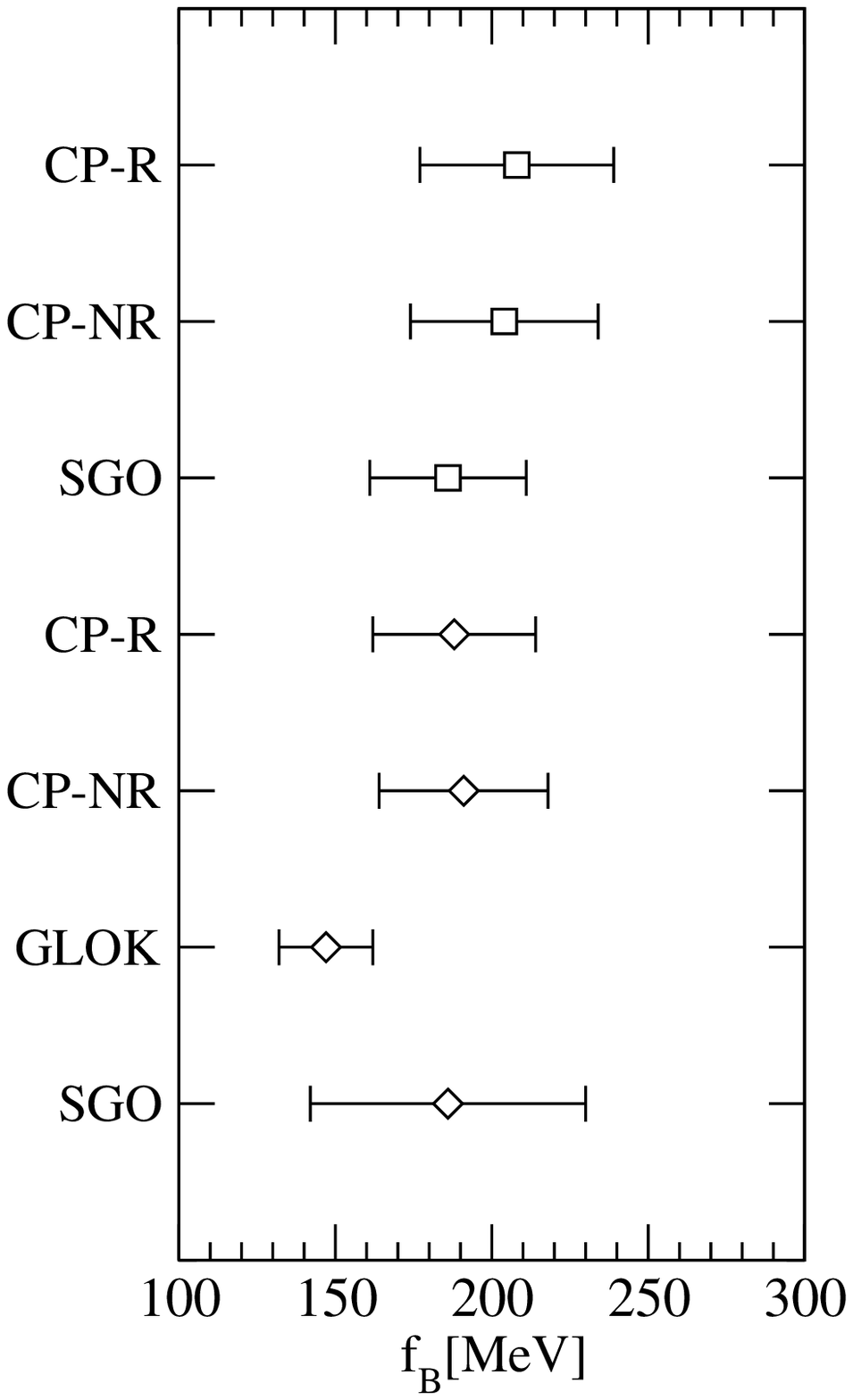}
\epsfysize=6cm \epsfbox{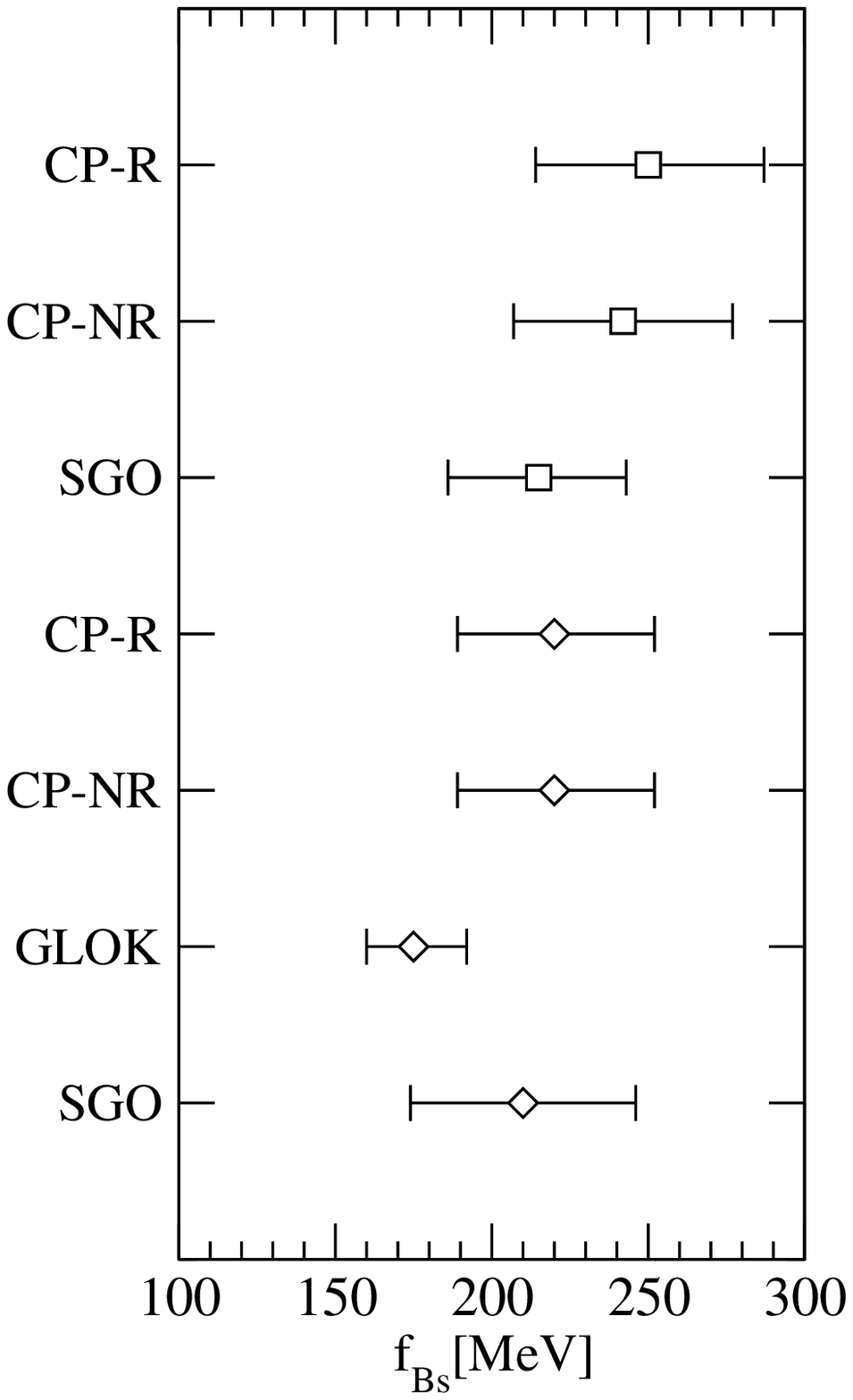}
}
\end{center}
\vspace{-0.5cm}
\caption{
Summary of own results on $f_B$ (left) and $f_{B_s}$ (right), from
Refs.~\protect\cite{sgo1997,collins1999} (SGO) \protect\cite{alikhan1998} (GLOK),  
\protect\cite{cppacs2001R}  (CP-R) and~\protect\cite{cppacs2001NR} (CP-NR). 
Diamonds denote quenched, squares $N_f = 2$ results.
}
\label{fig:fBcomp}
\end{figure}

Concerning the theoretical estimates of the discretization and perturbative errors used in the
NRQCD calculations one might object that it is not obvious which numerical value of $\lqcd$ 
should be used for the error estimate 
(e.g.\ quenched perturbative values from the determination of $\alpha_s$  are around $200-300$ MeV,
while generic momentum scales of QCD are given by hadron mass splittings, typically $\sim 500$ MeV).
Therefore we compare the for $\lqcd = 600$ MeV (used for the error estimate in the paper) and 400 MeV. 
The unquenched results in both formalisms are shown in Fig.~\ref{fig:fB_vs_a}.
Finally we summarize our results on $f_B$ and $f_{B_s}$ in Fig.~\ref{fig:fBcomp}.
A very important finding of the studies~\cite{cppacs2001NR,cppacs2001R} is that the $B$ and $B_s$ decay constants 
using NRQCD and Fermilab heavy quarks agree on the same set  of lattices on a range of lattice spacings.

\subsection{Comparison to other work}
We now compare our results to the ones from other lattice simulations. 
Lattice values for $f_B$ and $f_{B_s}$ since 1998 are listed in Table~\ref{tab:fB}.
The first error shown in the 
Table is statistical, the second is the systematical uncertainty, adding all the contributions 
quoted by the authors in quadrature. 

In the quenched approximation, a wealth of lattice results on $f_B$ is available, and we compare  
our results from NRQCD with two calculations using non-perturbative renormalization using the Wilson
gauge action, from the ALPHA~\cite{dellamorte2004,rolf2004,juettner2004} and the UKQCD
collaboration~\cite{bowler2001}.
\begin{figure}[thb]
\begin{center}
\epsfysize=6cm \epsfbox{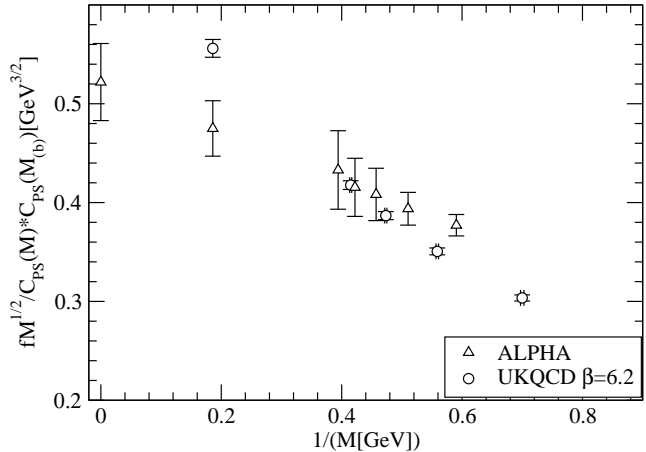}
\end{center}
\caption{Resummed matrix elements from non-perturbatively
$O(a)$ improved clover quarks. For the details of the resummation see the text.
The result at the $B_s$ mass is obtained by interpolation between charm and static heavy quarks 
for the ALPHA data~\protect\cite{rolf2004} or extrapolation from the charm for the UKQCD
data~\protect\cite{bowler2001}. The data in the Figure use $a$ set with $r_0$.}
\label{fig:Phi_NP} 
\end{figure}
The work of ALPHA~\cite{dellamorte2004,rolf2004,juettner2004} 
uses  the Schr\"odinger functional approach to simulate non-perturbatively
improved  clover charm and static quarks and obtains $f_{B_s}$ from an interpolation between the values
at infinite and at the charm mass. The results are continuum extrapolated. The scale is set using $r_0$. 

Ref.~\cite{bowler2001} uses two lattice spacings. The result from the finer lattice  is their
continuum estimate. An interpolation to a static value is not performed, since non-perturbative
renormalization was not available for the static action at the time of their calculation.
In Fig.~\ref{fig:Phi_NP} we compare the slopes of both nonperturbative calculations.
Both use a mass-independent renormalization scheme. 
The matrix elements themselves agree around the charm mass, their slope is slightly 
different. Ref.~\cite{bowler2001} uses a one-loop resummation
according to Eq.~(\ref{eq:leadinglog}), and \cite{rolf2004} uses the three-loop anomalous dimension
of the static current in terms of the renormalization group invariant quark masses. 
To calculate the anomalous dimensions for the plots, the value of $\lqcd = 238$ MeV is used.
Using the  three-loop anomalous dimension results in a slightly steeper running than 
using the 1-loop formula.
Adding results from finer lattices and larger masses to the calculation of \cite{bowler2001}  
would be useful to clarify the situation. 

\begin{figure}[thb]
\begin{center}
\epsfysize=6cm \epsfbox{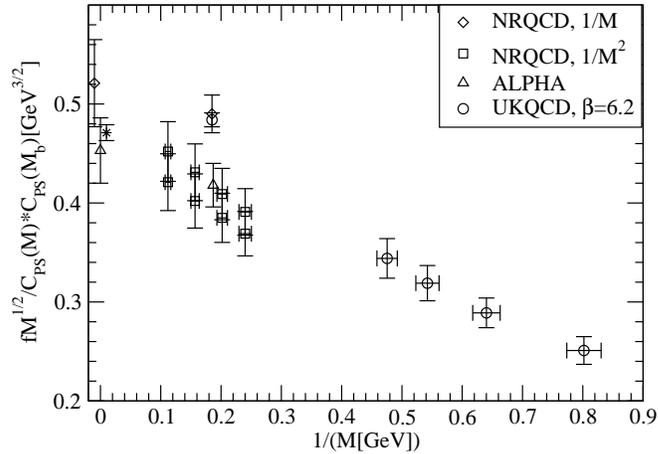}
\end{center}
\vspace{-0.5cm}
\caption{Comparison of NRQCD and perturbatively improved static decay matrix 
elements from \protect\cite{sgo1997,alikhan1998} with 
non-perturbatively improved results from~\protect\cite{bowler2001,rolf2004}. The non-perturbatively 
improved static value from $\beta = 6.0$ of \cite{dellamorte2004} is denoted by the star. The two data
sets from NRQCD at $O(1/M^2)$ correspond to 1-loop renormalization with $q^\ast = 1/a$ and $q^\ast = \pi/a$. }
\label{fig:fB_various}
\end{figure}

In Fig.~\ref{fig:fB_various} we compare quenched heavy-strange decay matrix elements from 
NRQCD~\cite{alikhan1998} with the non-perturbatively renormalized relativistic results.
The results have been rescaled to take the quenched
scale ambiguity into account. 
Since the lattice spacing of the NRQCD calculations is determined using the $\rho$ mass, the 
ALPHA value in the Figure is for comparison  decreased by $12\%$, which is the change they quote for $f_{B_s}$ if
$a$ is changed by $10\%$~\cite{rolf2004}, and their static result by $15\%$. A
static  value obtained from extrapolation of  the $1/M$ NRQCD results from \cite{sgo1997} to 
infinite mass at $\beta = 6.0$, and the non-perturbatively improved static value from 
ALPHA~\cite{dellamorte2004} at $\beta = 6.0$.
The difference of the NRQCD results at $O(1/M)$ and $O(1/M^2)$ is slightly less than 2 combined $\sigma$,
which we do not consider as significant since one of the ensembles is very small.
Ref.~\cite{ishikawa2000} compares  NRQCD results $O(1/M)$ and $O(1/M^2)$ on identical gauge field ensembles,
and finds very good agreement. The $1/M^2$ effects seem to be small. 
With a quadratic fit including results in a region around the charm mass, \cite{juettner2004} finds
$f_{B_s} = 198(8)$ MeV which is close to the published value of 205 MeV from~\cite{dellamorte2004}.

NRQCD results are obtained at a finite lattice spacing, and
using different gauge actions can lead to different discretization and perturbative errors.
In Fig.~\ref{fig:hugh_quenched_fits} we see that our results from renormalization group improved gluons 
and FNAL heavy quarks~\cite{cppacs2001R} extrapolate to a relatively high quenched value also on finer lattices.
As mentioned before, the NRQCD results agree at the $b$ mass with the  FNAL results.

The extrapolation to the static limit is higher than the nonperturbatively renormalized static result 
from the ALPHA collaboration using Wilson glue which can be taken as an indication of
remnant  discretization and perturbative errors. If the scale is set using $f_\pi$  instead of $m_\rho$,
the value decreases by $7\%$.

Results and global averages are listed in Table~\ref{tab:fB}. 
For the averaging, we employed the method used in the Particle Data Book for averaging the experimental 
results (see Ref.~\cite{pdg}, Introduction). The statistical and systematical errors were added in 
quadrature. Taking into account that sets of results have the same systematic error, the weights 
were computed and the averages obtained. The estimation of the systematic errors of all lattice results
is described in~\cite{alikhan2004}.
To briefly introduce the other calculations, 
Ref.~\cite{ishikawa2000} uses NRQCD at $O(1/M)$ and $O(1/M^2)$ with one-loop tadpole-improved clover 
light quarks and quenched 
Wilson glue, and~\cite{wing2003.stag} uses quenched NRQCD with an $O(a^2)$ tadpole improved gauge action and 
staggered light valence quarks.
Ref.~\cite{collins2001} uses quenched NRQCD at $O(1/M^2)$ and tadpole-improved and non-perturbatively $O(a)$
improved light clover quarks.  The quenched calculations of 
Refs.~\cite{abada2000,lin2001,becirevic2001} use Wilson gauge and
clover quarks at $a = 0.35-0.37$ GeV$^{-1}$ ($\beta = 6.2$)  simulated in the charm mass region, similar
to \cite{bowler2001}.
Ref.~\cite{abada2000} uses a non-perturbatively $O(a)$ improved clover quark action and a partly  
non-perturbative current renormalization.
Ref.~\cite{lin2001} uses tree-level tadpole-improved clover quarks without including $O(\alpha_s \times a)$
current correction terms.
Ref.~\cite{becirevic2001} uses nonperturbative $O(a)$ improvement except for a
perturbative value for the $O(\alpha_s am_q)$ quark mass correction to the renormalization constant.
Refs.~\cite{aoki1998,elkhadra1998,bernard2002} use heavy quarks in the Fermilab formalism and extrapolate
their results to $a \to 0$.
Refs.~\cite{guagnelli2002,dedivitiis2003} uses a step scaling function with the Schr\"odinger functional scheme
and clover heavy quarks. A part of  the renormalization factors is calculated nonperturbatively. 
 Their results are continuum extrapolated. 
Among the two-flavors results, Refs.~\cite{jlqcd2003,onogi2004} use NRQCD and Fermilab heavy quarks
respectively, at a fixed value of the gauge coupling.
\begin{table}[htb]
\begin{center}
\begin{tabular}{|lcl||cl|}
\hline
\multicolumn{5}{|c|}{Ratios of decay constants} \\
\multicolumn{1}{|c}{Ref.} &
\multicolumn{1}{c}{scale} &
\multicolumn{1}{c|}{$f_B^{N_f=2}/f_B^{N_f=0}$} &
\multicolumn{1}{|c}{Ref.} &
\multicolumn{1}{c|}{$f_{B_s}/f_{D_s}$} \\
\hline
\cite{bernard2002} & $f_\pi$  & 1.10(6) & \cite{cppacs2001R}($N_f=0$) &  0.88(1) \\                      
\cite{bernard2002} & $m_\rho$ & 1.19(6) & \cite{bernard2002}($N_f=0$)  & $0.891(12)(^{40}_{34})$ \\  
\cite{cppacs2001R} & $m_\rho$ & 1.11(6) & \cite{dellamorte2004} ($N_f=0$)  &     0.81(6)       \\
\cite{cppacs2001NR}& $m_\rho$ & 1.07(5) &  \cite{cppacs2001R}($N_f=2$)  & 0.94(6)  \\                 
\cite{cppacs2001NR}& $\Upsilon(\overline{P}-\overline{S})$ & 
0.97(5)  &  \cite{bernard2002}($N_f=2$) &  $0.922(13)(^{68}_{55})$ \\
\cite{maynard2002}($f_{D_s}$) & $r_0$   & 0.98(4) &   & \\
\hline
\end{tabular}
\end{center}
\caption{Ratios of decay constants. The first error is statistical, the second the systematical
errors given by the authors added in quadrature, where applicable.}
\label{tab:rat}
\end{table}

Refs.~\cite{aoki1998,elkhadra1998,bernard2002} use heavy quarks in the Fermilab formalism and extrapolate
their results to $a \to 0$.
The quenched calculations of Refs.~\cite{bowler2001,abada2000,lin2001,becirevic2001} use Wilson gauge and
clover quarks at $a = 0.35-0.37$ GeV$^{-1}$ ($\beta = 6.2$)  simulated at the charm quark mass. 
Ref.~\cite{lin2001} uses tree-level tadpole-improved clover quarks without including $O(\alpha_s × a)$
terms in the renormalization.
Ref.~\cite{abada2000} uses a non-perturbatively $O(a)$ improved clover quark action and a partly  non-perturbative
current renormalization.
Refs. \cite{bowler2001,becirevic2001} use nonperturbative $O(a)$ improvement except for a
perturbative value for the $O(\alpha_s am_q)$ quark mass correction to the renormalization constant.
Although different degrees of improvement are used, and the scaling behaviour is found to be different,
the results for $f_{D_s}$ agree at $\beta = 6.2$. 
A calculation from CP-PACS using the heavy quark formalism of \cite{aoki2002}
gives a preliminary result of  around $210$ MeV for $f_{B_s}$~\cite{ukawa2005}.

The unquenching effects on $f_B$ depend on the details of the calculation, e.g. 
how the scale is set. 
In Table~\ref{tab:rat}, ratios of decay constants from quenched and two-flavor simulations 
with the same gauge field and valence quark actions are listed. 
One finds an increase of $\sim 10\%$ with unquenching if $a$ is set with $f_\pi$,
$10-20\%$ if $a$ is set with $m_\rho$. On the other hand, Ref.~\cite{maynard2002} quotes no significant 
difference if the scale is set with $r_0$ for $f_{D_s}$, and
 Ref.~\cite{cppacs2001NR} finds no significant difference between quenched and two-flavor results
if the scale is set with the $\overline{P}-\overline{S}$ splitting of the $\Upsilon$, however 
on coarse lattices.
A new result with two flavors of domain-wall quarks on  the ratio $f_{B_S}/f_B$ of \cite{gadiyak2005} 
is somewhat higher than the CP-PACS results of \cite{cppacs2001R,cppacs2001NR}, but not statistically
significant.
%
\begin{table}[thb]
\begin{center}
\begin{tabular}{|l|c|ll|}
\hline
\hline
\multicolumn{1}{|l}{Ref.} &
\multicolumn{1}{c}{scale} &
\multicolumn{1}{c}{$f_B$[MeV]} & 
\multicolumn{1}{c|}{$f_{B_s}$[MeV]} \\
\hline
\hline
\multicolumn{4}{|c|}{{\bf Lattice}} \\
\hline
\multicolumn{4}{|c|}{$N_f = 0$} \\
\hline
\protect\cite{aoki1998}  & $m_\rho$     &  $173(4)(13)$ & $199(3)(14)$ \\
\protect\cite{elkhadra1998}  & $f_K$   &  $164(^{14}_{11})(8)$ & $185(^{13}_{8})(9)$ \\
\protect\cite{alikhan1998}   & $m_\rho$ &  $147(11)(^{13}_{16})$          & $175(8)(16)$ \\
\protect\cite{ishikawa2000} & $\sqrt{\sigma}=427$MeV & 170(5)(15) & 191(4)(17) \\
\protect\cite{abada2000} &$\frac{M_K^\ast}{M_K},M_K^\ast$  & $173(13)(^{34}_2)$ & $196(11)(^{42}_0)$\\
\protect\cite{bowler2001}   & $f_\pi$   & $195(6)(^{23}_{24})$ & $220(6)(^{23}_{28})$ \\
\protect\cite{lin2001}       & $f_K$    & $177(17)(22)$  & $204(12)(^{24}_{23})$ \\
\protect\cite{collins2001}  & $m_\rho$  &                      & 187(4)(15)    \\
\protect\cite{cppacs2001R} & $m_\rho$   & 188(3)(26)           & $220(2)(^{32}_{31})$ \\
\protect\cite{cppacs2001NR} & $m_\rho$  & 191(4)(27)           & 220(4)(31) \\
\protect\cite{becirevic2001} &$\frac{M_K^\ast}{M_K},M_K^\ast$ & $174(22)(^{8}_{0})$& $204(15)(^8_0)$\\
\protect\cite{bernard2002}  & $f_\pi$   & $173(6)(16)$ & $199(5)(^{23}_{22})$ \\
\protect\cite{guagnelli2002}    & $r_0$    & 170(11)(23)      & 192(9)(25)   \\
\protect\cite{dedivitiis2003}   & $r_0$    &                  & 192(6)(4) \\
\protect\cite{dellamorte2004} & $r_0$    &                  & 205(12) \\
\protect\cite{wing2003.stag} & $r_0$    &                     & 225(9)(34)      \\
        average1       &  &  $175(7)(^{48[21]}_{4})$        &     $198(5)(^{46[9]}_{16})$            \\
        average2      &  &                                 &     $201(6)(^{51[13]}_{13})$            \\
\hline
\multicolumn{4}{|c|}{$N_f = 2$} \\
\hline
\protect\cite{collins1999}  & $m_\rho$   & 186(5)(25)               & $215(3)(^{28}_{29})$ \\
\protect\cite{cppacs2001R} & $m_\rho$    & 208(10)(29)              & $250(10)(^{36}_{35})$ \\
\protect\cite{cppacs2001NR} & $m_\rho$   & 204(8)(29)               & 242(9)(34) \\
\protect\cite{bernard2002}  & $f_\pi$    & $190(7)(^{25}_{17})$    & $217(6)(^{36}_{28})$ \\
\protect\cite{jlqcd2003}    & $m_\rho$   & $191(10)(^{10}_{22})$   & $215(9)(^{14}_{13})$ \\
\protect\cite{onogi2004}    & $r_0=0.49$fm &   $181(7)(^{20}_{29})$   &  \\
average1               &            &    $190(10)(^{55[6]}_{13})$        & $226(15)(^{53[7]}_{2})$          \\
average2               &            &                                     & $226(15)(^{55[8]}_{1})$              \\
\hline
\multicolumn{4}{|c|}{$N_f = 2+1$} \\
\hline
\protect\cite{wing2003,gray2005}    & $\Upsilon (2S-1S), r_1$        &  216(9)(20)     & 260(7)(28) \\
\hline
\hline
\end{tabular}
\end{center}
\caption{$f_B$ and $f_{B_s}$ from the lattice. Statistical errors and systematical errors given by the authors 
are included, adding the systematical errors in quadrature. The method to set the scale is indicated in the 
second  column. The first error on the averages is due to the statistical and systematical errors of the individual 
results,
while the second error is from chiral extrapolation uncertainties and scale ambiguities as 
explained in the text. }
\label{tab:fB}
\end{table}

To investigate the effect of an additional dynamical strange quark, we compare our result to 
the published NRQCD lattice results with $2+1$ flavors of staggered sea quarks \cite{wing2003,gray2005}:
\ba
f_{B_s} & = & 260(7)(28) \mbox{ MeV}, \nonumber \\
f_B & = & 216(9)(20) \mbox{ MeV}, \label{eq:res}
\ea
where the first error is statistical and the second is from the systematic uncertainties added
in quadrature. Calculating at small quark masses, Ref.~\cite{gray2005} succeeds in achieving a 
precise result for the flavor breaking ratio of $1.20(3)$.
Calculations of the decay constants with the staggered $N_f=2+1$ configurations using 
FNAL \cite{bernard2004} heavy quarks are in further progress. 


So far there is no indication for large finite volume effects in the decay constants. 
Chiral perturbation theory predicts a 
finite size effect of $\leq 1 \%$ for lattice sizes $\sim 1.6$ fm and pion masses around
500 MeV for quenched QCD and much less than $1 \%$ for $N_f = 2$~\cite{arndt2004}. 
Ref.~\cite{bernard2002} estimates the finite volume uncertainty to be of the order of several percent;
the estimate is however from a comparison with very small volumes.

\begin{table}[thb]
\begin{center}
\begin{tabular}{|l|ll|}
\hline
\hline
\multicolumn{1}{|l}{Ref.} &
\multicolumn{1}{c}{$f_B$[MeV]} & 
\multicolumn{1}{c|}{$f_{B_s}$[MeV]} \\
\hline
\hline
\multicolumn{3}{|c|}{{\bf Sum rules}} \\
\hline
\cite{narison2001}  & 203(23)  & 236(30) \\
\cite{penin2002}    & 206(20)   &        \\
\cite{jamin2002}    & 210(19)  & 244(21) \\
\cite{braun1999}    & $180-190(30)$  &         \\
\hline
\hline
\multicolumn{3}{|c|}{{\bf Potential models}} \\
\hline
\cite{ebert2002}  & 178(15)   & 196(20)  \\
\hline
\end{tabular}
\end{center}
\caption{$f_B$ and $f_{B_s}$ from sum rule and potential model calculations.}
\label{tab:fBmodel}
\end{table}
We also use the experimental value for $f_{D_s}$ from unquenched lattice to $f_{B_s}$ and the lattice 
ratio $f_{B_s}/f_{D_s}$ from two-flavor calculations which work directly at the $b$ and $c$ quark masses 
without using extrapolations to calculate $f_{B_s}$. 
Taking the experimental value $f_{D_s} = 283(45)$ MeV using Eq.~(\ref{eq:fDs}) 
and $f_{B_s}/f_{D_s} = 0.93(^7_8)$ from the unquenched results from
Table~\ref{tab:rat}, one finds $f_{B_s} = 263(^{65}_{61})$ MeV, which is in agreement with the lattice
result of \cite{wing2003}.

Other recent review articles~\cite{ryan2002,lellouch2002,becirevic2003,wittig2003,hashimoto2004}
quote lattice estimates for $f_B$ and $f_{B_s}$ which are within errors in agreement with the
averages quoted in Table~\ref{tab:fB}.

Within errors the lattice results are also in agreement  with recent sum 
rule~\cite{narison2001,penin2002,jamin2002,braun1999} and potential model calculations~\cite{ebert2002},
see Tab.~\ref{tab:fBmodel}.
\section{SUMMARY AND CONCLUSIONS \label{sec:concl}}
This work presents a study of decay constants and the hadron spectrum in lattice QCD. 
The physics of hadrons with $b$ and light ($u,d,s$) quarks is of great relevance for the 
determination of parameters of the Standard Model, but both quarks are difficult to simulate because 
the light quark physics is sensitive to finite volume effects and the heavy quark physics
is sensitive to lattice spacing effects. Heavy quarks can be simulated  effectively
using NRQCD and HQET. The remaining difficulty is then to relate the light quarks simulated on the lattice 
to the physical
light $u$ and $d$ quark, since extrapolations over a more or less large range of masses are required. 
Chiral perturbation theory provides a useful description of hadron masses directly as a function of
the masses of the pseudoscalar mesons built from the quarks under consideration. 
The simulation parameters corresponding to the physical quark masses can be determined using
the light hadron spectrum. Of course the light hadron spectrum itself, in particular the nucleon mass, 
is one of the most basic quantities of which the lattice has to be able to show that they can be 
explained by QCD.

Light hadrons were studied in the quenched and two-flavor approximations using improved gluon and light 
quark actions. There are no indications of disagreement of the lattice results with experiment within the
estimated systematic errors.  Our results on the low-lying hadron spectrum  with  experiment 
agree surprisingly well with experiment already in the quenched approximation. Now further clarification
of the question whether there is really a  discrepancy between various improved fermion actions, would
be of interest.
Further one should note that quark model interpolating operators for the hadrons work very well.

Chiral extrapolations and finite volume effects of the nucleon mass find 
agreement with $O(p^4)$ chiral perturbation theory predictions at rather large quark masses, while  
chiral perturbation theory is assumed to be valid for pion masses up to $\sim 600$ MeV or less.
In particular in theories with dynamical quarks, lattice results at smaller quark masses would be of great 
interest to be able to make solid chiral extrapolations of hadron masses and matrix elements.

The agreement of the spectrum with experiment is an important result corroborating the reliability of the 
lattice calculations of the corresponding weak matrix elements. Most of the lattice results on the heavy-light spectrum
have been obtained using non-relativistic effective theories, in particular NRQCD. The agreement with experiment is 
a considerable success of the physical pictures used in these effective theories.
For example it implies that the strange-light quark mass difference determined from light physics is the same as in
heavy-light physics. 

For decay constants, comparison with results using different lattice discretizations and 
$0, 2$ and $2+1$ flavors of light sea quarks were made.  Effective theories, either those following the
ideas of NRQCD and the Fermilab method, or HQET together with interpolations to smaller masses are 
presently still indispensable in lattice calculations of $b$ hadrons. Using both methods next to each other
it should be possible to  obtain quite reliable estimates of hadronic matrix elements.

For heavy-strange decay constants an error not much larger than 10\% appears realistic. While the magnitude of the
error has not been reduced much compared to lattice errors quoted around 10 years ago, the error has become
significantly more reliable.
In unquenched calculations with heavy hadrons, effective theories will continue to play an
important role also in the coming years. While using NRQCD a continuum limit is not possible and 
renormalization constants are only calculated perturbatively, the scaling studies of previous calculations have 
shown that the error estimates are reliable. 
It would be important to support the simulations at the $b$ quark using non-relativistic effective theories 
also for dynamical quarks with  simulations of static and charm quarks using non-perturbative renormalization,
e.g.\ using the Schr\"odinger functional approach.
Since continuum extrapolations also involve uncertainties and since dynamical simulations on fine lattices
are presently still expensive, nonrelativistic methods such as NRQCD and the Fermilab method
are still important in precision calculations in $B$ physics.

Despite many lattice studies of a very broad range of observables there is no indication that 
QCD is not a consistent theory of the strong interactions. Comparison of 
of chiral perturbation theory in comparison with lattice data and experiment supports the present 
picture of chiral symmetry breaking in QCD. Increase of precision on a number of masses and matrix elements 
calculated on the lattice will be necessary to resolve the signatures of new physics. 
Important directions to investigate are unquenching and the chiral limit.
Further, the lattice has achieved a significant potential for calculations of many quantities in hadron physics
such as masses of excited hadrons, exotic states, a variety of form factors and structure functions, and the 
investigation of confinement and chiral symmetry. 

\subsection*{Acknowledgements}
I am grateful to D.~Becirevic, M.~Della Morte, 
D.~Ebert, A. El-Khadra, R.N.~Faustov, C.~Gattringer, V.O.~Galkin, M.~G\"ockeler, M.~G\"urtler, T.R.~Hemmert,
E.-M. Ilgenfritz, K.I.~Ishikawa, T.~Kaneko, A. Kronfeld, M.~Moch, C.~Morningstar, M.~Papinutto, 
G.~Schierholz, L.~Scorzato, J.~Shigemitsu, 
R.~Sommer, D.~Toussaint and J.~Zanotti for discussions. I thank D.~Ebert, M.~G\"ockeler,  
E.-M.~Ilgenfritz and M.~M\"uller-Preussker for a critical reading of the manuscript and invaluable 
comments, further I am grateful to W.~Bietenholz and A.~Walker-Loud for comments on the article.
I thank M. G\"urtler for communicating the hadron mass results of \cite{galletly2005}
to me, and J. Rolf for communicating to me the numerical data for $f\sqrt{M}$ from~\cite{rolf2004}. 
I am grateful to J.~Erdmenger for support and advice.
I thank R.~Burkhalter for a very useful comment about the manuscript.
I thank the group ``Theory of Elementary Particles/Phenomenology-Lattice Gauge Theories'' of the 
Humboldt University Berlin, the NIC/DESY Zeuthen and the Institute for Theoretical Physics of the
University Regensburg for their invaluable support and kind hospitality. 
Particular thanks goes to the group leaders M.~M\"uller-Preussker, K.~Jansen and A.~Sch\"afer.

I am grateful for a personal fellowship of the Deutsche Forschungsgemeinschaft and a fellowship for 
completion of the `Habilitation' from the Program for Equal Opportunities
for Women in Research and Academic Teaching of the City of Berlin. 

I would like to thank all my collaborators for many fruitful discussions and dedicated work, and
I thank the computer centers 
EPCC (Edinburgh), LANL (Los Alamos), LRZ (M\"unchen), NIC (J\"ulich) and NIC (Zeuthen),
NCSA (Urbana-Champaign), RCCP (Tsukuba) and SCRI (Tallahassee) for support.

\end{document}